\documentclass[11pt,letterpaper,twoside,openright]{book}
\makeatletter
\renewcommand{\@chapapp}{}% Not necessary...
\newenvironment{chapquote}[2][2em]
{\setlength{\@tempdima}{#1}%
	\def\chapquote@author{#2}%
	\parshape 1 \@tempdima \dimexpr\textwidth-2\@tempdima\relax%
	\itshape}
{\par\normalfont\hfill--\ \chapquote@author\hspace*{\@tempdima}\par\bigskip}
\makeatother

\usepackage{hyperref}
\usepackage[utf8]{inputenc}
\usepackage{epsfig}
\usepackage{graphicx, graphics}
\usepackage{amssymb,amsmath,amsfonts,amsthm}
\usepackage{fancybox,framed,tikz}
\usepackage{dsfont}
\usepackage{mathtools}
\usepackage{braket}
\usepackage{slashed}
\usepackage{bbold,bbding}
\usepackage{multirow}
\usepackage{verbatim}
\usepackage{scalerel}
\usepackage{fancybox,framed,tikz}
\usepackage[usestackEOL]{stackengine}
\usepackage[multiple]{footmisc}
\usepackage{rotating}
\usepackage{upgreek}
\usepackage{color}
\usepackage{euscript,stmaryrd}
\usepackage{caption}
\usepackage{subcaption}
\usepackage{tabularx}
\usepackage{longtable}
\usepackage{esint}
\usepackage{rotating}
\usepackage[sort]{cite}
\usepackage{float}

\usetikzlibrary{decorations.pathmorphing,patterns}
\usetikzlibrary{shapes.misc,arrows,decorations.markings}
\tikzset{cross/.style={cross out, draw=black, minimum size=2*(#1-\pgflinewidth), inner sep=0pt, outer sep=0pt},
	cross/.default={1pt}}
\stackMath

\tikzset{Witten diagram/.style={execute at begin picture={%
			\draw[blue ,fill=blue!05] circle[radius=\pgfkeysvalueof{/tikz/Witten/radius}];
			\path node (X){\phantom{X}};
		},baseline={(X.base)}},vertex/.style={circle,fill,inner sep=1.414pt,node
		contents={}},
	Witten/.cd,radius/.initial=1.414cm}
\newenvironment{wittendiagram}[1][]{\begin{tikzpicture}[Witten diagram,#1]}{\end{tikzpicture}}

\newcommand{\beq}{\begin{equation}}
	\newcommand{\eeq}{\end{equation}}
\newcommand{\be}{\begin{eqnarray}}
	\newcommand{\ee}{\end{eqnarray}}
\newcommand{\ba}{\begin{eqnarray}}
	\newcommand{\ea}{\end{eqnarray}}

\renewcommand{\a}{\alpha}
\renewcommand{\b}{\beta}
\renewcommand{\d}{\delta}
\newcommand{\g}{\gamma}

\newcommand{\D}{\Delta}
\newcommand{\e}{\epsilon}

\renewcommand{\k}{\kappa}

\renewcommand{\l}{\lambda}
\renewcommand{\L}{\Lambda}
\newcommand{\m}{\mu}

\newcommand{\n}{\nu}

\newcommand{\x}{\chi}

\newcommand{\s}{\sigma}

\renewcommand{\t}{\tau}

\newcommand{\y}{\upsilon}

\renewcommand{\O}{\Omega}

\newcommand{\wb}{\bar{w}}
\def \wb {{\bar{w}}}
\newcommand{\Xb}{\bar{X}}

\newcommand{\Li}{\textup{Li}}
\newcommand{\Tr}{\textup{Tr}}
\newcommand{\Str}{\textup{Str}}

\newcommand{\pa}{\partial}

\newcommand\To{\rule{0pt}{2.6ex}}       % Top strut
\newcommand\sbullet[1][.5]{\mathbin{\vcenter{\hbox{\scalebox{#1}{$\bullet$}}}}}
\def\ov{\over}
\newcommand{\del}{\partial}
\DeclarePairedDelimiter{\abs}{\lvert}{\rvert}
\newcommand{\tet}{\textstyle}

\definecolor{mypink1}{rgb}{0.958, 0.188, 0.478}

\newcommand{\sphere}{\textup{\textrm{S}}}

\newcommand{\cp}{\C \textup{\textrm{P}}}
\newcommand{\C}{\mathbb{C}}
\DeclareMathOperator{\vol}{vol}
\newcommand{\la}{\label}
\newcommand {\non}{\nonumber}
\newcommand{\mc}{\mathcal}
\DeclarePairedDelimiter{\cor}{\langle}{\rangle}

\newcommand{\mathsym}[1]{{}}
\begin{document}

\thispagestyle{empty} 
{}
	\vspace{ -3cm} \thispagestyle{empty} \vspace{-1cm}
\begin{flushright} 
	\footnotesize
	{HU-EP-23/58-RTG}
\end{flushright}%
\vspace{.1cm}
\begin{center}
	\Huge
	\textbf{Perturbative and non-perturbative analysis of defect correlators in AdS/CFT}\\
\vspace{2cm}
\normalsize 
\textbf{Dissertation}\\
 zur Erlangung des akademischen Grades \\
\vspace{.5cm}
\textbf{Doctor rerum naturalium (Dr. rer. nat.) }\\im Fach Physik \\ Spezialisierung: Theoretische Physik\\
\vspace{4mm}
eingereicht an der \\
Mathematisch-Naturwissenschaftlichen Fakultät\\
  der Humboldt-Universität zu Berlin  \par
  \vspace{6mm}
 von \par
 \vspace{6mm}
	\Large
\textbf{Gabriel James Stockton Bliard}
\normalsize \\MMath,   MA (Cantab)\par 
% (akademischer Grad, Vornamen, Name, Geburtsname, Geburtsdatum, Geburtsort)
\end{center}
\vspace{1cm}
Präsidentin der Humboldt-Universität zu Berlin \\
Prof. Dr. Julia von Blumenthal\par 
\vspace{4mm}
Dekanin der Mathematisch-Naturwissenschaftlichen Fakultät\\
 Prof. Dr. Caren Tischendorf\par 
 \vspace{1cm}
 \text{Gutascher}: \vspace{-8.7mm}
 \begin{flushleft}
\hspace{3cm}1. Prof. Dr. Valentina Forini \par \vspace{2mm}
\hspace{3cm}2. Prof. Dr. Nadav Drukker\par \vspace{2mm}
\hspace{3cm}3.  Prof. Dr. Thomas Klose 
\end{flushleft}
\vspace{1cm}
Dissertationsverteidigung: 18/09/2023
%Titel der Arbeit:\\
%{\centering{ Perturbative and non-perturbative analysis of defect correlators in AdS/CFT}}\\
%Dissertationzur Erlangung des akademischen Grades doctor rerum naturalium Doktor-Ingenieur Doktor philosophiae (Dr. rer. nat.) (Dr.-Ing.) (Dr. phil.)
%im Fach [Promotionsfach]: Spezialisierung:
%[Spezialisierung]\\
%eingereicht an der Mathematisch-Naturwissenschaftlichen Fakultät der Humboldt-Universität zu Berlin von\\
%akademischer Grad, Vornamen, Name, Geburtsname, Geburtsdatum, Geburtsort\\
%Präsidentin/Präsident der Humboldt-Universität zu Berlin Titel Vorname Name\\
%Dekanin/Dekan der Mathematisch-Naturwissenschaftlichen Fakultät Titel Vorname Name\\
\clearpage
\newpage 
\section*{Abstract}\label{abstract}
In this thesis, we consider two approaches to the study of correlation functions in one-dimensional defect Conformal Field Theories (dCFT$_1$), in particular those defined by 1/2-BPS Wilson line defects in the three- and four-dimensional superconformal theories relevant in the AdS/CFT correspondence.\par \vspace{3mm}
In the first approach, we use the analytic conformal bootstrap to evaluate two examples of defect correlators. 
We start with the four-point correlator of a set of 1/2-BPS protected operators forming the displacement supermultiplet inserted on the 1/2-BPS Wilson line in ABJM theory. This correlator is derived to the third order in a strong-coupling expansion and reproduces the explicit first-order Witten diagram calculations.  The CFT$_1$ data are then extracted from this correlator, and the operator mixing is solved at first order.  
%The first-order anomalous dimension does not contribute to mixing. 
Consequently, all-order results are derived for the part of the correlator with the highest logarithm power. This part uniquely determines the double-scaling limit of the correlator.  The second example is the five-point correlator of $1/2$-BPS operators inserted on the 1/2-BPS Wilson line in $\mc{N}=4$ super Yang-Mills. The superblocks are derived for all channels of the operator product expansion, and the five-point correlator is bootstrapped to first order in a strong coupling expansion. The CFT$_1$ data are then extracted, confirming that operator mixing does not affect the first-order anomalous dimension. \par \vspace{3mm}
The second approach considers the general structure of correlators in effective theories in AdS$_2$ and how this can best be used for the analytic conformal bootstrap program. The residue theorem is used to find all scalar $n$-point contact Witten diagrams for external operators of conformal dimensions $\Delta=1$ and $\Delta=2$ and to determine topological correlators of Yang–Mills in AdS. Effective theories in AdS$_2$ defined by an interaction Lagrangian with an arbitrary number of derivatives are then considered. A new formalism of Mellin amplitudes for 1d CFTs is introduced, and with it, the leading order of these effective theories is solved. Finally, the cusped Wilson line discretised action is presented as an alternative way to obtain non-perturbative data: through Lattice Field Theory. These developments aim to streamline the bootstrap process, from the Ansatz for a correlator to symmetry constraints to the non-perturbative physical input. \par \vspace{3mm}
\newpage
\section*{Zusammenfassung}
In dieser Arbeit betrachten wir zwei Ansätze zur Untersuchung von Korrelationsfunktionen in eindimensionalen und defektkonformen Feldtheorien, insbesondere solche, die durch die 1/2-BPS Wilson-Liniendefekte in den drei- und vierdimensionalen superkonformen Theorien, die in der AdS/CFT-Korrespondenz relevant sind, definiert sind. \par \vspace{3mm}
Im ersten Ansatz verwenden wir den analytischen konformen Bootstrap, um zwei Beispiele von Defektkorrelatoren in diesen Theorien auszuwerten. Wir beginnen mit dem Vierpunkt-Korrelator eines Satzes von 1/2-BPS geschützten Operatoren, die das Verschiebungssupermultiplett bilden, das auf der 1/2-BPS Wilson-Linie in der  ABJM-Theorie eingefügt ist. Dieser Korrelator wird bis zur dritten Ordnung in einer starken Kopplungsexpansion abgeleitet und reproduziert die expliziten Witten-Diagrammberechnungen erster Ordnung. Die CFT$_1$-Daten werden dann aus diesem Korrelator extrahiert und die Operatormischung erster Ordnung wird gelöst. Die anomale Dimension erster Ordnung trägt nicht zur Vermischung bei. Folglich werden für den Teil des Korrelators mit der höchsten logarithmischen Potenz Ergebnisse aller Ordnung abgeleitet. Dieser Teil bestimmt eindeutig die Double-Scaling-Grenze des Korrelators. Das zweite Beispiel ist der Fünfpunkt-Korrelator von $1/2$-BPS Operatoren, die auf der 1/2-BPS Wilson-Linie in $\mc{N}=4$-Super-Yang-Mills eingefügt sind. Die Superblöcke werden für alle Kanäle der Operatorproduktentwicklung abgeleitet und der Fünfpunkt-Korrelator wird in einer starken Kopplungsexpansion bis zur ersten Ordnung gebootstrapped. Die CFT$_1$-Daten werden dann extrahiert und bestätigen, dass die Operatormischung die anomale Dimension erster Ordnung nicht beeinflusst.\par \vspace{3mm}
Der zweite Ansatz betrachtet die allgemeine Struktur von Korrelatoren in effektiven Theorien in AdS$_2$ und wie diese am besten für das analytische konforme Bootstrap-Programm genutzt werden können. Der Residuensatz wird verwendet, um alle skalaren $n$-Punkt-Kontakt-Witten-Diagramme für externe Operatoren der konformen Dimensionen $\Delta=1$ und $\Delta=2$ zu finden und um topologische Korrelatoren von Yang-Mills in AdS zu bestimmen. Anschließend werden effektive Theorien in AdS$_2$, die durch einen Wechselwirkungslagrangian mit einer beliebigen Anzahl von Ableitungen definiert sind, betrachtet. Ein neuer Formalismus von Mellin-Amplituden für 1d-CFTs wird eingeführt, womit die führende Ordnung dieser effektiven Theorien gelöst wird. Schließlich wird die diskretisierte Wirkung der Cusped-Wilson-Linie als alternativer Weg zur Gewinnung nichtperturbativer Daten vorgestellt: durch die Gitterfeldtheorie. Diese Entwicklungen zielen darauf ab, den Bootstrap-Prozess zu rationalisieren: vom Ansatz für einen Korrelator über die Symmetriebeschränkungen bis hin zum physikalischen Input.
\clearpage
\newpage 

\section*{Statement of originality and independent work}\label{Sec:List of published works}
I declare that I have completed the thesis independently using only the aids and tools specified. I have not applied for a doctor’s degree in the doctoral subject elsewhere and do not hold a corresponding doctor’s degree. I have taken due note of the Faculty of Mathematics and Natural Sciences PhD Regulations, published in the Official Gazette of Humboldt-Universität zu Berlin no. 42/2018 on 11/07/2018.

\vspace{2cm}
{\Large \textbf{List of published works}}
	\begin{itemize}
	\item \cite{Bianchi:2020hsz} \href{https://doi.org/10.1007/JHEP08(2020)143}
		{L.~Bianchi, \underline{G.~Bliard}, V.~Forini, L.~Griguolo, and D.~Seminara,
			``Analytic bootstrap and Witten diagrams for the ABJM Wilson line as defect CFT$_1$'' JHEP 2020, 143 (2020). }
	\item \cite{Bianchi:2021piu}\href{https://journals.aps.org/prd/abstract/10.1103/PhysRevD.105.074507}
		{L.~Bianchi, \underline{G.~Bliard}, V.~Forini and G.~Peveri,
			``Mellin Amplitudes for 1d CFT,''\newline JHEP2021, 95 (2021). } 
		
	\item \cite{Bliard:2022kne}\href{https://doi.org/10.1007/JHEP08(2020)143}
		{\underline{G.~Bliard}, I.~Costa, V.~Forini and A.~Patella,
			``Lattice perturbation theory for the null cusp string'' Physical Review D 105, 074507 (2022)}

	\item \cite{Bliard:2022xsm} \href{https://iopscience.iop.org/article/10.1088/1751-8121/ac7f6b}
		{\underline{G.~Bliard},
			``Notes on $n$-point Witten diagrams in AdS$_2$'' J. Phys. A: Math. Theor. 55 325401 (2022)}
		
	\item \cite{Bliard:2022oof} \href{https://link.springer.com/article/10.1140/epjs/s11734-023-00769-w}{ \underline{G.~Bliard}, I.~Costa, V.~Forini,
	``Holography on the lattice: the string worldsheet perspective''  The European Physical Journal Special Topics volume 232, pages 339–353 (2023).}
\end{itemize}

This thesis is based on the publications above, written in equal parts with the co-authors listed whose work I gratefully acknowledge. It also includes results from work yet-to-be published \cite{Bliard-Ferrero,Bliard-Levine, Barrat-Bliard}.

\newpage
\section*{Acknowledgements}\label{Sec: Acknowledgements}
\vspace{-2mm}
This PhD has been nothing short of an epic journey in which I, and the world, have changed in countless ways. A journey is incomplete without a faithful companion, so I would like to start by thanking my wonderful wife, Olivia Brett, for supporting me during these four years, moving houses over ten times between five different countries, reading my papers, applications and this thesis, getting me discounted tickets for the Deutsche Oper and for saying yes to being my lifetime partner. \par 
This PhD is neither the beginning nor the end of my love for physics. I am very grateful to my parents for fostering a sense of wonder and curiosity and indulging me during late and cold nights of star gazing (with the Peterborough Astronomy Association!). Merci Papa de m'avoir fait découvrir le merveilleux monde qu'est la science, la chimie et les mathématiques et d'être resté curieux et jeune dans la tête. Thank you so much Mom for always being there and for preparing me for life's many challenges as a human and an traveller (and staying up late to reread this thesis). Thank you Uncle Steve for always having treated me as an adult and an equal, that meant and means more to me than you can imagine. Thank you Grandma for teaching me that I'm always growing in God's way and to all my family in Canada. Merci à toute ma famille Francaise d'avoir fait vivre notre patrimoine et notre terroir familial et de m'avoir soutenu pendent de nombreux moments. I have also been very fortunate to have had incredible teachers in life and throughout my relatively long educational journey: You were very patient with me and my provocative learning style. I believe that these mentors have truly formed the person I am. Some of my many mentors, teachers, lecturers, and coaches deserve special mention: Mr Gruzon, Mme Trezarieu, Mr Scat, Mr Devos, Sebastien Beaucourt, Mr Barré, JB Arnold, Andrew 'Grip' Watson as well as my Professors; Claudio Castelnovo, David Tong, Nick Dorey, Nick Manton, David Skinner, Malcom Perry and Ron Reid-Edwards.\par I would like to thank all the people I have met, worked with and learned from during my PhD. I want to start by thanking Valentina Forini, my supervisor, for enabling me to do this research through her PhD position, vision, network, and guidance. I would also like to thank Lorenzo Bianchi for being a mentor to me during this time. I am grateful for the discussions about physics and career with my other supervisors, Agostino Patella and Luca Griguolo. I have also been fortunate in my collaborations and have enjoyed working with all the above in addition to Domenico Seminara, Giulia Peveri, Ilaria Costa, Pietro Ferrero, Carlo Meneghelli, Martin Lagares, Diego Correa and Nat Levine. A little ode to all the unfinished (or unstarted) projects with Luigi Guerrini, Paolo Soresina, Davide Bonomi, Elia de Sabata, Ben Hoare,  and others.
\par Despite a raging Pandemic, I have managed to live in 5 different countries. I have many people to thank for making that possible:  Jenny Collard, for helping me with the heavy administration that goes with travel; the European Commission and the Europlex Grant for giving me access to the resources to travel this much - I believe that it was money well spent. Thank you to the Perimeter Institute for hosting me for the first two months of my PhD, and Karim and Nadia for housing me during this special time; the time spent in a physicist's heaven with blackboards galore and superstars walking the corridors was unforgettable. Thank you to City, University of London and the Mathematical Physics department there for having me for the next stint, the PhD seminar series, and many London conferences were a real highlight. Thank you again to my parents for having me during the first wave of the pandemic and having many apéro with  `roc blanc' with me.\par 
Desidero esprimere la mia sincera gratitudine all'Università di Parma, in particolare a Emil H. Leeb-Lundberg, Paolo Soresina, Luigi Guerrini, Jacopo Papalini, e ai professori Luca Griguolo, Francesco di Renzo e Marisa Bonini. Sono profondamente riconoscente per il sostegno che mi avete offerto durante un anno segnato dalla pandemia. Desidero inoltre rivolgere un sentito ringraziamento a Stefania e Alessandro, che sono stati come dei genitori per me e Olivia quando non ci è stato possibile tornare a casa durante le festività di Natale e Pasqua. Un ringraziamento speciale va all'enoteca Ombre Rosse per aver fornito le bottiglie perfette per il mio compleanno, e alla Torrefazione Gallo che ha reso la nostra permanenza qui come se fossimo a casa. Infine, vorrei esprimere la mia profonda gratitudine a Sarah e Giuseppe, che sono le persone migliori del mondo.
 %L'Università di Parma, Emil H. Leeb-Lundberg, Paolo Soresina, Luigi Guerrini, Jacopo Papalini, Francesco di Renzo e Marisa Bonini; vi rigrazio per avermi accolto un anno di Pandemia. Stefania e Allessandro, siete stati come Genitori per me e Olivia quando non siamo potuti tornare a casa per Natale e Pasqua, grazie dal profondo del cuore. Grazie al'Ombre rosse che ha trovato le bottiglie perfette per il mio compleanno e a Torrefazione Gallo ci che ha fatto sentire come abitanti del posto. Finalmente, grazie mille a Sarah, Giuseppe, che sono gli persone migliori del mondo.
\par 
Thank you Loïs, Grégoire and Malibu for making the last year of PhD the most enjoyable, and a huge thank you to Hélène and Christophe, who provided sanctuary during the critical writing time of the thesis, as well as Tristan and Styx ;) . I also cannot thank the Cuban Salsa Power Berlin community enough and the many climbing gyms around for keeping me sane during this write-up. Finally, I would like to thank the University of Humboldt-Berlin, and the countless people I have had the privilege to meet at the Mat-Nat Institute.\par\vspace{8mm}
I would like to thank the members of my PhD committee; Jan Plefka, Igor Sokolov, Lorenzo Bianchi and in particular the referees; Valentina Forini, Nadav Drukker and Thomas Klose, for their suggestions, questions and discussions during this process. 
\par 
I would like to thank all the people I have not mentioned above who have greatly impacted my life, academic and otherwise and to thank you, the reader, for opening this thesis.
\vfill\par 
The research received funding from the European Union’s Horizon 2020 research and innovation programme under the Marie Sklodowska-Curie grant agreement No 813942 ”Europlex” and from the Deutsche Forschungs- gemeinschaft (DFG, German Research Foundation) - Projektnummer 417533893/GRK2575 ”Rethinking Quantum Field Theory”.

\let\cleardoublepage\clearpage
\tableofcontents

\chapter{Introduction}\label{Sec:Introduction}
\begin{chapquote}{Edwin A. Abbott, \textit{Flatland: A Romance of Many Dimensions}}
	``I am no Woman," replied the small Line: ``I am the Monarch of the world. But thou, whence intrudest thou into my realm of Lineland?"
\end{chapquote}
This introduction will serve as a gateway into the world of holographic conformal defects by introducing the concepts of holography, conformal symmetry, and defects. The expert reader should feel free to skip to the overview of the thesis in subsection \ref{overview} since any essential or non-conventional information will be in the main text.\par \vspace{4mm}
Within the framework of Quantum Field Theory (QFT), Theoretical Physics has uncovered a powerful tool with which to model systems, from virus propagation \cite{DellaMorte:2020wlc,GRASSBERGER1983157} to economics \cite{Smolin:2009yk},  materials\cite{Kosterlitz:1973xp,Kane:2005zz}, knot theory \cite{Witten:1988hf} and the universe\cite{Hawking:1974rv}. For Lagrangian theories, each theory is a point in theory space specified by the fields (elements of the model) and how they interact. By varying these parameters, one explores this theory space. A specific variation plays a crucial role: the size/energy of the system. Limiting the energy range of a model and accounting for the effects of the higher energy fields leads to an effective theory at a lower energy scale. This change of scale triggers a variation of the theory, which is well-studied through the renormalisation group (RG) flow \cite{Wilson:1971dh,Wilson:1973jj}. The RG no longer flows at special points in theory space. In physical systems, these points correspond to critical phenomena where the correlation length of physical quantities diverges. 
These critical points have remarkable properties. The theory's stability under RG is extremely constraining and only allows certain configurations. The properties of critical points highlight the important aspects of physical theories: dimensionality, field content and symmetry. One of the major consequences of this is universality; models with the same field content and symmetry have the same behaviour at their critical point. This behaviour bridges the gap between seemingly different theories that are said to be in the same universality class. For example, by understanding the critical point of a system such as the Ising model, in which point-like magnets interact very simply, one can understand some properties of phase transitions of more complex materials \cite{Wilson:1973jj}.  In \cite{Polyakov:1970xd}, it was proposed that critical points enjoy a symmetry much larger than simple scale invariance: conformal invariance. This justified the study of Conformal Field Theories (CFTs) to describe these critical points. Therefore, our understanding of CFTs has far-reaching consequences in understanding various models relevant to the areas mentioned earlier. 
In theoretical physics, a complete theory should flow from a conformal point at small scales (ultraviolet fixed point) to one at large scales (infrared fixed point). Therefore, to understand the complete theory, it is paramount to study CFTs. Given the omnipresence of these conformal theories, it is quite fortunate that they possess additional symmetries which simplify computations. \par
In QFTs the central objects to calculate are correlation functions and scattering amplitudes, resulting from the way fields (or operators) interact through the Lagrangian of the theory. A modern approach is to have the correlators or amplitudes, not the Lagrangian, describe the theory.\footnote{Scattering amplitudes need asymptotic states, which are difficult to define in a CFT \cite{2003.07361}.} As a matter of fact, for local operators, the symmetries of the CFTs constrain the form of one-point, two-point and three-point correlators. The one-point correlator vanishes for non-unit operators, the two-point function is uniquely fixed from the conformal weight $\Delta$ of the operators, and the three-point function is also fixed from these weights up to an overall coefficient $c_{\Delta_1\Delta_2\Delta_3}$. The conformal weight $\Delta$ controls the behaviour of an operator under scaling of the spacetime coordinate and the coefficient $c_{\Delta_1\Delta_2\Delta_3}$ is defined by the three-point function. These form the CFT data of the theory. Furthermore, any two neighbouring operators inside a correlation function can be replaced by a series of single local operators.  This incredible tool, the operator product expansion (OPE), relates higher-point functions to three-point functions. Therefore, the entire theory can be specified in terms of the CFT data $\{\Delta,c_{\Delta_1\Delta_2\Delta_3}\}$. Despite the introduction of infinite sums through this OPE, this framework is compelling. Within this context, there is hope for solving models through analytical and numerical methods, providing rigorous and practical tools to study CFTs. This framework has led, for example, to the most accurate prediction of the critical exponents of the 3D Ising model through the numerical bootstrap  \cite{El-Showk:2012cjh}, far outperforming the most sophisticated Monte Carlo methods. \par
 %this correspondence is a duality of remarkable nature in many ways. First, it links fundamentally different theories: an inherently gravitational theory, String Theory in Anti-de-Sitter spacetime (AdS) is dual to a (super)conformal theory of non-abelian gauge fields - with no gravity at all - living at the boundary of AdS, therefore in one dimension less. The latter is a manifestation of the so-called holographic principle
Additionally, CFTs play a central role in the AdS/CFT correspondence, first postulated in the 1990s \cite{Maldacena:1997re,Gubser:1998bc,Witten:1998qj}. This correspondence is a duality which is remarkable in a number of ways. First, it links fundamentally different theories. On one side, one has an inherently gravitational theory: String Theory in Anti-de-Sitter spacetime in d+1 dimensions (AdS$_{\text{d+1}}$). One the other, one has a non Abelian gauge theory which has no gravity, living on the boundary of AdS in d-dimensions. This is a manifestation of the holographic principle\cite{Susskind:1994vu,tHooft:1993dmi}. Another remarkable feature is that this duality relates the non-perturbative regime of one theory to the perturbative one of its dual model. Computations in the strong-coupling regime of the gauge field theory are translated into perturbative, weak-coupling calculations of string theory. Conversely, the duality allows for the description of the non-perturbative string regime through the weakly coupled CFT description\cite{Gaberdiel:2018rqv}. Additionally, this duality has a wealth of non-perturbative evidence through supersymmetric localisation and integrability\cite{Kapustin:2009kz,Beisert:2010jr,Zarembo:2016bbk}. This string revolution, which celebrates its 25th birthday this year, has sparked a vigorous research program that has gained in breadth over the years, with applications to many aspects of theoretical (and even experimental \cite{Hartnoll:2008kx,Hubeny:2014bla}) physics.
Wilson loops are fundamental non-local observables of any gauge theory, and therefore play a central role in the context of AdS/CFT.  From the point of view of the CFT in which they are defined, straight or circular Wilson lines can be interpreted as defects~\cite{Andrei:2018die} which preserve a (one-dimensional) subgroup of the original conformal symmetry. Analysing such defect CFTs (dCFTs) allows one to probe the non-local observables of the embedding theory, thereby extending our knowledge beyond well-studied local data. The dual string worldsheet has AdS$_2$ geometry, and this AdS$_2/$dCFT$_1$ setup is an example of ``rigid'' holography~\cite{Giombi:2017cqn} since the AdS$_2$ metric is non-dynamical.  These AdS$_2/$dCFT$_1$ setups have recently been explored in a plethora of papers~\cite{Liendo:2016ymz, Giombi:2017cqn,Liendo:2018ukf,Giombi:2018hsx, Giombi:2018qox,  Bianchi:2020hsz, Cavaglia:2021bnz, Ferrero:2021bsb, Giombi:2022pas, Beccaria:2022bcr, Gimenez-Grau:2023fcy, Castiglioni:2023uus }.\vspace{3mm}
	
%which break part of the conformal symmetry. One can understand their dynamics as a theory on the defect. In this way, analysing these line defect CFTs (dCFT$_1$) allows us to probe the non-local observables of the embedding theory, thereby extending our knowledge beyond well-studied local data. On the other side of the duality, the defect is represented by a string whose geometry is AdS$_2$. These AdS$_2/$dCFT$_1$ setups have recently been explored in a plethora of papers \cite{Giombi:2022pas, Beccaria:2022bcr, Gimenez-Grau:2023fcy, Cavaglia:2021bnz,Castiglioni:2023uus,Giombi:2018hsx, Giombi:2017cqn, Giombi:2018qox, Bianchi:2020hsz, Liendo:2016ymz, Liendo:2018ukf, Ferrero:2021bsb} and constitute an example of rigid holography since there are no graviton propagating degrees of freedom on the `gravity' side of the duality.
\section{The Beauty of Things Conformal}\label{Intro Conformal}

\begin{chapquote}{Alexander Migdal}%, \textit{}}
``15 parameters!''
\end{chapquote}
This small overview of conformal symmetry aims at recalling some basic elements used in this thesis. For an introduction and more thorough description of CFTs, see \cite{DiFrancesco:1997nk,Ginsparg:1988ui}.
\subsection*{The Conformal Group}
Transformations are conformal when they preserve the angles (shapes) of the space, hence the name `\textit{con forma}'. There are four such types of transformations for conformal theories in $d>2$ dimensions:
\begin{itemize}
\item Translations $P_\mu$, which act by shifting coordinates
\item Lorentz transformations $M_{\mu \nu}$, which act as rotations (boosts in Minkowski)
\item Dilatation $D$, which acts by scaling coordinates
\item Special conformal transformations $K_\mu$
\end{itemize}
 \vspace{4mm}
The conformal group in $d-$dimensions is a Lie group $SO(d+1,1)$ whose algebra is given by the elements 
\begin{align}
D& &P_\mu &&K_\mu&&M_{\mu \nu}
\end{align}
satisfying the commutation relations
\begin{align}
&[D,P_\mu]= +i P_\mu\qquad[D,K_\mu]=-i K_\mu \qquad[K_\mu,P_\nu]=2i( \eta_{\mu \nu}D-M_{\mu \nu}) \\
&[K_\rho,M_{\mu \nu}]= i(\eta_{\rho \mu}K_\nu-\eta_{\rho \nu}K_\mu) \qquad [P_\rho,M_{\mu \nu}]= i(\eta_{\rho \mu}P_\nu-\eta_{\rho \nu}P_\mu) \\
&[M_{\mu \nu},M_{\rho \sigma}]=-i \left(\eta_{\mu \rho}M_{\nu \sigma}+\eta_{\nu \sigma}M_{\mu \rho}-\eta_{\mu \sigma}M_{\nu \rho}-\eta_{\nu \rho}M_{\mu \sigma}\right)
\end{align}
where $\eta_{\mu \nu}=\text{diag(-1,1,1,1)}$ is the Minkowski metric.
For most of the following, we will look at systems with extended supersymmetry but only one spatial dimension. As such, only the first line of these commutation relations will be used, corresponding to the generators of $SL(2,\mathbb{R})$, which can be written as\footnote{In practice we use the real Euclidean version (see \cite{Eberhardt:2020cxo}).}
\begin{align}
&[D,P]=  P \qquad[D,K]= -K \qquad[K,P]=2 D  \label{Eq: 1d CFT transformations}.
\end{align}

\subsubsection*{Addressing the triviality of 1d CFT}
The obvious element to address here is why look at conformal transformations in one dimension in the first place. In one dimension, conformal transformations are said to act trivially on the line; that is, if there are no angles in one dimension, what is the interest in `angle-preserving' transformations? Explicitely, any smooth coordinate transformation is conformal in one dimension. Furthermore, the quasi-entirety of the conformal literature considers two-dimensional or $d\geq 3$ systems. One of the reasons 1d CFTs have not been looked into much is because they are inherently non-local. When trying to define a local CFT$_1$, the Hamiltonian must vanish from the tracelessness of the energy-momentum tensor. Relaxing this assumption of locality (not requiring the presence of a conserved stress tensor) leads to one-dimensional CFTs that are invariant under the global conformal group $SL(2,\mathbb{R})$ (\ref{Eq: 1d CFT transformations}). These systems occur naturally when considering the embedding of a line in a higher-dimensional conformal theory and considering which symmetries are preserved by the line. Extended objects such as defects in conformal theories often preserve a remnant of this symmetry, so one can see conformal transformations acting on the full embedding space as one-dimensional conformal symmetry acting on the line. The coordinate on the line cannot be arbitrarily parametrised because of the embedding space.\par 
\subsubsection{Operators}
When defining a theory with conformal symmetry, the fields will be representations of this group. Since the dilatation operator $D$ can be identified with the Hamiltonian, the eigenvalue $\lambda_\phi$ (called conformal weight) of the diagonalised dilatation operator
\begin{align}
D \ket{\phi} = \lambda_\phi \ket{\phi}
\end{align}
can be identified with the energy of the eingenstate $\ket{\phi}$.\footnote{The terms field and operator will be used interchangeably when considering the eigenstate $\ket{\phi}$ because of the CFT state-operator correpondence. The identification of the Dilatation operator with the Hamiltonian of the theory comes from the radial quantisation of the theory, described for example in chapter 6 of  \cite{DiFrancesco:1997nk}.}
From the commutation in \eqref{Eq: 1d CFT transformations}, the special conformal transformation $K$ lowers the eigenvalue of a state and the translation $P$ raises it. 
As a consequence, since the energy spectrum should be bounded from below, there must be a highest weight state such that
\begin{align}
K \ket{\phi}_\Delta = 0.
\end{align}
This state is called primary and is labelled by its conformal weight $\Delta$. Each of these primary states has a tower of operators called \textit{descendants} under the action of the translation operator $P$. Thus all the states in the system can be created from the set of primary operators acting on the vacuum and the action of the translation generator $P$.

\subsubsection{Differential representation of conformal generators}
The translation $P$ allows one to relate an operator at the origin to one at a general point $x$
\begin{align}
	\phi(x) = \text{e}^{-Px}\phi(0)\text{e}^{Px}.
\end{align}
This procedure (known as the method of induced representations) allows the conformal transformations on a primary $\mathcal{O}_\Delta(x)$ of conformal dimension $\Delta$ evaluated at position $x$ to be written as differential operators
\begin{align}
	P &= -\partial_x \\
	K &= -2\Delta x-x^2 \partial_x\\
	D &= -\Delta -x \partial_x.
\end{align}
This parametrisation of the one-dimensional conformal algebra allows for an explicit understanding of how this symmetry constrains the correlators of one-, two-, and three-point functions
\begin{align}
&\langle \phi_\Delta(x)\rangle  = \delta_{\Delta,0}  \\ 
&\langle \phi_{\Delta_{1}}(x_1)\phi_{\Delta_2}(x_2)\rangle = \frac{\delta_{\Delta_1,\Delta_2}}{x_{12}^{2\Delta}}\\
&\langle \phi_{\Delta_{1}}(x_1)\phi_{\Delta_2}(x_2)\phi_{\Delta_3}(x_3)\rangle = \frac{c_{\Delta_1\Delta_2\Delta_3}}{(x_{12})^{\Delta_{123}}(x_{13})^{\Delta_{132}}(x_{23})^{\Delta_{231}}}\\
& \Delta_{ijk} = \Delta_i+\Delta_j-\Delta_k .
\end{align}
Here we use the convention
\begin{align}
	x_{ij} &= x_i-x_j& x_1&\leq x_2\leq x_3\leq ...\leq x_n.
\end{align} 
For more general $n$-point correlators, conformal transformations reduce the number of independent variables to $n \!-\!3$ conformally invariant cross-ratios
\begin{align}\label{Eq: cross-ratio}
\chi_i &= \frac{x_{1i}x_{n-1,n}}{x_{in}x_{1,n-1}}& 0<&\chi_i<1
\end{align}
where $x_{ij} = x_j-x_i$ are real numbers in $d=1$.
For equal conformal dimensions, the correlator is%\footnote{Note that the limit
%\begin{align}
%	\lim_{\epsilon\rightarrow 0} 	\langle \phi_\Delta(x_1)..\phi_\Delta(x_i)...\epsilon^{-2\Delta}\phi_\Delta(\epsilon^{-1})\rangle  = 	\langle \phi_\Delta(x_1)..\phi_\Delta(x_i)...\tilde{\phi}_\Delta(\infty)\rangle 
%\end{align}
%is well-defined. }
\begin{align}
\langle \phi_\Delta(x_1)..\phi_\Delta(x_i)...\phi_\Delta(x_n)\rangle = A(x_1,...,x_n)	\langle \phi_\Delta(0)..\phi_\Delta(\chi_i)...\phi_\Delta(1)\phi_\Delta(\infty)\rangle 
\end{align}
where
\begin{align}\label{Eq: prefactor}
A(x_1,...x_n) &=\left(\left(\frac{x_{1n}x_{n-1,n}}{x_{1,n-1}}\right)^{n-2} \prod _{j=1}^{n-1} x_{jn}^{-2}\right)^{\Delta }.
\end{align}
In particular, four-point correlators of scalar operators
\begin{align}\label{Eq: four point definition}
	\langle \phi_{\Delta}(x_1) \phi_{\Delta}(x_2) \phi_{\Delta}(x_3) \phi_{\Delta}(x_4)\rangle =\frac{1}{(x_{13}x_{24})^{2\Delta}} f(\chi)
\end{align} 
will be functions of a single conformal cross-ratio
\begin{align}\label{cross-ratio}
\chi = \frac{x_{12}x_{34}}{x_{13}x_{24}}. 
\end{align}
One might wonder why higher-point correlators are of any interest since the set $\{\Delta, c_{\Delta_1\Delta_2\Delta_3} \}$ determines the theory and all correlators. In practice, however, determining such a set is far from trivial, and working out perturbative higher-point correlators gives access to this information through the dynamics of the theory. Therefore, the OPE is also used to extract CFT data from correlation functions. 
%Therefore, the true power of conformal theories lies in the OPE. The higher-point correlators can be fixed entirely by the knowledge of the conformal data $\{\Delta,c_{123}\}$ defined above through the OPE. 
A simple example is the OPE of the four-point correlator of identical scalars in 1d in \ref{Eq: four point definition}, which will be relevant in the following sections.
This gives the block expansion 
\begin{align}\label{Eq: OPE}
f(\chi) =\sum_h c_{\Delta\Delta h}^2 \chi^{h-2\Delta} {}_2F_1(h,h,2h,\chi)
\end{align} 
where the conformal cross-ratio $\chi$, defined in \eqref{cross-ratio}, is a dimensionless quantity invariant under the conformal transformations listed above, and $h$ is the conformal weight of exchanged primary operators.\footnote{In the following, $\Delta_i$ will mainly be used for the weight of external primary operators and $h_i$ for exchanged primary operators. However, the use of $\Delta$ for the free theory exchanged dimension is standard and will also be used at times. To avoid confusion, it will be stated in the text whether $\Delta$ refers to an external or exchanged dimension.}\par \vspace{4mm}
Therefore, the knowledge of all two-point and three-point correlators (giving $h$ and $c_{\Delta \Delta h}$ respectively) is sufficient to construct four-point functions. This also applies to more higher-point correlators. Conversely, the knowledge of higher-point functions provides information about the conformal data.

\subsection*{Crossing and Polyakov Blocks}\label{subsec: CFT basics}
The block expansion in \eqref{Eq: OPE} can be done by doing the OPE expansion between the first two operators and between the second and third operators. The equality of these two sums is the crossing equation
\begin{align}
\sum_h c_h G_h(\chi) = \sum_h c_h G_h(1-\chi) .
\end{align}
The crossing equation is at the heart of the conformal bootstrap, both analytic and numerical, and remains one of the challenges moving forward.\footnote{The crossing equation still has not been solved in 4d\cite{Bissi:2022mrs}.}
\par
Within the context of the analytic bootstrap program for four-point functions, this crossing can be implemented directly into the Ansatz. The symmetries of the conformal blocks (similarity under $\chi\rightarrow \frac{\chi}{\chi-1}$), those of the correlator (crossing symmetry under $\chi\rightarrow1-\chi$), and those of the theory (e.g. Ward identities in \cite{Liendo:2018ukf}) highly constrain the form of four-point correlators. Complemented by a transcendentality Ansatz for the correlators \cite{Alday:2017zzv}, this provides a powerful way to compute perturbative correlators, as was done in \cite{Ferrero:2021bsb,Liendo:2018ukf,Ferrero:2019luz,Lemos:2016xke,Bianchi:2020hsz}. \par 
The bosonic four-point correlator in \eqref{Eq: four point definition} has symmetries under the permutations of the external operators. Given the ordering of the operators, the variable $\chi $ is naturally defined in the range $0<\chi<1$. Following the analysis of \cite{Mazac:2018qmi}, the bosonic symmetry can be used to define the function in \eqref{Eq: four point definition} on the entire real axis for the given prefactor\footnote{For fermions with $\Delta=n+\frac{1}{2}$, only the last line acquires a `$-$' sign because of the prefactor in \eqref{Eq: OPE}. }
\begin{align}\label{Eq: bosonic continuation}
f(\chi) = \begin{cases*}
	f^{(-)}(\chi)=(1-\chi)^{-2\Delta}f^{(0)}\left(\frac{\chi}{\chi-1}\right) &$\chi<0$\\
	f^{(0)}(\chi) & $0<\chi<1$\\
	f^{(+)}(\chi)=	\chi^{-2\Delta}f^{(0)} \left(\frac{1}{\chi}\right)& $\chi>1$
\end{cases*}
\end{align}
The resulting function has an explicit symmetry under crossing  
\begin{align}
\chi\rightarrow 1-\chi\qquad 0<\chi<1,
\end{align}
and braiding
\begin{align}
\chi\rightarrow \frac{\chi}{\chi-1}.
\end{align}
These two symmetries generate all the crossing symmetries from bosonic permutations.
In addition, these functions defined on a segment of the real line can be analytically continued outside their region of analyticity. For some functions (for example, those resulting from contact Witten diagrams), the analytic continuation of the function $f^{(0,\pm)}(\chi)$ outside its segment of definition matches the function $f(\chi)$. In this case, we speak of \textit{braiding symmetry}. This is linked to the vanishing of the double discontinuity, defined in  \cite{Mazac:2018qmi} as 
\begin{align}\label{Eq: Ddisc}
dDisc^{+}[\mathcal{G}(\chi)] = \mathcal{G}^{(0)}(\chi)-\frac{\mathcal{G}^{(+)}(\chi+i\epsilon)+\mathcal{G^{(+)}}(\chi-i \epsilon)}{2} \quad 0<\chi<1.
\end{align}
Unitarity arguments link this double discontinuity to the full correlator thanks to the inversion formula derived in \cite{Mazac:2018qmi} and provide a powerful tool constraining the correlators and, correspondingly, the OPE data.\par
The OPE expansion is the projection of the correlator on the conformal blocks basis. There is, however, another basis that is of some interest in this context: Polyakov blocks. These are defined to be crossing-symmetric, Regge-bounded\footnote{A Regge-bounded correlator $g(\chi)$ satisfies $\lim_{\zeta \rightarrow \infty} g(\frac{1}{2}+i \zeta) < C$ where $C $ is a constant. Thus, a correlator's large-$|\chi|$ behaviour in $d=1$ controls the Regge behaviour. Note that the identity has a constant contribution in this limit.} and to have the same expansion as the conformal blocks\footnote{See  section 6 of \cite{Mazac:2018qmi}.}
\begin{align}
\sum_h c_h G_h(\chi) = \sum_h c_h P_h(\chi).
\end{align}
Their existence in d=1 was motivated in \cite{Gopakumar:2016wkt,Gopakumar:2016cpb,Polyakov:1974gs} and proven in \cite{Mazac:2018ycv}. 
Additionally, they have the same double discontinuity as the conformal blocks and have a double zero at the position of two-particle operators ($h=2\Delta+2n$).
\begin{align}
P_{2\Delta+2n} &=0\\
\partial_nP_{2\Delta+n} &=\delta_{n,0}. 
\end{align}
These operators are usually called 'double-trace'. However, the use of the term double-trace in the context of 1d defect theories is abuse of language inherited from higher dimensional analysis. Given that the operators we will consider are insertions on a line, the trace is included in the definition of the correlator. The equivalent operators exchanged in GFF are instead 'two-particle' operators of the symbolic form $\phi\partial^n\phi$.
Consequently, they can be expressed as the sum of Witten exchange diagrams (see Appendix \ref{App: Polyakov}). 
Therefore, the links between CFT and AdS are deeply embedded so that any additional knowledge on the CFT side will help the understanding of topics such as quantum gravity, effective QFT in curved space, and integrability in curved space.
\section{AdS/CFT}
\begin{chapquote}{Dr Jeffrey Harvey, \textit{Strings 1998}}
	Ehhhh! Maldacena!
\end{chapquote}

The AdS/CFT correspondence is a remarkable duality stated 25 years ago \cite{Maldacena:1997re} that postulates the equivalence between a  gravitational theory and a lower-dimensional non-gravitational theory. One of the remarkable aspects of this duality is that the perturbative sector of one theory corresponds to the deeply non-perturbative sector of the other. For example, the perturbative Witten diagrams in AdS correspond to the strong coupling results of the CFT\cite{DHoker:1999kzh}, and weak coupling results in the CFT probe the very stringy regime of the AdS theory\cite{Gaberdiel:2018rqv}. \par 
The AdS/CFT dictionary relates the field theory partition function to the quantum gravity partition function. The correlators in the gauge theory then correspond to operators inserted on the boundary of the gravity dual. In the original construction \cite{Maldacena:1997re}, the rank $N$ of the gauge group $SU(N)$ then corresponds to the number of stacked D3-branes on which strings end (on the string theory side) and the couplings of the two theories are also related. In the low-energy limit of the theories, it is often useful to consider the large-$N$ limit with a fixed product of $N$ with the coupling of the field theory, called 't Hooft coupling $\lambda$ \cite{tHooft:1973alw}. The large $N$ limit, or planar limit, suppresses non-planar Feynman diagrams. In the two cases relevant for us, this is defined as
\begin{align}
	\lambda_{YM} &=  g_{YM}^2 N\\
	\lambda_{ABJM} &= \frac{k}{N}
\end{align}
%The first line corresponds to the ${\cal N}=4$ supersymmetric Yang-Mills theory (${\cal N}=4$ SYM)/Type IIB in AdS$_5\times \text{S}^5 $ duality.
 Above,   $g_{YM}^2$ is the inverse coupling in the Yang-Mills theory with gauge group $SU(N)$ and $k$ is the Chern-Simons level of ABJM with symmetry $U(N)\times U(N)$. \par 
 In the AdS$_5$/CFT$_4$ case, relating four-dimensional $\mc{N}=4$ super Yang-mills theory with gauge group $SU(N)$ and coupling $g_{YM}$ to type IIB string theory in AdS$_5\times$S$^{5}$ space with radius $R$, string tension $\alpha'$ and coupling $g_S$, one has
 \begin{align}
 	\frac{R^4}{\alpha'^{2}}&= g_{YM}^2 N=\lambda_{YM},&g_s&=\frac{4\pi\lambda_{YM}}{N},
 \end{align}
In the AdS$_4$/CFT$_3$ case, for which ABJM, the three-dimensional Chern-Simons theory with matter with gauge group $U(N)_k\times U(N)_{-k}$ is related to type ILIA string theory in AdS$_4\times \mathbb{C}$P$^3$ space with radius $R$, string tension $\alpha'$ and coupling $g_S$. Their couplings are related as
\begin{align}\label{Eq: Dictionary ABJM}
	\frac{R^4}{2\pi^2 \alpha'^2}&=\lambda_{ABJM} = \frac{N}{k},& g_S \sim \frac{\lambda_{ABJM}^{\frac{5}{4}}}{N}.
\end{align}
In the planar limit at fixed 't Hooft coupling $\lambda$, this corresponds to a vanishing string coupling ($g_s\rightarrow 0$) limit for both theories. In this duality, the fields in the boundary CFT are dual to bulk fields with the same quantum numbers and the mass in AdS is related to the conformal weight. In the AdS$_2$/CFT$_1$ case of interest here, this relates the mass of bosonic and fermionic excitations in AdS$_2$ to the conformal weight in the CFT as
\begin{align}
	m_B^2&=\Delta(\Delta-1)& m_F^2&=(\Delta-\frac{1}{2})^2.
\end{align}
At the centre of this correspondence sits the Wilson-Maldacena  loop \cite{Erickson:2000af, Correa:2012at, Maldacena:1997re, Witten:1998qj}, whose expectation value is the partition function of an AdS string which is anchored to the loop. In the strong coupling limit of the Wilson loop, the latter is then equal to the extremal surface of the classical string.\footnote{On the string side, this limit corresponds to the semiclassical saddle-point approximation.} In the defect CFT setup which will be of interest in this thesis, the local operator insertions on the Wilson loop correspond to fluctuations of the string. \par 
Below, the conformal bootstrap will provide an additional tool to study the strong-coupling regime of the CFT, which should agree with perturbative Witten diagram calculations in AdS. 
In the full AdS/CFT correspondence, the gravitational propagating degrees of freedom correspond to the local stress tensor at the boundary. In the AdS$_2$/CFT$_1$ case of interest here, there is no local CFT stress tensor operator and no gravitational propagating degrees of freedom. This is a special, non-gravitational case of the AdS/CFT correspondence. Some `gravitational flavour' is retained in this AdS$_2$/dCFT$_1$ correspondence \cite{Dubovsky:2012wk, Giombi:2022pas, Bliard-Levine} and it remains a physical and interesting setup as it is embedded within a larger holographic theory.

\subsection*{Perturbative Witten Diagrams}\label{Section Witten}
The power of the AdS/CFT correspondence lies in the fact that, in the strong-coupling regime of the field theory, observables can be computed perturbatively in the string picture. In the case of the defect configuration, the system is a string worldsheet whose boundary coincides with the defect. The classical solution of the string then encodes the information of the defect, while fluctuations orthogonal to the worldsheet describe deformations of this same defect. In the cases of the 1/2-BPS string in $\text{AdS}_5\times \text{S}^5$ and $\text{AdS}_4\times CP^3$, the geometry of these classical strings is $\text{AdS}_2$, and the fluctuations can be considered as a quantum field theory on $\text{AdS}_2$, with the dynamics of the theory being governed by the embedding of AdS$_2$.\footnote{The same applies to the cases of $\text{AdS}_3\times \text{S}^3\times \text{S}^3 \times \text{S}^1$ and AdS$_3\times \text{S}^3\times T^4$ \cite{Correa:2021sky, Bliard-Correa}.}\par

The first tool to compute correlators in AdS is Witten diagrammatics. This technique is equivalent to Feynman diagrams for QFT in flat space. The integrals related to the diagrams are well-studied for four-point functions at Next-to-Leading Order (NLO) \cite{DHoker:1999mqo} and have several useful representations in Mellin space \cite{Mack:2009mi,Fitzpatrick:2011ia} and in spectral representation \cite{Carmi:2018qzm}. Due to its low dimensionality, the AdS$_2$ case is simpler and allows for higher-order and higher-point computations. For example, four-point correlators are only a function of one variable (as opposed to two), and their functional form is simpler in AdS$_2$. In the publications leading up to this thesis \cite{Bliard:2022xsm, Bianchi:2020hsz,Bianchi:2021piu}, the formalism of Witten diagrams and the Mellin transform is adapted to this lower-dimensional case and applied to the 1/2-BPS Wilson line defect. \par  
In a general line defect holographic system, a bosonic operator of weight $\Delta$ inserted along the line is described by a fluctuation of mass $m^2=\Delta(\Delta-1)$ about the minimal surface of a string ending on the boundary of AdS space. When the minimal surface in question has an AdS$_2$ geometry, the fluctuations about this can be computed by considering effective fields propagating in a static AdS$_2$ space. 
We consider theories in Euclidean AdS$_2$, for which we use the Poincaré metric
\begin{align}
ds^2_{\text{AdS}_2} = \frac{dx^2+dz^2}{z^2}.
\end{align}
Scalar bosonic operators of conformal weight $\Delta$ are then dual to the boundary limit of these fluctuating fields in AdS$_2$
\begin{equation}
\tilde{\phi}_\Delta(t) = \lim\limits_{z\rightarrow 0} z^{-\Delta} \phi(z,t)
\end{equation}
where\footnote{For fermions, the map is $m^2 = (\Delta-\frac 12)^2$.}
\begin{align}
m^2 = \Delta(\Delta-1).
\end{align}
Using the common abuse of notation in the literature, we identify the boundary $\tilde{\phi}$ and bulk $\phi$ fields. These fields have a bulk-to-bulk propagator
\begin{align}\label{Eq: bulk to bulk}
G^{\Delta}_{ BB}(a,b) &=C_\Delta (2u)^{-1}{}_2F_1(\Delta,\Delta,2\Delta,-2u^{-1})&				u&=\frac{(z_a-z_b)^2+(x_a-x_b)^2}{2z_az_b},
\end{align}
which satisfies the AdS$_2$ equation of motion
\begin{align}
(\nabla_{AdS}^2-\Delta(\Delta-1)) G^{\Delta}_{ BB}(a,b)=z^2 \delta^{(2)}(a-b)\label{Eq: Equation of motion ads2},
\end{align}
and whose normalisation \cite{Freedman:1998tz,Fitzpatrick:2011ia} is
\begin{align}\label{Eq: normalisation C_Delta}
C_\Delta = \frac{\Gamma (\Delta )}{2 \sqrt{\pi } \Gamma \left(\Delta +\frac{1}{2}\right)}.
\end{align}
The bulk-to-boundary propagator corresponding to the $z\rightarrow 0$ limit of \eqref{Eq: bulk to bulk} is
\begin{align}\label{Eq: bulk-to-boundary prop}
K_\Delta(z,x;x_i)&=C_{\Delta}\tilde{K}_\Delta(z,x;x_i)\\
&=C_{\Delta}\left(\frac{z}{z^2+(x-x_i)^2}\right)^\Delta.
\end{align}
Due to the isometries of AdS$_2$, the boundary correlators will be conformal. 
This is best seen in the embedding coordinates\cite{Dirac:1936fq,Mack:1969rr, Ferrara:1973eg, Weinberg:2010fx,Penedones:2016voo}
\begin{align}
X^0&=R\frac{1+x^2+z^2}{2z}&X^1 &= R \frac{x}{z}&X^2&=R\frac{1-x^2-z^2}{2z}
\end{align}
where $X_A X^A=-R^2$. \par \vspace{2mm} The isometries in these coordinates are generated by the generators of $SO(2,1)$
\begin{align}
J_{AB} = -i\left( X_A \frac{\partial}{\partial X^B}-X_B \frac{\partial}{\partial X^A}\right).
\end{align} 
These are related to the usual conformal generators by
\begin{align}
J_{0,2}&=D&J_{1,0}-J_{1,2}&=P&J_{1,0}+J_{1,2}&=K.
\end{align}\par 
However, for the case of AdS$_2$, many simplifications can be seen in terms of the spacetime coordinates $(z,x)$ used in this thesis. Given an action (for example, the effective worldsheet theory on AdS$_2$ of \cite{Giombi:2017cqn}), boundary correlators are computed with Witten diagrams which are constructed just as Feynman diagrams are: with external points, propagators and vertices (see Figure \ref{Fig:contact-witten}). 
Contact diagrams are a subset of all possible contributions and are the building blocks for computing correlators. This can be used, for example, for local operators inserted along the 1/2-BPS Wilson line of ABJM in Chapter \ref{chapter: ABJM}.  
\section{Line Defects}\label{Intro: Line defects}
\begin{chapquote}{Leonard Cohen, \textit{Anthem}}
	There is a crack in everything.
\end{chapquote}

Any CFT can be fully described by the spectrum of local operators and the three-point coefficients $\{\Delta,c_{\Delta_1\Delta_2\Delta_3} \}.$ However, studying local operators through such methods misses an important class: defect operators. A particular class of defects are Wilson lines which are extended operators, defined as the traced, path-ordered integral of a connection along a contour. The two Wilson lines considered in this thesis are the 1/2-BPS Wilson lines in ABJM and in $\mc{N}=4$ SYM.\footnote{The construction of the supersymmetric Wilson loop in ABJM will be presented in section \ref{sec: ABJM Wilson line}.} The explicit expression of the latter is
\begin{align}
	\mc{W}_\mc{C}  = \frac{1}{N} \, \Tr \, \text{P}\, \text{exp}\int_\mc{C} dt (i A_\mu \dot{x}^\mu(t)+|\dot{x}(t)|n\cdot \Phi ).
\end{align}
where the contour $\mc{C}$ is parametrised by $x^\mu(t)$, the trace is taken in the fundamental representation, $\Phi^a$ is the scalar field in $\mc{N}=4$ SYM and the coupling is done through a unit vector $\vec{n}$ fixed in the  $\Phi^6$ direction by convention. \par \vspace{2mm}
In generic theories, the presence of a defect will break the overall (super) conformal symmetry to a subset. The defects are said to be conformal if the conformal transformations of the defect can be written in terms of those of the embedding coordinates.
% In general, the line will break the conformal group will break for a line defect as
%\begin{align}
%SO(d+1,1) \rightarrow  SO(d-1) \times SO(2,1).
%\end{align}
%If they conserve some conformal symmetry, t
The presence of the line breaks the conformal symmetry to a subgroup and brings additional sets of conformal data, the data intrinsic to the theory on the defect and that of the defect interacting with the rest of the space. The latter has been studied in \cite{Bianchi:2022sbz,Barrat:2022psm,Bianchi:2022ppi,deLeeuw:2017dkd,Mazac:2018biw} but will not be considered here. The data instrinsic to the defect will correspond to the defect operator data. For example, a straight line defect in a conformal theory will break translation invariance in the direction perpendicular to it. The breaking of the conservation of the energy-momentum tensor ($T^{\mu \nu}$) can be used to define the insertion of a defect operator intrinsic to the line: the displacement operator ($\mathbb{D}$). 
\begin{align}
\partial_\mu T^{\mu a} =\delta^{(d-1)}(x_{\perp}) \mathbb{D}^a(x_{\parallel}).
\end{align}
where $x_\perp$ is the coordinate perpendicular to the defect and $x_\parallel$ is the coordinate along the line (referred to in this paper as $x$ or $t$ depending on context). The symmetries broken by the defect no longer leave the system invariant. Instead, they deform the line and thus define insertions of local operators along it. The remaining symmetries can constrain correlators of these operators and, thus, the conformal data of the defect theory.
These defect insertions are generally not operators of the defect-free theory since they control line deformations. In particular, even though they can be expressed in terms of fields in the bulk theory Lagrangian at weak coupling, their correlators are not equivalent to the limit of a field in the bulk brought close to the line.
Given some local operators $\mathcal{O}_i(t_i)$, one can define the gauge invariant Wilson line with insertions as
\begin{equation}\label{eq: Path ordered Wilson line}
\mathcal{W}[\mathcal{O}_1(t_1)\mathcal{O}_2(t_2)\,...\,\mathcal{O}_n(t_n)]\equiv 
\Tr \left[\mathcal{W}_{t_i,t_1}\mathcal{O}_1(t_1)\mathcal{W}_{t_1,t_2}\mathcal{O}_2(t_2)\,...\,\mathcal{O}_n(t_n)\mathcal{W}_{t_n,t_f}\right],
\end{equation}
where $t$ parameterises an infinite straight line and $\mathcal{W}_{t_a,t_b}$ is the path-ordered exponential of a suitable connection that starts at position $x(t_a)$ and ends at position $x(t_b)$.  The  local operators $\mathcal{O}_i(t_i)$ are inserted between (untraced) Wilson lines and therefore are not invariant but have to transform in the adjoint representation of the gauge group. The (one-dimensional) defect  correlators are then defined as
\begin{equation}\label{eq: defect correlators}
\langle\mathcal{O}_1(t_1)\mathcal{O}_2(t_2)\,...\,\mathcal{O}_n(t_n)\rangle_{\mathcal{W}}\equiv
\frac{\langle\mathcal{W}[\mathcal{O}_1(t_1)\mathcal{O}_2(t_2)\,...\,\mathcal{O}_n(t_n)]\rangle}{\langle\mathcal{W}\rangle}\,\,.
\end{equation}
In particular, this is different from having a similarly adjoint bulk field $\Tr(\mc{O}_B)$ close to the line%\footnote{In this context, it is possible to have single and multi-trace operators as opposed to the untraced operator insertions.}
\begin{equation}
\lim_{x_{\perp \rightarrow 0}}\frac{\langle  \mathcal{P} \Tr(\mc{O}_B(t_i,\vec{x^i}_{\perp}))\Tr \left[\mathcal{W}_{t_i,t_1}\mathcal{O}_1(t_1)\mathcal{W}_{t_1,t_2}\mathcal{O}_2(t_2)...\mathcal{O}_n(t_n)\mathcal{W}_{t_n,t_f}\right],\rangle}{\langle\mathcal{W}\rangle}.
\end{equation}
Supersymmetric BPS Wilson lines provide one-dimensional supersymmetric defect field theories explicitly defined through the correlation functions of local operator insertions on the line~\cite{Drukker:2006xg}. From this perspective, $1/2-$BPS Wilson lines in the four-dimensional ${\cal N}=4$ supersymmetric Yang-Mills theory (SYM) have been actively studied in the last few years~\cite{ Giombi:2018qox,Giombi:2018hsx, Liendo:2018ukf, Ferrero:2021bsb}. The associated defect theory is conformal (dCFT). Further information has been gained by considering four-point correlators of certain protected operator insertions~\cite{Bianchi:2020hsz,Giombi:2017cqn,Liendo:2018ukf} whose two-point functions control the ${\cal N}=4$ SYM Bremsstrahlung function~\cite{Correa:2012at, Cavaglia:2022qpg}. This analysis places the study of defects at the heart of the AdS/CFT correspondence. 
\par \vspace{2mm}

\section{Analytic Conformal Bootstrap}
\begin{chapquote}{Erich Wolfgang Korngold, \textit{Das Wunder der Heliane } \ref{Heliane}}
Über mir wird Himmel sein,\\
und Sterne werden\\
sich ins Auge mir schmiegen\\
mütterlichen Lichts?
\end{chapquote}

The conformal data of a theory and, in this context, of the defect theory, can be obtained through a procedure known as the analytic conformal bootstrap. 
The conformal bootstrap is an ambitious project aiming to describe CFTs only by their symmetries and minimal physical input. This minimalist description is intellectually pleasing and provides tools to constrain correlators and conformal data\cite{Ferrero:2021bsb,Liendo:2018ukf, Alday:2017zzv, Bianchi:2020hsz}. For the strong-coupling results of holographic conformal defects, the method reviewed in depth in Chapter \ref{chapter: Analytic Bootstrap} is as follows:
\begin{enumerate}
\item Write an Ansatz with harmonic polylogarithms of  transcendentality increasing with each perturbative order. 
\item Impose constraints from the OPE
\item Input recursive \textit{unmixed} CFT data
\item Input physical data 
\end{enumerate} 

In systems with fewer symmetries, the analytic conformal bootstrap becomes more challenging due to the lack of constraints from the OPE, the absence of non-perturbative results and the greater mixing of operators. The mixing problem is especially present in 1d systems where, unlike in higher dimensions, single and double-trace operators are indistinguishable.
\newpage
\section{Overview of the Thesis}\label{overview}
\begin{chapquote}{Lewis Carroll, \textit{Alice's Adventures in Wonderland }}
``Begin at the beginning," The King said, very gravely,  ``and go on till you come to the end: then stop."
\end{chapquote}
Conformal defects provide an incredible playground in which to compute results. For line defects, Witten diagrams and the analytic conformal bootstrap provide ways to compute correlators of operators inserted on the line defect. Results can then be organised into simpler quantities using superspace formalism and the Mellin transform. This organisation is convenient for finding patterns and hints at which aspects of the correlators are most important and, ultimately, the simplest way to describe a theory. This thesis, beyond the study of correlators in holographic conformal defects, is the precursor to the solving of higher-point correlators, the finding of additional constraints for correlators, the computation of new data for theories such as AdS$_3$/CFT$_2$, and ultimately, the systemisation of one-dimensional defect analyses. By searching for additional symmetries, trends in different models, and discretised lattice models, there is a chance at solving 1d defects entirely. Given the prevalence of defects in CFT and the important role of CFT in modelling physical systems, understanding these defects is a unique opportunity to understand our world better.\par \vspace{2mm}
In this thesis, I present the analysis of these holographic line defects through the lens of the analytic conformal bootstrap. They stem from my contributions to the subject: \cite{Bliard:2022kne,Bliard:2022xsm,Bianchi:2021piu,Bianchi:2020hsz} and works yet to come \cite{Bliard-Ferrero,Bliard-Levine, Barrat-Bliard}. Chapter \ref{chapter: Analytic Bootstrap} introduces the different elements of the analytic conformal bootstrap for line defect CFTs and is dedicated to making the individual steps understandable. These steps are the writing of an Ansatz with harmonic polylogarithms of increasing  transcendentality, the imposing of constraints from the OPE, the recursive input of unmixed CFT data and the input physical data. Chapter \ref{chapter: ABJM} looks at the bootstrap of the four-point correlator of  insertions on the 1/2-BPS Wilson line dCFT. This supercorrelator can be written in terms of a function of a single cross-ratio and is fixed up to third-order perturbation about the strong-coupling free theory. The same quantity is computed and identified using perturbative Witten diagrams. The CFT data are extracted and the first order contribution is unmixed and found to be degenerate. As a consequence, the full non-perturbative correlator is computed in the double-scaling limit. In Chapter \ref{chapter N=4}, a five-point correlator in $\mc{N}=4$ SYM 1/2-BPS Wilson line dCFT is studied. The form of the superconformal blocks is derived and the first-order solution is bootstrapped. The CFT data are then extracted which supports the fact that the anomalous dimension also does not mix in this case. In Chapter \ref{Effective theories}, a range of effective theories are analysed in the context of strong coupling AdS$_2$ Witten diagrams, where $n$-point correlators are computed, and high-derivative interactions are solved effectively using the Mellin transform. Chapter \ref{Chap: Optimisation and limits} presents an analysis of the challenges that need to be overcome for all-order results to be obtained and some potential solutions, including lattice computations, are also presented. We conclude with an outlook in Chapter \ref{chap Outlook}. Following this, several technical appendices serve to lighten the main text and hopefully satisfy the curious reader.

\chapter{The Analytic Bootstrap for Conformal Line Defects}\label{chapter: Analytic Bootstrap}
\begin{chapquote}{Douglas Adams, \textit{The Hitchhiker's Guide to the Galaxy}}
	The ships hung in the sky in much the same way that bricks don't.
\end{chapquote}
\begin{figure}
	\centering
	\includegraphics[width=\linewidth]{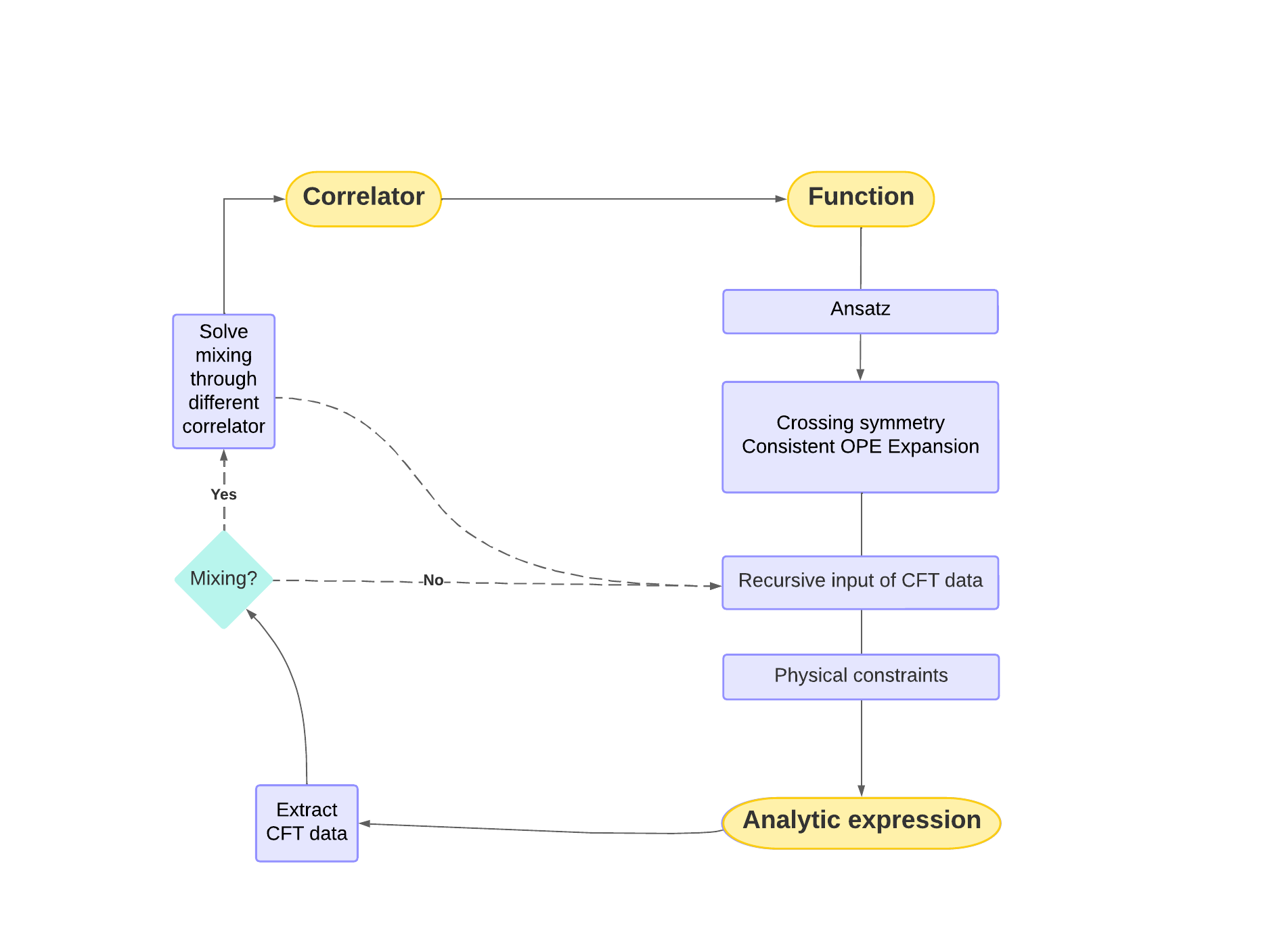}
	\caption{Flow diagram of the bootstrap process described in Chapter \ref{chapter: Analytic Bootstrap} starting from a correlator, written in terms of a minimal set of functions of the conformal cross-ratios. After formulating an appropriate ansatz for them, a series of constraints are then applied to the functions to fix them through the symmetries of the theory. Physical constraints then uniquely fix the solution. Conformal data can then be extracted from these functions and unmixed to use in the next-order bootstrap process.}
	\label{fig:bootstrap-flowchart}
\end{figure}
The conformal bootstrap is an ambitious project aiming to describe CFTs only by their symmetries and minimal physical input. Its main focus has shifted from the original goal in the 60s of describing the strong force \cite{Chew:1961ev} to the description of critical phenomena \cite{Polyakov:1970xd, Polyakov:1974gs} and the direct solving of known CFTs. It is an umbrella term for a topic which includes the numerical bootstrap\cite{El-Showk:2012cjh, El-Showk:2014dwa, Chester:2014fya} as well as the S-matrix bootstrap \cite{Paulos:2016fap, Fitzpatrick:2011hu} and the analytic bootstrap\cite{Aharony:2016dwx, Alday:2014tsa,Cornagliotto:2017dup, Cavaglia:2021bnz,Meneghelli:2022gps ,Ferrero:2021bsb}. Although most current bootstrap efforts work directly with CFTs, the original goals have not been forgotten, and there is some effort in the direction of the conformal window of QCD \cite{Nakayama:2014sba} and phase transitions\cite{Rychkov:2009ij}.  This Chapter covers all the steps of the analytic conformal bootstrap used in Chapters \ref{chapter: ABJM} and \ref{chapter N=4}. The overall concept is illustrated in Figure \ref{fig:bootstrap-flowchart}. First, the constraints from the OPE will be presented, and the perturbative structure  will be explained. Then, the symmetries from crossing and braiding will be analysed from the point of view of the OPE and the exchange of identical operators. These symmetries then inform the Ansatz. This functional form of the correlator is motivated by the perturbative analysis of the strong coupling perturbation of the OPE on the CFT side. Additionally, the AdS/CFT correspondence motivates this further from the diagrammatics on the weakly coupled string side. The recursive step of the analytic bootstrap is then explained, where the lower-order data are used as input for the higher-order correlators. A toy example illustrates one of the all-loop consequences, and the mixing problem is presented. Finally, several examples of  constraints used to fix the last terms in the correlator are presented. Once the correlator is fixed analytically, this expression gives CFT data. \par \vspace{3mm}

As the conformal booststrap can be quite technical in notation, assumptions and computations, the following subsections will cover some aspects that are central to the bootstrap effort. 
The non-perturbative constraints are the Operator Product Expansion (OPE) and crossing.
The OPE allows one to write any correlation function in terms of the CFT and of conformal blocs, fixed by the symmetry of the theory. Crossing and braiding are symmetries that relate different bloc expansions and as a consequence relate the correlator to itself or to another correlator. \par
We then impose constraints perturbatively. The function is written in terms of an Ansatz which can be justified for a few orders from perturbative computations. Then consequences of the bloc expansion above can be imposed perturbatively as a recursive input.\par 
Finally, we impose the physical constraints, which through non-perturbative results and constraints in the Regge limit select a unique solution in the best case scenario. 

\section{Operator Product Expansion}

\subsection{Scalar Blocks}
The Operator Product Expansion (OPE) is a convergent series of primary operators with increasing conformal weight, which encapsulate the behaviour of two primary operators.
\begin{align}
	\phi_{\Delta_1}(x_1)\phi_{\Delta_2}(x_2) = \sum_\Delta c_{\Delta_1\Delta_2\Delta}\mathcal{F}_{\Delta_1\Delta_2\Delta}(x_{21},\partial_{x_1})\mathcal{O}_\Delta(x_1)
\end{align}
where the sum is over the primary operators (those that are annihilated by the generator of special conformal transformations), the function $\mathcal{F}$ encodes the contribution from the conformal descendants (which have the form $\partial^{(l)} \mathcal{O}$), and the equality should be understood in the OPE sense; inserted in a correlator. Using this expansion iteratively allows one to relate any $n$-point function to three-point functions. There is, however, a conservation of difficulty, where one exchanges the number of insertions for infinite sums. Note that this type of expansion of multi-point operators is not unique to conformal theories. However, both the OPE's convergence and the fact that the constants are exactly the three-point coefficient are unique properties of CFTs. Thus the information from all $n$-point functions in a given CFT can be reduced to the spectrum $\{\Delta\}$ and the OPE coefficients $\{c_{\Delta_1\Delta_2\Delta_3}\}$, or CFT data $\{\Delta,c_{\Delta_1\Delta_2\Delta_3}\}$ of the theory.  Conversely, the study of $n-$ point functions gives the corresponding CFT data. \par
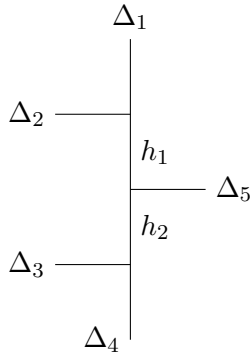
\begin{figure}[h]
	\centering
	\begin{tikzpicture}
		\draw[] (0,-2) -- (0,2);
		\draw[] (0,0) -- (1,0);
		\draw[] (0,1) -- (-1,1);
		\draw[] (0,-1) -- (-1,-1);
		\node[anchor=west] at (1,0) {$\Delta_5$};
		\node[anchor=east] at (-1,1) {$\Delta_2$};
		\node[anchor=east] at (-1,-1) {$\Delta_3$};
		\node[anchor=south] at (0,2) {$\Delta_1$};
		\node[anchor=east] at (0,-2) {$\Delta_4$};
		\node[anchor=west] at (0,.5) {$h_1$};
		\node[anchor=west] at (0,-.5) {$h_2$};
	\end{tikzpicture}
\caption{OPE diagram of a five-point correlator of scalars with external dimensions $\Delta_1,...,\Delta_5$ where the OPE is done in the (12) and (34) channels and the exchanged primaries have conformal weights $h_1$ and $h_2$.}
\label{diagram: 5pt block}
\end{figure}

 \vspace{2mm}
The Operator Product Expansion is inherently non-perturbative. However, one can expand the CFT data around a known point (for example, Generalised Free Fields, or GFFs) in a small parameter $\epsilon$ as done below in equation \eqref{Eq: Conformal block expansion} to extract perturbative CFT data. In one dimension, the OPE operator is given by (See appendix \ref{Appendix Conformal blocks} for more detail)
\begin{align}
	\mathcal{F}_{\Delta_1,\Delta_2,\Delta_3}(x_{21},\partial_{x_1}) = \sum_k \frac{(\Delta_{23,1})_k}{k! (2\Delta_3)_k} x^{k-\Delta_{12,3}}_{21}\partial_{x_1}^{(k)}
\end{align}
where the exchanged operator is evaluated in the coordinate of the derivative $\mathcal{O}_{\Delta_3}(x_1)$ and $\mathcal{O}_{\Delta_3}(x_2)$, respectively and 
\begin{align}
	\Delta_{i_1....i_n,\, \, j_1...j_m}&=\sum_i\Delta_i-\sum_j \Delta_j.
\end{align}
For the four-point function, the scalar OPE expansion gives the four-point conformal blocks, which are derived in several ways in Appendix \ref{Appendix Conformal blocks} and are explicitly
\begin{align}\label{4pt scalar block}
	\langle \phi_{\Delta_1}(x_1)\phi_{\Delta_2}(x_2)\phi_{\Delta_3}(x_3)\phi_{\Delta_4}(x_4)\rangle &=A(x_i)\sum_h c_{12\Delta}c_{34\Delta}\chi^h {}_2F_1(h+\Delta_{2,1},h+\Delta_{3,4},2h;\chi),
\end{align}
where
\begin{align}
	A(x_1,x_2,x_3,x_4) &=x_{12}^{-\Delta _1-\Delta_2}x_{34}^{-\Delta _3-\Delta_4} x_{13}^{\Delta _{4,3}} x_{14}^{\Delta _{23,14}} x_{24}^{\Delta _{1,2}}.
\end{align}
The same process can be iterated for higher-point functions and generates Appell functions, for example, the five-point function scalar block is\footnote{The definition of the Appell functions is reminded in \eqref{Eq: Appell Function} in Appendix \ref{App:five-point scalar bloc}. }
\begin{align}
	\langle \prod_{i=1}^{5} \phi_{\Delta_i}(\chi_i)  \rangle =& A_5(x_i)\chi_1^{-\Delta_1-\Delta_2} (1-\chi)^{-\Delta_3-\Delta_4}\sum_{h_1,h_2} c^3_{h_1,h_2} \chi_1^{h_1}(1-\chi_2)^{h_2}\times \\
	& F_2\left(
	h_1+h_2-\Delta_5,h_1-\Delta_{12},h_2+\Delta_{34};2h_1,2h_2\,|\,\chi_1,1-\chi_2\right)
\end{align}
where
\begin{align}
		A_5(x_i) = \left(\frac{x_{45}x_{15}}{x_{14}}\right)^{\Sigma-2\Delta_5}x_{15}^{-2\Delta_1}x_{25}^{-2\Delta_2}x_{35}^{-2\Delta_3}x_{45}^{-2\Delta_4},
\end{align}
and 
\begin{align}
	c^{3}_{h_1,h_2} = c_{\Delta_1\Delta_2h_1}c_{\Delta_3\Delta_4h_2}c_{h_1h_2\Delta_5}. 
\end{align}
These block expansions correspond to the exchange of conformal primaries of weight $h_i$ and can be summarised in OPE diagrams of the type shown in Figure \ref{diagram: 5pt block}.  However, since the exchanged operators are only labelled by their conformal weight, there is a risk of several operators contributing to the same term $h^{*}$; this is the mixing problem. \par 
The mixing problem has two aspects, one is perturbative (and problematic) and will be discussed later in subsection \ref{subsection: bootstrap mixing}, and the other is exact and due to supersymmetries. This can be treated quite effectively using superblocks. 
\subsection{Superblocks}
Conformal superblocks follow the same principle as scalar conformal blocks. However, in this case, the operators are superprimaries, so they are annihilated by the special superconformal transformations $\hat{S}$ and the special conformal transformations $\hat{K}$. In this way, the contribution of the conformal descendants is encapsulated by the expansion in spacetime cross-ratios and the contribution from superconformal descendants is encapsulated by the expansion in Gra\ss mann variables. The superblocks can be constructed by solving the differential equation corresponding to the superconformal Casimir (see \ref{Subsection: Superblocks ABJM}) or by expressing them as a sum of scalar blocks and using the constraints of $R$-symmetry selection rules and Ward identities (see section \ref{Section: superblocks}). 
Most superconformal theories also have an associated $R-$symmetry, so the correlators depend on additional terms. For example, the four-point function of the displacement multiplet inserted on the 1/2-BPS Wilson Line in $\mathcal{N}=4$ SYM is a function of variables that reduce to the following cross-ratios when the fermionic coordinates are turned off:
\begin{align}
	\chi &= \frac{x_{12}x_{34}}{x_{13}x_{24}}&\zeta_1 \zeta_2 &= \frac{y_{12}^2y_{34}^2}{y_{13}^2y_{24}^2}& (1-\zeta_1)(1-\zeta_2)&=\frac{y_{14}^2y_{23}^2}{y_{13}^2y_{24}^2}. 
\end{align}
However, the analysis of superconformal Ward identities from the consistency of the superspace as a coset constrains the correlator to be a function of a single  function $f(\chi)$\cite{Liendo:2018ukf, Liendo:2016ymz, Ferrero:2021bsb}. Additionally, these Ward identities provide an additional way to construct the superblocks (see Chapter \ref{chapter N=4}).
\subsection{Selection Rules}
In highly symmetric cases such as those studied in this thesis, the overall symmetry imposes sharper unitarity bounds than the conformal symmetry alone. These can be found by requiring the positivity of states when applying elements of the algebra, and examples in the literature are countless.\footnote{A useful reference when dealing with defect configurations are \cite{Agmon:2020pde} and \cite{Eberhardt:2020cxo}.} These constrain the correlator through the vanishing of the anomalous dimension for protected operators and the unitarity bound for the unprotected (long) operators. However, one of the universal bounds for all $1$d conformal correlators is the one from unitarity, which imposes
\begin{align}
	\Delta\geq0.
\end{align}
For the four-point function in \ref{Eq: four point definition}, for example, this fixes the boundary behaviour of the correlator in both the $\chi\rightarrow 0$ and $\chi\rightarrow1$ limits
\begin{align}
	f(\chi) &= \sum c_h \chi^h F_h(\chi) \\
	& = \sum \tilde{c}_h (1-\chi)^h F_h(1-\chi),
\end{align}
such that 
\begin{align}
	f(\chi) &=_{\chi\rightarrow 0}  o(\chi^0)\\ f(\chi)&=_{\chi\rightarrow 1} o((1-\chi)^0).
\end{align}
\section{Crossing and Braiding}

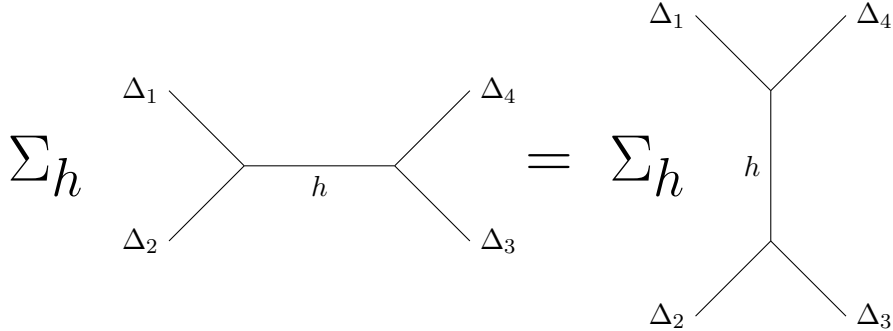
\begin{figure}[h]
	\centering
	\begin{tikzpicture}
		\hspace{-1cm}
		\draw[] (1,-1) -- (0,0);
		\draw[] (0,0) -- (1,1);
		\draw[] (0,0) -- (-2,0);
		\draw[] (-3,-1) -- (-2,0);
		\draw[] (-2,0) -- (-3,1);
		\node[anchor=west] at (1,-1) {$\Delta_3$};
		\node[anchor=west] at (1,1) {$\Delta_4$};
		\node[anchor=east] at (-3,-1) {$\Delta_2$};
		\node[anchor=east] at (-3,1) {$\Delta_1$};
		\node[anchor=north] at (-1,0) {$h$};
		\node[anchor=east] at (-4,0) {\Huge $ \Sigma_h$};
		\node[anchor=east] at (4,0) {\Huge $ \Sigma_h$};
		\node[anchor=east] at (2.5,0) {\Huge $ =$};
		\def\y{1}
		\draw[] (4,1+\y) -- (5,0+\y);
		\draw[] (5,0+\y) -- (6,1+\y);
		\draw[] (5,0+\y) -- (5,-2+\y);
		\draw[] (4,-3+\y) -- (5,-2+\y);
		\draw[] (5,-2+\y) -- (6,-3+\y);
		\node[anchor=east] at (4,1+\y) {$\Delta_1$};
		\node[anchor=west] at (6,1+\y) {$\Delta_4$};
		\node[anchor=east] at (4,-3+\y) {$\Delta_2$};
		\node[anchor=west] at (6,-3+\y) {$\Delta_3$};
		\node[anchor=east] at (5,-1+\y) {$h$};
	\end{tikzpicture}
	\caption{diagrammatic illustration of the crossing OPE. This is  understood as an OPE, not a Feynman diagram. The external legs are the operators considered in the correlator, and the internal vertex is the 'exchanged operator' from the OPE. This operator can be exchanged from the OPE between external operators in positions $1$ and $2$ or those in positions $1$ and $4$, hence the crossing. }
	\label{diagram: 4pt crossing }
\end{figure}

The crossing equation is one of the oldest elements of the conformal bootstrap and is ubiquitous in numerical and analytical methods. It is often graphically represented as in Figure \ref{diagram: 4pt crossing }.\par 
For the four-point function, crossing can be written as
\begin{align}
	\sum_h c^2_h G_h(\chi)= \sum_{\tilde{h}} \tilde{c}^2_{\tilde{h}}  G_{\tilde{h}}(1-\chi).
\end{align}
Remember that the operators exchanged in these two channels might be different. If additionally the operators exchanged are identical, this can be promoted to
\begin{align}
	\sum_h c^2_h G_h(\chi)= \sum_{h} c^2_{h}  G_{h}(1-\chi),
\end{align}
which is a functional equality 
\begin{align}
	f(\chi) = f(1-\chi)
\end{align}
depending on the conformal prefactor, there might be an additional $\chi-$dependant term. 
 In one dimension, this can be seen explicitly as a combination of the cyclic permutation invariance where the 1d line is mapped to a circle and the  $\mathbb{Z}_2$ time-reversal symmetry (called S-symmetry in \cite{Billo:2013jda}). In this case, there is an explicit symmetry between correlators
\begin{align}
	\langle 1234 \rangle = 	\langle 3214 \rangle 
\end{align}
which becomes a functional equality when operators $1$ and $3$ are identical. 
\newpage
\subsection{Braiding and Exchange of Identical Operators}
\begin{figure}[h]
	\centering
\begin{tikzpicture}
	\node[anchor=south east] at (-2,1.5) {$\chi$};
	\node[anchor=north west] at (2,-1.5) {$\frac{-\chi}{1-\chi}$};
	\node[anchor=south west] at (2,1.5) {$\frac{1-\chi}{-\chi}$};
	\node[anchor=south] at (0,2.5) {$1-\chi$};
	\node[anchor=north] at (0,-2.5) {$\frac{1}{1-\chi}$};
	\node[anchor=north east] at (-2,-1.5) {$\frac{1}{\chi}$};
	
	\draw[<->] (-2.3,1.5)--(-2.3,-1.5);
	\draw[<->] (0.5,-2.7)--(2,-1.9);
	\draw[<->] (0.5,2.7)--(2,1.9);
	
	\draw[double, <->] (-2,1.9)--(-.5,2.7);
	\draw[double,<->] (-.5,-2.7)--(-2,-1.9);
	\draw[double, <->] (2.3,-1.5)--(2.3,1.5);
	
	\draw[dashed, <->] (-2,1.5)--(2,-1.5);
	\draw[dashed, <->] (-2,-1.5)--(2,1.5);
	\draw[dashed, <->] (0,-2.4)--(0,2.4);
	
	\draw[<->] (4,2)--(5,2);
	\draw[double,<->] (4,0) -- (5,0);
	\draw[dashed,<->] (4,-2) --(5,-2);
	\node[anchor=west] at (5,2) {$x_2 \leftrightarrow x_3$};
	\node[anchor=west] at (5,0) {$x_2 \leftrightarrow x_4$};
	\node[anchor=west] at (5,-2) {$x_2 \leftrightarrow x_1$};
\end{tikzpicture}
\caption{diagram of the crossing and braiding for a four-point function of identical operators. The crossing and braiding, corresponding to the exchange of two operators, maps the cross-ratio $\chi$ to different cross-ratios corresponding to the function in a different region of analyticity. }
\label{crossing diagram}
\end{figure}
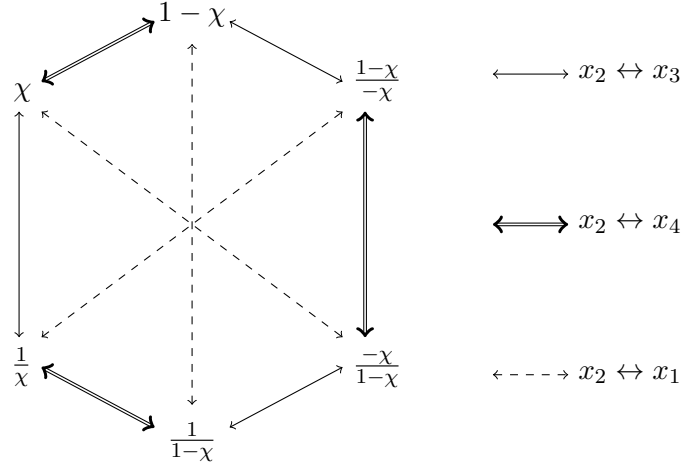

Another way to see this crossing is to consider the exchange of two identical operators (in positions 1 and 3 above). This is often problematic as it maps the conformal cross-ratios to an area outside its region of definition. Following the vocabulary of \cite{Liendo:2018ukf}, these mappings will be referred to as \textit{braiding} symmetries. For example, for the four-point function in 1d, since there is only one invariant cross-ratio, all the crossing symmetries can be summarised in the Figure \ref{crossing diagram}.\par 
However, the function $f(\chi)$ is only well-defined in the region $0<\chi<1$, so one requires an analytic continuation to compare the function to its braiding symmetric. This is not a problem in higher dimensions, as operators can cross without colliding. The way to consistently do this procedure is by considering the operator product expansion. The braiding is a partial symmetry of the four-point conformal blocks for external operators of equal dimensions:
\begin{align}
	f(\chi)&=\sum_h c^2_h \chi^{h-2\Delta}{}2F_1(h,h,2h;\chi)\\
	f(\frac{\chi}{\chi-1})&=\sum_h c^2_h \left(\frac{\chi}{\chi-1}\right)^{h-2\Delta} {}2F_1(h,h,2h;\frac{\chi}{\chi-1})\\
	&=(1-\chi)^{2\Delta}\sum_h c^2_h \left(-\chi \right)^{h-2\Delta} {}2F_1(h,h,2h;\chi)
\end{align}
where one of the hypergeometric identities was used (see e.g.  \cite{Gradshteyn:1943cpj}). 
When expanding perturbatively around even dimensions $h^{(0)}=2n \, , \,n\in \mathbb{N}$, we find the equality
\begin{align}
	f(\chi)|_{\log^l(\chi)} = (1-\chi)^{-2\Delta} f(\frac{\chi}{\chi-1})|_{\log^l(-\chi)}.
\end{align}
This happens in the case of Generalised Free Fields for example. 
This corresponds to equating the analytic continuation in equation \ref{Eq: bosonic continuation} to the original function by considering the function $\log(|\chi|)$ as a basis of logarithms, which also has motivation from the diagonal limit of higher dimensional theories. 
\subsection{Higher-$d$ Crossing Informing the Ansatz}
One way of gaining insight into $1d$ theories is with the diagonal limit of $d\geq2$ CFTs. For example, the four-point conformal invariant correlators known as $D-$functions have a vast literature in $d\geq2$ \cite{Zhou:2018sfz, Zhou:2020ptb, Dolan:2000ut,Dolan:2003hv} and reduce to the 1d D-functions (see Appendix \ref{app:Dfunctions}) in the diagonal limit. For example, the box integral, or $D_{1111}$\par
\begin{align}
	\Phi(\chi,\bar{\chi}) = \frac{1}{\chi-\bar{\chi}} \left(\log(\chi\bar{\chi})\log(\frac{1-\chi}{1-\bar{\chi}})+2\Li_2(\chi)-2\Li_2(\bar{\chi})\right).
\end{align}
Taking the limit $\chi\rightarrow \bar{\chi}$ we obtain 
\begin{align}
	\lim_{\bar{\chi} \rightarrow \chi} \Phi(\chi,\bar{\chi}) = -\frac{\log(\chi^2)}{1-\chi}-\frac{\log((1-\chi)^2)}{\chi},
\end{align}
which has both braiding and crossing up to the eventual replacement of the log terms depending on the expansion in the region of definition $0<\chi<1$. 
However, the higher dimensional crossing provides additional information beyond the behaviour of the logarithms; only a certain class of polylogarithms are allowed that are well defined under crossing. This restricts the starting Ansatz to a smaller basis, thus simplifying the bootstrap process (see section \ref{Section: Ansatz}). 
\section{Ansatz}\label{Section: Ansatz}
The form of the Ansatz closely follows the concepts developed for scattering amplitudes. However, the concepts of master integrals in AdS are less explored and the high-loop sector results are yet to be obtained.\footnote{This is especially true when compared to the eight loop calculations of \cite{Dixon:2022rse}. } 
Through explicit perturbative computations,  increasing loop orders at strong coupling have functions of  increasing transcendentality in the correlator, just as in scattering amplitudes. Contact diagrams with any number of external points have, at most, logarithmic terms \cite{Bliard:2022xsm}, and internal legs increase the transcendentality, usually by a factor of 2. As such, the Ansatz for a correlation function at $l^{th}$ order is given by \cite{Ferrero:2019luz, Ferrero:2021bsb, Liendo:2018ukf, Alday:2016njk, Alday:2015eya}:
\begin{align}
	f^{(l)}(\chi_i) = \sum_{t\leq 2l-1} r_t(\chi_i) T_t(\chi_i)
\end{align}
where $r_t(\chi_i)$ are rational functions with poles in the OPE limits of the correlators and $T_t(\chi_i)$ are harmonic polylogarithms (HPLs) of transcendentality up to $2l-1$ (Except for GFF, which is always rational). In practice, however, the transcendentality only increases by one at each increasing order for the cases studied (1/2-BPS Wilson lines in ABJM and $\mathcal{N}=4$ SYM). This could be from an additional symmetry which is not yet explored, such as the Yangian Symmetry\cite{Rigatos:2022eos}, or an Ansatz more naturally defined in Mellin space \cite{Rastelli:2016nze, Bianchi:2021piu}. This is also linked to the anomalous dimension being polynomial and not rational. When re-summing the $\gamma^{(1)}_h$ to obtain the prefactor of the leading power of logarithms, the resulting term is rational and not logarithmic, thus reducing the overall transcendentality of this term. \par 
Despite collision of operators preventing full crossing symmetry of the correlator, it is believed that, at strong coupling, the four-point correlators are single-valued in the region outside $0<\chi<1$. The condition that the functions be single-valued restricts the arguments of higher polylogarithms to be well defined under crossing and braiding. As such, for higher-loop four point correlators, the single-valuedness restricts a potentially large space of functions to those developed in \cite{Brown:2004ugm, Brown:2008um,Brown:2009qja} and used in a number of contexts, notably that of constraining higher-loop scattering amplitudes \cite{Dixon:2012yy,Chavez:2012kn}. 
\subsection{Perturbative Witten Diagrams}\label{Section Perturbative Witten diagrams}

Like Feynman diagrams, Witten diagrams provide a perturbative framework to compute correlators, but in $AdS$ space. Though there are many frameworks to make use of the isometries of the space to construct quantities similar to the momentum-space Feynman rules such as the Mellin formalism\cite{Paulos:2011ie, Penedones:2010ue, Fitzpatrick:2011ia, Mack:2009gy, Rastelli:2016nze}, momentum space in AdS \cite{Albayrak:2019asr, Albayrak:2020isk, Gillioz:2018mto, Bzowski:2019kwd} and spectral representation \cite{Carmi:2019ocp, Carmi:2021dsn, Carmi:2018qzm}, the explicit coordinate space methods are by far the most used\cite{Freedman:1998tz, Freedman:1998bj, DHoker:1999bve, DHoker:1999mqo, Zhou:2018sfz} and prove to have simplifications in the case of $\text{AdS}_2$. \par 
In the diagram, the external legs are depicted as points at the boundary, and the vertices in the bulk are integrated over the space (AdS$_2$).
\subsubsection{Witten diagrams}
The first tool to compute correlators in AdS is Witten diagrammatics. This technique is equivalent to Feynman diagrams for QFT in flat space. The integrals related to the diagrams are well-studied for four-point functions at Next-to-Leading Order (NLO) \cite{DHoker:1999mqo} and have several representations in Mellin space \cite{Mack:2009mi, Fitzpatrick:2011ia} and in spectral representation \cite{Carmi:2018qzm}. Due to its low dimensionality, the AdS$_2$ case is simpler and allows for higher-order and higher-point computations. For example, four-point correlators are only a function of one variable (as opposed to two), and their functional form is simpler in AdS$_2$. In the recent papers \cite{Bliard:2022xsm, Bianchi:2020hsz, Bianchi:2021piu}, the formalism of Witten diagrams and the Mellin transform is adapted to this lower-dimensional case and applied to the 1/2-BPS Wilson line defect. \par  \vspace{4mm}
In a general defect holographic system, a scalar bosonic operator of weight $\Delta$ inserted along the line is described by a fluctuation of mass $m^2=\Delta(\Delta-1)$ about the minimal surface of a string in AdS ending on the boundary of AdS space. When the minimal surface in question has an AdS$_2$ geometry, the fluctuations about this can be computed by considering effective fields propagating in a static AdS$_2$ space. The defect insertions are then dual to the boundary limit of these fluctuating fields
\begin{equation}
	\tilde{\phi}_\Delta(x) = \lim\limits_{z\rightarrow 0} z^{-\Delta} \phi(z,x).
\end{equation}
Using the common abuse of notation in the literature, we will denote $\tilde{\phi}$ as $\phi$. With this notation, the two-point function from two bulk points is
\begin{align}
	G_{\Delta_E}(z_1,t_1;z_2,t_2) &= C_\Delta \left(2u\right)^{-\Delta} {}_2F_1(\Delta,\Delta,2\Delta,-\frac{1}{2u})\\ u &= \frac{(z_1-z_2)^2+(x_1-x_2)^2}{2z_1 z_2}
\end{align}
which gives the Bulk-to-Boundary Propagator
\begin{align}
	K_{\Delta}(x_i; z,x) = C_\Delta \tilde{K}_{\Delta}(x_i; z,x) =C_\Delta \left(\frac{z}{z^2+(x-x_i)^2}\right)^\Delta
\end{align}
where $C_\Delta$ depends on the normalisation choice of the field, in this thesis, we use the canonical normalisation  \cite{Freedman:1998tz,Fitzpatrick:2011ia} 
\begin{align}\label{Eq: normalisation C_Delta}
	C_\Delta = \frac{\Gamma (\Delta )}{2 \sqrt{\pi } \Gamma \left(\Delta +\frac{1}{2}\right)}.
\end{align}
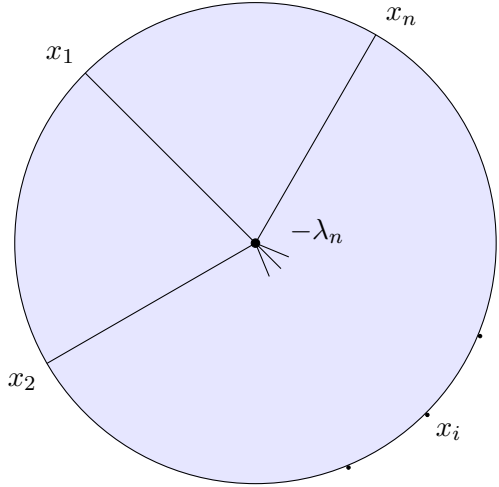
\begin{figure}[h]
	\centering
	\begin{tikzpicture}[scale=.8]
		\def\x{4}
		\filldraw[blue!10!white] (0,0) circle (\x cm);
		\draw[] (0,0) circle (\x cm);
		\node[anchor=west] at (0.1*\x,0.05*\x) {$-\lambda_n$};
		
		\draw[] (0,0) to (-0.707107*\x, 0.707107*\x);
		\node[anchor=south east] at (-0.707107*\x, 0.707107*\x){$x_1$};
		\draw[] (0,0) to  (-0.866025*\x, -0.5*\x);
		\node[anchor=north east] at (-0.866025*\x, -0.5*\x) {$x_2$};
		\draw[] (0,0) to  (0.5*\x,0.866025*\x);
		\node[anchor=south west] at (0.5*\x,0.866025*\x) {$x_n$};
		
		\draw[] (0,0) to  (.3 *0.191342*\x, -.3*0.46194*\x);
		\draw[] (0,0) to  (.3*0.353553*\x, -.3*0.353553*\x);
		\draw[] (0,0) to  (.3*0.46194*\x, -.3*0.191342*\x);
		
		\filldraw[]  (0,0) circle (0.5*\x pt);
		\filldraw[]  (0.191342*\x*2.02, -0.46194*\x*2.02) circle (0.2*\x pt);
		\filldraw[]  (0.353553*\x*2.02, -0.353553*\x*2.02) circle (0.2*\x pt);
		\filldraw[]  (0.46194*\x*2.02, -0.191342*\x*2.02) circle (0.2*\x pt);
		\node[anchor=north west] at (0.353553*\x*2, -0.353553*\x*2){$x_i$};
	\end{tikzpicture}
	\caption{Witten $n$-point contact diagram with a $\lambda_n\prod_i \phi_{\Delta_i}$ interaction and $n$ boundary insertions at positions $\{x_1,...,x_n\}$.}
	\label{Fig:contact-witten}
\end{figure}
Using these propagators, one can then compute correlators through Witten diagrams such as the one depicted in Figure \ref{Fig:contact-witten} by integrating over the bulk coordinate $z$. Explicitly, this corresponds to the integral
\begin{align}
	-\lambda_n\prod_{i=1}^{n} \left(C_{\Delta_i}\right) I_{\{\Delta_1,...,\Delta_n\}}(x_1,...,x_n) 
\end{align}
where
\begin{align}\label{Eq: contact integral 1}
	I_{\{\Delta_1,...,\Delta_n\}}(x_1,...,x_n) &= \int \frac{dzdx}{z^2} \prod_{i=1}^{n} \tilde{K}_{\Delta_i}(z,x;x_i),
\end{align}
where we have factored out the normalisation for later convenience.
The situation in AdS$_2$ is perfect for contour integration for the $x$-integral and partial fractions for the $z$-integral. This leads to the simple result from \cite{Bliard:2022xsm} for any number of massless operator insertions
\begin{align}
	I_{n,\Delta_i=1} = \frac{\pi}{(2i)^{n-2}}\sum_{i\neq j}\frac{x_{ij}^{n-4}}{\Pi_{k\neq i\neq j} x_{ik} x_{jk}}\ln\left(\frac{x_{ij}}{2i}\right).
\end{align}
These contact diagrams are a subset of all possible contributions and are the building blocks for computing correlators. 
 \subsubsection*{Contact diagrams}
For example, the contact diagram depicted in Figure \ref{Fig:contact-witten} corresponds to the integral \ref{Eq: contact integral 1} which was solved in \cite{Bliard:2022xsm} and presented in Chapter \ref{Effective theories} for all $n$ and $\Delta=1,2$. In the case of four points, there is a wealth of literature on these functions (referred to as $D-$functions) in this case and higher dimensions. The $D-$functions are then a special case of these integrals ($n=4$)
\begin{align} \label{D-function}
	\!\!\!\!\!\!\!\!
	D_{\Delta_1\Delta_2\Delta_3\Delta_4}(x_1,x_2,x_3,x_4) =\! \!\int \!\!\frac{dz dx}{z^{2}} 
	\tilde{K}_{\Delta_1}\!(z,x;x_1) \tilde{K}_{\Delta_2}\!(z,x;x_2) \tilde{K}_{\Delta_3}\!(z,x;x_3) \tilde{K}_{\Delta_4}\!(z,x;x_4).
\end{align}
Any $n-$point contact diagram in one dimension will, at most, give a logarithmic function. This classification of solutions in terms of functions of increasing transcendentality is important for the bootstrap procedure. \par 
For a one-dimensional boundary, the integration over the boundary coordinate ($x$ in this context) can be done through contour integration of the analytically continued variable\cite{Bliard:2022xsm}. 
We start by looking at $n$-point correlators of identical scalars with a simple contact interaction. These will serve as examples to demonstrate the simplifications occurring in this low-dimensional case and as building blocks for the massive contact diagrams and exchange diagrams. These correlators result from an interaction term $\lambda_n \phi^n$ in the bulk of AdS$_2$. They will be a function of $n-3$ independent cross-ratios due to the symmetry structure of CFT$_1$ or, equivalently, the isometry structure of AdS$_2$. These constitute the `master integrals' in AdS$_2$ for contact diagrams used in \cite{Bianchi:2020hsz,Giombi:2017cqn}. The contact diagram is illustrated in Figure \ref{Fig:contact-witten} and can be written as an integration over AdS$_2$ of the $n$ bulk-to-boundary propagators, leading to the connected tree-level correlator
\begin{align}
	\langle \phi_{\Delta_1}(x_1)...\phi_{\Delta_n}(x_n)\rangle_{conn}^{(1)} &= -\lambda_n \left(\Pi_{i=1}^n C_{\Delta_i} \right)I_{\Delta_1,...,\Delta_n}(x_1,...,x_n),
\end{align}
where we define the integral
\begin{align}\label{Eq: contact integral}
	I_{\Delta_1,...,\Delta_n}(x_1,...,x_n) =\int \frac{dx dz}{z^2}\Pi_{i=1}^n \left(\frac{z}{z^2+(x-x_i)^2}\right)^{\Delta_{i}}.
\end{align}
The simplifications of AdS$_2$ can be made explicit by evaluating the $x$-integral first with contour integration. This is especially effective for the massless case ($\Delta=1$) where the integrand of \eqref{Eq: contact integral} only has single poles. The general result for a massless $n$-point function is derived in subsection \ref{subsec:massless}:
\begin{align}\label{Eq: solution massless all n}
	I(x_i)&=\frac{\pi}{(2i)^{n-2}}\sum_{i\neq j}\frac{x_{ij}^{n-4}}{\Pi_{k\neq i\neq j}x_{ik}x_{kj}}\log\left(\frac{x_{ij}}{2i}\right),
\end{align}
For $\Delta=2$, a general result can also be derived (see Appendix \ref{App: Delta=2 derivation});
\begin{align}\label{Eq Delta=2 result}
	&I_{\Delta=2,n}=\sum_i\sum_{j\neq i}\frac{-\pi}{2(2i)^{2n-4}x_{ij}^2}\partial_{x_j}\left( \frac{x_{ji}^{2n-5}}{\prod_{k\neq j,k\neq i }x_{ik}^2x_{jk}^2}\log\frac{x_{ji}}{2i}\right)\nonumber \\
	&+\sum_i \sum_{j\neq i}\partial_{x_i}\frac{-\pi}{(2i)^{2n-2}x_{ij}^2}\partial_{x_j}\left( \frac{x_{ji}^{2n-4}}{\prod_{k\neq j,k\neq i}x_{ik}^2x_{jk}^2} \log\frac{x_{ji}}{2i}\right),
\end{align}
and higher weights can be derived from \textit{pinching}.
\subsubsection*{Pinching}\label{Sec: pinching}
One of the ways to relate correlators with a different number of points is through \textit{pinching}, that is, bringing an operator near another
\begin{align}
	\lim_{x_i\rightarrow x_{i+1}}\langle\phi(x_1)...\phi(x_i) \phi(x_{i+1})...\phi(x_n) \rangle .
\end{align}
One expects a divergence in this pinching limit from the operator product expansion. The contribution from the exchanged identity, in particular, leads to a power divergence
\begin{align}
	\lim_{\epsilon\rightarrow0} \langle \phi_\Delta(x_1)\phi_\Delta(x_1+\epsilon)\rangle  \sim \epsilon^{-2\Delta}.
\end{align}
Useful results can still be obtained through a similar limit relating not correlators but individual contact diagrams. In particular, the limit of the unit normalised propagators 
\begin{align}
	\lim_{x_2\rightarrow x_1} \, \, \tilde{K}_{\Delta_1}(x_1;x,z)\tilde{K}_{\Delta_1}(x_1;x,z) = \tilde{K}_{\Delta_1+\Delta_2}(x_1;x,z)
\end{align}
indicates that the pinching of operators should relate higher-point integrals to higher-weight integrals if this limit commutes with the integral.
\begin{figure}[h]
	\centering
	\begin{tikzpicture}[scale=.6]
		\def\x{4}
		\filldraw[blue!10!white] (0,0) circle (\x cm);
		\draw[] (0,0) circle (\x cm);

		\draw[] (0,0) to (\x, 0*\x);
		\node[anchor= west] at (\x, 0*\x){$x_5$};
		\draw[] (0,0) to (.5*\x, 0.866*\x);
		\node[anchor=south west] at (.5*\x, 0.866*\x){$x_6$};
		\draw[] (0,0) to (-.5*\x, 0.866*\x);
		\node[anchor=south east] at (-.5*\x, 0.866*\x){$x_1$};
		\draw[] (0,0) to (-1*\x, 0*\x);
		\node[anchor= east] at (-1*\x, 0*\x){$x_2$};
		\draw[] (0,0) to (-.5*\x, -0.866*\x);
		\node[anchor= north east] at (-.5*\x, -0.866*\x){$x_3$};
		\draw[] (0,0) to (.5*\x, -0.866*\x);
		\node[anchor=north west] at (.5*\x, -0.866*\x){$x_4$};
		
		\filldraw[]  (0,0) circle (0.7*\x pt);
	\end{tikzpicture} \hspace{1cm}
	\begin{tikzpicture}[scale=.6]
		\def\x{4}
		\filldraw[blue!10!white] (0,0) circle (\x cm);
		\draw[] (0,0) circle (\x cm);

		\draw[] (0,0) to (.5*\x, 0.866*\x);
		\node[anchor=south west] at (.5*\x, 0.866*\x){$x_{6}$};
		\draw[] (0,0) to (-.5*\x, 0.866*\x);
		\node[anchor=south east] at (-.5*\x, 0.866*\x){$x_1$};
		\draw[very thick] (0,0) to (-1*\x, 0*\x);
		\node[anchor=east] at (-1*\x, 0*\x){$x_2$};
		\draw[very thick] (0,0) to (\x, 0*\x);
		\node[anchor= west] at (\x, 0*\x){$x_5$};
		
		\filldraw[]  (0,0) circle (0.7*\x pt);
	\end{tikzpicture} 
	\caption{Pinching of the $I_{\Delta=1, n=6}$ integral to $I_{[1,2,2,1]}$ where the thick lines represent $\Delta=2$ bulk-to-boundary propagators.}
	\label{Fig: pinching}
\end{figure}
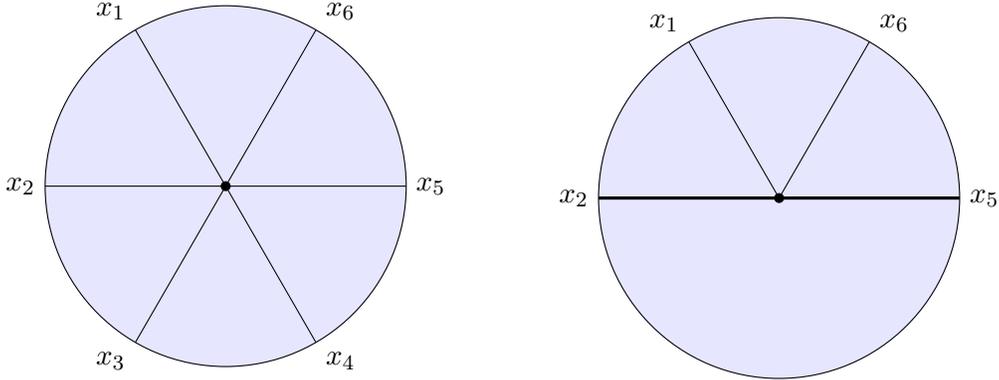\par
One expects this commuting between the limit and the integral to break down as soon as one encounters divergences. Therefore, one can take the following limit in the absence of divergences, 
\begin{align}
	\lim_{x_{i+1}\rightarrow x_{i}} I_{[\Delta_1,...,\Delta_n]}(x_1,...,x_n) = I_{[\Delta_1,...,\Delta_i+\Delta_{i+1},\bar{\Delta}_{i+1},...,\Delta_n]}(x_1,...,\bar{x}_{i+1},...,x_n),
\end{align}
where $\bar{x}_{i+1}$ denotes the absence of an operator at position $x_{i+1}$. The process can be iterated to form massive contact $n$-point diagrams from the basis of massless contact diagrams. 
This can, in principle, be done for all values of $\Delta$ and $n$. A list of examples for scalars of differing dimensions is given in Appendix \ref{App: list of correlators}, agreeing with numerical integration and known results. This provides a non-trivial check of the 6-point function and a way to evaluate the four-point correlator of massive fields. 
When divergences are present in the individual diagrams, these might cancel in the full correlator and, if not, need to be regularised. Divergences appear, for example,  in the pinching from a four- to a three-point function
\begin{align}
	\lim_{x_2\rightarrow x_1} <\phi(x_1)\phi(x_2)\phi(x_3)\phi(x_4)> &= 	\lim_{x_2\rightarrow x_1}\left( \frac{1}{x_{13}^2x_{24}^2} \left( \frac{-2\log(\chi)}{1-\chi}-\frac{2\log(1-\chi)}{\chi}\right)\right)\nonumber\\
	&=\frac{1}{x_{13}^2x_{14}^2}\lim_{\epsilon\rightarrow 0}\left(-2\log(\epsilon)-2\right),
\end{align}
where the pinched cross-ratio $\chi \rightarrow 0$ generates a divergence in the pinched correlator. 
In some physical systems, the cancellation of such divergences can occur thanks to the symmetries of the theory. For example, in the pinching of $\mathcal{N}=4$ fields in \cite{Barrat:2021tpn}, the contraction of the R-symmetry indices with a null vector ensures that the protected operators form a chiral ring, and the powers of a single protected operator are still protected. In the generic case where the divergences are retained, these do not necessarily match the corresponding correlator.

From the examples shown in the Appendix \ref{App: list of correlators}, the class of scalar contact diagrams follows this general property. In particular, since the dictionary of $D$-functions is well known, this provides a non-trivial check of the higher-point functions. One would be tempted to apply this pinching to the formal expression \eqref{Eq: massless} to have an independent derivation of the $\Delta=2$ case in \eqref{Eq Delta=2 result}. However, the pinching has to be done after the sum to have consistent limits, which impedes deriving the $\Delta=2$ case from the massless one.\par With this caveat in mind, the $n$-point contact diagrams of massless scalars can generate all $n$-point contact diagrams.\par
An simplification which occurs in this setting is that one can bring the final point to $\infty$ before beginning the procedure detailed above.  In such a situation, the integral
\begin{align}
	\lim_{x_n\rightarrow \infty}I_{\Delta_1,\Delta_2,...\Delta_n}(x_1,..,x_n) = \int dz dx z^{\Sigma-2}\prod_{i=1}^{n-1}\left(\frac{1}{z^2+(x-x_i)^2}\right)^{\Delta_i}
\end{align}
has a large $|z|$ behaviour as $\sim z^{2\Delta_n-\Sigma-2}$ and converges in the $z-$coordinate if
\begin{align}
	\Sigma-2\Delta_n > -1
\end{align}
where $\Sigma = \sum_{i}\Delta_i$. In this situation, the coordinate $x_n$ can be ignored and the pinching can be evaluated in a straightforward manner by considering a variation of \ref{Eq: massless}\footnote{For other cases, we can easily map it to a convergent case through cyclic permutations of the external coordinates. }
\begin{align}
	I_{\Delta_1,...,\Delta_n}(\chi_i)&=\lim_{\epsilon\rightarrow0}\text{Pinch}\frac{\pi}{(2i)^{n-2}}\sum_{i\neq j}\frac{x_{ij}^{\Sigma-4}}{\Pi_{k\neq i\neq j}x_{ik}x_{kj}}\log\left(\frac{x_{ij}}{2i}\right),\\
	\text{Pinch} &= x_{k+\sum_{p=0}^l\Delta_p}\rightarrow \chi_l+(k-1) \epsilon
\end{align}
where the pinching limit is in point-splitting formalism, $x_i=\{x_1,...,x_{\Sigma-\Delta_n}\}$, $k$ ranges from $1$ to $\Delta_{l+1}$ and $l$ ranges from $0$ to $n-2$. To condense the notation, we used the conventions
\begin{align}
	\chi_0&=0&\chi_{n-2}&=1&\Delta_0&=0
\end{align}
A mathematica notebook is attached to \cite{Bliard:2022xsm} which evaluates this pinching along with giving the appropriate conformal and numerical prefactor. This is relatively efficient at computing $n$-point contact diagrams of arbitrary weights. 

\subsubsection*{Exchange Diagrams and Polyakov Blocks }
The first extension to this class of integrals is to consider multiple bulk integrations. This happens when there are loops and exchanges in the corresponding Witten diagram. For some exchange diagrams, the various integrals can be related to \eqref{Eq: contact integral 1} through the action of a differential operator.
For example, the four-point single-exchange diagram can be found by solving the differential equation
\begin{align}
	(C_{(34)}^{(2)}-m^2_E)	J(x_1,x_2,x_3,x_4) &=\int \frac{dz_a dx_a}{z_a^2} \Pi_{i=1}^4K_{\Delta_\phi}(z_a,x_a;x_i)\\
	&= \tilde{I}_{\Delta_\phi ,4}(z),
\end{align}
where $C^{(2)}_{34}$ is the quadratic Casimir acting on the external legs 3 and 4, $m_E$ is the mass of the exchanged operator, and the full integral is given by
\begin{align}
	J(x_1...x_4) &= \int \frac{dz_adx_a}{z_a^2} \int \frac{dz_bdx_b}{z_b^2}G^{\Delta_E}_{BB}(a,b)\prod_{i=1}^{2}\left(K_{\Delta_\phi}(z_a,x_a;x_i)\right) \prod_{i=3}^{4}\left( K_{\Delta_\phi}(z_b,x_b;x_i)\right).
\end{align}
The simple structure of one-dimensional conformal correlators allows us to write the result explicitly in position space for the case $\Delta=\Delta_E=1$, see below in \eqref{Eq: First polyakov block}.

\bigskip
These quantities have been computed in Mellin space\cite{Mack:2009mi,Penedones:2010ue,Rastelli:2016nze,Fitzpatrick:2011ia,Ferrero:2019luz}, where Witten diagrams have a natural language. In particular, contact diagrams with no derivatives are given by constant truncated Mellin amplitudes, and exchange diagrams have poles in Mellin space.\footnote{For an introduction to the subject, a useful resource is \cite{Penedones:2016voo}. }
However, there are several caveats to these results. Firstly, the Mellin and anti-Mellin transforms are not trivial computations, so knowledge of the Mellin amplitude does not imply that of the position space correlator and vice-versa. Additionally, the generality of such results prevents the use of the simplifications occurring in $d=1$.  Furthermore, many spurious Mellin variables do not correspond to a cross-ratio in position space when the number of external legs is large enough ($n>d+2$). This is already relevant for the four-point $d=1$ correlator. In one dimension, several attempts were made to use the Mellin transform using the higher-dimensional formalism \cite{Ferrero:2019luz} or developing a one-dimensional formalism \cite{Bianchi:2021piu}. Using as a guide the principle that contact Witten diagrams correspond to constant Mellin amplitudes, the results in subsection \ref{subsec:massless} may provide a starting point in generalising the one-dimensional Mellin formalism developed in \cite{Bianchi:2021piu}. There, results for contact diagrams with general external dimensions $\Delta$ were derived so that, in combination with the insights of the present study, results for all $(n,\Delta)$ may be achievable.

\par 
Exchange diagrams immediately increase the difficulty of the calculation by adding a bulk integration and a bulk-to-bulk propagator. For example, the $t$-channel exchange diagram is given by the integral
\begin{align}\label{Eq: Exchange integral}
J(x_1,...,x_4) &= \int \frac{d^2z_a}{z_{a,0}^2}I(z_a,x_2,x_3)K_{\Delta_1}(x_1,z_a)K_{\Delta_4}(x_4,z_a)\\
I(w,x_2,x_3) &= \int \frac{d^2z}{z_0^2}G_{\Delta_E}(w,z)K_{\Delta_2}(x_2,z)K_{\Delta_3}(x_3,z),
\end{align}
where $K_\Delta$ and $G_{\Delta_E}$ are defined in equations \eqref{Eq: bulk-to-boundary prop} and $z_a= (z_{a,0},z_{a,1})$ are bulk coordinates. 
Using the isometries of AdS space, the exchange diagrams can be related to contact diagrams \cite{DHoker:1999mqo,Rastelli:2017udc} such as those presented in subsection \ref{subsec:massless}. This toy example is slightly different from the case in \cite{DHoker:1999mqo} since the sum of contact diagrams does not truncate but allows for an explicit example of a non-vanishing Polyakov block. Since both bulk integrations in the exchange Witten diagram integral have conformal symmetry, the solution is invariant under the action of symmetry generators on the legs attached to each of the bulk points
\begin{align}
(\mathbf{\mathcal{L}}_a+\vec{L}_2+\vec{L}_3)I(z_a,x_a;x_2,x_3) = 0.
\end{align}
This allows one to relate the quadratic Casimir acting on the external legs to the Laplacian acting on the corresponding bulk point (see Appendix \ref{App: quad Casimir} for details)
\begin{align}\label{Eq: Casimir exchange}
(C_{(23)}^{(2)}-m^2_E)I(z_a,x_a;x_2,x_3) &=  \int \frac{d^2z}{z_0^2}\left((\nabla_{a}^2-m_E^2)G_{\Delta_E}(w,z)\right)K_{\Delta_2}(x_2,z)K_{\Delta_3}(x_3,z).
\end{align}
Given that the bulk-to-bulk propagator satisfies the equation of motion, this term in \eqref{Eq: Casimir exchange} reduces to a delta function, thus reducing the number of integrals.
The problem is then more tractable since the double AdS$_2$ bulk integrations are replaced by a differential equation relating the answer to the known case of contact diagrams, which is a single bulk integral. \par Let us now consider a $\lambda \phi^3$ interaction in AdS$_2$ which gives a non-vanishing three-point function and an exchange diagram for the four-point function. For four-point correlators, the exchange diagram integral is solved explicitly in the $s$, $t$ and $u$ channels
\begin{align}
<\phi(x_1)\phi(x_2)\phi(x_3)\phi(x_4)> &=\frac{\lambda^2C_\Delta^4C_{\Delta_E}}{(x_{13}x_{24})^2}\left(f_t^{\Delta,\Delta_E}(\chi) +f_s^{\Delta,\Delta_E}(\chi) +f_u^{\Delta,\Delta_E}(\chi) \right),
\end{align}
where 
%$\lambda$ is the coupling constant for the $\phi^3$ interaction and 
$C_\Delta$ is the normalisation defined in equation \eqref{Eq: normalisation C_Delta}. For example, in the t channel, 
\begin{align}\label{Eq: Casimir eq 1}
(C_{(23)}^{(2)}-\Delta_E(\Delta_E-1))	 \frac{1}{(x_{13}x_{24})^{2\Delta_\phi}}  f_t^{\Delta,\Delta_E}(\chi)&= \frac{1}{(x_{13}x_{24})^{2\Delta_\phi}} I_{\Delta,n=4}(\chi).
\end{align}
$C_{(23)}^{(2)}$ is the quadratic Casimir acting on the external points 2 and 3, $\Delta_E$ is the conformal dimension of the exchanged operator, and $I_{\Delta,n=4}$ is the contact integral defined in \eqref{Eq: contact integral}. $f_t^{\Delta,\Delta_E}$ corresponds to the Witten exchange diagram in the $t$-channel ($J(x_1,...,x_4)$ in equation \eqref{Eq: Exchange integral}).  In one dimension, for $\Delta_\phi=\Delta_{exch}=1$, this is solved by
\begin{align}
f_{t}^{(1,1)} (\chi)&=\frac{\pi}{4}\frac{c_1+c_2 \log (\chi^2)+6 \text{Li}_3(\chi)- \text{Li}_2(\chi) \log (\chi^2)}{(\chi-1)^2}.
\end{align}
The same can be done in the other channels
%; in the $s$-channel, we have 
%\begin{align}
%f_{s}^{(1,1)}(\chi)&=\frac{\pi}{4}\frac{c_1+c_2\log \left((1-\chi)^2\right)+6 \text{Li}_3(1-\chi)- \text{Li}_2(1-\chi) \log\left( (1-\chi)^2\right)}{\chi^2}\label{Eq: fs1}\\
%&= f_t^{(1,1)}(1-\chi).
%\end{align}
%In the $u$-channel, we have 
%\begin{align}
%f_u^{(1,1)}(\chi)&= \frac{\pi}{4}\left(c_3+6 \text{Li}_3\left(\frac{\chi}{\chi-1}\right)- \text{Li}_2\left(\frac{\chi}{\chi-1}\right) \log \left((\frac{\chi}{1-\chi})^2\right)+c_4 \log \left((\frac{\chi}{1-\chi})^2\right)\right)\nonumber \\
%&= (1-\chi)^{-2}f_t(\frac{\chi}{\chi-1}). 
%\end{align} 
%The symmetry of the three channels is clear: The 
where the $s$ and $t$ channels are related by $\chi\rightarrow 1-\chi$ crossing, which identifies their integration constants. The solution, which is crossing-symmetric and makes the OPE expansion consistent,
% has the integration coefficients equal to
%\begin{align}
%c_1=c_3 &= 6\zeta(3)\\
%c_2 = c_4 &= -\frac{\pi^2}{6} .
%\end{align}
%Additionally, this solution 
has the mildest Regge growth. We can then define the correlator from the sum of the exchanges in the different channels
\begin{align}
<\phi(x_1)\phi(x_2)\phi(x_3)\phi(x_4)> &=\frac{\lambda^2}{\pi^5(x_{13}x_{24})^2}\left(f_t^{\Delta,\Delta_E}(\chi) +f_s^{\Delta,\Delta_E}(\chi) +f_u^{\Delta,\Delta_E}(\chi) \right).
\end{align}
The sum of exchanged Witten diagrams is related to the Polyakov block \cite{Mazac:2018qmi}. For an exchanged weight $\Delta=1$ and external weights $\Delta_\phi =1$,
\begin{align}\label{Eq: First polyakov block}
P_{1,1}^{(0)} (\chi)&= \frac{4}{\pi}\left(f_u^{(1,1)}(\chi) +f_t^{(1,1)}(\chi) +f_s^{(1,1)}(\chi) \right).
\end{align}
The $u$ and $t$ channels evaluated in the $\frac{\chi}{1-\chi}$ variable are well defined on the analytic continuation to the interval $0<\chi<1$
\begin{align}
f_u(\chi) = (1-\chi)^{-2}f_t(\frac{\chi}{\chi-1})\\
f_t(\chi) = (1-\chi)^{-2}f_u(\frac{\chi}{\chi-1}).
\end{align}
Due to this `pseudo-braiding' and crossing properties of this analytically continued function, the double discontinuity defined in \cite{Mazac:2018qmi} and reviewed in \eqref{Eq: Ddisc} can be evaluated quite easily as
\begin{align}
dDisc^{(+)}[P_{(1,1)}(\chi)] = \frac{2 (\chi \, \, {}_2F_1(1,1,2;\chi))}{\chi^2}.
\end{align}
This is the discontinuity in the $s$-channel of the corresponding Polyakov block \cite{Mazac:2018qmi}. Additionally, the bosonic continuation defined via \eqref{Eq: bosonic continuation} is fully symmetric under $s\rightarrow t$ and $s\rightarrow u$ and is Regge-bounded. These all provide a consistency check that $P^{(0)}_{1,1}$ is indeed the Polyakov block\footnote{I thank Pietro Ferrero for sharing results done via the conformal bootstrap relating to \cite{Ferrero:2019luz} allowing for a verification of this result.}  (defined on $0<\chi<1$) with external weight $\Delta=1$ and exchanged weight $\Delta_E=1$
\begin{align}\label{Eq: Exchange solution Delta=1}
P_{1,1}^{(0)} (\chi)&=\text{Li}_2\left(\frac{\chi}{\chi-1}\right) \log \left(\frac{\chi^2}{(\chi-1)^2}\right)-6 \text{Li}_3\left(\frac{\chi}{\chi-1}\right)-\frac{1}{6} \pi ^2 \log \left(\frac{\chi^2}{(\chi-1)^2}\right)+6 \zeta (3) \nonumber \\
&+ \frac{\text{Li}_2(1-\chi) \log \left((\chi-1)^2\right)-6 \text{Li}_3(1-\chi)-\frac{1}{6} \pi ^2 \log \left((\chi-1)^2\right)+6 \zeta (3)}{\chi^2}\nonumber \\
&+\frac{\text{Li}_2(\chi) \log \left(\chi^2\right)-6 \text{Li}_3(\chi)-\frac{1}{6} \pi ^2 \log \left(\chi^2\right)+6 \zeta (3)}{(\chi-1)^2} .
\end{align}
However, not all exchange diagrams have transcendentality 3, already in \cite{DHoker:1999kzh}, it was observed that the sum of diagrams reduces the order of transcendentality of the result. As we will see from the bootstrap result, a similar simplification occurs, even at higher loops. Sometimes, even the individual diagrams have simpler expressions, such as in the exchange diagram for the vertex $\lambda \phi_{\Delta=2}$ which is present in the perturbative computations of the 1/2-BPS string minimal surface in $\text{AdS}_3 \times \text{S}^3 \times \text{S}^3 \times \text{S}^1$ \cite{Bliard-Correa}. \par 
The Polyakov blocks for higher-valued exchanged or external weights can be computed (see \cite{Bliard:2022xsm}). Along with constraints from the double discontinuity and a suitable Ansatz, this method might provide a way to compute all Polyakov blocks  $P_{1,\Delta_E}^{(0)} (\chi)$.\par 
Therefore, given the possible vertices in the strong-coupling description, the second-order solution of four-point line-defect correlators has functions up to transcendentality three, increasing by two orders in the transcendentality at each order in perturbation. However, one expects a milder behaviour in physical systems. 

\subsection{Constraints from Crossing/ Anomalous Dimension}\label{constraints ansatz}
In practice, in both cases, the 1/2-BPS Wilson Line in ABJM \cite{Bianchi:2020hsz, Bliard-Ferrero} and in $\mathcal{N}=4$ SYM\cite{Ferrero:2021bsb}, the transcendentality only increases by a factor of one at each order in perturbation at strong coupling.\footnote{This is not the case at weak coupling. \cite{Cavaglia:2022qpg}} One of the origins of this feature can be seen in the first-order anomalous dimension which in both cases is polynomial in the exchanged weight and proportional to the quadratic Casimir eigenvalue\cite{Giombi:2017cqn}. The consequence of this is that when computing the higher powers of logarithms corresponding to 
\begin{align}
	f(\chi)|_{\log^l(\chi)} = \sum_h (\gamma_h^{(1)})^l c_h G_h(\chi). 
\end{align}
One can evaluate this directly by applying the differential Casimir operator to the GFF solution
\begin{align}
	\sum_h (\gamma_h^{(1)})^l c^{(0)}_h G_h(\chi) &= (C^{(2)}_{12})^l \sum_h c^{(0)}_h G_h(\chi),
\end{align}
which can only be rational. If, on the other hand, the anomalous dimension was a rational function of the exchanged weight, for example, inversely proportional to the Casimir eigenvalue, then the resulting functions would increase in transcendentality leading to non-rational functions multiplying the highest power of logarithms. Crossing symmetry is another factor that protects the solutions from having higher transcendental terms. The leading power of logarithms will be related to other parts of the function through crossing and braiding. The condition that the functions be single-valued restricts the arguments of higher polylogarithms to be well defined under crossing and braiding. As such, for higher-loop four point correlators, the single-valuedness restricts a potentially large space of functions to those developed in \cite{Brown:2004ugm,Brown:2004ugm, Brown:2008um,Brown:2009qja} and used in a number of contexts, notably that of constraining higher-loop scattering amplitudes \cite{Dixon:2012yy,Chavez:2012kn}. 

\section{Recursive Input}
The core principle of the analytic conformal bootstrap is the recursion step which happens by reinserting lower-loop data to find information about the next order in perturbation theory. This step is common for several methods, including the double discontinuity \cite{Bianchi:2022ppi, Mazac:2018mdx, Caron-Huot:2020adz, Meltzer:2019nbs, Correia:2020xtr}. 
\subsection{Input of $\gamma^{(l<L)}$}
If we write our function with a basis of powers of logarithms determined by the behaviour near the OPE limit $z\rightarrow 0$
\begin{align}
	f^{(l)}(\chi) = \sum_k \log^k(\chi) f^{(l)}_k(\chi),
\end{align}
then all of the terms except for $ f^{(l)}_0(\chi) $ and $ f^{(l)}_1(\chi)$ can be reconstructed from the lower order CFT data. Below are listed the perturbative expansions of the OPE of a four-point function where the conformal block is $G_h(\chi) =\chi^h F_h(\chi)$ :

\begin{align}\label{Eq: Conformal block expansion}
	f(\chi)&= \sum_\Delta c_\Delta \chi^\Delta F_\Delta (\chi)\\
	f^{(0)}(\chi) &= \sum_h c_h \chi^h F_h(\chi)\\
	f^{(1)}(\chi) &= \log(\chi) \sum_h c_h^{(0)} \gamma_h^{(1)} \chi^h F_h(\chi)+\sum_h \chi^{h} \left(c_h^{(0)} \gamma_h^{(1)} \partial_h+c_h^{(1)} \right)F_h(\chi)\\
	\vspace{2mm}
	f^{(2)}(\chi) &= \log^2(\chi) \sum_h \frac{1}{2} c_h^{(0)} \left(\gamma_h^{(1)}\right)^2 \chi^{h} F_h(\chi) \\
	&+\log(\chi)\sum_h \chi^{h} \left(c_h^{(0)} \left(\gamma_h^{(1)}\right)^2 \partial_h+c_h^{(1)} \gamma_h^{(1)}+c_h^{(0)} \gamma_h^{(2)} \right)F_h(\chi)\non \\
	&+\sum_h \chi^{h} \left(\frac{1}{2}c_h^{(0)} \left(\gamma_h^{(1)}\right)^2\partial_h^{2} + (c_h^{(1)} \gamma_h^{(1)}+c_h^{(0)} \gamma_h^{(2)})\partial_h + c_h^{(2)} \right) F_h(\chi)\non\\ 
	\vspace{2mm}
	f^{(3)}(\chi) &= \log^3(\chi)\sum_h \frac{1}{6} c_h^{(0)} \left(\gamma_h^{(1)}\right)^3 \chi^{h} F_h(\chi)\\
	&+\log^2(\chi) \sum_h  \frac{1}{2} \gamma_h^{(1)} \chi^{h} \left(c_h^{(0)} \left(\gamma_h^{(1)}\right)^2 \partial_h+c_h^{(1)} \gamma_h^{(1)}+2 c_h^{(0)} \gamma_h^{(2)}\right) F_h(\chi)\non \\
	&+\log(\chi)\sum_h \chi^{h} \left(\frac{1}{2} c_h^{(0)} \left( \gamma_h^{(1)}\right)^3 \partial_h^{2}+  \gamma_h^{(1)}(c_h^{(1)} \gamma_h^{(1)}+2 c_h^{(0)} \gamma_h^{(2)}) \partial_h +\right. \non \\
	&\hspace{5cm}\left.+ c_h^{(2)} \gamma_h^{(1)}+c_h^{(1)} \gamma_h^{(2)}+c_h^{(0)} \gamma_h^{(3)} \right) F_h(\chi)\non\\
	&+\sum_h\frac{1}{6} \chi^{h} \left(c_h^{(0)} \left(\gamma_h^{(1)}\right)^3 \partial_h^{3}+3 \gamma_h^{(1)} (c_h^{(1)} \gamma_h^{(1)}+2 c_h^{(0)} \gamma_h^{(2)})\partial_h^{2}+ \right. \non \\
	&\hspace{5cm}\left.+6 (c_h^{(2)} \gamma_h^{(1)}+c_h^{(1)} \gamma_h^{(2)}+c_h^{(0)} \gamma_h^{(3)}) \partial_h+6 c_h^{(3)} \right) F_h(\chi), \non
\end{align}
where we have expanded the terms
\begin{align}
	f(\chi) &= \sum_k \e^k f^{(k)}(\chi)& h &= h^{(0)} +\sum_k \e^k \gamma^{(k)}_h & c_\Delta &= \sum_k \e^k c_h^{(k)}.
\end{align}
Notice that only the $ f^{(l)}_0(\chi) $ and $ f^{(l)}_1(\chi)$ terms depend on CFT data from the given order in perturbation. Moreover, this analysis can be done in any OPE channel, which solves for a large part of the correlator. However, this simplicity is deceiving as even the part from the known CFT data can be hard to re-sum due to \textit{mixing} (see \ref{subsection: bootstrap mixing})
\subsection{All-Loop Consequences}
Given that in the best-case scenario, there is a sector of the correlator which can be known at all perturbative orders, it is a natural question to ask if information can be extracted from this. There is a specific limit (the double scaling limit), which selects precisely the highest logarithm Power. This is useful when looking at quantities such as the reparametrisation chaos and out of time ordered correlators (OTOC) such as in \cite{Giombi:2022pas}. In this situation, one looks at four-point correlators in the region of definition $\chi>1$ given by the braiding symmetric
\begin{align}
	\tilde{f}(\chi)|_{\chi>1} =\chi^{2\Delta}f(\frac{1}{\chi}).
\end{align} 
This function's double scaling limit selects the highest power of logarithms accessible from first-order data. 
For example, let us consider a fundamental scalar field of conformal weight $\Delta=1$ and the tree-level anomalous dimension of exchanged operators given by their eigenvalue under the quadratic Casimir
\begin{align}
	\gamma^{(1)}_\Delta&= \epsilon \Delta(\Delta-1).
\end{align} 
The GFF result is given by
\begin{align}
	\langle \phi(t_1)\phi(t_2)\phi(t_3)\phi(t_4)\rangle  =\frac{1}{(t_{12}t_{34})^2} \left( 1+\chi^2+\left(\frac{\chi}{1-\chi}\right)^2\right). 
\end{align}
The leading powers of logarithms can be found using the expansion in \eqref{Eq: Conformal block expansion} using the scalar conformal blocks \eqref{4pt scalar block}. The only necessary terms are those proportional to $\log(1-\chi)^l$ which are 
\begin{align}
	f^{(0)}_{\log^0(1-\chi)}&=\frac{\chi^2}{(1-\chi)^2}+\chi^2+1\\
	f^{(1)}_{\log^1(1-\chi)}&=\e \frac{2 \chi \left(2 \chi^4-8 \chi^3+12 \chi^2-7 \chi+4\right)}{(\chi-1)^2}\\
	f^{(2)}_{\log^2(1-\chi)}&=\e^2\frac{2 \left(9 \chi^6-44 \chi^5+87 \chi^4-88 \chi^3+49 \chi^2-6 \chi+2\right)}{(\chi-1)^2}\\
	f^{(3)}_{\log^3(1-\chi)}&=\e^3 \frac{4 \left(72 \chi^7-405 \chi^6+952 \chi^5-1203 \chi^4+872 \chi^3-350 \chi^2+90 \chi-1\right)}{3 (\chi-1)^2}. 
\end{align}
This access to certain parts of the function gives information about certain limits of the correlator. In higher dimensions, it is easier to find such limits due to the independence of the cross-ratios. In 1d this is achieved with the double scaling limit, explained in \cite{Giombi:2022pas}. When studying operators on the 1d CFT boundary of AdS$_2$, this corresponds to placing them along the line and mapping the line to a circle.\footnote{The points are equally spaced with coordinates $x_n= n\frac{\pi}{2}+i t\qquad n=\{1,2,3,4\}$. } This corresponds to the cross-ratio
\begin{align}
	\chi = \frac{2}{1-i\sinh(t)}.
\end{align}
The next step is to take the $t\rightarrow \infty$ limit of the function valid in the region $\chi>1$
\begin{align}
	\tilde{f}_{\chi^+}&= \frac{1}{\chi^{2\Delta}}f_{\chi^0}(\frac{1}{\chi}).
\end{align}
The term multiplying the $\log(1-\chi)$ has the same transformation from braiding seen above
\begin{align}
		\tilde{f}_{\chi^+}|{}_{\log(1-\chi)}(\chi)&= \frac{1}{\chi^{2\Delta}} f_{\chi^0}|{}_{\log(1-\chi)}(\frac{1}{\chi}).
\end{align}
In the late time limit, the $\log(1-\chi)$ terms give a $2\pi i$ contribution, and the inverse powers $1/\chi$ give an exponential $\text{e}^t/4i$ contribution. Keeping $\kappa = \e \text{e}^t$ constant in the late time limit gives the out-of-time correlator (OTOC) as in \cite{Giombi:2022pas} for this toy model (up to corrections of order $O\left(\kappa^n \times \text{e}^{-t}\right)$)
\begin{align}
	f_{DS}(\kappa) &= \sum _{n=1}^{\infty } (n+1) \left(\frac{\pi }{2}\right)^n \kappa ^n \Gamma (n+2)\\
	\kappa &= \e \exp(t). 
\end{align}
This is a divergent series, which we can regularise by considering the Borel transform but it should not be taken too seriously as it was an unphysical toy example.\footnote{See \cite{Dorigoni:2014hea} or \cite{Marino:2015yie} for an introduction to resurgence and non-perturbative methods in QFT.} \par 
The Borel transform is defined as usual for a series expansion as
\begin{align}
	F(t) &= \sum_n F_n t^n\\
	\mc{B}F(t) &= \sum_n \frac{F_n t^n}{n!}\\
	F_B(a) &= \int_0^\infty dt e^{-t} 	\mc{B}F(ta) 
\end{align}
One of the main objections to such all-loop results is that they only hold in the absence of operator mixing. However, we will see that the analogous result for ABJM holds as the first-order anomalous dimension does not mix. 
\subsection{Mixing} \label{subsection: bootstrap mixing}
Mixing, or the mixing problem, is the bane of all perturbative bootstrap situations, but makes most problems more interesting. It is an inherently perturbative issue as it is generated by the degeneracy of operators in non-interacting theories.\footnote{An extensive analysis and review of this issue can be found in \cite{Henriksson:2020jwk,Alday:2019clp}.} Let us consider the point at GFF for a single scalar of dimension $\Delta=1$. For example, the operators in the same column in Table \ref{mixing Table} have the same classical conformal dimension. \par 
\begin{figure}[h]
\centering
\begin{tabular}{|c|c|c|c|c|}
	\hline
	$\Delta_{\text{exchanged}}$&  $\Delta= 2$ &  $\Delta=3$& $\Delta=4$ & $\Delta=5$  \\
		\hline
	Exchanged Operator&  $\phi^2$ &  $\phi\partial\phi$&  $\phi \partial^2 \phi$&  $\phi \partial^3 \phi$  \\
	$(\Delta_\phi=1)$&  & $\phi^3$ & $\phi^2 \partial \phi$ & $\phi^2 \partial^2\phi$ \\
	&  &  &  $\phi^4 $& $\phi^3 \partial \phi$  \\
	&  &  &  & $\phi^5$ \\
	\hline
\end{tabular}
\caption{Table of different possible operators being exchanged at a certain level $\Delta$, there are additional operators of the same form which are not related by integration by parts. These illustrate the concept of mixing and are usually not actual eigenfunctions of the dilatation operator. }
\label{mixing Table}
\end{figure}
Therefore, it is necessary to bear in mind that when writing the OPE expansion, each weight $h$ has several terms
\begin{align}
	\sum_h c_h G_h \delta_{h,h^*} = \sum_{h=h^*}c_h G_h,
\end{align}
and in particular, that when extracting CFT data $\{c_h, \gamma^{(i)}\}$, one is actually computing \begin{align}
	\{\sum_{h^*=h}c_h,\langle\gamma_h\rangle_{c_h}\}.
\end{align}
The clearest example is seen for the highest logarithm powers of $\log(\chi)$ and $\log(1-\chi)$, which can be expressed in terms of the OPE expansions in the corresponding channels
\begin{align}
	f^{(l)}(\chi)|_{\log(\chi)^l} = \frac{1}{2}\sum_h c_h^2 \left(\gamma_h^{(1)}\right)^l G_{h}(\chi). 
\end{align}
Given that, perturbatively, several operators can contribute to the term with free dimension $\Delta$, we are looking at 
\begin{align}
	c_\Delta^2 \left(\gamma_\Delta^{(1)}\right)^l  = \sum_{h^{(0)}=\Delta}	c_h^2 \left(\gamma_h^{(1)}\right)^l  
\end{align}
so one cannot generically use the results from the first order, which give 
\begin{align}
	\sum_{h^{(0)}=\Delta}	c_h^2\gamma_h^{(1)}.
\end{align}
In the best-case scenario, there is no mixing between the operators; this can happen if there is a symmetry which sets most of the OPE to zero or if the degeneracy of the operators is not lifted. This absence of mixing occurs at first-order in both ABJM and $\mathcal{N}=4$ SYM
\begin{align}
	\sum_{h^{(0)}=\Delta}	c_h^2\gamma_h^{(1)} = \gamma_\Delta^{(1)}\sum_{h^{(0)}=\Delta}	c_h^2. 
\end{align}
In this case, we write that 
\begin{align}
	<(\gamma^{(1)})^2> = <\gamma^{(1)}>^2
\end{align}
where the average is over the different operators, weighted by the OPE coefficients. In this case, the operators do not mix, and the highest power of logarithms in the correlator can always be fixed. If this is not the case, the anomalous dimension will need to be diagonalised, requiring the study of a (potentially infinite) family of correlators.\footnote{This is done in \cite{Ferrero:2021bsb} to solve the fourth-order mixing.}  When considering this mixing, one can organise the families of operators quite naturally in terms of the length of the operator, as in the number of free fundamental fields composing it.  In this language, the mixing of two operators of different lengths can be understood in terms of particle production. In this sense, the absence of mixing would correspond to the absence of particle production, which is a statement of integrability.\footnote{The author thanks Miguel Paulos for enlightening discussions concerning this.}

\section{Final Constraints}
At this stage of the conformal bootstrap, where the crossing, the (super-)symmetry, and the CFT data from previous orders have constrained the Ansatz, the correlator is usually fixed up to a few elements (possibly an infinite family). The solutions organise themselves naturally in terms of a family with heavy operators growing  increasingly rapidly in the large-$\Delta$ limit. Constraining the growth of heavy operators in this Regge limit selects a finite number of terms which then can be fixed by other non-perturbative inputs. 

\subsection{Regge Behaviour and Perturbation}
At this point, the infinite family of solutions can be organised in terms of the behaviour of the anomalous dimension of the exchanged operators at large conformal weights. 
One criterium is to choose which term to keep by choosing the mildest behaviour at large-$\Delta$. This is justified in higher dimensions from the analysis of double-trace operators \cite{Cornalba:2007fs,Heemskerk:2009pn, Fitzpatrick:2011dm}. However, there is no distinction in 1d between single- and multi-trace operators. In Regge Bounded/superbounded theory, the equivalent requirement is that the anomalous dimension goes as 
\begin{align}
	\gamma^{(i)}&\sim O(1) &\gamma^{(i)}&\sim o(1)
\end{align}
respectively. However, this bound is too strict and does not yield solutions. This can easily be seen on the strong perturbation side in AdS$_2$, as derivative interactions give a faster growth in this limit so that the first order solution already has quadratic growth $\gamma_\Delta^{(1)]}\sim \Delta^2$. As a consequence, the Regge limit can be relaxed to the large-$\Delta$ behaviour
\begin{align}\label{Eq: Regge Behaviour of gamma}
	\gamma^{(l)} \sim_{\Delta\rightarrow \infty } \Delta^{l+1}.
\end{align}
Given this relaxed condition, it is natural to ask whether the perturbation is still valid in the expansion
\begin{align}\label{Eq: Weight expansion}
	\Delta = \Delta^{(0)}+\e \gamma^{(1)}+\e^2 \gamma^{(2)}+\e^3 \gamma^{(3)}+...
\end{align}
since the anomalous dimension grows faster than the classical dimension. In particular, at a fixed value of $\e$, there will be a lightest operator $\mc{O}_{\Delta^*}$ such that 
\begin{align}
	\e \gamma_\Delta^{(1)} \sim \Delta
\end{align}
so that the expansion \eqref{Eq: Weight expansion} for anomalous dimensions satisfying the bounds \eqref{Eq: Regge Behaviour of gamma} will have all of its terms being of the same order. This would break the perturbation. \par 
The perturbative expansion still works because the OPE expansion is dominated by lighter operators so that for a fixed small $\e$, one can truncate the sum to approximate the solution. Furthermore, the behaviour of ever-increasing powers of $\Delta$ in the anomalous dimension is reminiscent of many power series where the large-$\Delta$ limit is bounded and well-defined. 

\subsection{Flat-Space Limit}\label{flat-space limit}
There are many definitions of the flat-space limit in the context of AdS/CFT configurations \cite{Fitzpatrick:2011hu,Komatsu:2020sag, Paulos:2016fap,  Polchinski:1999ry, Costa:2012cb, Fitzpatrick:2011dm, Susskind:1998vk, Raju:2012zr, Alday:2017vkk, Fitzpatrick:2012cg, Mazac:2018mdx, Levine:2023ywq} notably because of the appeal of S-matrices. These exist in terms of Mellin space representation \cite{Fitzpatrick:2011hu, Penedones:2010ue}, and many in terms of very massive excitations with finite mass in the flat-space limit.
 To have some intuition into this defect configuration, we will give the following example of a flat-space limit from \cite{Bliard-Levine} for elementary excitations that become massless in the flat-space limit. The goal is to match the bosonic action to the free open string in flat space.
  In this context, the flat-space Mellin amplitude has not yet been developed. One must have a consistent 1d Mellin amplitude and flat-space limit definition to use results from the analytic bootstrap.\par 
  Consider the bosonic part of the action describing an open string (the example is given for Type IIB in AdS$^5\times \text{S}^5$, as studied in \cite{Giombi:2017cqn}) whose minimal surface is AdS$_2$.\footnote{This is the case of the 1/2-BPS defect configuration of Type IIB in AdS$^5\times \text{S}^5$ /$\mc{N}=4$ SYM and Type IIA in AdS$_4\times \C P^3$/ABJM. } This can be described by the Nambu-Goto action
  \begin{align}
  	S_{NG} = \alpha \int d^2 \sigma \sqrt{-|\left(\frac{1+\tfrac 1 4 x^2}{1-\tfrac 1 4 x^2}\right) g^{\text{AdS}_2^R}_{\mu \nu}(\sigma)+\frac{\partial_\mu x^i \partial_\nu x^i}{(1-\tfrac 1 4 x^2)^2}+\frac{\partial_\mu y^a \partial_\nu y^a}{(1+\tfrac 1 4 y^2)^2}|}.
  \end{align}
In the case of Witten diagrams, one factors out the Radius $R$ and uses the Green's Functions on AdS$_2^{R=1}$, which can be seen when writing this radius explicitly
  \begin{align}
	S_{NG} = \alpha \int \frac{d^2 \sigma}{R^2} \sqrt{-|\left(\frac{1+\tfrac 1 4 x^2}{1-\tfrac 1 4 x^2}\right) R^2 g^{\text{AdS}_2^{1}}_{\mu \nu}(\sigma)+R^2 \frac{\partial_\mu x^i \partial_\nu x^i}{(1-\tfrac 1 4 x^2)^2}+R^2\frac{\partial_\mu y^a \partial_\nu y^a}{(1+\tfrac 1 4 y^2)^2}|}.
\end{align}
Instead of this, one can rescale the excitations $(x^i,y^a)\rightarrow\frac{1}{R}(x^i,y^a)$ before taking the limit $R\rightarrow \infty$\par 
  \begin{align}
	S_{NG} &= \alpha \int \frac{d^2 \sigma}{R^2}\sqrt{-\text{det}(g_{\mu \nu})^{\text{AdS}_2^R}}  \sqrt{|\left(\frac{1+\tfrac{1}{4R^2}  x^2}{1-\tfrac{1}{4R^2} x^2}\right)^2\mathbb{1}+g^{\mu \nu}_{\text{AdS}_2^R}\left(\frac{\partial_\mu x^i \partial_\nu x^i}{(1-\tfrac{1}{4R^2} x^2)^2}+\frac{\partial_\mu y^a \partial_\nu y^a}{(1+\tfrac{1}{4R^2} y^2)^2}\right)}\non\\
	&\rightarrow_{R\rightarrow\infty} \lambda \int d^2 \sigma \sqrt{-\text{det}(\eta_{\mu \nu})}  \sqrt{\mathbb{1}+\eta^{\mu \nu} \left( \partial_\mu x^i \partial_\nu x^i+\partial_\mu y^a \partial_\nu y^a \right)} \label{Flat-space action}.
\end{align}
This, as expected, matches the NG action of an open, gauge-fixed string in $10d$ flat space. \par 
A few non-trivial conclusions can be drawn from this simple limit.
In the flat-space limit, the interaction terms of the gauge-fixed action are isotropic, meaning that the difference between $AdS$ and $S^5$ terms become irrelevant so that all fields have the same behaviour. Conversely, the large-$\Delta$ behaviour of the anomalous dimension in the original theory is determined by the flat-space Lagrangian \cite{Fitzpatrick:2011dm}. In the flat-space limit, only the maximal derivative terms survive, meaning that when expanding \eqref{Flat-space action}, there are only terms with as many derivatives as fields. In turn, the high-derivative interaction dominates the large-$\Delta$ behaviour of the anomalous dimension \cite{Bianchi:2021piu, Heemskerk:2009pn}. This will be used in the bootstrap in Chapter \ref{chapter: ABJM}, and there is hope to use results in the flat-space limit to fix the all-order large-$\Delta$ behaviour of the anomalous dimension \cite{Dubovsky:2012wk,Bliard-Levine}. This would not only rigorously prove the Regge-limit \eqref{Eq: Regge Behaviour of gamma} but would be an additional input for the bootstrap process. 

\subsection{Non-Perturbative Input}
In any bootstrap problem, additional physical input is needed to relate the small expansion parameter $\e$ to a parameter in the theory, such as the 't Hooft coupling $\lambda$. In Chapters \ref{chapter: ABJM}~and~\ref{chapter N=4}, two methods will be presented for this input. For less symmetric cases, possible solutions are presented in Chapter \ref{Chap: Optimisation and limits}. 

\section{CFT Data} \label{CFT Data}
In the eyes of the conformal bootstrap, the CFT data is the theory, such that every CFT can be specified by a set of operators labelled by their conformal weight and the three-point functions between the different operators. The CFT data generates correlators and vice versa. 
\subsection{Inverting the OPE / Inner Product}
The perturbative OPE in \eqref{Eq: Conformal block expansion} can be inverted through properties of the blocks which are solutions to the Casimir eigenvalue equation.
Let $F_{1\lambda}$ and $F_{2 \lambda}$ be the two linearly independent solutions to the $2^{\text{nd}}$ order eigenvalue problem
\begin{equation}\label{Eq: Eigenvalue equation}
	\mathcal{L} F_\lambda = c(\lambda)F_{\lambda}.
\end{equation}
We expect the set of eigenfunctions $\{F_\lambda\}$ to form a orthonormal basis with respect to an inner product. Let us define the weighted inner product as 
\begin{equation}\label{Eq: weighted inner product}
	\langle F_{\lambda_1},F_{\lambda_2} \rangle_w = \oint_\mathcal{C} \frac{d\chi}{2 \pi i}w(\chi) F_{\lambda_1}(\chi) F_{\lambda_2}(\chi),
\end{equation}
where the curve $\mathcal{C}$ is an anti-clockwise circle near 0.
We choose our inner product such that the operator $\mathcal{L}$ is self-adjoint
\begin{equation}\label{Eq: Self adjoint}
	\langle F_{\lambda_1},\mathcal{L}F_{\lambda_2} \rangle_w  = \langle \mathcal{L}F_{\lambda_1},F_{\lambda_2} \rangle_w .
\end{equation}
For this to be satisfied, we must have the following
\begin{align}\label{Eq: Self adjoint inner product}
	\langle F_{\lambda_1},\mathcal{L}F_{\lambda_2} \rangle_w &=\langle \mathcal{L}^{-1}w F_{\lambda_1},F_{\lambda_2} \rangle_1\\
	&=\langle w \mathcal{L} F_{\lambda_1},F_{\lambda_2} \rangle_1.
\end{align}
We write this differential operator equation as
\begin{equation}\label{Eq: Self adjoint weight}
	\mathcal{L}^{-1} w-w \mathcal{L} =0
\end{equation}
where $\mathcal{L}^{-1}$ is defined by integrating by parts the inner product and the equation
\begin{equation}
	\langle f_1,\mathcal{L}f_2\rangle =\langle \mathcal{L}^{-1}f_1,f_2 \rangle_1.
\end{equation}
This method allows us to define the correct inner product and find the appropriate orthogonality relations. For example in the simple scalar bloc, the eigenvalue equation given by following eigenvalues $\lambda_h$ and eigenfunctions $F_h$ of the quadratic Casimir operator $\mathcal{L}$
\begin{align}
	\mathcal{L} &= -\chi^2(1-\chi)\partial_\chi^2-\chi^2\partial_\chi \\
	F_h (\chi) &= \chi^h {}_2F_1(h,h,2h,\chi) \\
	\lambda_h &= h(h-1)\\
	\mathcal{L}^{-1} &= -\chi^2(1-\chi)\partial_\chi^2 +(4\chi-5\chi^2)\partial_\chi+(2-4\chi).
\end{align}
The equation \eqref{Eq: Self adjoint weight} is then solved by the weight\footnote{$w$ is normalised to obtain a normal basis.}
\begin{equation}
	w(\chi) = \frac{1}{\chi^2},
\end{equation}
and from equation \eqref{Eq: Self adjoint inner product}, we have
\begin{align}
	\langle F_{\lambda_1},\mathcal{L}F_{\lambda_2} \rangle_w & = \lambda_2(\lambda_2-1) \langle F_{\lambda_1},F_{\lambda_2} \rangle_w \\
	&= \lambda_1(\lambda_1-1)\langle F_{\lambda_1},F_{\lambda_2} \rangle_w.
\end{align}{}
The inner product, orthonormal basis, and $h$-projectors are then given by
\begin{align}
		\langle F_{\lambda_1},F_{\lambda_2} \rangle_w &= \oint_\mathcal{C} \frac{d\chi}{2 \pi i}\frac{1}{\chi^2} F_{\lambda_1}(\chi) F_{\lambda_2}(\chi)= \delta_{h,1-h'} \\
		F_h &= {}_2F_1(h,h,2h,\chi) \\
		F_{h \perp} &= F_{1-h}.
\end{align}
This inversion can be done to find the anomalous dimension from the logarithm terms and the OPE coefficients from the free theory. However, in practice, the most convenient is often to expand the blocks as a power series and solve for each coefficient.  \par Once the OPE is inverted, the average CFT data are found. Unmixing the contributions from different operators requires carefully analysing which operators mix and the comparison of several correlators. A careful analysis of this mixing in \cite{Ferrero:2021bsb} to compute a 4th-order correlator. In Chapter~\ref{chapter: ABJM}, the first order mixing will also be solved to compute the 3rd-order perturbative result and the all-order double scaling limit of the same correlator.
\section*{Conclusion}
This chapter described all the elements and guiding principles of the analytic conformal bootstrap; the constraints from the OPE, crossing, Ansatz, lower order CFT data, and physical input. Additionally, the constraints that they impose on the correlator were justified. This constitutes a toolbox with which to tackle most line dCFT analytic bootstrap problems.  The next two chapters will apply these principles and cover the explicit bootstrap of defect correlators.

\chapter{Bootstrapping the 1/2-BPS Wilson Line in ABJM}\label{chapter: ABJM}
\begin{chapquote}{Dante Alighieri, \textit{Inferno: Canto 3} \ref{Dante 1} }
	Per altra via, per altri porti\\
	verrai a piaggia, non qui, per passare:\\
	più lieve legno convien che ti porti.
\end{chapquote}
%\begin{chapquote}{Dante Alighieri,  \textit{De Monarchia, Libro I}\ref{Dante 2}}
%In ogni azione il primo scopo di chi agisce è di rivelare la propria immagine.
%\end{chapquote}

This section presents the bootstrap of the four-point functions of the displacement multiplet in the defect CFT defined by the 1/2-BPS Wilson line, thus exploring the spectrum at strong coupling to three perturbative orders. \par 
Wilson loops are fundamental non-local observables of any gauge theory and admit a representation in terms of Lagrangian fields employed in the weak coupling description. However, at strong coupling, the perturbative description in terms of these fields breaks down and one must resort to other methods to compute quantities. For example, through holography, their properties are naturally encoded into the degrees of freedom of a semiclassical open string when a gauge/gravity description is available \cite{Maldacena:1998im,Rey:1998ik,Drukker:1999zq}. Wilson lines are a prototypical example of a defect in QFT and could support a defect field theory characterising their dynamical behaviour. In the supersymmetric case, BPS Wilson lines provide one-dimensional supersymmetric defect field theories, explicitly defined through the correlation functions of local operator insertions on the contour~\cite{Drukker:2006xg}. From this perspective, 1/2-BPS Wilson lines in the four-dimensional ${\cal N}=4$ supersymmetric Yang-Mills theory (SYM) have been actively studied in the last few years~\cite{Giombi:2018qox,Giombi:2018hsx, Giombi:2017cqn}. In this case, the associated defect field theory is conformal (dCFT). Correlation functions can be generated through a ``wavy line" procedure and studied at weak coupling~\cite{Cooke:2017qgm} using general results for Wilson loops~\cite{Bassetto:2008yf}. \par 
Further information has been gained by considering four-point correlators of certain protected operator insertions~\cite{Giombi:2017cqn} whose two-point functions control the ${\cal N}=4$ SYM Bremsstrahlung function~\cite{Correa:2012at}. These correlators are evaluated at strong coupling by studying the relevant AdS/CFT dual string sigma model. The latter corresponds to an effective field theory in $\text{AdS}_2$, and correlation functions can be computed using standard Witten diagrams~\cite{Giombi:2017cqn}. The conformal bootstrap has also been applied to the computation of the same four-point functions~\cite{Liendo:2018ukf, Ferrero:2021bsb}, recovering and extending the analytical results at strong coupling and studying the finite-coupling regime numerically. The same approach yielded comparable results in a less supersymmetric scenario~\cite{Gimenez-Grau:2019hez}.   More generally, line defects provide a useful and physically interesting laboratory for the application of the analytical techniques developed in the context of one-dimensional CFTs~\cite{Mazac:2016qev,Maldacena:2016hyu,Simmons-Duffin:2017nub,Mazac:2018mdx,Mazac:2018ycv,Ferrero:2019luz}.
Finite-coupling results on defect conformal data were also obtained using integrability via the quantum spectral curve technology~\cite{Grabner:2020nis, Cavaglia:2022qpg}. 

\section{ABJM and Wilson Line Defect}\label{sec: ABJM Wilson line}

In this section, we will present the study of the dCFT associated with the 1/2-BPS Wilson line in ${\cal N}=6$ super Chern-Simons theory with matter (ABJM)~\cite{Aharony:2008ug}. These correlators of line insertions were bootstrapped in \cite{Bianchi:2020hsz, Bliard-Ferrero}.  
The structure of Wilson loops in ABJM theory is richer than in ${\cal N}=4$ SYM~\cite{Drukker:2019bev}, admitting different realizations through the fundamental Lagrangian fields~\cite{Drukker:2008zx,Chen:2008bp,Drukker:2009hy}, sometimes leading to the same quantum expectations values through a cohomological equivalence~\cite{Drukker:2009hy}. The relevant defect field theories should be able to fully distinguish them, possibly describing different Wilson lines in terms of marginal deformations~\cite{Correa:2019rdk}. Further motivations are the potential existence of topological sectors that could be associated with new supersymmetric localization procedures and the relations with integrability that might lead to an alternative derivation of the  $h(\lambda)$ function of ABJM~\cite{Gromov:2014eha} (see also~\cite{Bianchi:2014ada}) recently achieved in \cite{Correa:2023lsm}. 
The construction of supersymmetric Wilson loops in ABJ(M) theory ~\cite{Drukker:2009hy, Cardinali:2012ru} is notably more intricate than in the four-dimensional relative $\mathcal{N}=4$ SYM~\cite{Maldacena:1998im,Rey:1998ik,Zarembo:2002an,Drukker:2007qr}. For instance, when exploring the dynamics of the 1/2-BPS heavy massive particles obtained via the Higgsing procedure~\cite{Lee:2010hk}, one discovers that they are coupled not only to bosons (as occurs in $D=4$) but to fermions as well. Then the low-energy theory of these particles turns out to possess a $U(N|N)$ supergauge invariance instead of the {\it smaller} but  {\it expected} $U(N)\times \hat U(N)$ gauge symmetry. Therefore the (locally) 1/2-BPS Wilson loop operator  must be realized as the holonomy  of a  superconnection $\mathcal{L}(t)$ living in $U(N|N)$ ~\cite{Drukker:2009hy, Lee:2010hk, Cardinali:2012ru}:
\begin{equation}\label{WL}
	\mathcal{W}=\Str\left[P\exp\left(-i\oint dt \mathcal{L}(t) \right) \mathcal{T}\right].
\end{equation}
This superconnection can be written in terms of the ABJ(M) fields
\begin{equation}\label{superconnection}
	\mathcal{L}=\begin{pmatrix} A_\mu \dot{x}^\mu-\frac{2\pi i}{k} |\dot x| M_J{}^I C_I \bar C^J & -i\sqrt{\frac{2\pi}{k}}|\dot{x}|\eta_I \bar \psi^I\\
		-i\sqrt{\frac{2\pi}{k}}|\dot{x}| \psi_I \bar \eta^I & \hat A_\mu \dot x^\mu-\frac{2\pi i}{k} |\dot x| \hat{M}^I{}_J  \bar C^J C_I
	\end{pmatrix}.
\end{equation}
Here, $x^{\mu}(t)$ parametrizes the contour while the matrices $M_J{}^I(t)$ and $\hat M^I{}_J(t)$ and the spinors $\eta_I(t)$ and $\bar \eta^I(t)$ are local couplings,  determined in terms of the circuit $x^{\mu}(t)$ by the requirement of preserving some of the supercharges. 
The invariance for this type of loop operator does not follow  from imposing  $\d_\text{susy}\mathcal{L}=0$  as usual, but   the weaker condition
\begin{equation}\label{susyvarL}
	\d_\text{susy}\mathcal{L}=\mathcal{D}_{t} \mathcal{G}=\pa_{t} \mathcal{G}+i[\mathcal{L},\mathcal{G}] 
\end{equation}
where $\mathcal{G}$ is a $u(N|N)$ supermatrix. Namely, the action of supersymmetry on the connection $\mathcal{L}(t)$ can be cast as an infinitesimal supergauge transformation of   $U(N|N)$ \cite{Drukker:2009hy,Lee:2010hk,Cardinali:2012ru}. This directly implies a vanishing variation for the (super)traced Wilson loop, provided that 
$\mathcal{G}$ is periodic along the contour.  When $\mathcal{G}$ is not exactly periodic, one can correct the  lack of periodicity   either by inserting a twist supermatrix $\mathcal{T}$ in the supertrace  (see \eqref{WL}) ~\cite{Cardinali:2012ru} or by adding to $\mathcal{L}(t)$ a background connection living on the contour  ~\cite{Drukker:2019bev}. The explicit form of either  $\mathcal{T}$  or the background connection is irrelevant for the subsequent 
analysis. \par \vspace{3mm}
We will focus on the straight-line case.   The line is located at  $x^2=x^3=0$, and the couplings  in \eqref{superconnection} are
given by
\be
\renewcommand*{\arraystretch}{0.8}
\!\!\!\!\!\! M_J{}^I =\hat{M}^I{}_J  =\left(~\begin{array}{cccc}
	\!\!\!-1 	&0	&0	&0\\
	0	&1		&0	&0\\
	0		&0				&1	&0\\
	0		&0				&0	&1
\end{array}\right) \, , \quad \mathrm{and}\quad
\eta_I^\a=
\left( \begin{array}{c}
	1\\
	0\\
	0\\
	0
\end{array}\right)_{\!\! I}\!\!
\eta^+, \ \ 
\bar\eta^I_\a=i
\begin{pmatrix}
	1&
	0&
	0&
	0
\end{pmatrix}^I\!\!
\bar\eta_+
\ee
where $\eta^+=\bar\eta_+^T=\begin{pmatrix}1 & 1\end{pmatrix}$.  These couplings break the original symmetry  OSp$(6|4)$ to
SU$(1,1|3)$: in particular the original bosonic subsector of the supergroup containing the Euclidean conformal group in three-dimensions Sp$(4)\simeq$ SO$(1, 4)$ and the R-symmetry group SO$(6)\simeq$ SU$(4)$ reduces to  SU$(1,1) \times$  SU$(3)_R \times$ U$(1)_{J_0}$. The first factor SU$(1,1)\simeq$ SO$(2,1)$ is simply the conformal algebra in one dimension;   SU$(3)_R$ is the residual R-symmetry group,  and the U$(1)_{J_0}$ factor is a recombination of the rotations around the line and a broken R-symmetry diagonal generator. The preserved generators are given in Appendix \ref{algebra}.  The structure of the residual supergroup implies that we have a super-dCFT living along the straight line. Its operators  are characterised by a set of four quantum numbers $[\Delta, j_0, j_1, j_2]$ associated with the four Cartan generators of the above bosonic subalgebra.\footnote{$\Delta$ is the conformal dimension, $j_0$ is the  U$(1)_{J_0}$ charge and $j_1$ and $j_2$ are the SU$(3)_R$ labels.} The structure of short and long multiplets representing this subalgebra has been studied thoroughly in~\cite{Bianchi:2017ozk}, and we review it in Appendix~\ref{representations}. Recent development in $1/3$-BPS lines gives additional applications to the results below since the 1/2- and 1/3-BPS lines share some features for operator insertions \cite{Drukker:2022txy}.

\section{Line Operators and Correlators}
This  study will concentrate mainly on  defect correlation functions (see \eqref{eq: defect correlators}) associated with the components of a short multiplet -the displacement multiplet- which plays a fundamental role in any supersymmetric dCFT. It contains the displacement operator $\mathbb{D}(t)$ that is supported on every conformal defect~\cite{Billo1} and describes  infinitesimal deformations of the defect profile, as well as other operators associated with the broken R-symmetries and supercharges. Furthermore, since the operator is chiral, it preserves half of the dCFT's supercharges, making the superblock expansion accessible. For more general operators, such as the 1/3-BPS operators studied in \cite{Bliard-Ferrero}, additional constraints from the superconformal Ward identities are needed to constrain the superblock expansion and the full super correlator. \par 
We will consider three different complementary realizations of this protected multiplet that are useful to probe its properties and to calculate its correlation functions, depending on the coupling regime and the computational method. A first,  more general, realization of the displacement multiplet is obtained using a superfield $\Phi$ associated with a short multiplet of SU$(1,1|3)$. A representation theory analysis shows that this superfield is neutral under SU$(3)$ and annihilated by half of the supercharges (and thus chiral or antichiral). In four dimensions, the analogous superfield representation for the displacement multiplet has been derived in~\cite{Liendo:2016ymz}, resulting in a superfield charged under the residual $R$-symmetry group. A natural strong-coupling realization  is  provided by the AdS/CFT correspondence. In ABJM theory, a 1/2-BPS Wilson line is dual to the fundamental open string living in $\text{AdS}_4\times \cp^3$, with the appropriate boundary conditions on the straight contour. The minimal surface spanned by the string encodes its vacuum expectation value at leading order in the strong-coupling expansion. Fluctuations around the minimal surface solution nicely organize  in an AdS$_2$ multiplet of transverse modes~\cite{Forini:2012bb, Correa:2014aga}, whose components precisely match the quantum numbers of the displacement multiplet. This correspondence is summarised in Table \ref{tabledisp}. A third realization is obtained by inserting field operators constructed explicitly from the elementary fields appearing in the ABJM Lagrangian. In our case, the identification is trickier than in the ${\cal N}=4$ SYM case: the 1/2-BPS Wilson line in ABJM is constructed as the holonomy of a superconnection living on a $U(N|N)$ superalgebra~\cite{Drukker:2009hy, Cardinali:2012ru}, and our multiplet should correspond to the insertion of appropriate supermatrices. Perturbative weak-coupling results for the correlation functions can be obtained in this framework with Feynman diagrams. We construct the supermultiplet explicitly in terms of supermatrices, taking into account the fact that the action of the relevant supercharges is deformed by the presence of the Wilson line itself~\footnote{A basic difference between 1/2-BPS Wilson lines in ${\cal N}=4$ SYM and ABJM is that in the first case, the relevant connection is invariant under supersymmetry. In contrast, in the second one, it undergoes a supergauge $U(N|N)$ transformation \cite{Drukker:2009hy, Cardinali:2012ru}. This fact will be important in deriving the correct field-theoretical representation of the displacement supermultiplet.}. 
\begin{table}
	\begin{center}
		\begin{tabular}{l c c c}
			\sc Grading &\sc Operator & $\D$ & $m^2$\\
			\hline
			Fermion & $\mathbb{F}(t)$ & $\frac12$ & 0\\
			Boson   & $\mathbb{O}^a(t)$ & $1$ & 0\\
			Fermion & $\mathbb{\Lambda}_a(t)$ & $\frac32$ & 1\\
			Boson   & $\mathbb{D}(t)$ & $2$ & 2\\
			\hline
		\end{tabular}
		\caption{Components of the displacement supermultiplet with their scaling dimensions and the masses of the dual $\text{AdS}_2$ string excitations. The mass is obtained through the AdS/CFT dictionary $m^2=\D(\D-1)$ for the bosons and $m^2=(\Delta-\frac12)^2$ for the fermions.}
		\label{tabledisp}
	\end{center}
\end{table}

\subsection{Symmetry Breaking and Superconformal Algebra}
The $1/2$ BPS Wilson line in ABJM breaks the three-dimensional $\mathcal{N}=6$ superconformal algebra $osp(6|4)$ to the one-dimensional $\mathcal{N}=6$ superconformal algebra $su(1,1|3)$. This algebra is generated by  $\{D, P, K, R_a^b,J;Q^a,\bar{Q}_a,S^a,\bar{S}_a\}$, where $a=1,2,3$ and the generators can be identified with $osp(6|4)$ generators (see Appendix A of \cite{Bianchi:2020hsz} and C of \cite{Bianchi:2017ozk}). The bosonic generators  $\{D,P,K\}$ form the one-dimensional conformal algebra $su(1,1)\sim so(2,1)$, the traceless generators ${R_a}^b$ ($a,b=1,2,3$) form SU$(3)$, and $J_0$ forms the $u(1)$ R-symmetry. Their commutation relations are
\begin{align}
	[P,K]&=-2 D\,,& ~~ [D,P]&=P\,, &  [D,K] &=-K & [{R_{a}}^{b},{R_{c}}^{d}]&=\delta_{a}^{d} {R_{c}}^{b} -\delta^{b}_{c} {R_{a}}^{d}\,,
\end{align}
and $J_0$ commutes with all of them.
%%%%
The anticommutation relations for the fermionic generators $Q^a$, $\overline{Q}_a$ and the corresponding superconformal charges  $S^a, ~\overline{S}_a$, $a=1,2,3$ are
\begin{align}
	\{Q^a,\bar Q_b\}&=2 \delta^a_b P\,, &  \{S^a,\bar{S}_b\}&=2\delta^a_b K \\  
	\{Q^a,\bar S_b\}&= 2\delta_b^a ( D+\tfrac13 J_0)-2 {R_b}^a & \{\bar Q_a, S^b\}&= 2\delta_b^a(D-\tfrac13 J_0)+2 {R_a}^b \,,
	\label{anticommQS}
\end{align}
and the non-vanishing mixed commutators are \small
\begin{align}
	[D,Q^a]&=\frac12 Q^a & [D,\overline{Q}_a]&=\frac12 \overline{Q}_a &  [K,Q^a]&=S^a & [K,\overline{Q}_a]&= \overline{S}_a\\
	[D,S^a]&=-\frac12 S^a & [D,\overline{S}_a]&=-\frac12 \overline{S}_a &  [P,S^a]&=-Q^a & [P,\overline{S}_a]&=- \overline{Q}_a\\
	[{R_a}^b,Q^c]&=\delta_a^c Q^b-\tfrac13 \delta_a^b Q^c & [{R_a}^b, \overline{Q}_c]&=-\delta_c^b \overline{Q}_a+\tfrac13 \delta_a^b \overline{Q}_c & [J_0,Q^a]&=\tfrac12 Q^a & [J_0,\overline{Q}_a]&=-\tfrac12 \overline{Q}_a\\
	[{R_a}^b,S^c]&=\delta_a^c S^b-\tfrac13 \delta_a^b S^c & [{R_a}^b,\overline{S}_c]&=-\delta_c^b \overline{S}_a+\tfrac13 \delta_a^b \overline{S}_c & [J_0,S^a]&=\tfrac12 S^a & [J_0,\overline{S}_a]&=-\tfrac12 \overline{S}_a.
\end{align}\normalsize
The quadratic Casimir is
\be
\label{Casimir}
{\bf C}^{(2)}=D^2-\frac{1}{2}\,\{K,P\}+\frac{1}{3}J_0^2-\frac{1}{2}{R_a}^b\,{R_b}^a+\frac{1}{4}\,[\overline{S}_a,Q^a]+\frac{1}{4}\,[S^a,\overline{Q}_a]~,
\ee
and, when acting on a highest weight state $[\Delta,j_0,j_1,j_2]$  of $su(1,1|3)$, it has eigenvalue
\be\label{Casimireigenv}
{\bf c_2} =   \Delta (\Delta+2)+\frac{j_0^2}{3}\,-\frac{1}{3} (j_1^2 + j_1 j_2 + j_2^2).
\ee
%Similar superspaces have already been used in the literature to study 1-BPS correlators in different superconformal setups
To compute the superconformal blocks, we must represent the above algebra as an action on superconformal primaries. For this purpose, we use the superspace $(t, \theta_a, \bar\theta^a)$ introduced above.\footnote{The natural superspace to study correlation functions in the bulk was introduced in~\cite{Liendo:2015cgi}, and the setup adopted here can be seen as the reduction of the one considered there.}  We can then write the differential action of the $\mathcal{N}=6$ generators  as
% ~\footnote{One way to derive~\eqref{superspace}-\eqref{superspace-end} is e.g. via a straightforward extension of $su(1,1|1)$ case, obtained as the global part   of the  $\mathcal{N}=2$ superconformal algebra in two dimensions  when restricted to the holomorphic part, see Refs.~\cite{hep-th/9208069, P. Di Vecchia J. L. Petersen and H. B. Zheng, 
		%G. Mussardo G. Sotkov and M. Stanishkov}.}
\begin{align}\label{generators-diff}
	P&=-\del_{t} \\
	D&=\textstyle{-t \del_t-\frac12 \theta_a \del^{a} -\frac12 \bar \theta^a \bar \del_a - \Delta}\\
	K&=-t^2 \del_{t}-(t+\theta\bar\theta) \theta_a \del^{a}-(t-\theta \bar\theta) \bar \theta^a \bar \del_a-(\theta \bar \theta)^2 \del_{t} -2\,t\, \Delta+\frac23 j_0\,\theta\bar \theta\\
	Q^a&=\del^{a}-\bar \theta^a \del_t\\
	\bar Q_a &=\bar \del_{a}- \theta_a \del_t\\
	S^a&=(t+\theta \bar \theta)\del^a-(t-\theta \bar \theta) \bar \theta^a \del_{t}-2\bar \theta^a \bar \theta^b \bar \del_b-2( \Delta +\frac13 j_0 ) \bar\theta^a\\
	\bar S_a&=(t-\theta \bar \theta)\bar \del_a-(t+\theta \bar \theta)  \theta_a \del_{t}-2 \theta_a \theta_b \del^b-2( \Delta -\frac13 j_0 )\theta_a\\
	J_0&=-\frac12 \theta_a \del^a+\frac12 \bar \theta^a \bar \del_a+j_0\\\label{generators-diff-end}
	R_a{}^b&=-\theta_a \del^b+\bar \theta^b \bar \del_a+\textstyle{\frac{1}{3}}\,\delta_a^b\, (\theta_c\del^c-\bar\theta^c\bar\del_c)
\end{align}
where $\del^a=\frac{\del}{\del \theta_a}$, $\bar \del_a=\frac{\del}{\del \bar \theta^a}$, $\theta \bar \theta =\theta_a \bar \theta^a$ and we neglect the SU$(3)$ charges $j_1$ and $j_2$ because we will only be  interested in neutral superfields. 
The superspace is also equipped with super covariant derivatives
\be\label{supercovder}
\mathsf{D}^a=\del^{a}+\bar \theta^a \del_t\,,\qquad\qquad  \bar{\mathsf{D}}_a= \bar{\del}_{a}+ \theta_a \del_t\,.
\ee

\subsection{Displacement Multiplet}\label{Sec: ABJM Displacement}
This section is devoted to defining and constructing the displacement super multiplet for a line defect given by the $1/2$ BPS Wilson line in ABJM. For completeness, we start by recapitulating some basic facts  about this theory. The gauge sector consists of two gauge fields $A_\mu$ and  $\hat A_\mu$ belonging respectively to  the adjoint of $U(N)$ and $\hat U(N)$. The matter sector instead contains the complex scalar fields  $C_I$, $\bar C^I$, and the fermions $\psi_I$ and $\bar \psi^I$. The fields $(C_I, \bar \psi^I)$ transforms in the bifundamental  $(N,\bar N)$ while the couple $(\bar C^I,  \psi_I)$ lives in the $(\bar N,N)$. The additional capital index $I = 1,2,3,4$ labels the (anti)fundamental representation of the R-symmetry group SU$(4)$.  The kinetic term for the gauge fields consists of two Chern-Simon actions of opposite level $(k,-k)$, while the ones for scalars and fermions  take the standard form in terms of the usual  covariant derivatives. To ensure super-conformality, the action is also endowed with a suitable  sextic scalar potential  and $\psi^2 C^2$ Yukawa-type interactions explicitly spelt out in~\cite{Aharony:2008ug, Benna:2008zy}.\par 
An infinitesimal variation of the Wilson line \eqref{WL} translates into an operator insertion according to the identity
\begin{equation}\label{insert}
	\frac{\braket{(\delta\mathcal{W}) \dots}}{\braket{\mathcal{W}}}=-i\int_{\mathcal{C}} dt\,\langle \delta \mathcal{L}(t)\dots \rangle_{\mathcal{W}},
\end{equation}
where on the left-hand side (l.h.s.) we are considering an arbitrary correlator with the deformed Wilson line while on the right-hand side (r.h.s.) we are using the definition \eqref{eq: defect correlators}. Notice that $\delta$ could be any generator of a symmetry broken by the Wilson line. When $\delta$ is the action of a broken fermionic  or bosonic charge inside the parent superalgebra, the resulting operators  can be related to  an element of the 
super-multiplet of the 
displacement operator. Consider, for instance, the  six broken R-symmetry generators $J_1{}^a\equiv\mathsf{J}^a$ and $J_a{}^1\equiv\bar{\mathsf{J}}_a$ (see Appendix \ref{algebra}). Their action will yield  six bosonic defect operators by the symbolic action
\begin{align}\label{O}
	[\mathsf{J}^a, \mathcal{W}]&\equiv i\delta_{\mathsf{J}^a} \mathcal{W}= \int dt~ \mathcal{W}[\mathbb{O}^a(t)]  & [\bar{\mathsf{J}}_a, \mathcal{W}]&\equiv i\delta_{\bar{\mathsf{J}}_a} \mathcal{W}= \int dt~ \mathcal{W}[\bar{\mathbb{O}}_a(t)],
\end{align}
where we indicate by $\mathcal{W}[\mathbb{O}^a(t)]$ an operator that is inserted on the Wilson line. The equations \eqref{O} are Ward identities and need to be thought of as inserted in some correlation function. The defect operator $\mathbb{O}^a(t)$ has  conformal dimension  $\Delta=1$ since  the line defect is dimensionless and the dilatations commute with $\mathsf{J}^a$. The $U(1)_{J_0}$ charge is $2$ and it can be read from the commutator $[J_0,\mathsf{J}^a]=-2[ J_1{}^1,\mathsf{J}^a]=2 \mathsf{J}^a$. Finally, it transforms in the fundamental representation of SU$(3)$. Namely, this operator is characterised by the following set of four quantum numbers $[1,2,1,0].$  Similarly for $\bar{\mathbb{O}}_a(t)$  we have $[1,-2,0,1].$ \par \vspace{4mm}
The six broken supercharges $Q^{1a}_-\equiv i \bar{\mathsf{Q}}^a$ and $Q^{ab}_{{+}}=\e^{abc}\mathsf{Q}_c$ define six fermionic defect operators
\begin{align}
	\label{La}
	[\mathsf{Q}_a, \mathcal{W}]&=\int dt ~\mathcal{W}
	[\mathbb{\L}_a(t) ] &  [\bar{\mathsf{Q}}^a, \mathcal{W}]&=\int dt ~\mathcal{W}
	[ \bar{\mathbb{\L}}^a(t) ].
\end{align}
The quantum numbers of the line defect and the commutation relations of the broken and preserved charges fix the quantum numbers of the defect operators.  We find  $[\frac32,\frac52,0,1]$ for $\mathbb{\L}_a(t)$   and $[\frac32,-\frac52,1,0]$ for
$ \bar{\mathbb{\L}}^a(t)$. 
Finally,  we  can consider the two  broken translations $\mathsf{P}$ and $\bar{\mathsf{P}}$ in the directions orthogonal to the defect. They define the  displacement operators
\begin{align}\label{D}
	[\mathsf{P}, \mathcal{W}]&=\int dt  \mathcal{W}[\mathbb{D}(t)] & [\bar{\mathsf{P}}, \mathcal{W}]&=\int dt  \mathcal{W}[\bar{\mathbb{D}}(t)]
\end{align}
with charges $[2,3,0,0]$ and $[2,-3,0,0]$ respectively. 
The above construction does not uniquely determine the above set of operators. Since they are defined as objects inserted in the defect, and we integrate over the position along the loop, we can add  a total derivative with respect to $t$  to the integrand  without altering the result.
The set of defect operators obtained through the action of  the broken charges organizes itself as a supermultiplet. The action of the preserved supercharges $Q^a=Q^{1a}_{+}$ and  $\bar Q_a=i\,\frac12 \epsilon_{abc} Q_{-}^{bc} $ on them is essentially dictated by the commutation relations between  broken and preserved charges. For instance, to compute $[Q^a,\mathbb{O}^b]$ one has simply to consider the commutator $[Q^a,\mathsf{J}^b]$ acting on $\mathcal{W}$  and we get
\begin{align}
	[Q^a,\mathsf{J}^b]&=Q^{ab}_+=\e^{abc}\mathsf{Q}_c  &\Rightarrow& & [Q^a,\mathbb{O}^b]&=\e^{abc}\mathbb{\L}_c.
\end{align}
From the commutation relation of  $Q_a$ with the broken supercharges and translation,  we immediately find
\be
\{Q^a,\mathbb{\L}_b\}&=-2 \d^a_b \mathbb{D}\qquad \text{and}\qquad [Q^a,\mathbb{D}]=0. 
\ee
Fixing the action of $\bar Q_a$ on these defect operators requires more attention. The naïve application of the above procedure would give zero since $\bar Q_a$ commutes with all the broken charges. This is  inconsistent with  $\{Q^a,\bar Q_b\}=2 \d^a_b P$, where $P$ generates the translations along the line.  However, as stressed above, any result obtained from \eqref{O}, \eqref{La}, and  \eqref{D} is
defined up to  a derivative with respect to $t$, which yields zero when integrated.  To fix the form of  these derivatives in the commutation relations, it is more convenient to use the super-Jacobi identities.  For instance, we can determine $[\bar Q_c,\mathbb{D}]$ as follows\footnote{We choose to represent $P$ as $-\partial_t$ and then $[P,\Phi]=\partial_t\Phi$, see the discussion in \cite{Fortin:2011nq}.}
\begin{align}
	&0=\{\bar Q_c,[Q^a,\mathbb{D}]\}-\{Q^a,[\mathbb{D},\bar Q_c]\}+ [\mathbb{D},\{\bar Q_c,Q^a\}]=	Q^a,[\bar Q_c,\mathbb{D}]+\partial_t \mathbb{\Lambda}_c\non\\
&\Rightarrow\qquad[\bar Q_c,\mathbb{D}]=-\partial_t \mathbb{\Lambda}_c.
\end{align}
where we assumed that the action of the translation $P$ is realised by  $-\partial_t$ to be consistent with \eqref{generators-diff}.
Similarly we can show that  $ \{\bar{Q}_a,\mathbb{\L}_b\}=-2 \e_{abc}\pa_{t}\mathbb{O}^c$. Since we do not have an operator of lower dimension, it would be natural to set $[\bar Q_c,\mathbb{O}^b]=0$. However, this choice is inconsistent with the super-Jacobi identity
\begin{align}
	\label{SJac}
	0=&\{\bar Q_c,[Q^a, \mathbb{O}^b]\}-\{Q^a,[\mathbb{O}^b,\bar Q_c]\}+ [\mathbb{O}^b,\{\bar Q_c,Q^a\}]=-2\delta_c^{b}\partial_t\mathbb{O}^{a}
	+\{Q^a,[\bar Q_c,\mathbb{O}^b]\}.
\end{align}
The consistency of \eqref{SJac}  suggests the existence of an additional fermionic operator $\mathbb{F}$, that is a singlet under SU$(3)$,
which obeys the anticommutation relation:
\begin{align}
	\{Q^a,\mathbb{F}\}&=\mathbb{O}^a, 
\end{align}
which in turn implies $[\bar Q_c,\mathbb{O}^b]=2 \delta^b_c \partial_t \mathbb{F}$.  The Jacobi identity for this new field immediately shows that   $ \{\bar Q_a,\mathbb{F}\}$ can be consistently chosen to vanish. Furthermore, $\mathbb{F}$ has the correct quantum numbers to be the superprimary of the chiral multiplet $\bar{\mathcal{B}}^{\frac12}_{\frac32,0,0}$ (see Appendix \ref{representations}) with the structure 
\begin{align} \bar{\mathcal{B}}^{\frac12}_{\frac32,0,0}: &\nonumber\\ &
	\begin{tikzpicture}
		\draw[->]  (-5,5)--(-4.7,4.7);
		\node[above] at (-5.5,5) {$[\frac12,\frac32,0,0]$};
		\draw[->]  (-4,4)--(-3.7,3.7);
		\node[above] at (-4.5,4) {$[1,2,1,0]$};
		\draw[->]  (-3,3)--(-2.7,2.7);
		\node[above] at (-3.5,3) {$[\frac32,\frac52,0,1]$};
		\node[above] at (-2.5,2) {$[2,3,0,0]$};
	\end{tikzpicture}
	\label{displacement}
\end{align}
where this supermultiplet  is realized in terms of the fundamental fields of  ABJM theory \cite{Bianchi:2020hsz}. 
An analogous action of the supercharges can be derived for the conjugated operators leading to the barred version of the commutation relations in Table \ref{tab:susy-displacement}.

\begin{table}[h]
	\caption*{\sc Summary of supersymmetry transformations}
	\centering
	{\renewcommand{\arraystretch}{1.5}%
		\begin{tabular}{ l c c l }
			\hline
			$\{Q^a,\mathbb{F}\}=\mathbb{O}^a $ &  &  &$ \{\bar Q_a,\mathbb{F}\}=0$ \\ 
			$[Q^a,\mathbb{O}^b]=\e^{abc}\mathbb{\L}_c $ &  &  &$[\bar Q_a,\mathbb{O}^b]=2 \delta^b_a \partial_t \mathbb{F}$ \\  
			$\{Q^a,\mathbb{\L}_b\}=-2 \d^a_b \mathbb{D}$ &  &  &$ \{\bar{Q}_a,\mathbb{\L}_b\}=-2 \e_{abc}\pa_{t}\mathbb{O}^c$ \\
			$[Q^a,\mathbb{D}]=0$& & & $[\bar Q_c,\mathbb{D}]=-\partial_t \mathbb{\Lambda}_c$\\
			\hline
	\end{tabular}}
	\medskip
	\caption{\small Summary of the supersymmetry transformations of the displacement supermultiplet of a 1/2-BPS line-defect in  $\mathcal{N}=6$ supersymmetric theories in $d=3$.}
	\label{tab:susy-displacement}
\end{table}

The displacement supermultiplet \eqref{displacement} appearing in the dCFT$_1$ living along the Wilson line
should match  the string transverse excitations via AdS/CFT. 
The expansion of the Green-Schwarz string action   around the minimal surface solution results in a multiplet of fluctuations transverse to the string~\cite{Forini:2012bb, Correa:2014aga}  whose components match precisely the  quantum numbers of the operators in~\eqref{displacement},  once the two standard relations between AdS$_{2}$  masses and the corresponding CFT$_1$ operator dimensions  are taken into account, $m^2=\Delta (\Delta-1)$ for scalars and $\Delta=\frac{1}{2}+|m|$  for spinors~\cite{Henningson:1998cd,Mueck:1998iz}.  In particular,  in the scalar fluctuation sector, one finds one  massive  (with $m^2=2$)  complex scalar field $X$ in AdS$_4$, corresponding to the $\Delta=2$ displacement operator $\mathbb{D}$, and  three  massless  \emph{complex} scalar fluctuations $w^a, ~a=1,2,3$ in $\cp^3$, corresponding to the $\Delta=1$ operators $\mathbb{O}^a,  ~a=1,2,3$. In the fermionic sector, there are two massless fermions which should correspond  to the $\Delta=\frac{1}{2}$ fermionic superprimary $\mathbb{F}$ of the multiplet and its conjugate, as well as six massive fermions (of which three with mass $m_F=1$ and three with $m_F=-1$) corresponding to the $\Delta=\frac{3}{2}$ fermionic operator $\mathbb\Lambda_a$ and its conjugate.

\subsection{Chiral Correlators}\label{Sec: chiral correlators}
The representation theory analysis of the supergroup SU$(1,1|3)$ reviewed in the previous section shows that the displacement operator belongs to a chiral multiplet with R-charge $j_0=\frac32$ and dimension $\Delta=\frac12$ (the conjugate operator belongs to an antichiral multiplet with opposite R-charge). For this section, we will consider a chiral multiplet with arbitrary R-charge $j_0$ and dimension $\D=\frac{j_0}{3}$. A chiral superfield $\Phi_{j_0}$ should respect the chirality condition
\begin{equation}\label{chiralitycondition}
	\bar{\mathsf{D}}_a \Phi_{j_0}=0\,,\qquad 
\end{equation}
for every value of $a$.
By defining the chiral coordinate $y=t+\theta_a \bar \theta^a$, such that $\bar{\mathsf{D}}_a y = 0$, one simply has the component expansion

\begin{align}\label{chiralsuperfield}
	\Phi_{j_0}(y,\theta)=\phi(y)+\theta_a \psi^a(y)-\frac12 \theta_a \theta_b \,\e^{abc}\, \eta_c(y) +\frac13 \theta_a \theta_b \theta_c \,\e^{abc} \xi(y)\,,
\end{align} 
where the numerical coefficients of each  component are fixed by the consistency between the action of the supercharges on the superfield and the commutation relations in Table~\ref{tab:susy-displacement}.  \\
Similarly, the antichiral field is expanded as
\begin{align}\label{antichiralsuperfield}
	\bar\Phi_{j_0}(y,\theta)=\bar\phi(y)+\bar\theta^a \bar\psi_a(y)+\frac12 \bar\theta^a \bar\theta^b \,\e_{abc}\, \bar\eta^c(y) -\frac13 \bar\theta^a \bar\theta^b \bar\theta^c \,\e_{abc}\, \bar\xi(y)\,,
\end{align} 
where the coefficients in the expansion are determined by the Ward identities relating the two-point functions of different super descendants.\footnote{Or equivalently, one can act explicitly on the superfield with the supercharges since the commutations with the component fields were computed using the same Ward identities. Writing the superfield as \begin{equation}
		\Phi(y) =  \mathbb{F}+c_1 \theta^a \mathbb{O}_a+\frac{1}{2}c_2\theta^a \theta^b \epsilon_{abc} \Lambda^c +\frac{1}{3}c_3\theta^3 \mathbb{D},
	\end{equation} one obtains the commutation relations \begin{align}
		&\{ Q_a,\mathbb{F}\} =  c_1 \mathbb{O}_a \\
		&[Q_a, \mathbb{O}_b]  = -\frac{c_2}{c_1}\epsilon_{abc} \Lambda^c\\
		& \{ Q_a, \Lambda^c \}=2\frac{c_3}{c_2}\delta_a^c \mathbb{D}\\
		&[Q_a,\mathbb{D} ]=0,
	\end{align} which when compared to \eqref{chiralsuperfield} give the correct factors. } Using the commutation relations from above, we can relate the constants from the two-point functions
$$\langle \mathbb{F}(0) \bar{\mathbb{F}}(t) \rangle  = \frac{c_{\mathbb{F}}}{t}\quad \quad \langle \mathbb{O}_a(0) \bar{\mathbb{O}}^b(t) \rangle  = \frac{\delta_a^b c_{\mathbb{O}}}{t^2}\quad \quad \langle \Lambda(0)^a \bar{\Lambda}_b(t) \rangle  = \frac{\delta^a_b c_{\Lambda}}{t^3}\quad \quad \langle \mathbb{D}(0) \bar{\mathbb{D}}(t) \rangle  = \frac{c_{\mathbb{D}}}{t^4}\quad \quad $$
by using the vanishing of some of the two-point correlators such as $\langle  \mathbb{F} \bar{\mathbb{O}}_a\rangle $. Acting on it with Q  and using the previous commutation relations gives:
\begin{align}
	& \quad \quad Q_a \langle  \mathbb{F} \bar{\mathbb{O}}^b\rangle  =0 \\
	&\Leftrightarrow  \langle \{Q_a,\mathbb{F}\} \bar{\mathbb{O}}^b\rangle  + \langle \mathbb{F} [Q_a,\bar{\mathbb{O}}^b]\rangle =0 \\
	&\Leftrightarrow  \langle \mathbb{O}_a \bar{\mathbb{O}}^b \rangle + 2 \delta_a^b \langle \mathbb{F}\partial_t\bar{\mathbb{F}}\rangle =0\\
	& \Leftrightarrow \frac{\delta_a^b c_{\mathbb{O}}}{\tau^2} +2\delta_a^b \partial_t \frac{c_{\mathbb{F}}}{\tau} = 0 \\
	&\Leftrightarrow c_{\mathbb{O}} = +2 c_{\mathbb{F}}.
\end{align}
Doing the same process on $ \langle \mathbb{O}_a \bar{\Lambda}_b \rangle \quad \langle \Lambda^b \bar{\mathbb{D}}\rangle$ gives the equalities:
\begin{align}
	c_\Lambda = 4 c_\mathbb{O}\\
	2c_\mathbb{D} = 3c_\Lambda.
\end{align}

The two-point function of a chiral and antichiral superfield must be expressed in terms of the chiral distance ($\bar{\mathsf{D}}_i\braket{i\bar j}=\mathsf{D}_j\braket{i\bar j}=(Q_i+Q_j)\,\braket{i\bar j}=(\bar Q_i+\bar Q_j)\,\braket{i\bar j}=0$)
\be\label{distance}
\braket{i\bar j}=y_i-y_j-2{\theta_{a}}_i \bar {\theta^a}_j\,,% \qquad\qquad \braket{\bar i j}=y_i-y_j-2\bar\theta_{i a} \theta^a_j
\ee  
and can be written as
\be\label{two-point-superfield}
\braket{\Phi_{j_0}(y_1,\theta_1) \bar \Phi_{-j_0}(y_2,\bar \theta_2)}=\frac{c_{\Phi_{j_0}}}{\braket{1\bar 2}^{\frac{2j_0}{3}}}\,.
\ee
For the case of the displacement supermultiplet, the normalization factor $C_{\Phi}$ has an important physical interpretation. As shown in subsection \ref{Sec: ABJM Displacement}, most of the components in the displacement multiplet are obtained by the action of a broken symmetry generator. Such generators have a natural normalization in the bulk theory. Therefore the Ward identities \eqref{D} fix the physical normalization of the displacement operator, making its two-point function an important piece of dCFT data. There is concrete evidence that in the presence of a superconformal defect, this coefficient is related to the one-point function of the stress tensor operator \cite{Lewkowycz:2013laa,Bianchi:2018zpb,Bianchi:2019sxz}. Moreover, for the case of the Wilson line, this coefficient is particularly important as it computes the energy emitted by an accelerating heavy probe in a conformal field theory \cite{Correa:2012at,Fiol:2012sg}, often called Bremsstrahlung function. The relation with the two-point function of the displacement operator in the context of superconformal theories in three and four dimensions has allowed for the exact computation of this quantity in a variety of examples~\cite{Correa:2012at,Fiol:2012sg,Lewkowycz:2013laa,Forini:2012bb,Correa:2014aga,Bianchi:2014laa,Fiol:2015spa,Mitev:2015oty,Bianchi:2017ozk,Bianchi:2017svd,Bianchi:2018scb,Bianchi:2018zpb,Fiol:2019woe,Bianchi:2019dlw}. In the present context we have
\be\label{Bremss}
C_{\Phi}(\lambda)=2 B_{1/2}(\lambda)\,,
\ee
where $B_{1/2}(\lambda)$ is the Bremsstrahlung function associated to the 1/2-BPS Wilson line in ABJM theory (see \cite{Bianchi:2014laa,Bianchi:2017ozk}). 
From the Gra\ss mann expansion of the two-point function \eqref{two-point-superfield}, we extract the following two-point functions
\begin{align}\label{2-p-insertions}
	\langle \mathbb{F}(t_1)\bar{\mathbb{F}}(t_2) \rangle  &= \frac{C_\Phi}{t_{12}} &
	\langle \mathbb{O}^a(t_1)\bar{\mathbb{O}}_b(t_2) \rangle  &= \frac{2 \,C_\Phi\,\delta^a_b}{t_{12}^{2}} \\
	\langle \mathbb{\Lambda}_a(t_1)\bar{\mathbb\Lambda}^b(t_2) \rangle & = \frac{8 \,C_\Phi\,\delta_a^b}{t_{12}^{3}}&
	\langle \mathbb{D}(t_1)\bar{\mathbb{D}}(t_2) \rangle  &= \frac{12\, C_\Phi}{t_{12}^{4}}
\end{align}
which is consistent with the component expansion above. \par \vspace{4mm}
The four-point function of 1/2-BPS (chiral) operators will, in general, be a function of four spacetime coordinates and twelve Gra\ss manian coordinates subject to the constraints from three bosonic generators $(P,K,D)$ and twelve Gra\ss mannian generators ($Q^a,\bar{Q}_a,S^a, \bar{S}_a$) of $su(1,1|3)$. By simple counting, one expects the correlator to depend on a single super cross-ratio. For the two different four-point function configurations, we have
\begin{align}\label{four-point-1}
	\!\! \braket{\Phi_{j_0}(y_1,\theta_1) \bar \Phi_{-j_0}(y_2,\bar \theta_2)\Phi_{j_0}(y_3,\theta_3) \bar \Phi_{-j_0}(y_4,\bar \theta_4)}&=
	%\braket{\Phi_{j_0}(y_1,\theta_1) \bar \Phi_{-j_0}(y_2,\bar \theta_2)}\braket{\Phi_{j_0}(y_3,\theta_3) \bar \Phi_{-j_0}(y_4,\bar \theta_4)}
	\frac{C^2_{\Phi_{j_0}}}{\braket{1\bar 2}^{\frac{2j_0}{3}}\braket{3\bar 4}^{\frac{2j_0}{3}}}\,f (\mathcal{Z})\,, \\
	\!\! \braket{\Phi_{j_0}(y_1,\theta_1) \bar \Phi_{-j_0}(y_2,\bar \theta_2) \bar \Phi_{-j_0}(y_3,\bar \theta_3)  \Phi_{j_0}(y_4, \theta_4)}&=
	%\braket{\Phi_{j_0}(y_1,\theta_1) \bar \Phi(y_2,\bar \theta_2)}\braket{\bar \Phi(y_3,\bar \theta_3) \Phi(y_4, \theta_4)}
	- \,\frac{C^2_{\Phi_{j_0}}}{\braket{1\bar 2}^{\frac{2j_0}{3}}\braket{4\bar 3}^{\frac{2j_0}{3}}}\,h (\mathcal{X})\, , \label{four-point-2}
\end{align}
where
\begin{align} \label{supercrossratios}
	\mathcal{Z}&=\frac{\braket{1\bar 2}\braket{3 \bar 4}}{\braket{1\bar 4}\braket{3{\bar 2}}}  & \mathcal{X}&=-\frac{\braket{1\bar 2}\braket{4  \bar 3}}{\braket{1\bar 3}\braket{2 \bar 4}}
\end{align}
are the two superconformal cross-ratios corresponding to the two different correlators. These two invariants are built out of the chiral distance~\eqref{distance}. In the general case, one may have a set of additional superconformal invariants which are nilpotent due to their Gra\ss mann nature. For long-multiplet four-point functions,  using such nilpotent invariants guarantees a finite truncation in the superspace expansion  (see~\cite{Cornagliotto:2017dup}). 
However, none of these invariants is compatible with the chirality condition~\eqref{chiralitycondition}. The absence of nilpotent invariants is expected for correlators of 1/2-BPS operators and, in general, for four-point functions containing two chiral and two anti-chiral operators~\cite{Fitzpatrick:2014oza}. It is important to notice that both correlators \eqref{four-point-1} and \eqref{four-point-2} are ordered such that $t_1<t_2<t_3<t_4$. These two correlators would be related by crossing in higher dimensions, but this is not the case in one dimension. To make this more concrete, let us consider the bosonic part of the cross-ratios \eqref{supercrossratios}
\begin{align}\label{z-chi}
	z&=\frac{t_{12}t_{34}}{t_{14}t_{32}} & \chi&=\frac{t_{12}t_{34}}{t_{13}t_{24}}
\end{align}
where $z$ is the bosonic part of $\mathcal{Z}$ and $\chi$ is the bosonic part of $\mathcal{X}$. With our ordering we have $z<0$ and $0<\chi<1$. The transformation
\be\label{z-chi-rel}
z=\frac{\chi}{\chi-1}\,
\ee
relates the two cross-ratios.
The absence of nilpotent invariants implies that the superprimary correlators
\begin{align}\label{four-point-1-primary}
	\!\! \braket{\phi(t_1) \bar \phi(t_2)\phi(t_3) \bar \phi(t_4)}&=\braket{\phi(t_1) \bar \phi(t_2)}\braket{\phi(t_3) \bar \phi(t_4)}\,f (z)\,, \\
	\!\! \braket{\phi(t_1) \bar \phi(t_2)\bar \phi(t_3) \phi(t_4)}&=\braket{\phi(t_1) \bar \phi(t_2)}\braket{\bar \phi(t_3) \phi(t_4)}\,h (\chi)\, , \label{four-point-2-primary}
\end{align}
fully determine the four-point functions of the whole superconformal multiplet, i.e. the correlators of superconformal descendants can be obtained by the action of differential operators on $f(z)$. Of course, one is free to express the function $f(z)$ in terms of the cross-ratio $\chi$ by considering $f(\frac{\chi}{\chi-1})$. One can also take the analytic continuation of $f(z)$ for $0<z<1$, which would naively establish a relation between \eqref{four-point-1-primary} and \eqref{four-point-2-primary}.\footnote{$f(z)$ has branch cut singularities at coincident points, i.e. $z=0,1,\infty$.} Nevertheless, this is not the case in one dimension, and $h(\chi)$ is not the analytic continuation of $f(z)$. Still, we will see in subsection \ref{superblocks} that a relationship between these two functions exists through their $s$-channel block expansion.

Just as in the case of the two-point function, the correlators of the super descendants can be obtained by expanding the Gra\ss mann coordinates of the four-point correlator in superspace. 
%\allowdisplaybreaks
\begin{align}\label{corrF}
	&\!\!\!\!\! \braket{\mathbb{F}(t_1)\bar{\mathbb{F}}(t_2)\mathbb{F}(t_3)\bar{\mathbb{F}}(t_4)}=\frac{C_{\Phi}^2}{t_{12} t_{34}}f(z)\\\nonumber
	&\!\!\!\!\! \braket{\mathbb{O}^{a_1}(t_1)\bar{\mathbb{O}}_{a_2}(t_2)\mathbb{O}^{a_3}(t_3)\bar{\mathbb{O}}_{a_4}(t_4)}=\frac{4 C_{\Phi}^2}{t_{12}^2 t_{34}^2}\,\Big[\delta^{a_1}_{a_2} \delta^{a_3}_{a_4}\, \big(\,f(z)-z f'(z)+z^2f''(z) \big)\\\label{corrO}
	&\qquad\qquad\qquad\qquad\qquad\qquad\qquad \qquad  -\delta^{a_1}_{a_4} \delta^{a_3}_{a_2}\,\big(z^2f'(z)+z^3 f''(z)\big)\,\Big]
	\\\nonumber
	&\!\!\!\!\! \braket{\mathbb{D}(t_1)\bar{\mathbb{D}}(t_2)\mathbb{D}(t_3)\bar{\mathbb{D}}(t_4)}=\frac{(12 C_{\Phi})^2}{t_{12}^4 t_{34}^4}\, 
	\frac{1}{36}\Big[36 f(z)-36 (z^4+z) f'(z) +18 z^2 (-14 z^3+3 z^2+1) f''(z)\\\nonumber
	&\qquad\qquad\qquad\qquad  -6 z^3 \left(55 z^3-39 z^2+3 z+1\right) f^{(3)}(z)-3 z^4 \left(46 z^3-63 z^2+18 z-1\right) f^{(4)}(z)\\\label{corrD}
	& \qquad\qquad\qquad \qquad -3 (z-1)^2 z^5 (7 z-1) f^{(5)}(z) -(z-1)^3 z^6 f^{(6)}(z) 
	\,\Big]\,\\\nonumber
	&\!\!\!\!\! \braket{\mathbb{D}(t_1)\bar{\mathbb{D}}(t_2)\mathbb{O}^{a_3}(t_3)\bar{\mathbb{O}}_{a_4}(t_4)}=\frac{24 C_{\Phi}^2}{t_{12}^4t_{34}^2}\delta^{a_3}_{a_4}\frac16\Big[\,(\!1\!-z)\, z^4 f^{(4)}\!(z)-(3 z+\!1\!) \,z^3 f^{(3)}\!(z)\\\label{corrMIXED}
	&\qquad\qquad\qquad \qquad\qquad\qquad\qquad\qquad+3 z^2\, f''(z)-6 z f'(z)+6 f(z)\,\Big]\,,
\end{align}
where for each correlator, we factorised the (squared) two-point function contribution~\eqref{2-p-insertions}  arising from the double OPE. 
In the next section, we will use analytic bootstrap techniques to evaluate the function $f(z)$ perturbatively to third-order at strong coupling. 
In section~\ref{sec:sigmamodel}, the superspace analysis is used to find $f(z)$ from the correlators of superconformal descendants evaluated directly at strong coupling using Witten diagrammatics.  The consistency of these various correlators with the superspace analysis provides a non-trivial check of the relations above.

\subsection{Symmetry and Superblocks}\label{Subsection: Superblocks ABJM}

\subsubsection*{Selection rules for the 1/2-BPS operators}
Even though we only consider half-BPS multiplets as external operators, more general multiplets can be exchanged when an OPE is applied to the correlator.  
In our case,  there are  two qualitatively different OPE channels to consider, depending on whether we take the chiral-antichiral OPE $\Phi\times \bar\Phi$  or the chiral-chiral OPE $\Phi \times\Phi$. Each has selection rules for superconformal representations appearing in these two channels and corresponding superconformal blocks. We start from the chiral-antichiral channel. In~\cite{Bianchi:2018scb}, all the selection rules for the 1/6-BPS defect theory were derived,  and it was found that only the identity and long multiplets can appear in the chiral-antichiral OPE for $su(1,1|1)$, a.k.a. $\mathcal{N}=2$ supersymmetry. For the $\mathcal{N}=6$ case of interest here, every pair $\{Q^a,\bar Q_a\}$ at fixed $a$ generates a $su(1,1|1)$ subgroup of $su(1,1|3)$, and a chiral multiplet of $su(1,1|3)$ is also a chiral multiplet of all the three $su(1,1|1)$ subgroups. This implies that the operators appearing in the $\Phi \times \bar \Phi$ OPE %for a $su(1,1|3)$ chiral, 
must belong to long multiplets of all the $su(1,1|1)$ subgroups, i.e. they must not be annihilated by any supercharge. Furthermore, three-point functions with one chiral, one antichiral and one long multiplet are non-vanishing only when the superprimary R-charges sum to zero, namely a long multiplet enters only when its superprimary can be exchanged. We conclude that
\begin{align}
	\bar{\mathcal{B}}_{j_0} \times \mathcal{B}_{-j_0} \sim \mathcal{I} + \mathcal{A}^{\Delta}_{0,0,0}\,,
\end{align}
where $\mathcal{I}$ is the identity, $\mathcal{B}_{-j_0}$ is the chiral multiplet, and $\mathcal{A}^{\Delta}_{0,0,0}$ is an uncharged long multiplet.\footnote{We use the notation of Appendix~\ref{app:reprs} for the $su(1,1|3)$ supermultiplets.} 
In this case, the unitarity bound~\eqref{unitaritybounds} is $\Delta\geq0$, and every positive dimension is allowed for long operators.

For the chiral-chiral channel, the situation is richer. We will borrow an argument from \cite{Poland:2010wg} to extract the selection rules. Consider the chiral superprimary operator $\phi$. chirality gives
\begin{align}
	[\bar Q_a, \phi(t)]&=0 & [\bar S_a, \phi(t)]&=0
\end{align}
for any $a$ and any $t$. The first condition is simply the definition of chirality. In contrast, the second one comes from the requirement that $\phi$ is a superprimary at the origin together with the commutation relation $[P,\bar S_a]=- \bar Q_a$. These two conditions imply  that any operator $\mathcal{O}$ appearing in the $\phi \times \phi$ OPE must respect
\begin{align}
	[\bar Q_a, \mathcal{O}(t)]&=0 & [\bar S_a, \mathcal{O}(t)]&=0
\end{align}
for any $a$. It is then immediate to realize that the only superprimary operator which is allowed to appear is a chiral operator of dimension $\Delta_{\text{exc}}=\frac{2j_0}{3}$. All other multiplets will contribute with a single super descendant (and all its conformal descendants) generated by the repeated action of $\bar{Q}_a$. Concretely, a long multiplet will contribute with the operator generated by $\bar{Q}^3 O$ (here $\bar{Q}^3=\e^{abc}\bar Q_a \bar Q_b \bar Q_c)$, where $O$ is the superprimary. The complete analysis yields
\begin{align}
	\bar{\mathcal{B}}_{j_0} \times \bar{\mathcal{B}}_{j_0} \sim \bar{\mathcal{B}}_{2j_0} + \bar{\mathcal{B}}_{2j_0+\frac12,1} + \bar{\mathcal{B}}_{2j_0+1,0,1}+ \mathcal{A}^{\Delta}_{2j_0+\frac32,0,0} .
\end{align}
In particular, in the OPE of the superprimary operator $\phi$, every supermultiplet contributes with a single conformal family, whose conformal primary has quantum numbers $[\D,2j_0,0,0]$. The dimension is fixed in terms of $j_0$ for short multiples. In Table \ref{selrulestab}, we summarize the schematic form of the only relevant super descendant operators and make explicit their conformal dimension, which can be easily obtained from the associated superprimary. The dimension of the long multiplet is unfixed, but it should respect the unitarity bound  \eqref{unitaritybounds}. For this specific case, we find that the dimension of the superprimary must be $\D> \frac{2j_0}{3}+\frac12$, while the relevant super descendant, obtained by acting with all the $\bar{Q}$'s, must have dimension\footnote{Here the equality is excluded because in that case, the long multiplet decomposes as in Table \ref{longmultdec} and the relevant super descendant falls back into the $\bar{\mathcal{B}}_{2j_0+1,0,1}$ multiplet.} 
\begin{equation}\label{boundexchange}
	\D_{\text{exc}}^{\text{long}}> \frac{2j_0}{3}+2 \, .
\end{equation}
This bound will be very important in the following, where we will focus on the $j_0=\frac32$ case. In summary, the $\phi \times \phi$ spectrum admits three protected conformal primaries of dimensions $\frac{2j_0}{3}$, $\frac{2j_0}{3}+1$ and $\frac{2j_0}{3}+2$ and infinitely many operators with unprotected dimensions strictly higher than the protected ones. 
\begin{table}[htbp]
	\begin{center}
		\begin{tabular}{|l|c|c|}
			\hline
			Multiplet & Exc. & $\D_{\text{exc}}$  \\
			\hline
			$\bar{\mathcal{B}}_{2j_0}$ & $O$ & $\frac{2j_0}{3}$  \\
			$\bar{\mathcal{B}}_{2j_0+\frac12,1} $ & $ \bar{Q}O $ & $ \frac{2j_0}{3}+1 $ \\
			$\bar{\mathcal{B}}_{2j_0+1,0,1} $ &  $\bar{Q}^2O $ & $ \frac{2j_0}{3}+2 $ \\
			$ \mathcal{A}^{\D}_{2j_0+\frac32,0,0} $ & $ \bar{Q}^3O $ & $ \D+\frac32 $ \\
			\hline
		\end{tabular}
	\end{center}
	\caption{The multiplets contributing to the chiral-chiral OPE with a schematic representation of the only superconformal descendant (but conformal primary) contributing to the OPE (in this Table $O$ indicates the superprimary of each multiplet).}
	\label{selrulestab}
\end{table} 

\subsubsection{Superblocks}\label{superblocks}
We now derive the superconformal blocks associated with the two channels. In the chiral-chiral channel, each supermultiplet only contributes with a single conformal family. Therefore one only needs to select the $sl(2)$ conformal blocks with the appropriate dimensions \cite{Dolan:2011dv}.
\begin{align}\label{sl2block}
	g_h(\chi)=\chi^h \, _2F_1(h,h;2 h;\chi).
\end{align}
Each conformal primary listed in Table \ref{selrulestab} contributes with an $sl(2)$ block with $h=\D_{\text{exc}}$. Let us consider the specific correlator of interest here\footnote{The correlator \eqref{four-point-1-primary} does not admit an expansion in this channel since one cannot take point $\phi(t_1)$ close to $\phi(t_3)$.} 
\begin{align}\label{tildeh}
	\braket{\phi (t_1) \bar{\phi} (t_2) \bar{\phi} (t_3) \phi (t_4)}= \frac{C_{\Phi}^2}{t_{14}^{\frac{2j_0}{3}} t_{23}^{\frac{2j_0}{3}}} \hat{h}(\chi)
\end{align}
where, comparing with \eqref{four-point-2-primary} we defined $\hat{h}(\chi)=(\frac{1-\chi}{\chi})^{\frac{2j_0}{3}}h(\chi)$. The $\phi\times \phi$ channel corresponds to the $\chi\to1$ limit and we have
\begin{align}\label{tildehexpansion}
	\!\!\!\!
	\hat{h}(\chi)=\mathsf{c}_{\frac{2j_0}{3}} g_{\frac{2j_0}{3}}(1-\chi)+ \mathsf{c}_{\frac{2j_0}{3}+1} g_{\frac{2j_0}{3}+1}(1-\chi)+\mathsf{c}_{\frac{2j_0}{3}+2} g_{\frac{2j_0}{3}+2}(1-\chi)+\sum_{\Delta} \mathsf{c}_{\Delta} g_{\Delta}(1-\chi)
\end{align}
where the sum runs over unprotected operators with dimension $\Delta > \frac{2j_0}{3}+2$, and $\mathsf{c}_\Delta$'s are the squared moduli of the OPE coefficients.\footnote{We use a different font to distinguish the coefficients from the chiral-chiral and the chiral-antichiral channel.}\par \vspace{4mm}
In the chiral-antichiral channel, long multiplets with quantum numbers $[\Delta,0,0,0]$ contribute with all the allowed super descendants, and we can decompose the four-point correlation function in terms of superconformal blocks. The latter are always expressed as a linear combination of ordinary $sl(2)$ blocks with shifted dimensions. Conformal blocks can be seen as eigenfunctions of the Casimir differential operator~\cite{Dolan:2011dv}. Analogously, superconformal blocks can be computed by considering the differential equation generated by the action of the superconformal Casimir~\cite{Fitzpatrick:2014oza}. We start by the $s$-channel OPE expansion of the superspace four-point function \eqref{four-point-1}, insert a resolution of the identity between points $t_2$ and $t_3$, and act on each term of the sum with the quadratic Casimir ${\bf C}_1+{\bf C}_2$, where ${\bf C}_i$ is the differential operator~\eqref{Casimir} acting on the supercoordinates $(t_i,\theta_{ai},\bar \theta^a_i)$. This leads to the block expansion
\be\label{superblock}
f(z)=1+\sum_\Delta\,c_\Delta\,G_\Delta(z)
\ee
where the sum runs over the dimensions  $\Delta>0$ of the superprimary operators exchanged in the chiral-antichiral channel, and each conformal block satisfies the differential equation
\be\label{eigenfunction}
\big(-z^2(z-1)\partial_z^2 -z(z-3)\partial_z \big)\,G_\Delta (z)  = \Delta(\Delta+2)\,G_\Delta (z) \,,
\ee   
where $\Delta(\Delta+2)$ is the Casimir eigenvalue~\eqref{Casimireigenv} with  zero R-charges. The equation above is solved by the hypergeometric function~\footnote{The appearance of a minus sign in~\eqref{superblock} is due to our use of $z$, which takes real negative values, rather than the more standard $\chi$, see~\eqref{z-chi}.}
\be\label{solsuperblock}
G_\Delta(z)=  (-z)^\Delta {}_2F_1(\Delta,\Delta,2\Delta+3; z)\,.
\ee
$G_\Delta(z)$ can be decomposed in terms of a finite sum of $sl(2)$ blocks \eqref{sl2block}.\footnote{Here we use $\hat{g}(z)=g(\frac{z}{z-1})=(-z)^\Delta \, _2F_1(\Delta,\Delta;2 \Delta;z)$, which implements the change of variable from $\chi$ to $z$.}\par 
\begin{align}
	G_\Delta(z)&=\hat{g}_\Delta(z)+ \frac{3 \Delta}{2 (2 \Delta+3)}   \hat{g}_{\Delta+1}(z)+ \frac{3 \Delta^2 (\Delta+1)}{4 (2\Delta+3)(2\Delta+1)(\Delta+2)}\hat{g}_{\Delta+2}(z) \nonumber  \\ 
	&+\frac{\Delta^2 (\Delta+1)}{8 (2 \Delta+3)^2 (2 \Delta+5)}\hat{g}_{\Delta+3}(z) \label{expansion-conformal-blocks}
\end{align}
To find a connection between the two correlators \eqref{four-point-1-primary} and \eqref{four-point-2-primary}, it is important to consider the $s$-channel expansion of \eqref{four-point-2-primary}
\begin{align}\label{hexpansion}
	h(\chi)&=1+\sum_\Delta\,\tilde{c}_\Delta\,\tilde{G}_\Delta(\chi)
\end{align}
where 
\begin{align}
	\tilde{G}_\Delta(\chi)=\chi^\Delta {}_2F_1(\Delta,\Delta,2\Delta+3; \chi)\,.
\end{align}
The similarity between this block expansion and \eqref{superblock} is apparent, especially when considering the relation between $\tilde c_\Delta$ and $c_\Delta$. Denoting by $\mathcal{O}_\Delta$ the exchanged operator of dimension $\Delta$, the expressions of $c_\Delta$ and $\tilde c_\Delta$ are given by
\begin{align}
	c_\Delta&= f_{\phi \bar \phi \mathcal{O}_\Delta} f_{\phi \bar \phi \mathcal{O}_\Delta} & \tilde c_\Delta=f_{\phi \bar \phi \mathcal{O}_\Delta}f_{\bar \phi  \phi \mathcal{O}_\Delta}
\end{align}
where $f_{\phi \bar \phi \mathcal{O}_\Delta}$ is the three-point coefficient determining $\braket{\phi \bar \phi \mathcal{O}_\Delta}$. In one-dimensional CFT, the OPE coefficients depend on the signature of the permutation. In particular, despite there being no continuous group of rotation, there is a $\mathbb{Z}_2$ parity transformation $t\to -t$.\footnote{This symmetry was called S-parity in \cite{Billo:2013jda}.} operators are charged under this symmetry, and one has
\begin{align}
	\braket{\mathcal{O}_1(t_1)\mathcal{O}_2(t_2)\mathcal{O}_3(t_3)}=(-1)^{T_1+T_2+T_3} \braket{\mathcal{O}_3(-t_3)\mathcal{O}_2(-t_2)\mathcal{O}_1(-t_1)}
\end{align}
where $T_1$, $T_2$ and $T_3$ are the charges of the operators under parity. For our example, if $\phi$ is a bosonic operator, we have
\begin{align}
	f_{\phi \bar \phi \mathcal{O}_\Delta}= (-1)^{T_{\mathcal{O}}} f_{\bar \phi  \phi \mathcal{O}_\Delta},
\end{align}
whereas for a fermionic $\phi$
\begin{align}
	f_{\phi \bar \phi \mathcal{O}_\Delta}=(-1)^{1+T_{\mathcal{O}}} f_{\bar \phi  \phi \mathcal{O}_\Delta}.
\end{align}
Therefore the two coefficients $c_\Delta$ and $\tilde c_\Delta$, for a fermionic $\phi$ are related by
\begin{align}\label{ctildec}
	c_\Delta=(-1)^{T_{\mathcal{O}}+1} \tilde c_\Delta.
\end{align}
As an example, which will be useful in the following, operators of the schematic form $\mathcal{O}_n=\phi\pa_t^n\bar \phi$ have charge $T_{\mathcal{O}}=n$. In section \ref{Subsec: ABJM bootstrap}, these relations will allow us to establish a precise connection between the correlators \eqref{four-point-1-primary} and \eqref{four-point-2-primary} in a perturbative expansion around the free theory result.

\section{Diagrammatic Results}\label{sec:sigmamodel}
In this section, we use the superconformal symmetry to relate the four-point correlation function of the massless bosonic excitations $\{w^a \wb_a\}$ to that of the fundamental operator $\{\mathbb{F}, \bar{\mathbb{F}}\}$.
The Type IIA background $\text{AdS}_4 \times\cp^3$ is defined by
\begin{gather}
	\label{total_metric}
	ds^2=R^2\left ( ds^2_{\textrm{AdS}_4}+ 4 ds^2_{\cp^3}\right) \,,
	\qquad
	e^\phi= \frac{2 R}{ k} \,,
	\qquad
	R^2 \equiv {\widetilde{R}^3\over 4 k}= \frac{k^2}{4} e^{2\phi} \,,
	\\
	F_2 = 2\,k\,J_{\cp^3} \,,
	\qquad
	F_4=\frac{3}{8} R^3\, \vol(\text{AdS}_4) \,,
\end{gather}
where $R$ is the $\text{AdS}_4$ radius, $\phi$ the dilaton, $k$  results from the compactification of the original M-theory on an $\text{AdS}_4\times \sphere^7/Z_k$ background.\footnote{$k$ coincides in the dual theory with the  Chern-Simons  level number~\cite{Aharony:2008ug}.} Above, $F_2$, $F_4$ are the two-form and four-form field strengths with  $J_{\cp^3}$ the K\"ahler form on $\cp^3$. As they only play a role in the fermionic part of the Lagrangian we only report them here for completeness.\footnote{At first order in perturbation, it is sufficient to consider the bosonic part of the action to compute bosonic correlators.}   
Using the Poincar\'e patch for the AdS$_4$ metric
\be 
ds^2_{\text{AdS}_4} = \frac{dz^2 + dx^r dx^r }{z^2} \,,
\ee
where $x^r=(x^0,x^1,x^2)$ parametrize the Euclidean three-dimensional  boundary of $\text{AdS}_4$ and $z$ is the radial coordinate,
the bosonic part of the superstring  action in AdS$_4 \times \,\mathbb{CP}^3$   reads 
\begin{equation}\label{2.1} 
	S_B =\frac{1}{2} T   \int d^2\s \sqrt{h}\,   h^{\m\n} \Big[  \frac{1}{z^2}  \left(\partial_\m x^r\partial_\n x^r+\partial_\m z\partial_\n z\right)
	+   4\,G_{MN}^{\cp^3}\,{\partial_\m Y^M\partial_\n Y^N } \Big]  \ .
\end{equation}
Here, $\sigma^\mu= (t,s) $ are  Euclidean world-sheet  coordinates, and $T$ is the effective string tension. In its original ``dictionary''  proposal~\cite{Aharony:2008ug} it is related to the effective 't Hooft 
coupling $\lambda$ of the dual $\mathcal{N}=6$ superconformal Chern-Simons theory (realized in the limit of $k$ and $N$ large with  their ratio fixed) by equation \ref{Eq: Dictionary ABJM} which we remind here:
\be\label{T}
T= 
\frac{R^2}{2\pi\alpha'}
=\sqrt{\frac{\lambda}{2}}\,,
\qquad\qquad
\lambda=\frac{N}{k}\,.
\ee
As we are interested in the leading and tree-level orders in perturbation theory, we may disregard the corrections to the effective string tension $T$  due to the geometry of the background~\cite{Aharony:2008ug,Bergman:2009zh}, which start at order $\frac{1}{\sqrt{\lambda}}$~\cite{Bianchi:2014ada}.
The classical solution to \eqref{2.1}, which is relevant here, is the minimal surface  corresponding to the straight Wilson line at the boundary  
\begin{align} \label{solution}   
z&=s&x^0&=t&x^{i=1,2}&=0&Y^{M}&=0.
\end{align}
This is the embedding of AdS$_2$ in the AdS$_4$ background of the %AdS$_5$ 
solution of~\cite{Drukker:2000ep,Giombi:2017cqn}. 
The induced AdS$_2$ metric is
\be
g_{\mu\nu} d\sigma^\mu d\sigma^\nu=   {1\ov s^2} ( dt^2 + ds^2)\,.
\ee
We will consider correlators of small fluctuations  of  ``transverse"  string coordinates near this minimal surface.\footnote{The transverse coordinates are $x^i, \, i=1,2$ and the $\cp^3$ coordinates.} The bosonic symmetry of the defect conformal field theory associated with the 1/2-BPS Wilson line is SU$(1,1) \times $ SU$(3) \times $U$(1)_{J_0}$, and it turns out to be the manifest symmetry of the bosonic string action \eqref{2.1}. 
The SU$(1,1) \simeq$ SO$(2,1)$  symmetry 
%(corresponding to dilatations, translation and special conformal transformation along the line, namely   the conformal   symmetry at the corresponding  1d boundary theory) 
can be made manifest by fixing  a static gauge where $z$ and $x^0$  do not fluctuate. AdS$_2$ is embedded~\cite{Giombi:2017cqn} into AdS$_4$  using the following parametrization
\be \label{2.3} 
ds^2_4 =\frac{ (1+\frac{1}{2} |X|^2)^2}{(1-\frac{1}{2} |X|^2)^2} ds^2_{2}  + \frac{dX d\bar X}{(1-\frac{1}{2} |X|^2)^2} \ , \ \ \ \ \ \  \ \ \ \ \ \
ds^2_2 =\frac{1}{z^2} (dx_0^2+dz^2)      \ . 
\ee
The complex combination  $X=\frac{1}{\sqrt{2}}(x^1+i x^2)$  was introduced in terms of the transverse AdS coordinates $x^i$~. $X$ and $\bar X$ have opposite charge under  $U(1)_{J_0}$.   
Finally, adopting the following parametrization of the $\cp^3$ metric
\be
ds^2_{\cp^3}=\frac{d\bar{w}_a\,dw^a}{1+|w|^2}-\frac{d\bar{w}_a\,w^a\,d{w}^b\,\bar w_b}{(1+|w|^2)^2} \,, \qquad |w|^2=\bar{w}_a \,w^a\,,\qquad a,b=1,2,3\,,
\ee 
the preserved SU$(3)$ subgroup of the SU$(4)$ global symmetry of $\cp^3$ is manifest.  

The  Nambu-Goto action with fixed static gauge reads then
reads
%\be \label{2.4} 
%S_B =  T \int d^2\s   \sqrt{\det \Big[   \frac{ (1+\fo  x^2)^2}{(1-\fo x^2)^2} \,  g_{\m\n} (\s)  + \frac{\del_\m x^i \del_\n x^i  }{(1-\fo x^2)^2}
	% +    g_{MN}\,\partial_\m z^M\partial_\n z^N  \Big] } \equiv T \int d^2\s   \sqrt{ g}\  L_B   \ , 
%\ee
%
\begin{equation}\label{stringaction}
	S_B = T \int d^2\sigma \sqrt{\det\Big[ \frac{(1+1/2|X|^2)^2}{(1-1/2|X|^2)^2} g_{\mu \nu} + 2 \frac{\del_\mu X \del_\nu \bar{X}}{(1-1/2|X|^2)^2} +4 \Big(\,\frac{\del_\mu \bar{w}_a \del_\nu w^a}{1+\abs{w}^2}-\frac{\del_\mu \bar{w}_aw^a \bar{w}_b \del_\nu w^b}{(1+\abs{w}^2)^2}\Big)\,\Big]} .
\end{equation}
\vspace{2mm}
Above, $ g_{\m\n}= {1\ov s^2} \delta_{\m\n}$ is the  background AdS$_2$ metric. Along the lines of~\cite{Giombi:2017cqn}, \eqref{stringaction} can be interpreted as the action of a  straight fundamental   string in AdS$_4 \times \cp^3$ stretched from the boundary towards the  AdS$_4$ center (so, stretched along  $z$), as well as the action for a $2d$  ``bulk'' field  theory of  1+3 complex   scalars in AdS$_2$ geometry with SO$(2,1) \times [$U$(1) \times  $SU$(3)]$ as manifest  symmetry. From the AdS/CFT point of view, this second interpretation leads to a CFT$_1$  dual  living at  the $z=s=0$ boundary, namely the  defect CFT defined by operator insertions on the straight Wilson line. \par \vspace{4mm}
Expanding the action above in powers of  $X$ and $w^a$ one gets
\begin{align}   
	S_B \equiv& T \int d^2\s   \sqrt{g}\,  L_B \,,\qquad L_B =\,  L_2  + L_{4X}   + L_{2X,2w} + L_{4w}   + ...   \, \label{lagr} \\
	L_2=&\tet   g^{\m\n}\pa_\m X \del_\n \bar{X}  +  2 |X|^2 + g^{\m\n}\del_\m w^a \del_\n \wb_a\ ,  \la{2.6} 
	\\[2pt]
	L_{4X} =&\ \tet  2|X|^4+|X|^2  \,(g^{\m\n}\del_\m X \del_\n \bar{X}) -\frac{1}{2}\,(g^{\m\n}\del_\m X \del_\n X) \,(g^{\rho \k}\del_\rho \Xb \del_\k \Xb) \,
	\label{2.7}            
	\\[2pt]\nonumber
	L_{2X,2w}=&\ \tet   (g^{\m\n}\del_\m X \del_\n \Xb  )\,(g^{\rho\kappa} \del_\rho w^a \del_\kappa  \wb_a) 
	-  (g^{\m\n} \del_\m X \del_\n w^a) \; (g^{\rho\kappa} \del_\rho \Xb \del_\kappa  \wb_a)\\
	&   -  (g^{\m\n} \del_\m \Xb \del_\n w^a) \; (g^{\rho\kappa} \del_\rho X \del_\kappa  \wb_a)\,  
	\\[2pt]\nonumber
	L_{4w} =&\ - \tet \frac{1}{2} \tet (w^a \wb_a) (g^{\m\n} \del_\m w^b \del_\n \wb_b) -  \frac{1}{2} \tet (w^a \wb_b) (g^{\m\n} \del_\m w^b \del_\n \wb_a) 
	+   \frac{1}{2}\,(g^{\m\n}\del_\m w^a \del_\nu \wb_a)^2\\
	&\tet-\frac{1}{2}  (g^{\m\n} \del_\m w^a \del_\n \wb_b) \; (g^{\rho\kappa} \del_\rho \wb_a \del_\kappa  w^b)
	-\frac{1}{2}  (g^{\m\n} \del_\m w^a \del_\n w^b) \; (g^{\rho\kappa} \del_\rho \wb_a \del_\kappa  \wb_b)\ . \label{lagrend}
\end{align}
There is one  massive ($X$ with $m^2=2$) and three massless  ($w^a, ~a=1,2,3$) complex scalar fields propagating in AdS$_2$ that correspond to the bosonic elementary CFT$_1$ insertions represented in the displacement supermultiplet.\footnote{Respectively, these are the dual fields to the $\Delta=2$ displacement operator $\mathbb{D}$ and to the $\Delta=1$ operators $\mathbb{O}^a,  ~a=1,2,3$.}  As written above in subsection~\ref{Sec: ABJM Displacement}, one must also consider the fermionic fluctuations to obtain the AdS/CFT dual of the full displacement supermultiplet. The fermionic spectrum has been worked out at quadratic level in~\cite{Forini:2012bb, Correa:2014aga}. It consists of two massless  and six massive fermions (of which three with mass $m_F=1$ and three with $m_F=-1$), which should correspond, respectively,   to the $\Delta=\frac{1}{2}$ fermionic superprimary $\mathbb{F}$ of the multiplet and its conjugate and to the $\Delta=\frac{3}{2}$ fermionic operators $\mathbb\Lambda_a$ and their conjugates.  Expanding the full Type IIA Green-Schwarz action in $\text{AdS}_4 \times\cp^3$ background~\cite{Gomis:2008jt,Uvarov:2009hf} around the solution~\eqref{solution} up to quartic order in fermions would yield  the interaction vertices from which to evaluate directly, via Witten diagrams, the four-point functions of fermionic fluctuations. Below we will limit our analysis to the direct calculation of bosonic four-point functions from the vertices in \eqref{lagr} above and compare with the bootstrap results through the superspace analysis of section \ref{Sec: chiral correlators}.\footnote{The computation for the massless fluctutations is below and the mixed and massive correlators can be found in Appendix \ref{App: AdS correlators}.} However, since the the function $f(z)$ is the four-point correlator \eqref{corrF}  of the fermionic superprimary, \emph{all} four-point functions can be found  from the Gra\ss mann-expansion of the correlator.\par \vspace{4mm}
%We emphasize, however, that in so doing, we will evaluate directly  the function $f(z)$, which governs the four-point correlator \eqref{corrF}  of the fermionic superprimary  $\mathbb{F}$ - and thus \emph{all} four-point functions - as the unique solution of the differential equations in \eqref{corrF}-\eqref{corrMIXED}, arising from the Gra\ss mann-expansion of the correlator for the four chiral fields in superspace. 
Below, we will use these vertices of the  AdS$_2$ bulk theory to compute the corresponding tree-level Witten diagrams in AdS$_2$, with bulk-to-boundary propagators ending at points ${t_n}$ on the boundary.
As in the AdS$_5\times \text{S}^5$ case, no cubic terms appear in the bosonic Lagrangian above, so at this level of perturbation theory, the correlation functions are only a sum of four-point ``contact"  diagrams with four bulk-to-boundary propagators.

\subsection*{Four-point Function of Massless Fluctuations in $\cp^3$}

\begin{figure}
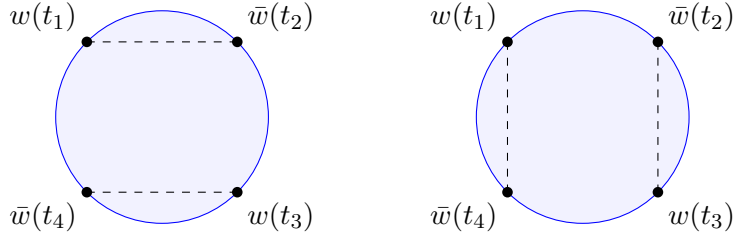

	\centering
	\begin{wittendiagram}
		\draw[dashed] (-1,1) node[vertex] -- (1,1) node[vertex];
		\draw[dashed] (-1,-1) node[vertex] -- (1,-1) node[vertex];
		\node[anchor=south east] at (-1,1) {$w(t_1)$};
		\node[anchor=south west] at (1,1) {$\bar{w}(t_2)$};
		\node[anchor=north west] at (1,-1) {$w(t_3)$};
		\node[anchor=north east] at (-1,-1) {$\bar{w}(t_4)$};
	\end{wittendiagram} \quad \quad \quad \begin{wittendiagram}
		\draw[dashed] (-1,1) node[vertex] -- (-1,-1) node[vertex];
		\draw[dashed] (1,1) node[vertex] -- (1,-1) node[vertex];
		\node[anchor=south east] at (-1,1) {$w(t_1)$};
		\node[anchor=south west] at (1,1) {$\bar{w}(t_2)$};
		\node[anchor=north west] at (1,-1) {$w(t_3)$};
		\node[anchor=north east] at (-1,-1) {$\bar{w}(t_4)$};
	\end{wittendiagram}
	\caption{Witten diagram for the disconnected contribution to the four-point function~\eqref{4-p-w}.}
	\label{fig:Witten-disconn}
\end{figure}

Here we compute the tree-level four-point Witten diagram of the $\cp^3$ fluctuations $w,\wb$  appearing in the AdS$_2$ action in \eqref{lagr}-\eqref{lagrend}. 
As discussed above, these are AdS/CFT dual to the scalar operator insertions $\mathbb{O}^a, \bar{\mathbb{O}}_a, ~a=1,2,3$ with protected dimension $\Delta=1$. Due to the $SO(2, 1)$ conformal invariance, the four-point function is expected to take the form
\begin{equation}\label{4-p-w}
	\langle w^{a_1}(t_1)\,\wb_{a_2}(t_2)\,w^{a_3}(t_3)\,\wb_{a_4}(t_4)\rangle  =  \frac{\big[C_{w}(\lambda)\big]^2}{t_{12}^2 t_{34}^2}
	G^{a_1\,a_3}_{a_2\,a_4}(\chi) .
\end{equation}
Here, $\chi$ is the conformally invariant cross-ratio defined in~\eqref{z-chi}, and  we used for the two-point function
\begin{equation}\label{two-point-w}
	\langle w^{a_1}(t_1)\wb_{a_2}(t_2)\rangle = \delta^{a_1}_{a_2}\frac{C_{w}(\lambda)}{t_{12}^2}\,.
\end{equation}
The function $G^{a_1\,a_3}_{a_2\,a_4}(\chi)$ in~\eqref{4-p-w} does not depend on the normalization of the $w^a$ fields, and thus on~$C_{w}(\lambda)$.  One can of course choose $C_w(\lambda)\equiv 4 B_{1/2}(\lambda)$, so as to realise a direct identification of $w^a$ with the $\mathbb{O}^a$ in view of~\eqref{2-p-insertions}.\footnote{This is the formal choice of~\cite{Giombi:2017cqn}, where the analogue relation is to the $\mathcal{N}=4$ SYM Bremsstrahlung function~\cite{Correa:2012at,Forini:2010ek,Drukker:2011za}. see also a related discussion in~\cite{Drukker:2020swu}.} This would correspond to an overall rescaling of the fields.\footnote{At tree level, this amounts to the rescaling $w^a\rightarrow \sqrt{2T} w^a$. This can be found from the leading strong coupling value of the Bremsstrahlung function $B_{1/2}(\lambda)=\frac{\sqrt{2\lambda}}{4\pi}\equiv \frac{T}{2\pi}$, and the canonical choice~\eqref{Eq: bulk-to-boundary prop} of the propagator.}
By perturbatively evaluating the two-point function~\eqref{two-point-w} (calculating loop corrections to the boundary-to-boundary propagator) one should then be able to verify that the elementary excitations $w^a$ are protected, as well as reproduce the strong coupling expansion of the corresponding 1/2-BPS Bremsstrahlung function~\eqref{Bremss}.The disconnected part of the four-point function~\eqref{4-p-w} originates from Wick contractions illustrated in Figure~\ref{fig:Witten-disconn}, and is 
\begin{align}\nonumber
	\cor{w^{a_1}(t_1)\,\wb_{a_2}(t_2)\,w^{a_3}(t_3)\,\wb_{a_4}(t_4)}_\text{disconn.} &= \big[C_w(\lambda)\big]^2\,\Big[\frac{\delta^{a_1}_{a_2}\delta^{a_3}_{a_4}}{t_{12}^2 t_{34}^2}+\frac{\delta^{a_1}_{a_4}\delta^{a_3}_{a_2}}{t_{14}^2 t_{23}^2}\Big]  
	\\\label{w-disconn}
	&  =\frac{\big[C_{w}(\lambda)\big]^2}{t_{12}^2 t_{34}^2}\,\Big[\delta^{a_1}_{a_2}\delta^{a_3}_{a_4}+\frac{\chi^2}{(1-\chi)^2} \,\delta^{a_1}_{a_4}\delta^{a_3}_{a_2}\,\Big]\,.
\end{align}

\begin{figure}
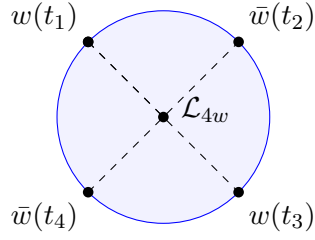

	\centering
	\begin{wittendiagram}
		\draw[dashed] (-1,1) node[vertex] -- (1,-1) node[vertex];
		\draw[dashed] (-1,1) node[vertex] -- (0,0) node[vertex];
		\draw[dashed] (-1,-1) node[vertex] -- (1,1) node[vertex];
		\node[anchor=south east] at (-1,1) {$w(t_1)$};
		\node[anchor=south west] at (1,1) {$\bar{w}(t_2)$};
		\node[anchor=north west] at (1,-1) {$w(t_3)$};
		\node[anchor=north east] at (-1,-1) {$\bar{w}(t_4)$};
		\node[anchor=west] at (0.1,0.1) {$\mathcal{L}_{4w}$};
	\end{wittendiagram}
	\caption{Witten diagram for the connected contribution to the four-point function~\eqref{4-p-w}.}
	\label{fig:Witten-conn}
\end{figure}

The first connected contribution to the four-point function comes from the   tree-level connected Witten diagrams illustrated in Figure \ref{fig:Witten-conn}. These are obtained from the four-point interaction vertices  $L_{4w}$ in~\eqref{lagrend} with four bulk-to-boundary propagators attached and therefore are subleading in $1/T$.  

Then the connected correlator reads 
\begin{equation}
	\langle w^{a_1}(t_1)\,\wb_{a_2}(t_2)\,w^{a_3}(t_3)\,\wb_{a_4}(t_4)\rangle_\text{conn} = \frac{1}{T}\,\big(\mathcal{C}_{\Delta=1}\big)^4 \, \Big[\mathcal{Q}_1\, \delta^{a_1}_{a_2} \delta^{a_3}_{a_4}+\mathcal{Q}_2\,\delta^{a_1}_{a_4} \delta^{a_3}_{a_2}\Big]
\end{equation}
where ${\cal C}_{\Delta=1}={1\ov \pi}$ and $\mathcal{Q}^{a_1\,a_2}_{a_3\,a_4}$ is built out of the  $D$-functions\footnote{Appendix~\ref{app:Dfunctions} contains a list of such $D$-functions.}~\cite{Liu:1998ty,DHoker:1999kzh, Dolan:2003hv}:
\begin{eqnarray}\nonumber
	\mathcal{Q}_1&=& \,3D_{1111}+t_{12}^2 D_{2211}+t_{34}^2 D_{1122}-2t_{24}^2 D_{1212}-2t_{13}^2 D_{2121}\\
	&&  -3 t_{23}^2 \,D_{1221} - 3 t_{14}^2 D_{2 1 1 2}+4 (t_{14}^2 t_{23}^2 + t_{13}^2 t_{24}^2 - t_{12}^2 t_{34}^2) D_{2222}\,\\\nonumber
	\mathcal{Q}_2&=&\,3D_{1111}+t_{14}^2 D_{21 1 2} + t_{23}^2 D_{1 2 2 1} -  2 t_{13}^2 D_{2 1 2 1} -2 t_{24}^2 D_{12 1 2}\\
	&&   -  3 t_{12}^2 D_{2 2 1 1} - 3 t_{34}^2  D_{1 1 2 2} +
	4 (  t_{12}^2 t_{34}^2+t_{13}^2 t_{24}^2-t_{14}^2 t_{23}^2 ) D_{2 2 2 2}\,.
\end{eqnarray}
This leads to the expression
\be\label{w-conn}
\!\!\!\!\!\!    \cor{w^{a_1}(t_1)\,\wb_{a_2}(t_2)\,w^{a_3}(t_3)\,\wb_{a_4}(t_4)}_\text{conn.} &=&\frac{1}{4\pi T}\,\frac{\big[\mathcal{C}_{\Delta=1}\big]^2}{ t_{12}^2 t_{34}^2}\,\Big[\delta^{a_1}_{a_2} \delta^{a_3}_{a_4}\,G_1(\chi)
+\delta^{a_1}_{a_4} \delta^{a_3}_{a_2}\,G_2(\chi)\Big]\,
\ee
where, according to~\eqref{4-p-w},  we factored out $[\mathcal{C}_{\Delta=1}\big]^2$, and
\begin{eqnarray} \label{G1}
	G_1(\chi) &=&-3+\frac{1}{ (\chi-1 )}-\frac{\chi ^2 }{(1-\chi )^2}\log \chi +\left(1-\frac{4}{\chi   }\right) \log (1-\chi )\\\label{G2}
	G_2(\chi) &=&-\frac{\chi  (3 \chi +1)}{ (1-\chi )^2}+\frac{\chi ^2 (\chi +3)}{(\chi -1)^3}\,\log\chi-\log (1-\chi ).
\end{eqnarray}
 Identifying these to the differential equation in~\eqref{corrO} leads to the following system of second-order differential equations\footnote{
 	The functions above can be expressed in terms of the invariant $z$ using \eqref{z-chi-rel}.}
\begin{align} \label{system-f}
	f(z)-zf'(z)+z^2f''(z) &= 1+ \,\frac{1}{4\pi T}\big[-4+z-z^2\,\log (-z)+\big(z^2-\frac{4}{z}+3\big)\,\log(1-z)\,\big]\\
	-z^2f'(z)-z^3 f''(z) &=  z^2\,+\frac{1}{4\pi T}\,\big[\,z-4\,z^2-  z^2 (3-4z)\,\log (-z)+(1 + 3 z^2 - 4 z^3) \,\log (1-z)\,\big]\,
\end{align}
whose unique solution agrees with the bootstrap solution~\eqref{Eq ABJM f1sol}
\be\label{fstring}
f(z)=1-z+\frac{1}{4\pi T}\,\big[z-1+z (3-z) \log (-z)-\frac{(1-z)^3 }{z}\log (1-z)\big]\,.
\ee
One can now  repeat the analysis for the correlator $\cor{w^{a_1}(t_1)\,\wb_{a_2}(t_2)\,\wb^{a_3}(t_3)\,w_{a_4}(t_4)}$. The result coincides with the one obtained using on  the $f(z)$ above with the replacement $z\to \chi$, and neglecting the imaginary part of the logarithm.  This perfectly agrees with what was observed in the bootstrap analysis of section~\ref{Subsec: ABJM bootstrap}. 
In Appendix \ref{App: AdS correlators}, the $f(z)$ evaluated above also solves the corresponding differential equations  for the correlators of massive and mixed worldsheet excitations once the normalization factors defining the corresponding two-point functions are identified with the ones of their field theory dual. 

\section{Setting up the Bootstrap}
In the following, the four-point function of chiral 1/2-BPS operators is bootstrapped the quantity of interest is
\begin{align}\label{four-point-1}
	\!\! \braket{\Phi_{j_0}(y_1,\theta_1) \bar \Phi_{-j_0}(y_2,\bar \theta_2)\Phi_{j_0}(y_3,\theta_3) \bar \Phi_{-j_0}(y_4,\bar \theta_4)}&=
	%\braket{\Phi_{j_0}(y_1,\theta_1) \bar \Phi_{-j_0}(y_2,\bar \theta_2)}\braket{\Phi_{j_0}(y_3,\theta_3) \bar \Phi_{-j_0}(y_4,\bar \theta_4)}
	\frac{C^2_{\Phi_{j_0}}}{\braket{1\bar 2}^{\frac{2j_0}{3}}\braket{3\bar 4}^{\frac{2j_0}{3}}}\,f (\mathcal{Z})\,, \\
	\!\! \braket{\Phi_{j_0}(y_1,\theta_1) \bar \Phi_{-j_0}(y_2,\bar \theta_2) \bar \Phi_{-j_0}(y_3,\bar \theta_3)  \Phi_{j_0}(y_4, \theta_4)}&=
	%\braket{\Phi_{j_0}(y_1,\theta_1) \bar \Phi(y_2,\bar \theta_2)}\braket{\bar \Phi(y_3,\bar \theta_3) \Phi(y_4, \theta_4)}
	- \,\frac{C^2_{\Phi_{j_0}}}{\braket{1\bar 2}^{\frac{2j_0}{3}}\braket{4\bar 3}^{\frac{2j_0}{3}}}\,h (\mathcal{X})\, , \label{four-point-2}
\end{align}
where the supercross-ratios reduce to $\chi$ and $z=\frac{\chi}{\chi-1}$ when the fermionic coordinates are set to 0. From this point on, all the analysis can be done in terms of bosonic quantities of a single cross-ratio $\chi$ where the constraints come from the full superspace analysis above. As such, we consider the bosonic part of this super correlator obtained by setting 
\be 
\theta^a = \bar{\theta}_a=0
\ee
which corresponds to the four-point function of the lowest weight field in the supermultiplet
\begin{align}
	\langle \mathbb{F}(t_1)\bar{\mathbb{F}}(t_2)\mathbb{F}(t_3)\bar{\mathbb{F}}(t_4)\rangle &= \frac{C_{\mathbb{F}}^2}{t_{12}t_{34}}f(\chi)\\
	\langle \mathbb{F}(t_1)\bar{\mathbb{F}}(t_2)\bar{\mathbb{F}}(t_3)\mathbb{F}(t_4)\rangle &= \frac{C_{\mathbb{F}}^2}{t_{12}t_{34}}h(\chi).
\end{align}
These functions have an OPE corresponding to the bosonic part of those described in subsection \ref{superblocks}, which are
\begin{align}
	f(\chi) &= 1+\sum_h c_h \chi^h {}_2F_1(h,h+3,2h+3;\chi)\\
	& = \left(\frac{\chi}{1-\chi}\right)\left(1+ \sum_h c_h (1-\chi)^h {}_2F_1(h,h+3,2h+3;1-\chi)\right),
\end{align}
and 
\begin{align}
	h(\chi) &= \sum_h \tilde{c}_h \chi^\Delta {}_2F_1(h,h,2h+3;\chi)\\
	&=\frac{\chi}{1-\chi}\left(\sum_{\Delta_S=\{1,2,3\}}\bold{a}_{\Delta}g_{\Delta}(1-\chi)+\sum_{\Delta_L>3}\bold{a}_\Delta g_{\Delta}(1-\chi) \right),\\
	g_\Delta&=\chi^\Delta {}_2F_1(\Delta,\Delta, 2\Delta;\chi).
\end{align}
Above, the bounds for the long operators and the presence of the short operators is explained above and given in equation \eqref{boundexchange}.\footnote{In our case, $j_0=\frac{3}{2}$ and the short operators of dimension $\Delta=1$ and $\Delta=3$ cannot appear due to the fermionic nature of the excitations at strong coupling.}
Additionally, from the S-parity of the correlators, we can relate $a_\Delta$ to $\tilde{a}_{\Delta}$ through the charges of the operators under S-parity. In practice, this means that 
\begin{align}
	a_\Delta = (-1)^{\Delta^{(0)}}\tilde{a}_\Delta.
\end{align}
This allows us to relate the perturbative expansions of $f(\chi)$ and $h(\chi)$ in the same way as the usual 1d braiding in \cite{Liendo:2018ukf,Ferrero:2019luz,Ferrero:2021bsb}. Combined with the crossing relations, this gives the following braiding/crossing symmetries:
\begin{align}
	f(\chi) &= \frac{\chi}{1-\chi}f(1-\chi)\label{crossing symmetry},\\
	h^{(i)}(\chi) |_{\log(\chi)^{(l)}}&= f^{(i)}(\frac{-\chi}{1-\chi})|_{\log(-\chi)^{(l)}}  \label{braiding symmetry},\\
	h^{(i)}(\chi)|_{\log(1-\chi)^{(l)}} &= -\chi h(\frac{1}{\chi}) |_{\log(\chi-1)^{(l)}}\label{cross-braiding symmetry}.
\end{align}
For convenience, we will write the braiding relations using ``$\simeq$", for example
\begin{align}
	h(\chi)&\simeq -\chi h(\frac{1}{\chi}) .
\end{align}

\subsection{Ansatz}\label{Section ABJM Ansatz}
At strong coupling, the perturbative correlators can be obtained by computing contributions from Witten diagrams in AdS$_2$ see  \cite{Giombi:2017cqn, Bliard:2022xsm}. Just as in flat space, there is a natural organisation in terms of functions of increasing transcendentality where for the four-point function, the disconnected diagram is rational, the contact diagrams are logarithmic, and the exchange diagrams have polylogarithms up to $Li_3$. Therefore, we begin with the 1d HPL Ansatz as in \cite{Ferrero:2019luz,Ferrero:2021bsb,Liendo:2018ukf}
\begin{align}\label{Ansatz ABJM 1}
	f^{(i)}(\chi) = \sum_{p_i} r(\chi) T_{p_i}(\chi)
\end{align}
where $r_p$ are rational functions and $T_p$ are Harmonic polylogarithms of transcendentality $p_i<2i-1$. For the third-order bootstrap, the Ansatz should be defined more rigorously, as braiding for polylogarithms is a little more subtle. In this case and in $\mc{N}=4$ SYM \cite{Ferrero:2021bsb, Ferrero:2019luz}, the polylogarithmic part of the solution can be constrained further by requiring the correct bosonic symmetry considerations, analogous to the function found by considering a special subset of functions.\footnote{One can repeat the procedure as above with simple polylogarithms. However, braiding relations become more intricate and this only leads to more terms to fix with the integrated correlators.}\footnote{The author thanks Pietro Ferrero for an introduction to these concepts.} These obey special crossing properties such that the result is always real and single valued, no matter what bosonic crossing is considered. These are the functions:
\begin{equation}
	LL_n(\chi) = \sum_{s=0}^{n-1} \frac{(-\log(|\chi|))^s}{s!} Li_{n-s}(\chi) +\frac{\left(-\log(|\chi|)\right)^{n-1}}{n!}\log(|1-\chi|)
\end{equation}
and the crossing and braiding relations are 
\begin{align}
	&LL_3(\chi )=LL_3(\tfrac{1}{\chi})\\
	&LL_3(\tfrac{\chi }{\chi -1})+LL_3(1-\chi )+LL_3(\chi )=\zeta (3).
\end{align}
\subsection*{Growth of the Anomalous Dimension}
The growth of the anomalous dimension of heavy exchanged operators is bounded and related to the relevance of the bulk interactions on the string worldsheet\cite{Heemskerk:2009pn,Fitzpatrick:2011dm}. The condition that $\gamma^{(l)}_\Delta$ be bounded in the large-$\Delta$ limit is relaxed to 
\begin{align}\label{Eq: ABJM growth gamma}
	\gamma_\Delta^{(l)} \sim_{\Delta\rightarrow \infty} \Delta^{l+1}
\end{align}
where $l$ is the order in perturbation theory and $\Delta$ is the weight of the exchanged operator. In this Regge limit, the leading order should be the same for all operators. The reason behind this \cite{Bliard-Levine} is that the large-$\Delta$ behaviour or the anomalous dimension is dominated by the flat space contribution where only isotropic terms contribute; therefore, excitations in different directions are equivalent. \par 

\subsection{Integrated Correlator Conditions}\label{Section: Integrated correlator}
In the case of $\mathcal{N}=4$ SYM, the topological sector known non-perturbatively through localisation allows this integration constant to be fixed. The analysis in \cite{Bianchi:2020hsz} shows us that such a sector does not exist for the displacement supermultiplet we are considering. Instead, one can use the integral relations which relate the integrated four-point correlators of the displacement multiplet to the Bremsstrahlung function of the corresponding Wilson line. A similar relation was used in \cite{Cavaglia:2022qpg} in addition to the quantum spectral curve to bootstrap the weak coupling spectrum. The relevant relation derived in \cite{Drukker:2022pxk} is
\begin{align}\label{Eq: integral eq}
	2\left(\int f(\chi)-\frac{h(\chi)}{\chi(1-\chi)}\right) = \frac{1}{4 B_{1/2}(\lambda)}.
\end{align}
This relation provides the physical input to match the expansion parameter with the coupling constant. It also constrains the type of functions in the correlator through the convergence of the integral. For example, terms that violate the large-$\Delta$ bound in \eqref{Eq: ABJM growth gamma} lead to power divergences in the integral \eqref{Eq: integral eq}, as do terms that are not in the minimal set described above in \ref{Section ABJM Ansatz}. \par \vspace{2mm}
The exploration of additional integrated correlator constraints such as in \cite{Cavaglia:2022yvv} would be essential to computing higher-order correlators. 
\newpage
\subsection*{Mixing}\label{Sec: mixing}
At first order, since the relevant vertices at strong coupling have four legs, the solving of the mixing only needs to include two-particle  and four-particle operators.\footnote{See \ref{ABJM solving mixing} for a full detailed analysis of the mixing.} In particular, the protected operator $\mathcal{O}_2 = \mathbb{F}\partial \mathbb{F} $ exchanges both two-particle and four-particle operators in free theory.\footnote{Note that this is a conformal primary, but a not a superprimary; it is a super descendent  of $\bar{\mathcal{B}}_{\frac72,1} \sim \mathbb{F} w^a$. } 
The strong coupling free theory result is obtained by Wick contracting the elementary fields\footnote{In this case, the elementary field is $\mathbb{F}$. This differs from the generalised free field result since $\mathcal{O}_2$ is a composite operator, not an elementary excitation. In the free theory, length-two and length-four operators will be exchanged.}
\begin{align}
	\langle \mathcal{O}_2 \bar{\mathcal{O}}_2 \bar{\mathcal{O}}_2\mathcal{O}_2 \rangle  &= \lim_{\{x_2,x_4,x_6,x_8\}\rightarrow\{x_1,x_3,x_5,x_7\}}\partial_{x_2}\partial_{x_4}\partial_{x_6}\partial_{x_8}\langle \mathbb{F}\mathbb{F}\bar{\mathbb{F}}\bar{\mathbb{F}}\bar{\mathbb{F}}\bar{\mathbb{F}} \mathbb{F}\mathbb{F}\rangle,
\end{align}
which gives
\begin{align}
	f^{(0)}_{\mathcal{O}_2}(\chi)&= \frac{1}{(1-\chi)^4}&	h^{(0)}_{\mathcal{O}_2}(\chi)&=(1-\chi)^4 .
\end{align}
In particular, even though the unitarity bound is $h_{L}>4$ in the charged channel, the lightest operator exchanged in the free theory has dimension $h_L = 8$. This weight is the same as the operator $\mathbb{F}\partial\mathbb{F}\partial^2\mathbb{F}\partial^3\mathbb{F}$.\par
Using the same constraints as for $\mathbb{F}$, the specific ones for the chiral ring detailed in see{Section: chiral ring}, and the first-order CFT data from the four-point correlator of the displacement multiplet, the first-order perturbation can be fixed entirely, giving\footnote{In particular, the anomalous dimension of $\mathbb{F}\bar{\mathbb{F}}$ can be found from both these correlators as can the OPE coefficient $c_{\mathcal{O}_2,\mc{O}_2 \mathbb{F}\mathbb{F}}$of $f(\chi)$.}
\begin{align}	
	f^{(1)}_{\mathcal{O}_2}(\chi)=4\epsilon\frac{(\chi-4) (\chi-1)^2 \log (1-\chi)-\chi (-4 \chi+(\chi+3) \chi \log (\chi)+4)}{ (\chi-1)^5 \chi}
\end{align}
where 
\begin{align}
	\epsilon = -\frac{1}{2\pi \sqrt{2\lambda}}
\end{align}
is the strong-coupling expansion parameter and $\lambda$ the 't Hooft coupling for ABJM. 
This gives the following CFT data:
\begin{align}
	\tilde{c}^{(0)}_h& =
	\frac{\sqrt{\pi } 2^{-2h -3} (h+1)^2 (h+2) (h  (h +2)+9) \Gamma (h+3)}{9 \Gamma \left(h +\frac{3}{2}\right)}\\
	\mathbb{c}^{(0)}_h& = \frac{\sqrt{\pi } 4^{-h } \left((-1)^{h }+1\right) (h -6) (h -4) (h -2) (h +1) (h +3) (h +5) \Gamma (h )}{9 \Gamma \left(h -\frac{1}{2}\right)}\\
	\gamma_s^{(1)} &= \epsilon h(h+2)\\
	\gamma_t^{(1)} &= \epsilon (h(h-1)-20).
\end{align}
In particular, this means that both the two-particle and the four-particle operators with vanishing $U(1)$ charge have anomalous dimensions proportional to the quadratic Casimir and we have 
\begin{align}
	\langle(\gamma^{(1)})^n\rangle =\langle\gamma^{(1)}\rangle ^n.
\end{align}

\section{Bootstrap Results}\label{Subsec: ABJM bootstrap}

In this section, by imposing symmetries and consistency with the OPE expansion, we extract the strong-coupling corrections to the four-point function of the displacement supermultiplet. We then extract the anomalous dimensions and the OPE coefficients from this expression. 
Let us start from the expressions of the four-point functions of the superprimary \eqref{four-point-1-primary} and \eqref{four-point-2-primary}, which for this specific case, are
\begin{align}
	\braket{\mathbb{F}(t_1)\bar{\mathbb{F}}(t_2)\mathbb{F}(t_3)\bar{\mathbb{F}}(t_4)}&=\frac{C_{\Phi}^2}{t_{12} t_{34}} f(\chi)\\
	\braket{\mathbb{F}(t_1)\bar{\mathbb{F}}(t_2)\bar{\mathbb{F}}(t_3)\mathbb{F}(t_4)}&=\frac{C_{\Phi}^2}{t_{12} t_{34}} h(\chi)
\end{align}
where for the bosonic part, we have decided to use the same cross-ratio; when considering the full superspace correlator, one must revert to the formalism in \eqref{four-point-1}.
\subsection{Leading Order Result}
The leading order result at strong coupling can be easily obtained from Wick contractions and is
\begin{align}
	f^{(0)}(\chi)&=1-\frac{\chi}{\chi-1} & 
	h^{(0)}(\chi)&=1-\chi. 
\end{align}
Looking at the $s$-channel expansion of the two correlators \eqref{superblock} and \eqref{hexpansion} and using the well-known fact that in free theory, the only exchanged operators are of the schematic form $[\mathbb{F}\bar{\mathbb{F}}]_n\sim\mathbb{F}\pa_t^n \bar{\mathbb{F}}$ with dimension $h_n=1+n$. The mismatch factor $(-1)^{1+n}$ between \eqref{hexpansion} and \eqref{superblock} is compensated by an identical factor from the relation~\eqref{ctildec} for the OPE coefficients, which in the case of the operators $[\mathbb{F}\bar{\mathbb{F}}]_n$ is
\begin{align}\label{c0andctilde0}
	c^{(0)}_n=(-1)^{1+n} \tilde c^{(0)}_n.
\end{align}
As a result, in free theory, the two expansions \eqref{superblock} and \eqref{hexpansion} are identical up to the braiding  $\chi \leftrightarrow \frac{\chi}{\chi-1}$. The explicit form of the OPE coefficient at leading order is extracted
\begin{align}
	c^{(0)}_n=\frac{\sqrt{\pi } 2^{-2 n-3} \Gamma (n+4)}{(n+1)\Gamma \left(n+\frac{5}{2}\right)}.
\end{align}
The same can be done in the chiral-chiral channel
\begin{align}
	\mathsf{c}^{(0)}_n&=\frac{\sqrt{\pi } 2^{1-2 n} \Gamma (n+1)}{\Gamma \left(n+\frac{1}{2}\right)}  & &n \text{ odd}\\
	\mathsf{c}^{(0)}_n&=0  & &n \text{ even}.
\end{align}
As expected, in the OPE of two identical fermions, only operators of the schematic form $[\mathbb{F}\pa^n\mathbb{F}]_n$ with \emph{odd} values of $n$ appear. 
In the following, a first-order perturbation of this result is considered.

\subsection{Next-to-Leading Order}
We are interested in finding the first-order strong coupling correction to the correlator. We expand the function $\tilde f(\chi)$ as 
\begin{align}\label{fhat}
	\hat{f}(\chi)&=\hat{f}^{(0)}(\chi)+\e\, \hat{f}^{(1)}(\chi) & h(\chi)&=h^{(0)}(\chi)+\e\, h^{(1)}(\chi)
\end{align}
where $\e$ is a small parameter whose precise relation with the string tension cannot be predicted by symmetry considerations. Following~\cite{Liendo:2018ukf, Ferrero:2019luz}, we start with the following Ansatz for the first order correction to $\hat{f}$ and $\hat{h}$
\begin{align}\label{Ansatz}
	f^{(1)}(\chi)&=\chi \left(r(\chi) \log(1-\chi)+r(1-\chi) \log \chi+q(\chi)\right)\\
	\hat{h}^{(1)}(\chi)&=-r\left(\tfrac{1}{1-\chi}\right)\log \chi + \left[r\left(\tfrac{\chi}{\chi-1}\right)+r\left(\tfrac{1}{1-\chi}\right)\right]\log (1-\chi)-q\left(\tfrac{\chi}{1-\chi}\right),
\end{align}
where $r(\chi)$ and $q(\chi)$ are rational functions and 
\begin{align}\label{qcross}
	q(\chi)=q(1-\chi).
\end{align}
The OPE expansions constrain the behaviour of both functions as $\chi\rightarrow 0$ and $\chi \rightarrow 1$
\begin{align}
	&f(\chi) = 1+O(\chi)\\
	&h(\chi) = O(\chi)\\
	&\left(\frac{1-\chi}{\chi}\right)h(\chi)= O((1-\chi)) \\
&\left(\frac{1-\chi}{\chi}\right) h(\chi)|_{\log(1-\chi)} = O((1-\chi)^4).
\end{align}
Since the identity operator receives no contributions, the rational function $r(1-\chi)$ is regular at $\chi=0$. Since it should only have poles for colliding operators ($\chi\rightarrow \{0,1,\infty\}$) one can expand it as a Laurent series
\begin{align}
	r^{(1)}(1-\chi) = \sum_{k } r_k (1-\chi)^k.
\end{align}
The large-$\Delta$ behaviour of the anomalous dimension
\begin{align}
	\gamma_\Delta^{(1)} \sim_{\Delta\rightarrow \infty} \Delta^2
\end{align} 
in all OPE channels is extremely constraining and sets almost all coefficients $r_k$ to 0 except for $k=-2$ and $k=-1$. The boundary behaviour from the OPE shown above is then sufficient to fix the function up to an overall constant which can be reabsorbed in the expansion parameter $\e$
\begin{align}\label{Eq ABJM f1sol}
	\epsilon f^{(1)}(\chi) &= \epsilon  \left(\left(\frac{1}{\chi }+2\right) \log (1-\chi )+\frac{-\chi +(3-2 \chi ) \chi  \log (\chi )+1}{(\chi -1)^2}\right)\\
	\epsilon h^{(1)}(\chi)&=-\frac{\epsilon  \left((\chi -1)^3 \log (1-\chi )+\chi  (\chi -((\chi -3) \chi  \log (\chi ))-1)\right)}{\chi }.
\end{align}
This parameter can be fixed using the integrated correlator constraint in \ref{Section: Integrated correlator}
\begin{align}
	&2\e \left(\int f^{(1)}(\chi)-\frac{h^{(1)}(\chi)}{\chi(1-\chi)}\right) = \frac{\pi }{\sqrt{2 \lambda }}&\Rightarrow \qquad \e &= -\frac{1 }{2\pi \sqrt{2 \lambda }}.
\end{align}
The remaining configurations of the fields can be found through crossing and braiding
\begin{align}
	<\mathbb{F}\bar{\mathbb{F}}\mathbb{F}\bar{\mathbb{F}}>&= <\bar{\mathbb{F}}\mathbb{F}\bar{\mathbb{F}}\mathbb{F}> \simeq \frac{1}{(t_{12}t_{34})}h(\frac{\chi}{\chi-1})\\
	<\mathbb{F}\bar{\mathbb{F}}\bar{\mathbb{F}}\mathbb{F}>&= <\bar{\mathbb{F}}\mathbb{F}\mathbb{F}\bar{\mathbb{F}}>=\frac{1}{(t_{12}t_{34})}h(\chi)\\
	<\bar{\mathbb{F}}\bar{\mathbb{F}}\mathbb{F}\mathbb{F}>&= <\mathbb{F}\mathbb{F}\bar{\mathbb{F}}\bar{\mathbb{F}}>=\frac{1}{(t_{12}t_{34})}\chi h(1-\chi).
\end{align}
\subsection{Leading Logarithms and Double Scaling Limit}\label{double scaling limit}
Given the absence of mixing for $\gamma^{(1)}$, the leading logarithm powers are entirely fixed by tree-level data presented in section \ref{Subsec CFT data} and are:
\begin{align}
	&h^{(0)}|_{\log(\chi)^0} =1-\chi , \\
	&h^{(1)}|_{\log(\chi)^1} =(\chi -3) \chi \\
	&h^{(2)}|_{\log(\chi)^2} =\frac{1}{2} \chi  ((11-4 \chi ) \chi -9)\\
	&h^{(3)}|_{\log(\chi)^3} =\frac{1}{6} \chi  (2 \chi -1) (\chi  (18 \chi -43)+27) \\
	&h^{(4)}|_{\log(\chi)^4} =\frac{1}{24}\chi  (\chi  (803-4 \chi  (18 \chi  (8 \chi -25)+487))-81),\\
	\text{and}\qquad &\\
	&h^{(0)}|_{\log(1-\chi)^0} =1-\chi\\
	&h^{(1)}|_{\log(1-\chi)^1} = -\frac{(\chi -1)^3}{\chi } \\
	&h^{(2)}|_{\log(1-\chi)^2} = -\frac{(\chi -1)^3 \left(2 \chi ^2+\chi +2\right)}{\chi ^2}\\
	&h^{(3)}|_{\log(1-\chi)^3} = -\frac{2 (\chi -1)^3 (9 \chi ^4+2 \chi ^3+3 \chi ^2+2 \chi +9)}{3 \chi ^3}\\
	&h^{(4)}|_{\log(1-\chi)^4} =-\frac{(\chi -1)^3 (72 \chi ^6-18 \chi ^5+6 \chi ^4+5 \chi ^3+6 \chi ^2-18 \chi +72)}{3 \chi ^4}.
\end{align}
Since these are generated by acting with the superconformal Casimir in the different channels, they are straightforward to generate at high orders. So one can look at the double-scaling limit \cite{Giombi:2022pas} of these correlators in the out-of-time-ordered configuration. To match their  conventions, we will look at the $\log(1-\chi)$ term of $h(\chi)$ and $h(1-\chi)$ related to the correlators given above. This should incorporate all the different configurations, as only two OPE limits are available. In this limit, the leading term will be given by the leading power of $\log(1-\chi)$ of the transformed function 
\begin{align}
	\tilde{h}(\chi) &= \chi h\left(\frac{1}{\chi}\right). 
\end{align}
Note that, given the braiding properties of $h(\chi)$, this simply corresponds to a continuation of $h(\chi)$ with the replacement $\log(1-\chi)\rightarrow \log(\chi-1)$. We therefore drop the tilde notation in the following for convenience. 
Taking the limit $t\rightarrow \infty$ of the function evaluated in the cross-ratio
\begin{align}
	\chi = \frac{2}{1-i \sinh (t)},
\end{align} 
while keeping $\kappa = \frac{\text{e}^t}{4  \sqrt{2\lambda }}$ fixed gives this double-scaling limit.
Organising the correlator as 
\begin{align}
	h(\chi) &= \sum_n \epsilon^n h^{n}(\chi)\\
	h^{n}|_{\log^l(\chi-1)} &= \frac{1}{\chi^l}\sum_{k}a_{k}^{n,l} \chi^k
\end{align}
Given, the OPE is naturally written in terms of 
\begin{align}
	\hat{h}(\chi) = \frac{1-\chi}{\chi}h(\chi)
\end{align}
we will use it in the following. 
The fact that the leading anomalous dimension is proportional to the eigenvalue of the Quadratic Casimir allows us to write the leading power of logarithms as
\begin{align}
	\gamma^{(1)}_\Delta &= \epsilon \lambda_\Delta\\
	\hat{h}^{n}|_{\log^n(\chi-1)} &= \frac{\left(\mathcal{C}^{(2)}\right)^n}{n!}\hat{h}^{(0)}(\chi)
\end{align}
where $\mathcal{C}^{(2)}$ is the differential expression of the quadratic Casimir in the $\chi\rightarrow 1$ channel of $h(\chi)$, and the subleading power of logarithms as 
\begin{align}
	\hat{h}^{n}|_{\log^{n-1}(\chi-1)} &= \frac{\left(\mathcal{C}^{(2)}\right)^{n-2}}{(n-2)!}\left(\hat{h}^{(2)}|_{\log(1-\chi)}(\chi)\right)-\frac{\chi}{1-\chi}\frac{(n-2)\left(\mathcal{C}^{(2)}\right)^{n-1}}{(n-1)!}\left(\hat{h}^{(1)}|_{\log(1-\chi)^0}(\chi)\right).
\end{align}
These expression give recursion relations between the the $a_{k}^{n,l}$ coefficients for $l=n$ and $l=n-1$ in terms of the lower perturbative orders.
%This gives for both channels (up to an overall sign)
%\begin{align}
%	h_{\text{DS}}(\lambda,\text{e}^{t}) =1-\frac{e^t}{4 \sqrt{2} \sqrt{\lambda }}+\frac{e^{2 t}}{16 \lambda }-\frac{3 e^{3 t}}{64 \sqrt{2} \lambda ^{3/2}}+\frac{3 e^{4 t}}{128 \lambda ^2}-\frac{15 e^{5 t}}{512 \sqrt{2} \lambda ^{5/2}}+\frac{45 e^{6 t}}{2048 \lambda ^3}+O(\left(\text{e}^{t}\lambda^{-\tfrac 12}\right)^7)
%\end{align}
%where we have neglected exponentially suppressed terms such as $\kappa^n \text{e}^{-t}$ coming from the non-maximal powers of logarithms. 
This gives the series
\begin{align}
	h_{\text{DS}}(\kappa ) =\sum_n (-\kappa)^n \Gamma (n+1)+O(\text{e}^{-t}).
\end{align}
where the subleading terms give a correction to the Lyapunov exponent:
\begin{align}
	\kappa = \frac{\text{e}^{t(1-\frac{1}{4\pi \sqrt{2\l}})}}{4  \sqrt{2\lambda }} 
\end{align}
This series is Borel resummable where the Borel transform and Borel resummation are\footnote{see \cite{Dorigoni:2014hea} or \cite{Marino:2015yie} for an introduction to resurgence and non-perturbative methods in QFT.}
\begin{align}
	\hat{h}_{\text{DS}}(\zeta)&=\frac{1}{1+\zeta} &h_{\text{DS}}(\mu)&=\kappa^{-1}U\left(1,1,\kappa^{-1}\right),
\end{align}
Here $U$ is a confluent hypergeometric function of the second kind and one can expand it around $\kappa^{-1}=0$ which gives a series in $\sqrt{\lambda}$ as opposed to $\frac{1}{\sqrt{\lambda}}$. This matches equation (3.20) in  \cite{Giombi:2022pas} if we identify $\Delta_V =\Delta_W= \frac{1}{2}$, corresponding to the conformal weight of the superconformal primary, and $\kappa = \frac{\text{e}^{(1-\frac{1}{4\pi \sqrt{2\l}})t}}{4  \sqrt{2\lambda }} $. Note that we also observe a correction to the Lyapunov exponent at this order. 
This formalism in terms of conformal blocs is a reorganising of the perturbative series, where the (partially unmixed) conformal data is in reality contained in all perturbative orders as the leading powers of logarithms. In this way, one can access a limit of the non-perturbative correlator simply with first-order CFT data and exponentially suppressed terms with second-order data.
\subsection{Second Order Bootstrap}
The next order has an Ansatz with functions up to transcendentality three:
\begin{align}
	h^{(2)}(\chi)&= q^{(2)}(\chi)+r_1^{(2)}(\chi)\log(\chi)+r_2^{(2)}(\chi)\log(1-\chi)+\\
	&s_1^{(2)}(\chi) \log(1-\chi)\log(\chi)+s_2^{(2)}(\chi)\log^2(\chi)+s_3^{(2)}(\chi)\log^2(1-\chi)+s_4^{(2)}(\chi)(\text{Li}_2(\chi )-\text{Li}_2(1-\chi ))\nonumber 
\end{align}
Polylogarithm identities were used to eliminate redundant terms, and the polynomiality of $\gamma^{(1)}$ was used to discount non-rational functions multiplying $\log^2$ in both OPE limits. Applying crossing/braiding to the known leading logarithms eliminates all transcendentality-three terms and fixes $s_1^{(2)}(\chi)$. 
As with the first order bootstrap, the growth of the anomalous dimension imposes conditions on the rational functions $\{q,r_i,s_4\}$ such that we have a finite sum of terms which are then fixed by the boundary conditions set by the OPE.\footnote{It was noted here that relaxing this condition to $\gamma^{(2)}\sim \Delta^4$ only leads to a single additional term which leads to a power divergence in the integrated correlator. }  This fixes the function up to a constant $a$
\begin{align}
	h^{(2)}(\chi) &= \e^2 \left(-\frac{(\chi -1)^3 \left(2 \chi ^2+\chi +2\right) \log ^2(1-\chi )}{\chi ^2}-\frac{1}{2} \chi  (4 \chi ^2-11 \chi +9) \log ^2(\chi )+\right. \non\\
	&\left.+\frac{(8 \chi ^4-21 \chi ^3+19 \chi ^2-\chi -1) \log (1-\chi ) \log (\chi )}{2 \chi }+\frac{1}{2} (\chi  (2 \chi -3)-1) \log (\chi )+\right. \non \\
	&\left.-\frac{(\chi -1) \left(\chi ^2+1\right) \log (1-\chi )}{\chi }+\chi -1\right) +\e a h^{(1)}(\chi),
\end{align}
and equivalently
\begin{align}
	f^{(2)}&= \e^2 \chi \left( \frac{\left(-2 \chi ^5+7 \chi ^4-9 \chi ^3\right) \log ^2(\chi )}{2 (\chi -1)^3 \chi ^3} -\frac{(1-2 \chi )^2 \log (1-\chi ) \log (\chi )}{4 (\chi -1)^2\chi^2}+\right.  \\
	&\left. +\frac{\left(-2 \chi ^5+7 \chi ^4-6 \chi ^3+\chi ^2\right) \log (\chi )}{2 (\chi -1)^3 \chi^3}+ \frac{1}{(\chi -1) \chi }+\chi \leftrightarrow1-\chi\right) +\e a f^{(1)}(\chi) .\non
\end{align}
The integrated correlator condition fixes this final free parameter to $a=0$.\footnote{Note that the resulting function only has terms up to transcendentality two and only products of logarithms.}
\subsection{Third Order Bootstrap}\label{Bootstrapping ABJM order 3}
For the next order correlator, the $\log^2$ terms can also be computed without the preoccupation of mixing.\footnote{There is no mixing from $\gamma^{(1)}$ and the corresponding perturbative OPE expansion term is linear in $\gamma^{(2)}$.}
\begin{align}
	h^{(3)}|_{\log^2} = \gamma^{(1)} h^{(2)}|_{\log}-\frac{1}{2}\left( \gamma^{(1)}\right)^2h^{(2)}|_{\log^0}
\end{align}
where this is to be understood in the sense of the terms in the OPE expansion; the anomalous dimension must be summed over the weights of the primaries. The precise evaluation of these terms is a little more involved than for the leading power of logarithms but is well-defined in terms of known CFT data. \par 
Given the Ansatz described above in \eqref{Ansatz ABJM 1}, the OPE constraints and the Regge bound fix all terms up to 2. One of them can be fixed by noticing that the leading large-$\Delta$ behaviour of the anomalous dimension $\gamma^{(3)}_\Delta$ should be the same for the $\mathbb{F}\times \mathbb{F}$ channel and for the $\mathbb{F}\times \bar{\mathbb{F}}$ channel from flat-space limit considerations (see section \ref{flat-space limit}) 
\begin{align}
	\gamma_\Delta^{(3)} &\rightarrow_{\Delta\rightarrow \infty}\left(  \frac{1}{6} \left(\pi ^2-30\right) \epsilon^3+\frac{1}{4}x_{2,1}\right) \Delta^4 \\
	\upgamma_\Delta^{(3)}&\rightarrow_{\Delta \rightarrow \infty} \left( +\frac{1}{6} \left(\pi ^2-30\right) \epsilon^3 -\frac{1}{4}x_{2,1}\right)\Delta^4.
\end{align}
The final term can be fixed with the integrated correlator giving the result
\begin{align}
	&f^{(3)}(\chi) = \e^3 \chi \left( \frac{\chi  \left(2 \chi  \left(-2 \chi ^2+\chi +7\right)-5\right)+1}{2 (\chi -1)^3 \chi ^2}LL_3(\chi) -\frac{(\chi +1) (\chi  (2 \chi -11)+27) }{6 (\chi -1)^4}\log^3(\chi) \right. \non\\
	&\left.+\frac{(\chi  ((17-6 \chi ) \chi -12)+9)  ^3}{(\chi -1)^3}\log^2(\chi)+\frac{\left(14 \chi ^4-37 \chi ^3+23 \chi ^2-5 \chi +1\right) }{6 (\chi -1)^3 \chi ^2}\log ^2(\chi ) \log (1-\chi )\right. \non \\
	&\left. +\frac{\left(12 \chi ^4-24 \chi ^3+5 \chi ^2+7 \chi -3\right)}{2 (\chi -1)^2 \chi ^2} \log (\chi ) \log (1-\chi )+ \frac{\left(-12 (\chi -3) \chi +\pi ^2 (2 \chi -3) \chi -21\right) }{6 (\chi -1)^2 \chi }\log(\chi)\right. \non \\
	&\left. +\frac{\left(12 \chi ^4-24 \chi ^3-9 \chi ^2+21 \chi -8\right) \zeta (3)}{4 (\chi -1)^3 \chi ^3}+\frac{\left(\pi ^2-12\right)}{12 (\chi -1) \chi } +\chi \rightarrow 1-\chi\right).
\end{align}
Where the only presence of polylogarithms is through the single-valued function $LL_3(\chi)$ presented in \ref{Section ABJM Ansatz}. 
\section{CFT Data}\label{Subsec CFT data}
Before computing the values of anomalous dimensions and OPE coefficients, we need to comment on the class of operators one expects to appear in this context. When classifying operators that can be exchanged in a given channel, one is interested in the eigenstates of the dilatation operator. When perturbing the leading order result, one can think of building operators out of the fundamental fields in the worldsheet Lagrangian (see section \ref{sec:sigmamodel}), which are in one-to-one correspondence with the components of the super displacement multiplet. Two-particle operators of the schematic form $O_n\sim\mathbb{F}\pa^n\bar{\mathbb{F}}$ will have a three-point function $\braket{\mathbb{F}\bar {\mathbb{F}} O_n}$ that is leading compared to higher particle operators. Nevertheless, these two-particle  operators are not well-defined eigenstates of the dilatation operator. A simple example is the mixing between two- and four-particle operators $\mathbb{F}\pa^2 \bar{\mathbb{F}}$ and $\mathbb{F} \bar{\mathbb{F}}\mathbb{F} \bar{\mathbb{F}}$. Only a linear combination of these two operators will be an eigenstate of the one-loop dilatation operator, but both of them are allowed to appear in the $\mathbb{F}\times \bar{\mathbb{F}}$ OPE.\footnote{We would like to thank Pietro Ferrero, Shota Komatsu and Carlo Meneghelli for very useful discussions on this point.} Of course, the situation becomes increasingly more intricate for heavier operators. Therefore, we conclude that any operator with a two-particle contribution will appear in the leading order OPE. Therefore, the anomalous dimension we extract is a linear combination of the anomalous dimensions of these operators, weighted by their OPE coefficients. Given that part of this degeneracy is lifted by decomposing in terms of superconformal blocks instead of ordinary conformal blocks, there are a few cases where we can be certain that a single long multiplet with a given dimension can appear. These are the $n=0$ and $n=1$ case for the $\mathbb{F}\times \bar{\mathbb{F}}$ channel associated to the operators $\mathbb{F}\bar{\mathbb{F}}$ and $\mathbb{F}\pa\bar{\mathbb{F}}$ and the $n=1$ case in the $\mathbb{F} \times \mathbb{F}$ channel associated to $\mathbb{F}\pa\mathbb{F}$, which is protected. The mixing was solved at the first order in the previous section. Still, to solve the mixing beyond the first order for heavier operators, one must study a larger class of correlators, particularly the 1/3 BPS operators studied in \cite{Bliard-Ferrero}, a task still underway. \par \vspace{4mm}
In the chiral-anti-chiral OPE channel, the superconformal bloc basis, Casimir differential equation, eigenvalue and orthogonality relation needed for the superblock expansion are
\begin{align}\label{orthogonality}
	&G_n = \chi^n {}_2F_1(n,n,2n+3;\chi) \\
	&\mathcal{L}=-\chi^2(\chi-1)\partial_\chi^2 -\chi(\chi-3)\partial_\chi\\
	&\lambda_h = h(h+3)\\
	&\oint \frac{d\chi}{2 \pi i} \frac{\chi}{(1-\chi)^3} G_{1+n}(\chi) G_{-3-n'}(\chi)= \delta_{n,n'}\,,
\end{align}
where the contour circles $0$  counterclockwise. 
Expanding the blocks as in \eqref{Eq: Conformal block expansion} and inverting the expansion, we find the following CFT data. The anomalous dimension is
{\allowdisplaybreaks
\begin{align}
	\gamma^{(1)}_\Delta &= -\e \Delta  (\Delta +2)\\
	\gamma^{(2)}_\Delta&= \epsilon^2 \left( \frac{1}{2} \left(4 \Delta ^3+4 \Delta ^2-9 \Delta +\frac{5}{\Delta +1}-\frac{4}{\Delta +2}+\frac{4}{\Delta }+3\right)+\left(\Delta  (\Delta +2)-4\right)H_{\Delta+1} \right. \\
	&\left.-(-1)^{\Delta} \frac{3+(1+\Delta)^4}{\Delta  (\Delta +1) (\Delta +2)}\right)\non\\
	\hspace{-2mm}
	\gamma^{(3)}_{\Delta, even}&= \epsilon ^3 \left(\frac{1}{6} \left(-3 \Delta ^2-6 \Delta +4\right) \left(-3 \psi ^{(1)}_{\frac{\Delta }{2}}+3 \psi ^{(1)}_{\frac{\Delta +1}{2}}-\pi ^2\right)+\left(-\Delta ^2-2 \Delta +4\right) \left(H_{\Delta -1}\right){}^2\right. \\
	&\left. -\frac{5 \Delta ^7+19 \Delta ^6+13 \Delta ^5-15 \Delta ^4-26 \Delta ^3-3 \Delta ^2+48 \Delta +16}{\Delta  (\Delta +1) (\Delta +2)}-\frac{1}{6} \left(\pi ^2-3\right) \Delta  (\Delta +2) \right. \non\\
	&\left. -\frac{\left(4 \Delta ^6+16 \Delta ^5+3 \Delta ^4-30 \Delta ^3-12 \Delta ^2-8 \Delta +8\right) H_{\Delta -1}}{\Delta  (\Delta +1) (\Delta +2)}+\Delta  (\Delta +2) \left(\Delta ^2+2 \Delta -4\right) H_{\Delta -1}^{(2)}\right) \non
	\end{align}}
\begin{align}
	\hspace{-6mm}\gamma^{(3)}_{\Delta, odd}=\epsilon ^3& \left(\frac{1}{6} \left(3 \Delta ^2+6 \Delta -4\right) \left(-3 \psi ^{(1)}_{\frac{\Delta }{2}}+3 \psi ^{(1)}_{\frac{\Delta +1}{2}}+\pi ^2\right)+\Delta  (\Delta +2) \left(\Delta ^2+2 \Delta -4\right) H_{\Delta -1}^{(2)}\right. \\
	&\left. -\frac{\Delta  \left(5 \Delta ^5+19 \Delta ^4+19 \Delta ^3+7 \Delta ^2+2 \Delta -7\right)}{(\Delta +1) (\Delta +2)}-\frac{1}{6} \left(\pi ^2-3\right) \Delta  (\Delta +2)\right. \non\\
	&\left. -\frac{\left(4 \Delta ^6+16 \Delta ^5+11 \Delta ^4+2 \Delta ^3+36 \Delta ^2+24 \Delta -24\right) H_{\Delta -1}}{\Delta  (\Delta +1) (\Delta +2)}+\left(-\Delta ^2-2 \Delta +4\right) \left(H_{\Delta -1}\right){}^2\right). \non
\end{align}
The OPE coefficients are
\begin{align}
	c_\Delta^{(0)}&= \frac{\sqrt{\pi }  2^{-2 \Delta -1} (\Delta +1) (\Delta +2) \Gamma (\Delta )}{\Gamma \left(\Delta +\frac{3}{2}\right)} \\
	c_\Delta^{(1)}&= \partial_\Delta \left( c_\Delta^{(0)} \gamma^{(1)}_\Delta  \right)\\
	c_\Delta^{(2)}&= \partial_\Delta \left( c_\Delta^{(0)}\upgamma^{(2)}_\Delta+c_\Delta^{(1)}\upgamma^{(1)}_\Delta \right)-\partial^2_{\Delta}\left(c_\Delta^{(0)}(\upgamma^{(1)}_\Delta)^2\right) \\
	&+ \tfrac12 \e^2 c_\Delta^{(0)}\left(\Delta ^4+4 \Delta ^3+3 \Delta ^2-2 \Delta +4\right) \left(\psi ^{(1)}_{\frac{\Delta +1}{2}}-\psi ^{(1)}_{\frac{\Delta }{2}+2}\right)\\
	&+\tfrac 12  \e^2 c_\Delta^{(0)}\left( \frac{1+(-1)^\Delta}{2}\left(-6 \Delta  (\Delta +1)-\frac{2}{(\Delta +1)^2}+6\right)+\right. \non \\
	&\qquad\left. +\frac{1+(-1)^\Delta}{2}\left(\frac{3 (\Delta +1) \Delta ^3-4 \Delta +8}{\Delta ^2}+\frac{11}{(\Delta +1)^2}+\frac{4}{\Delta +2}+\frac{8}{(\Delta +2)^2}-3\right)\right).
\end{align}
A similar analysis can be carried out in the chiral-chiral channel, with conformal scalar blocks. This gives
\begin{align}
	\upgamma^{(1)}_\Delta &= -\e (\Delta  (\Delta -1)-2) \\
	\upgamma^{(2)}_\Delta&= \epsilon^2 \left( 2 \Delta ^3-7 \Delta ^2-\Delta -\frac{4}{\Delta +1}-\frac{2}{\Delta }+9+((\Delta -1) \Delta +2)H_{\Delta+1}\right)\\
	\upgamma^{(3)}_\Delta&=\epsilon ^3 \left(21-5 \Delta ^4+26 \Delta ^3-\frac{63 \Delta ^2}{2}-\frac{27 \Delta }{2}-\frac{1}{6} \left(\pi ^2-3\right) (\Delta -2) (\Delta +1)\right. \non \\
	&\left.+\frac{1}{12} \left(-5 \Delta ^2+5 \Delta +6\right) \left(3 \psi ^{(1)}_{\frac{\Delta }{2}}-3 \psi ^{(1)}_{\frac{\Delta +1}{2}}+\pi ^2\right)+\left(-\Delta ^2+\Delta -2\right) \left(H_{\Delta -1}\right){}^2 \right. \non \\
	&\left. -2 \left(2 \Delta ^3-7 \Delta ^2+5 \Delta -4\right) H_{\Delta -1}+(\Delta -2) (\Delta +1) \left(\Delta ^2-\Delta +2\right) H_{\Delta -1}^{(2)}\right),
\end{align}
where the anomalous dimension vanishes for $\Delta=2$, as expected. 
The OPE coefficients are
\begin{align}
	\mathsf{c}_\Delta^{(0)}&= \frac{\sqrt{\pi } 2^{3-2 \Delta } \Gamma (\Delta )}{\Gamma \left(\Delta -\frac{1}{2}\right)} \\
	\mathsf{c}_\Delta^{(1)}&= \partial_\Delta \left( \mathsf{c}_\Delta^{(0)} \upgamma^{(1)}_\Delta  \right)\\
	\mathsf{c}_{\Delta}^{(2)}&= \partial_\Delta \left( \mathsf{c}_\Delta^{(0)}\upgamma^{(2)}_\Delta+\mathsf{c}_\Delta^{(1)}\upgamma^{(1)}_\Delta \right)-\partial^2_{\Delta}\left(\mathsf{c}_\Delta^{(0)}(\upgamma^{(1)}_\Delta)^2\right) \\
	&+ \e^2 \mathsf{c}_\Delta^{(0)}\left((\Delta -3) (\Delta +2)-\frac{1}{2} \left(\Delta ^4-2 \Delta ^3-4 \Delta ^2+5 \Delta +2\right) \left(\psi ^{(1)}_{\frac{\Delta }{2}}-\psi ^{(1)}_{\frac{\Delta +1}{2}}\right)\right)\non 
\end{align}

The functional form of the 3rd-order OPE coefficients is neither compact nor illuminating, so it will not be listed here.  However, the weight of the lightest uncharged operator and the OPE coefficient from the protected operator $\mathsf{c}_{\Delta=2}$ is not affected by mixing and is the following
\begin{align}
	\Delta_{\mathbb{F}\bar{\mathbb{F}}} &= 2+\frac{2 \sqrt{2}}{\pi  \sqrt{\lambda }}+\frac{65}{24 \pi ^2 \lambda }+\frac{179+8 \pi ^2}{96 \sqrt{2} \pi ^3 \lambda ^{3/2}}+O\left(\lambda ^{-2}\right),\\
	c^2_{\mathbb{F}\mathbb{F}(\bar{\mathbb{F}}\partial\bar{\mathbb{F}})} &=1+\frac{3}{2 \pi  \sqrt{2 \lambda }}+ \frac{12 \zeta (3)+\pi ^2-28}{2 (2 \pi  \sqrt{2 \lambda })^3}+O\left(\lambda ^{-2}\right).
\end{align}

\section{Comments}
This series of analyses and results for the ABJM 1/2-BPS Wilson line dCFT is extensive and leads to several conclusions. These will be listed here. The first concerns the structure of the supermultiplet and of the supercorrelator, the second is the obtaining of results through the analytic bootstrap. The third result is the addition of the integrated correlator which frees the bootstrap result from needing an input from the string-theory description. The fourth is the explicit computation of worldsheet fluctuation correlators on the string side of the AdS/CFT duality, these match the bootstrap results and are consistent with the superspace analyis.  The fifth is the analysis of the mixing at first-order which not only enables the higher-order perturbative bootstrap, but also a part of the all-order solution. This constitutes the sixth and final conclusion, which is the double-scaling limit of the OTOC, which is an all-loop result. \par \vspace{4mm}
The first important result of this analysis concerns the general structure of the four-point functions of the displacement supermultiplet. The superfield formalism, together with the underlying superconformal symmetry, enables us to determine all the non-vanishing correlators for a given ordering of the external superfields in terms of a single function $f(\chi)$ of the relevant conformal cross-ratio $\chi=\frac{t_{12} t_{34}}{t_{13} t_{24}}$,  as well as fixing the superconformal block expansion, essential to bootstrap process. 
This function appears directly in the correlator of the superconformal primary $\mathbb{F}(t)$ and its conjugate $\bar{\mathbb{F}}(t)$ \begin{align}\label{corrFintro}
	\braket{\mathbb{F}(t_1)\bar{\mathbb{F}}(t_2)\mathbb{F}(t_3)\bar{\mathbb{F}}(t_4)}&=\frac{C_{\Phi}^2}{t_{12} t_{34}} f(z)\,.
\end{align}
Above, $C_{\Phi}$ is the normalization of the superfield two-point function and has a physical interpretation in terms of the Bremsstrahlung function, discussed in subsection~\ref{Sec: ABJM Displacement}.
In one-dimensional CFTs, correlators come with a specific ordering, and one can take OPEs only for neighbouring operators. 
Therefore, crossing symmetry implies that the exchange $1\leftrightarrow 3$ is a symmetry of the correlator~\eqref{corrFintro}, whereas $1\leftrightarrow 2$ is not. This means that different orderings correspond to different functions of the cross-ratio. However, the OPE expansion relates these functions, often through braiding symmetry. 
Having encoded all the information into this function, the analytic bootstrap process is sufficient to compute $f(\chi)$ up to the third order in perturbation around the generalized free field theory result obtained by Wick contractions. 
The two OPEs of this correlator correspond to the OPE channels, charged and uncharged under $U(1)$. In the chiral-chiral OPE, each superconformal primary only contributes one operator in the tower of super descendants. Therefore, the scalar blocks are sufficient. In this channel, short (protected) multiplets appear in the block expansion, constraining the long operators' dimension in the same channel. In the chiral-antichiral OPE (uncharged), every descendent contributes. Hence, the superconformal block is a sum of a finite number of conformal blocks corresponding to the contribution from the superconformal descendants as seen in \eqref{expansion-conformal-blocks}. The explicit form of the superblocks is explicitly obtained by diagonalizing the superconformal Casimir operator. \par \vspace{2mm}
This constitutes the second important series of results. The symmetry considerations and a minimal input are sufficient to perturbatively find the strong coupling correlator up to three orders, which is beyond the current capabilities in perturbative diagrammatics and is the first prediction at this order.  Using the constraints from the conformal bootstrap detailed in the previous section, the symmetries constrain the four-point function up to the third order. \par \vspace{2mm}
One of the new elements in this analysis compared to \cite{Bianchi:2020hsz}, beyond the additional two orders in perturbation, was the ability to fix the bootstrap parameter through the integrated correlators in \cite{Drukker:2022pxk}. The parameter $\epsilon$ controls the expansion around generalized free field theory, and it will be interpreted as the inverse string tension  $T$ appearing in the effective AdS$_2$ sigma model. The precise relation between the $\epsilon$ parameter and the string tension $T$ can be established by comparing the explicit Witten diagram computation to the bootstrap result. However, integrated correlator results \cite{Drukker:2022pxk} allow for an explicit fixing of this coefficient by comparison to the ABJM Bremsstrahlung function\cite{Bianchi:2018scb,Bianchi:2017ozk, Bianchi:2017svd, Drukker:2019bev} which is also sufficient to fix remaining integration constant up to third-order and to justify some of the bootstrap requirements. \par \vspace{2mm}
The third element of this analysis concerns the mixing of operators. At leading order (generalized free field theory), the operators exchanged in the OPE channels are ``two-particle'' operators of the schematic form $\mathbb{F}\partial^{n}_t\bar{\mathbb{F}}$ in the chiral-antichiral channel, and $\mathbb{F}\partial^{n}_t{\mathbb{F}}$ (with odd $n$) in the chiral-chiral channel.\footnote{At strong coupling these operators should represent worldsheet bound states, made of two of the corresponding fluctuations, as discussed in~\cite{Giombi:2017cqn}.} The CFT data can then be extracted from these analytic expressions of the correlator. At first order, the correction to the CFT data corresponds to the leading corrections to the classical dimension of the two-particle  operators defined above (or their supersymmetric generalization). However, at higher-order, two-particle operators generally mix with multi-particle operators, and the results should correspond to linear combinations of the actual anomalous dimensions weighted by OPE coefficients, except for a handful of operators for which there is no mixing. \par \vspace{2mm}
A fourth element concerns the strong coupling perturbative results. As mentioned above, the worldsheet fluctuations around the  (AdS$_2$) minimal surface corresponding to the 1/2-BPS Wilson line are in direct correspondence with the components of the displacement supermultiplet. Through AdS/CFT,  correlators of these  AdS$_2$ fields evaluated at the boundary correspond to correlation functions of the dual defect operators~\cite{Giombi:2017cqn}. Their boundary-to-boundary propagator is free for large string tension $T$, leading to a generalized free field theory result for their four-point function.  The $1/T$ expansion for the Nambu-Goto string action  involves non-trivial bulk interactions, and the associated boundary correlators are evaluated via AdS$_2$ Witten diagrams~\footnote{We remark that, compared to the usual $1/N$ expansion in Witten diagrams for higher-dimensional AdS/CFT, we are expanding the large-$N$ string sigma-model in inverse powers of the string tension. see related discussion in~\cite{Giombi:2017cqn}.}.  Having derived the effective Lagrangian governing the interactions of the AdS$_2$, the computation of all bosonic correlators confirms the result of $f^{(0)}(\chi)$ and $f^{(1)}$, obtained by bootstrap. The analogous computation for higher order quantities quickly becomes unpractical as the number of terms already reaches $\sim 10^4$ for $f^{(2)}$ hence the power of the conformal bootstrap. Another issue in explicit computations is regularization subtleties in AdS$_2$ models with derivative interactions (see discussion in~\cite{Menotti:2005fk,Menotti:2006tc,Beccaria:2019stp,Beccaria:2019mev,Beccaria:2019dju,Beccaria:2020qtk}). However, independent verifications could be done through integrability along the lines of \cite{Grabner:2020nis}. \par \vspace{2mm}
A non-perturbative result stemming from this analysis is the double scaling limit of the correlator, which can be found at all orders from  unmixed first-order CFT data and gives
\begin{align}
	h_{\text{DS}}(\mu)&=\kappa^{-1}U\left(1,1,\kappa^{-1}\right)
\end{align}
where $\kappa = \frac{\text{e}^t}{4  \sqrt{2\lambda }} $ and $U$ is a confluent hypergeometric function of the second kind, which matches equation (3.20) in  \cite{Giombi:2022pas} for the fermionic displacement superfield with $\Delta=\frac 12$ and a correction to the Lyupanov exponent.\par \vspace{2mm}

\vspace{6mm}
A natural development of this work would be to solve the mixing at this order to push the bootstrap process to higher orders and find the unmixed spectrum. This could be done along the lines of \cite{Ferrero:2021bsb} by looking at mixed four-point functions of protected operators and long (unprotected) operators. Another method is to look at higher-point functions. The next section will present this in the case of $\mc{N}=4$ SYM. \par 
An interesting direction concerns topological sectors, hopefully amenable to localization, These could  appear considering other supermultiplets \cite{Bliard-Ferrero}. A fascinating case is that of the 1/3 BPS operators, of which $\mathbb{F}\mathbb{O}^a$ is an example; they are charged under R-symmetry and seem to have similar properties to the 1/2-BPS operators on the Wilson line defect in $\mc{N}=4$ SYM studied in\cite{Giombi:2017cqn, Liendo:2018ukf, Ferrero:2021bsb}. The R-symmetry offers the potential of constructing topological operators and computing exact correlation fucntions through localisation~\cite{Giombi:2009ds, Giombi:2018qox,Giombi:2018hsx}.  \par 
For less symmetric systems where all-order data may not be available to fix the bootstrap expansion parameter, finite-coupling data for the correlators of interest here may also be obtained with lattice field theory methods applied to the string worldsheet, discretizing the Lagrangian of~\cite{Uvarov:2009nk} expanded around the  minimal surface corresponding to the 1/2-BPS line, and using Monte Carlo techniques on the lines of~\cite{Bianchi:2016cyv, Forini:2016sot, Forini:2017mpu,Forini:2017ene, Bianchi:2019ygz} for the correlators of the worldsheet excitations.

\chapter{Bootstrapping Higher-Point Functions in $\mc{N}=4$ SYM dCFT}\label{chapter N=4}
\begin{chapquote}{Jacques Brel, \textit{La valse a mille temps} \ref{Brel}}
	Une valse à mille temps\\
	Offre seule aux amants\\
	Trois cent trente-trois fois le temps\\
	De bâtir un roman
\end{chapquote}

The 1/2-BPS Wilson line defect in $\mathcal{N}=$4 SYM  has extensively been studied at weak \cite{Kiryu:2018phb} and strong coupling \cite{Giombi:2017cqn} using Feyman diagrammatics \cite{Barrat:2021tpn}, Witten diagrams \cite{Giombi:2017cqn}, the analytic bootstrap \cite{Liendo:2016ymz,Liendo:2018ukf,Ferrero:2021bsb} and integrability through the quantum spectral curve \cite{Cavaglia:2021bnz}. 
The extensive work on this defect motivates the search for more complex quantities to compute, such as higher $n$-point functions.\footnote{At the time of writing and submitting the Thesis, the important advances made in \cite{Giombi:2023zte} had not yet been published} In \cite{Barrat:2021tpn}, the correlators at weak coupling of operator insertions were computed through a recursion relation relating 1-loop lower point functions to 1-loop $n-$point functions. At strong-coupling, such a relation would be trivial at the same loop order since the corresponding Witten diagrams are disconnected. As such, $n$-point functions factorise into products of lower point functions until the perturbative order of the first connected diagram.\footnote{The first connected diagram appears at the $\tfrac n2-1^{\text{st}}$ order for massless excitations.} However, the analysis of higher-point correlators\cite{Barrat-Bliard}, even at GFF or at the next order, can be illuminating in terms of resolving the mixing and finding the free OPE data. \par \vspace{3mm}
In this Chapter, the non-vanishing five-point correlator\footnote{The correlator we study corresponds to a pinching limit of the six-point function of the displacement multiplet.} of 1/2-BPS insertions on the 1/2-BPS Wilson line is studied. The superblocks are obtained by requiring the multipoint Ward Identities in \cite{Barrat:2021tpn} to be satisfied. The CFT data for the free theory is obtained, and the first-order solution is bootstrapped, which confirms the statement in \cite{Ferrero:2021bsb} that the anomalous dimension does not mix at first order. 
\newpage
\section{Symmetry and Operators}
The 1/2-BPS Wilson in $\mc{N}=4$ SYM breaks the overall symmetry to $OSP(4^*|4)$. The displacement multiplet preserves half the supersymmetries of the line and is the lightest 1/2-BPS operators in the family $\mc{D}_k$ \cite{Liendo:2016ymz} where $k$ labels the R-symmetry ($SP(4)$) irrep of the superconformal primary. The structure of the multiplets $\mc{D}_1$ is \cite{Liendo:2018ukf}. 
\begin{align}
	\mc{D}_1:\qquad [0,1]^{s=0}_{\Delta=1} \rightarrow [1,0]^{s= \tfrac 12}_{\Delta=\tfrac 32} \rightarrow [0,0]^{s=1}_{\Delta=2}
\end{align}
where the displacement operator is a superconformal descendent of the lowest weight operator $[0,1]$. The family of operators can be obtained by considering the chiral ring generated by $\mc{D}_1$ and can be obtained explicitly by pinching several of these operators together, a process used in \cite{Barrat:2021tpn} to relate higher-point functions to higher-weight four-point functions at weak coupling. 
In the following, we will consider the four-point correlator
\begin{align}
	\langle \mc{D}_1 \mc{D}_1 \mc{D}_1 \mc{D}_1 \mc{D}_2 \rangle = \tilde{\mc{A}}(y_i, x_i, \theta_i)
\end{align}
Where the R-symmetry dependence is encapsulated in the cross-ratios $\{r_1,r_2,s_1,s_2,t\}$ obtained by contracting the free indices of the operators with a null vector $y_a$. This correlator is a function of the superspace coordinates \cite{Liendo:2016ymz}
\begin{align}
X= 	\begin{pmatrix}
		x \e^{ab} & \theta^{a \beta} \\
		\theta^{b \alpha} & y^{(\alpha \beta)}
	\end{pmatrix}
\end{align}
which are useful to derive the Ward Identities satisfied by the correlator and to find the correlation functions of all the superconformal descendants through differential equations such as above in \eqref{corrF}. However, if we instead consider the bosonic part of this correlation function, it will be a function of two spacetime cross-ratios and 5 R-symmetry cross-ratios
\begin{equation}
	\langle 11112\rangle  = \frac{(u_1\cdot u_4) (u_2\cdot u_5) (u_3\cdot u_5)}{\tau_{14}^2 \tau_{25}^2 \tau_{35}^2}
	\mc{A}(\chi_1,\chi_2; r_1,r_2,s_1,s_2,t)\,,
\end{equation}
where
\begin{align}
	\chi_1 &= \frac{\tau_{12} \tau_{45}}{\tau_{14} \tau_{25}}\,& \chi_2 &= \frac{\tau_{13} \tau_{45}}{\tau_{14} \tau_{35}} \\
	r_1 =&\, \zeta_1 \zeta_2 = \frac{y_{12}^2 y_{45}^2}{y_{14}^2 y_{25}^2}\,, &s_1 &= (1-\zeta_1)(1- \zeta_2) = \frac{y_{15}^2 y_{24}^2}{y_{14}^2 y_{25}^2}\,, \\
	r_2 =&\, \eta_1 \eta_2 = \frac{y_{13}^2 y_{45}^2}{y_{14}^2 y_{35}^2}\,, & s_2 &= (1-\eta_1)(1- \eta_2) = \frac{y_{15}^2 y_{34}^2}{y_{14}^2 y_{35}^2}\,, \\
	t =&\, \frac{y_{15}^2 y_{23}^2 y_{45}^2}{y_{14}^2 y_{25}^2 y_{35}^2}\,,
\end{align}
These should be related to the bosonic pieces of the eigenvalues of the superMatrix cross-ratios. The correlation function above can be written explicitly as
\begin{equation}
	\mc{A} = F_0 + \frac{r_1}{\chi_1^2} F_1 + \frac{s_1}{(1-\chi_1)^2} F_2 + \frac{r_2}{\chi_2^2} F_3 + \frac{s_2}{(1-\chi_2)^2} F_4 + \frac{t}{\chi_{12}^2} F_5\,,
	\label{eq:5ptRchannels}
\end{equation}
where each of the $F_i$ are functions of the two spacetime cross-ratios $\{\chi_1,\chi_2\}$. 
\section{Ward Identities}
An extension to the Ward identities in \cite{Liendo:2016ymz} for higher-point correlation functions was conjectured in \cite{Barrat:2021tpn} as :
\begin{align}\label{Eq: WardId}
	\sum_{k=1}^{n-3} \left(\frac{1}{2}\partial_{\chi_k}+\alpha_k\partial_{r_k}-(1-\alpha_k)\partial_{s_k} \right)\mathcal{A}_{\Delta_1...\Delta_n}\raisebox{-.2cm}{$\Biggr|$}_{\raisebox{+.2cm}{$\begin{subarray}{l}
				r_i\rightarrow \alpha_i \chi_i \\s_i\rightarrow (1-\alpha_i)(1-\chi_i)\\t_{ij}\rightarrow (\alpha_i-\alpha_j)(\chi_i-\chi_j)
			\end{subarray}$}} =  0
\end{align}
\par 
where
\begin{align}
	\chi_{i-1} &=\frac{x_{1i}x_{n-1,n}}{x_{in}x_{1,n-1}}\\
	r_{i-1}&=\frac{(u_{1}\cdot u_{i})(u_{n-1}\cdot u_{n})}{(u_{i}\cdot u_{n} )(u_{1}\cdot u_{n-1})}\\
	s_{i-1}&=\frac{(u_{1}\cdot u_{n})(u_{i}\cdot u_{n-1})}{(u_{i}\cdot u_{n} )(u_{1}\cdot u_{n-1})}\\
	t_{i-1,j-1} &=\frac{(u_{i}\cdot u_{j})(u_{1}\cdot u_{n})(u_{n-1}\cdot u_{n})}{(u_{i}\cdot u_{n} )(u_{j}\cdot u_{n} )(u_{1}\cdot u_{n-1})},
\end{align}
This was found under the minimal assumptions the conformal prefactors should vanish under its action, that it should be linear,  have single derivatives and be related to the topological sector. This identity was verified for various correlators at weak and strong coupling \cite{Bliard:2022xsm}. 

This Ward identity applied to the five-point correlator reduces the problem of 6 independent functions of two variables to one of three functions and a constant which corresponds to the topological sector
\begin{align}\label{definition N=4 f-functions}
		\mc{A} (\chi_1,\chi_2; \zeta_1, \zeta_2, \eta_1, \eta_2, t) = \mathds{F} + \sum_{i=1,2,3} \mathds{D}_i f_i (\chi_1, \chi_2)\,,
\end{align}
where we have defined the differential operators
\begin{align}
	\mathds{D}_{1} :=&\, (v_1 + v_2) + v_1 v_2 (\partial_{\chi_1} + \partial_{\chi_2})\,, \\
	\mathds{D}_{2} :=&\, (w_1 + w_2) + w_1 w_2 (\partial_{\chi_1} + \partial_{\chi_2})\,, \\
	\mathds{D}_{3} :=&\, z + \chi_{12}(v_1 v_2 - w_1 w_2) (\partial_{\chi_1} + \partial_{\chi_2})\,,\\
		v_i =& \chi_1-\zeta_i\qquad w_i=\chi_2-\eta_i \qquad z=\chi_{12}^2-t.
\end{align}
with the use of more convenient variables in the topological limit.
The functions are related by 
\begin{align}
	f^0 &= \sum_{i=0}^5 F_i (\chi_1,\chi_2)\\
	f_1 &= -\frac{F_1 (\chi_1,\chi_2)}{\chi_1}+ \frac{F_2 (\chi_1,\chi_2)}{1-\chi_1}\\
	f_2 &= - \frac{F_3 (\chi_1,\chi_2)}{\chi_2}+ \frac{F_4 (\chi_1,\chi_2)}{1-\chi_2}\\
	f_3 &= -\frac{F_5 (\chi_1,\chi_2)}{(\chi_1-\chi_2)^2}
\end{align}
The different solutions to ward identities correspond to the different blocks exchanged in the relevant OPE limits. Therefore, the Ward Identities should be enough to fully constrain the form of the superconformal blocks. This analysis is shown below.
\section{Superconformal Blocks}\label{Section: superblocks}
\begin{figure}[h]
	\centering
	\begin{tikzpicture}
		\draw[] (0,-2) -- (0,2);
		\draw[] (0,0) -- (1,0);
		\draw[] (0,1) -- (-1,1);
		\draw[] (0,-1) -- (-1,-1);
		\node[anchor=west] at (1,0) {$\mc{D}_2(X_5)$};
		\node[anchor=east] at (-1,1) {$\mc{D}_1(X_3)$};
		\node[anchor=east] at (-1,-1) {$\mc{D}_1(X_2)$};
		\node[anchor=south] at (0,2) {$\mc{D}_1(X_4)$};
		\node[anchor=east] at (0,-2) {$\mc{D}_1(X_1)$};
		\node[anchor=west] at (0,.5) {$\mathcal{I}+\mc{D}_2+\mathcal{L}_{h_1}$};
		\node[anchor=west] at (0,-.5) {$\mathcal{I}+\mc{D}_2+\mathcal{L}_{h_2}$};
	\end{tikzpicture}
	\caption{OPE diagram for the five-point function $\langle \mc{D}_1 \mc{D}_1 \mc{D}_1 \mc{D}_1 \mc{D}_2 \rangle $ for the symmetric OPE channel, where the same operators are exchanged from (12) and (34).}
	\label{Figure: OPE diagram symmetric}
\end{figure}
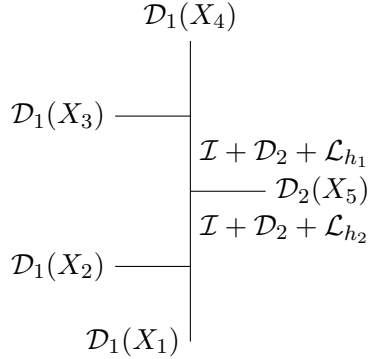\par \vspace{2mm}
The R-symmetry selection rules \cite{Liendo:2016ymz} and the Ward Identities \cite{Barrat:2021tpn} fix the conformal blocks entirely. In the symmetric channel, which corresponds to the OPE between points $(12)$ and $(34)$, the operators exchanged are 
\begin{align}
	\mc{D}_1 \times \mc{D}_1  \sim \mathcal{I}+\mc{D}_2+\sum_h \mathcal{L}^{h>1}_{0,[0,0]}
\end{align}
which can be summarized in the OPE diagram in Figure \ref{Figure: OPE diagram symmetric}\par
In this expansion, the non-vanishing blocks are given by
\begin{align}
	\mc{A} =  &c_{011} c_{022}c_{112}\left(\mathcal{G}_{\mathcal{I},\mathcal{D}_2}+\mathcal{G}_{\mathcal{D}_2,\mathcal{I}}\right) +c_{112}^2c_{222}\mathcal{G}_{\mathcal{D}_2,\mathcal{D}_2}\nonumber\\
	&+\sum_{h_2}c_{112}c_{11h_2}c_{22h_2}\mathcal{G}_{\mathcal{D}_2,\mathcal{L}_{h_2}}+\sum_{h_1}c_{112}c_{11h_1}c_{22h_1}\mathcal{G}_{\mathcal{L}_{h_1},\mathcal{D}_2} \nonumber \\
	&+\sum_{h_1,h_2}c_{11h_1}c_{11h_2}c_{h_1h_22}\mathcal{G}_{\mathcal{L}^{h_1}_{0,[0,0]},\mathcal{L}^{h_2}_{0,[0,0]}}
\end{align}
where the OPE coefficient $c_{ijk}$ are labelled by the conformal weights of the Short/Long operators and the superblocks $\mathcal{G}_{A,B}$ are functions of the spacetime cross-ratios as well as the R-symmetry cross-ratios. Their explicit form will be presented below. They represent the sum of the conformal blocks from the descendants of the exchanged fields. The multiplets for the exchanged operators can be found in from \cite{Ferrero:2021bsb}. Each conformal block will be weighted by an R-symmetry block corresponding to the representations of the given descendent of operators being exchanged. In this case, the R-symmetry selection rules between these and the remaining operator ($\mc{D}_2$) will restrict the possible terms. Additionally, only the uncharged terms under the $SO(3)$ transverse spin will contribute. 
The relevant scalar blocks are (see \ref{App:five-point scalar bloc})
\begin{align}\label{Eq: scalar five-point block}
	g_{h_1,h_2} =\chi_1^{h_1-2}(1-\chi_2)^{h_2-2}F_2(h_1+h_2-2,h_1,h_2,2h_1,2h_2;\chi_1,1-\chi_2)
\end{align}
where $F_2$ is the Appell Hypergeometric function.
Since the correlator can be expressed in terms of a constant topological term and three functions, it is useful to write the expression of the blocks in terms of this decomposition. This leads to the results in the following sections, where the  superblocks have the same complexity as the scalar blocks leading to efficient expansions and relatively elegant expressions. Additionally, the vanishing of the topological term for superblocks of long operators is a non-trivial check of the validity of these expressions. 
In this channel, the topological limit is given by 
\begin{align}
	f^0 = c_{\mc{D}_2,\mc{I}}+c_{\mc{I},\mc{D}_2}+c_{\mc{D}_2,\mc{D}_2}
\end{align}

The process of fixing these correlators is somewhat intricate and long, in addition to being needed for over ten different superblocks (counting those in the asymmetric channel), so the derivation will be illustrated with the simplest non-trivial superblock $\mathcal{G}_{\mathcal{D}_2,\mathcal{D}_2}$. The process was repeated with all the blocks listed in the Appendices  \ref{App: superblocks symmetric} and \ref{App: superblocks asymmetric} where the superconformal blocks are all fully constrained by the Ward Identity above, including $\mathcal{G}_{\mathcal{L}^{h_1}_{0,[0,0]},\mathcal{L}^{h_2}_{0,[0,0]}}$ and $\mathcal{G}_{\mathcal{L}^{h_1}_{0,[0,0]},\mathcal{L}^{h_2}_{0,[0,1]}}$ which are each a sum of over 50 scalar conformal blocks. This reveals the true power of these multipoint Ward identities, which fix all the unknown constants. \par 
\subsection*{$\mathcal{G}_{\mathcal{D}_2,\mathcal{D}_2}$}
The superblock $\mathcal{G}_{\mathcal{D}_2,\mathcal{D}_2}$ includes the contributions from the superprimary $[0,2]^{\Delta=2}_0$ as well as all its super descendants with vanishing transverse spin (charge under the $SO(3)$). These are 
\begin{align}
	&[0,2]^2_0 &&[2,0]_0^3&&[0,0]^4_0
\end{align}
where, in the following, we will drop the transverse spin label as it will always vanish. There is an additional selection rule from the R-symmetry being applied at the remaining OPE vertex $\mc{D}_2(5)$ in \ref{Figure: OPE diagram symmetric} where the superconformal primary $[0,2]^2$ imposes the constraint that the operator exchanged on the right must be in the product of the two on the left, i.e 
\begin{align}
	\begin{matrix}
		[0,2]\\ [2,0]\\ [0,0]
	\end{matrix} 
\subset [0,2] \otimes 
 \begin{matrix}
	[0,2] \\ [2,0] \\ [0,0]
\end{matrix} 
\end{align}
Applying these selection rules, we obtain the following decomposition
\begin{align}
	\mathcal{G}_{\mathcal{D}_2,\mathcal{D}_2}& = \beta_1 h_{[0,0][0,2]} g_{4,2}+ \beta_2 h_{[0,2][0,0]} g_{2,4}+ \beta_3 h_{[0,2][2,0]} g_{2,3}\non \\
	&+ \beta_4h_{[2,0][0,2]} g_{3,2}+ \beta_5 h_{[0,2][0,2]} g_{2,2}+ \beta_6 h_{[2,0][2,0]} g_{3,3}
\end{align}
where $g_{h_1,h_2}$ are the scalar conformal blocks in \eqref{Eq: scalar five-point block}, $h_{[a_1,b_1],[a_2,b_2]}$ are the R-symmetry blocks, solving the $SO(5)$ quadratic Casimir equation and $\beta_i$ are unknown constants to be determined up to an overall factor. The Ward Identities constrain the coefficients up to an overall constant which is normalised to  $\beta_5=\frac{-1}{5}$ for the lowest conformal weight $g_{2,2}$ to match the normalisation conventions.
\begin{align}
	\beta_1 &= \frac{3}{70}\beta_5&\beta_2&= -\frac{3}{70}\beta_5&\beta_3&= -\frac{1}{4}\beta_5&\beta_4&= \frac{1}{4}\beta_5&\beta_6&=\frac{3}{40}\beta_5
\end{align}
 This gives a fully fixed superconformal block which depends on the two spacetime and 5 R-symmetry cross-ratios. In turn, this block can be decomposed like the correlator in terms of $F_i$ and $f_i$ functions. In terms of the Ward identity channels, this gives
\begin{align}\label{Eq: G22 bloc}
	f^0_{\mc{D}_2,\mc{D}_2}& = 1\\
	f^1_{\mc{D}_2,\mc{D}_2}& =\sum_{k,l}\frac{ (k+1) (l+1) (2 (k+1) l+3 k+2)}{\nu_2^2 \chi_1^2 (k+l+1)}G_{\mc{D}_2,\mc{D}_2} ^{k,l}\\
	f^2_{\mc{D}_2,\mc{D}_2}& =-\sum_{k,l}\frac{ (k+1) (l+1) (2 k (l+1)+3 l+2)}{\nu_2^2 \chi_1^2 (k+l+1)}G_{\mc{D}_2,\mc{D}_2} ^{k,l}\\
	f^3_{\mc{D}_2,\mc{D}_2}&=- \sum_{k,l}\frac{(k+1) (k+2) (l+1) (l+2)}{2\nu_2^2\chi_1^2}	G_{\mc{D}_2,\mc{D}_2} ^{k,l}
\end{align}
One can define the bloc
\begin{align}
	G_{\mc{D}_2,\mc{D}_2} ^{k,l}= \frac{ 3 (2)_{k+l}}{(5)_k (5)_l  }\chi_1^{k+2} \nu_2^{l+2}\\
	\sum_{k,l}	G_{\mc{D}_2,\mc{D}_2} ^{k,l} = 3 \chi_1^{2} \nu_2^{2}F_2\left(
	2,1,1;5,5\,|\,\chi_1,\nu_2\right)
\end{align}
and the function $	f^3_{\mc{D}_2,\mc{D}_2}$ is then the derivative of this Appell function. 
What is remarkable in these explicit forms of the blocks above \eqref{Eq: G22 bloc} and listed in the Appendix \ref{App: superblocks symmetric} is that the superconformal blocks are simple functions of the cross-ratios (essentially of the same form as the Appell functions of the scalar conformal blocks), even in the case where the functions came from the sum of 50 terms.\footnote{This being said, one must be careful of which variable to use in the OPE expansion as an awkward choice can lead to quite cumbersome expressions.}
\subsection*{Superconformal Blocks in the Asymmetric Channel}
In the other channel, where we have the OPE expansion 
\begin{align}
	\mc{D}_1 \times \mc{D}_2 \sim \mc{D}_1+\mc{D}_3+\sum_h \mathcal{L}^{h}_{0,[0,1]}
\end{align}
in addition to the one above. This can be seen in the diagram
\begin{figure}[h]
\centering
\begin{tikzpicture}
	\draw[] (-1,1) -- (-1,2);
	\draw[] (5,1) -- (-2,1);
	\draw[] (1.5,1) -- (1.5,2);
	\draw[] (4,1) -- (4,2);
	\node[anchor=west] at (5,1) {$\mc{D}_2(X_5)$};
	\node[anchor=east] at (-2,1) {$\mc{D}_1(X_1)$};
	\node[anchor=south] at (4,2) {$\mc{D}_1(X_4)$};
	\node[anchor=south] at (-1,2) {$\mc{D}_1(X_2)$};
	\node[anchor=south] at (1.5,2) {$\mc{D}_1(X_3)$};
	\node[anchor=west] at (-1.5,.5) {$\mathcal{I}+\mc{D}_2+\mathcal{L}^{h_1}_{0,[0,0]}$};
	\node[anchor=west] at (1.8,.5) {$\mc{D}_1+\mc{D}_3+\mathcal{L}^{h_2}_{0,[0,1]}$};
\end{tikzpicture}
\caption{OPE diagram for the five-point function $\langle \mc{D}_1 \mc{D}_1 \mc{D}_1 \mc{D}_1 \mc{D}_2 \rangle $ for the Asymmetric OPE channel, where the same operators are exchanged from (12) and (45).}
\label{Figure: OPE diagram asymmetric}
\end{figure}
The OPE limits in this diagram can be seen at the vertices, and we can choose variables that vanish in these limits; for example, it will be useful to write the results in terms of 
\begin{align}
	\nu_1&=\frac{\chi_1-\chi_2}{1-\chi_2}& \nu_2& = 1-\chi_2
\end{align}
In this expansion, the non-vanishing blocks are given by
\begin{align}
	\mc{A} =  &c_{110}^2 c_{112}\mathcal{G}_{\mathcal{I},\mathcal{D}_1} +c_{112}^3\mathcal{G}_{\mathcal{D}_2,\mathcal{D}_1}+c_{112}c_{123}^2\mc{G}_{\mc{D}_2,\mc{D}_3}+\sum_{h_1} c_{112}c_{11h_1}^2\mc{G}_{\mc{L}^{h_1}_{0,[0,0]},\mc{D}_1} \\
	&+\sum_{h_1} c_{11h_1}c_{13h_1}c_{123}\mc{G}_{\mc{L}^{h_1}_{0,[0,0]},\mc{D}_3} +\sum_{h_2} c_{112}c_{12h_2}^2\mc{G}_{\mc{D}_2,\mc{L}^{h_2}_{0,[0,1]}}+\sum_{h_1,h_2}c_{11h_1}c_{12h_2}c_{1h_1h_2} \mc{G}_{\mc{L}^{h_1}_{0,[0,0]},\mc{L}^{h_2}_{0,[0,1]}} \non
\end{align}

In particular, the topological limit is given by 
\begin{align}
	g^0 = c_{\mc{D}_1,\mc{I}}+c_{\mc{D}_2,\mc{D}_1}+c_{\mc{D}_2,\mc{D}_3}
\end{align}
\section{Free Theory Data}\label{N=4 free theory data}
We can find the free theory OPE coefficients in both channels from the block expansion above. 
The Free result is obtained by Wick contracting the elementary fields and gives\footnote{To do this, we compute the six-point GFF result and pinch points 5 and 6.}
\begin{align}
\mc{A} = \sqrt{2} \left(\frac{r_1}{\chi_1^2}+\frac{r_2}{\chi_2^2}+\frac{s_1}{(\chi_1-1)^2}+\frac{s_2}{(\chi_2-1)^2}+\frac{t}{(\chi_1-\chi_2)^2}+1\right)
\end{align}
which gives the $F_i = \sqrt{2}$ and 
\begin{align}
	f_{0}^{(0)}&=6 \sqrt{2}\\
	f_1^{(0)}&=\frac{\sqrt{2} (1-2 \chi_1)}{(\chi_1-1) \chi_1}\\
	f_2^{(0)}&=\frac{\sqrt{2} (1-2 \chi_2)}{(\chi_2-1) \chi_2} \\
	f_3^{(0)}&=-\frac{\sqrt{2}}{(\chi_1-\chi_2)^2}
\end{align}
For the symmetric channel, this gives us the following:
\begin{align}
	&c_{\mc{I},\mc{D}_2}=c_{\mc{D}_2,\mc{I}} = c_{112}=\sqrt{2}\\
	&c_{\mc{D}_2,\mc{D}_2} = c_{112}^2c_{222} = 4\sqrt{2}\\
	&c_{\mc{L}^{\Delta}_{0,[0,0]},\mc{D}_2} = c_{\mc{D}_2, \mc{L}^{\Delta}_{0,[0,0]}} =c_{112}c_{11\Delta}c_{22\Delta}=\left(\frac{1+(-1)^\Delta}{2}\right) \frac{\sqrt{\pi } 2^{-2 \Delta +\frac{1}{2}} (\Delta -1) \Gamma (\Delta +3)}{\Gamma \left(\Delta +\frac{3}{2}\right)}\\
	&c_{\mc{L}^{\Delta_1}_{0,[0,0]},\mc{L}^{\Delta_2}_{0,[0,0]}}=c_{11\Delta_1}c_{11\Delta_2}c_{2\Delta_1\Delta_2}= -\frac{\left(\pi  \, 2^{-2 \Delta_1-2 \Delta_2-\frac{3}{2}} (\Delta_1)_3 (\Delta_2)_3 \Gamma (\Delta_1+\Delta_2)\right)}{\Gamma \left(\Delta_1+\frac{3}{2}\right) \Gamma \left(\Delta_2+\frac{3}{2}\right)}
\end{align}
where the last coefficient is vanishing for $\Delta_1$ and $\Delta_2$ odd.
For the asymmetric channel, this gives:
\begin{align}
	&c_{\mc{I}, \mc{D}_1}=c_{110}^2c_{112}= \sqrt{2}\\
	&c_{\mc{D}_2, \mc{D}_1}=c_{112}^3 =  2\sqrt{2}\\
	&c_{\mc{D}_2, \mc{D}_3}=c_{112}c_{123}^2 =  3\sqrt{2}\\
	&c_{\mc{L}^{\Delta}_{0,[0,0]}, \mc{D}_1}	=c_{112}c_{11\Delta}^2=\left(\frac{1+(-1)^{\Delta}}{2}\right) \frac{\sqrt{\pi } 2^{-2 \Delta-\frac{1}{2}} (\Delta-1) \Gamma (\Delta+3)}{\Gamma \left(\Delta+\frac{3}{2}\right)}\\
	&c_{\mc{D}_2,\mc{L}^{\Delta}_{0,[0,1]}}= c_{112}c_{12\Delta}^2=-\frac{\sqrt{\pi } 2^{-2 \Delta -\frac{7}{2}} (\Delta -2) \left((-1)^{\Delta } (\Delta +1) (\Delta +2)-12\right) \Gamma (\Delta +4)}{3 (\Delta +1) \Gamma \left(\Delta +\frac{3}{2}\right)}\\
	&c_{\mc{L}^{\Delta_1}_{0,[0,0]},\mc{L}^{\Delta_2}_{0,[0,1]}}=c_{11\Delta_1}c_{12\Delta_2}c_{1\Delta_1\Delta_2}\\
	&=\frac{1+(-1)^{\Delta_1}}{2}\frac{2\sqrt{2}(\Delta_1-1) \Delta_1 (\Delta_1+1) (\Delta_2-\Delta_1) \Gamma (\Delta_1+3) \Gamma (\Delta_2-1) \Gamma (\Delta_1+\Delta_2+4)}{\Gamma (2 \Delta_1+3) \Gamma (2 \Delta_2+3)}\non \\
	& \text{for} \quad \Delta_2>\Delta_1 \quad\text{otherwise}\non \\ &c_{\mc{L}^{\Delta_1}_{0,[0,0]},\mc{L}^{\Delta_2}_{0,[0,1]}}=0 
\end{align}
This gives the OPE coefficients for the protected operators
\begin{align}
	c_{112} &= \sqrt{2}\\
	c_{222} &= 2\sqrt{2}\\
	c_{123}&=\sqrt{3}
\end{align}
which agree with \cite{Giombi:2018qox}
Those of the uncharged long operators 
\begin{align}
	c_{11\Delta} &= \left(\frac{1+(-1)^{\Delta}}{2}\right) \sqrt{ \frac{\sqrt{\pi } 2^{-2 \Delta-1} (\Delta-1) \Gamma (\Delta+3)}{\Gamma \left(\Delta+\frac{3}{2}\right)}}\\
	c_{22\Delta} &=2 c_{11\Delta}\\
	c_{2\Delta_1,\Delta_2}&= \sqrt{\frac{2\Delta_1^2 (\Delta_1+1) (\Delta_1+2)\Delta_2^2 (\Delta_2+1) (\Delta_2+2) \Gamma (\Delta_1+\Delta_2)^2}{(\Delta_1-1) (\Delta_2-1) \Gamma (2\Delta_1+2) \Gamma (2\Delta_2+2)}}\qquad \{\Delta_1,\Delta_2\}\, \text{even}
\end{align}
where the last expression is only valid when both $\Delta_1$ and $\Delta_2$ are even given the form of the OPE expansion. \par  
For the charged channel, just as for the protected operators, the Ward Identities only constrain the blocks up to a constant. Unlike the protected operators, no topological sector fixes this constant. We thus impose a minimal condition of reality and choose a normalisation factor of $-\frac{(-1)^{\Delta +1} \left((-1)^{\Delta } (\Delta +1) (\Delta +2)-12\right)}{\Delta +1}$ for the $\mc{G}_{\mc{D}_2,\mc{L}^\Delta_{0,[0,1]}}$ block which gives the OPE coefficient for the charged long and mixed uncharged/charged long, respectively:
\begin{align}
	c_{12\Delta} &= \sqrt{\frac{\sqrt{\pi } 2^{-2 \Delta -4} (\Delta -2) \Gamma (\Delta +4)}{3 \Gamma \left(\Delta +\frac{3}{2}\right)}}\\
	c_{1\Delta_1\Delta_2}&= \sqrt{\frac{12 \Delta_1 \Delta_2 (\Delta_2-1) (\Delta_1-\Delta_2)^2 \left((\Delta_2-1)_4\right){}^2 \Gamma (\Delta_1+\Delta_2+4)^2}{\Gamma (2 \Delta_1+2) \Gamma (2 \Delta_2+2) \left((\Delta_1-1)_3\right){}^2 (\Delta_1-1)_4 (\Delta_2-2)_6}}\qquad \Delta_1\, \, \text{even and }\, \Delta_2>\Delta_1
\end{align}
One can already recognise patterns observed in four-point functions, such as the absence of uncharged long operators of odd dimensions. This is, however, not the case for the charged long operators $\mc{L}^{\Delta_2}_{0,[0,1]}$. For the final OPE coefficient, we observe a bound which seems to come from the OPE
\begin{align}
	\mc{D}_1 \times \mc{L}^{\Delta_1}_{0,[0,0]} \sim \sum_{\Delta_2>\Delta_1}  \mc{L}^{\Delta_2}_{0,[0,1]} +...
\end{align}
\section{Conformal Bootstrap}\label{N=4 conformal bootstrap}
We will focus on bootstrapping the  function$f_{3}$ since it has both braiding and crossing symmetry and deducing the rest of the functions from there. The crossing is
\begin{align}
	f_{3}(\chi_1,\chi_2)&=f_{3}(1-\chi_2,1-\chi_1)
\end{align}
and the braiding is 
\begin{align}
	f_{3}(\chi_1,\chi_2)&\simeq f_{3}(\chi_2,\chi_1). 
\end{align}
For the first order functions, we start with the Ansatz which satisfies this crossing and braiding
\begin{align}
	f_{3}^{(1)}(\chi_1,\chi_2)& = p(\chi_1,\chi_2)+r_1(\chi_1,\chi_2)\log(\chi_1)+r_1(\chi_2,\chi_1)\log(\chi_2) +r_1(1-\chi_1,1-\chi_2)\log(1-\chi_1)\non \\
	&+r_1(1-\chi_2,1-\chi_1)\log(1-\chi_2)+r_5(\chi_1,\chi_2)\log(\chi_2-\chi_1)
\end{align}
where
\begin{align}
	p(\chi_1,\chi_2) &= p(\chi_2,\chi_1)\\
	p(\chi_1,\chi_2) &= p(1-\chi_2,1-\chi_1)\\
	r_5(\chi_1,\chi_2) &= r_5(\chi_2,\chi_1)\\
	r_5(\chi_1,\chi_2) &= r_5(1-\chi_2,1-\chi_1)
\end{align}
The easiest way to see the effect of crossing and braiding is for the individual R-symmetry channels where the relevant transformations are:
\begin{align}
	F_0(\chi_1,\chi_2)&\simeq F_5\left(\frac{\chi_2-1}{\chi_2-\chi_1},\frac{\chi_2}{\chi_2-\chi_1}\right)\\
	F_1(\chi_1,\chi_2)&\simeq F_5(\frac{\chi_2-\chi_1}{\chi_2-1},\frac{\chi_2}{\chi_2-1})\\
	F_2(\chi_1,\chi_2) &\simeq F_5(\frac{\chi_1}{\chi_2},\frac{1}{\chi_2})\\
	F_3(\chi_1,\chi_3) &\simeq F_5(\frac{\chi_1}{\chi_1-1},\frac{\chi_2-\chi_1}{1-\chi_1})\\
	F_4(\chi_1,\chi_2)&\simeq F_5(\frac{1}{\chi_1},\frac{\chi_2}{\chi_1})
\end{align}
So there are only three rational functions to fix.  The rational functions can only have poles when operators collide; that is 
\begin{align}
	\{\chi_1,\chi_2,1-\chi_1,1-\chi_2,\chi_2-\chi_1, \frac{1}{\chi_1},\frac{1}{\chi_2}\}\rightarrow 0
\end{align}
Additionally, the OPE expansion dictates the behaviour of the overall function as well as the logarithms in the OPE limits; this gives the following boundary conditions
\begin{align}
	r_1(\chi_1,\chi_2)& \sim O(\chi_1^1 ,(1-\chi_2)^0)\\
	r_5(\chi_1,\chi_2) &\sim O((1-\chi_2)^{-2},(\chi_1-\chi_2)^0)\\
	f_3(\chi_1,\chi_2)&\sim a_{\mc{D}_1,\mc{I}}(\chi_1-\chi_2)^{-2}+ O(\chi_1^0,(1-\chi_2)^0,(\chi_1-\chi_2)^{0})
\end{align}
Additionally, the anomalous dimension's growth constrains the poles' maximal order in the other variables. For example, $r_1(\chi_1,\chi_2)$ cannot have poles in $\{\chi_1,1-\chi_2\}$ so one can parametrise it as
\begin{align}
	r_1(\chi_1)&=\frac{\chi_1 \sum_{\{k,l\}=0} r_{1,kl} \chi_1^k \chi_2^l}{\chi_2^{N_{r_1}}(\chi_2-\chi_1)^{M_{r_1}}}
\end{align} 
where the growth of the anomalous dimension fixes 
\begin{align}
	N_{r_1} &= l+2&
	M_{r_1} &=l
\end{align}
Likewise, one can use the crossing symmetry to best parametrise 
\begin{align}
	r_5(\chi_1,\chi_2)&= \left(\frac{\tilde{r_5}(\chi_1,\chi_2)}{\chi_1^{N_{r_5}}}+\frac{\tilde{r_5}(\chi_2,\chi_1)}{\chi_2^{N_{r_5}}}+\frac{\tilde{r_5}(1-\chi_1,1-\chi_2)}{(1-\chi_1)^{N_{r_5}}}+\frac{\tilde{r_5}(1-\chi_2,1-\chi_1)}{(1-\chi_2)^{N_{r_5}}} \right)
\end{align}
where the growth of the anomalous dimension sets
\begin{align}
	N_{r_5} = l+1
\end{align}
Possible cancellations between the rational functions and the logarithms in the OPE limits mean that the boundary conditions for $p_1$ are relaxed from those of $f_3$ to 
\begin{align}
	p_1(\chi_1,\chi_2)&\sim O(\chi_1^{-1},(1-\chi_2)^{-1},(\chi_1-\chi_2)^{-2})
\end{align}
so we can parameterise $p_1$ accordingly as
\begin{align}
	p_1&= \frac{p_{1,1}}{(\chi_2-\chi_1)^2}+ \frac{\sum_{k,l} p_{k,l} \chi_1^k,\chi_2^l}{\chi_1 \chi_2 (1-\chi_1)(1-\chi_2)}
\end{align}
The crossing symmetry (which still relates some coefficients of $p_1$), the boundary conditions of $f_1,f_2,f_3$ from the OPE expansion, as well as the constraint that $f_0$ is a constant are all extremely constraining along with the Regge behaviour fixes all the coefficients up to an overall constant and $p_{1,1}$. At this point, two physical inputs are needed, known from integrability: The OPE coefficient of the protected operator. 
\begin{align}
	c_{\mc{D}_1\mc{D}_1\mc{D}_2} = 2-\frac{3}{2\sqrt{2\lambda}}-\frac{9}{16 \sqrt{2}\lambda}+O(\lambda^{-3/2})
\end{align}
and the value of the topological sector at large N
\begin{align}
	f_0 = \frac{6 \mathbb{I}_2^2}{\lambda\mathbb{I}_1^2} \, \frac{2(\mathbb{I}_1-2)(\mathbb{I}_1+28)+\lambda(2\mathbb{I}_1-19)}{\sqrt{3\lambda-(\mathbb{I}_1-2)(\mathbb{I}_1+10)}}
\end{align}
where 
\begin{align}
	\mathbb{I}_i = \frac{\sqrt{\lambda}I_0(\sqrt{\lambda})}{I_a(\sqrt{\lambda})}
\end{align}
and $I_a$ are modified Bessel functions of the first kind.  The explicit expansion at strong coupling is
\begin{align}
	f_0 &= 6\sqrt{2}-\frac{33}{\sqrt{2\lambda}}+\frac{189}{8\sqrt{2}\lambda}+O(\lambda^{-3/2})
\end{align}
This fixes the free rational functions to 
\begin{align}
	r_1(\chi_1) &= \frac{\sqrt{2}}{\sqrt{\lambda}}\frac{\chi_1 \left(\chi_1^2- 3\chi_1\chi_2+4\chi_2^2\right)}{\chi_2^2 (\chi_2-\chi_1)^3}\\
	r_5(\chi_1-\chi_2)&=\frac{\sqrt{2}}{\sqrt{\lambda}} \left(\frac{1}{\chi_1^2}+\frac{1}{(\chi_1-1)^2}+\frac{1}{\chi_2^2}+\frac{1}{(\chi_2-1)^2}\right)\\
	p(\chi_1,\chi_2) &=\frac{1}{2 \sqrt{2} \sqrt{\lambda }}\left(\frac{4}{(\chi_1-1) (\chi_2-1)}+\frac{4}{\chi_1 \chi_2}+\frac{19}{(\chi_1-\chi_2)^2}\right)
\end{align}
This gives a degenerate anomalous dimension in all the channels, independent of multiplying OPE coefficients. This solves the mixing between all these operators
\begin{align}
	\gamma^{(1)}_\Delta = \frac{-1}{2\sqrt{\lambda}} \Delta (\Delta+3)
\end{align}
\section{Comments}
The low dimensionality of this system, along with the high symmetry of the insertions, allows for a simple analysis of a presumably complex quantity: the five-point correlator of operator insertions on the Wilson line. The system is set up to allow the recursive analytic conformal bootstrap to be implemented, and the first-order bootstrap solution is presented where the higher-order solutions will be found in \cite{Barrat-Bliard}. The structure of the superconformal blocks is a non-trivial result in itself, which enables the bootstrap process for the defect correlators but could potentially also inform more general setups such as the bulk-defect blocks presented in section 3.4 of \cite{Liendo:2016ymz}. The bootstrap of the five-point quantity means that the CFT one can access is much greater and includes the OPE coefficients with 2 long operators. The analysis of the six-point function in \cite{Barrat-Bliard} will also shed light on the full long OPE coefficient $c_{\Delta_1\Delta_2\Delta_3}$. 

\chapter{Effective Theories in AdS$_2$}\label{Effective theories}
\begin{chapquote}{Frank Herbert, \textit{Dune}}
	``Deep in the human unconscious is a pervasive need for a logical universe that makes sense, But the real universe is always one step beyond logic.” - Princess Irulan
\end{chapquote}
A common trait of the theories above, as well as other defect theories (such as the 1/2-BPS minimal surface string solution in $\text{AdS}_3\times \text{S}^3 \times \text{S}^3\times \text{S}^1$ \cite{Correa:2021sky} and in $\text{AdS}_3\times \text{S}^3 \times T^4$), is that they have an effective theory description in $\text{AdS}_2$. For this reason, it is beneficial to understand features of QFT in $\text{AdS}_2$ in general; such as generic diagrams, properties from different interaction vertices, and tools to compute and interpret quantities within this framework effectively. \par 
As introduced in subsection \ref{Section Perturbative Witten diagrams}, boundary (and bulk) correlators in $\text{AdS}_2$ are computed using Witten diagrams. These have the same structure as Feynman diagrams. However, the difference in metric does bring complications when it comes to propagators but simplifications when it comes to the symmetries of the correlators. The conformal symmetry can help solve some of these diagrams, particularly involving exchange diagrams. Another method is to use the embedding space formalism to map the problem to flat-space coordinates, though the bulk propagators remain difficult. Yet another method uses the ``dimension-independent representation of Conformal Theories"\cite{Mack:2009mi} provided by the Mellin transform. However, this description is redundant for one-dimensional defects and does not allow for the Mellin transforming of analytically-obtained results. This Chapter will present two methods specific to one-dimensional defects  addressing these issues. The first is a tool for Witten diagrams in $\text{AdS}_2$ that uses the one-dimensional boundary to simplify direct spacetime computations of contact diagrams. The second is a Mellin formalism tailored to one-dimensional systems that provides a useful tool for studying effective theories in $\text{AdS}_2$. 
\section{Explicit Computations in AdS$_2$}\label{Section AdS2 Witten diagrams}
 Perturbative computations in Anti-de-Sitter space are done through Witten diagrams, whose structure has been studied extensively \cite{Witten:1998qj,DHoker:1999bve,DHoker:1999kzh,DHoker:1999mqo,Freedman:1998bj,Rastelli:2016nze,Zhou:2018sfz,Zhou:2020ptb,Dolan:2000ut,Penedones:2010ue,Fitzpatrick:2011ia}. Yet, the complexity relative to their flat-space counterpart is still a roadblock to perturbative analysis. There is still a search for the full equivalent of Feynman rules \cite{Paulos:2011ie}. As such, the most efficient methods for perturbative correlators  are through the conformal bootstrap \cite{Bissi:2022mrs, Gopakumar:2016cpb}. However, explicit computations remain a reliable way to make progress in perturbation theory and can provide some insight into assumptions that may simplify the bootstrap process.  First-order four-point correlators with quartic interactions in the strong coupling limit can be written in terms of $D$-functions \cite{DHoker:1999kzh}, which are four-point Witten contact diagrams. At higher order, with loops and exchanges corresponding to additional integrated bulk points, some diagrams can be related to contact integrals through differential equations \cite{DHoker:1999mqo,Rastelli:2017udc}. As such, the $n$-point $D$-functions are used beyond the first order and can be seen as a starting point to build `master integrals' for Witten diagrams. AdS$_2$ is a perfect place to look at these integrals as it provides a simple framework with relevance in its own right (e.g. in defect theories) and corresponds to a diagonal limit ($\chi= \bar{\chi}$ for four points) of its higher-dimensional counterparts.\par 
\vspace{.2cm}
Boundary correlators in AdS$_2$ enjoy a one-dimensional conformal symmetry. In higher dimensions, the conformal symmetry simplifies perturbative computations. However, the structure of AdS$_2$ provides a framework in which another elementary method can be used to compute perturbative quantities: the residue theorem. Using contour integration for one of the AdS$_2$ integrals, the contact diagram in $\lambda_n \phi_{\Delta}^n$ theory for $n$ scalars of low conformal weights is remarkably simple, leading to the results for the integral \ref{Eq: contact integral 1}
\begin{align}
	I_{\Delta=1,n}(x_i)&=\frac{\pi}{(2i)^{n-2}}\sum_{i\neq j}\frac{x_{ij}^{n-4}}{\Pi_{k\neq i\neq j}x_{ik}x_{kj}}\log\left(\frac{x_{ij}}{2i}\right),\\
	&I_{\Delta=2,n}(x_i)=\sum_i\sum_{j\neq i}\frac{-\pi}{2(2i)^{2n-4}x_{ij}^2}\partial_{x_j}\left( \frac{x_{ji}^{2n-5}}{\prod_{k\neq j,k\neq i }x_{ik}^2x_{jk}^2}\log\frac{x_{ji}}{2i}\right)\nonumber \\
	&+\sum_i \sum_{j\neq i}\partial_{x_i}\frac{-\pi}{(2i)^{2n-2}x_{ij}^2}\partial_{x_j}\left( \frac{x_{ji}^{2n-4}}{\prod_{k\neq j,k\neq i}x_{ik}^2x_{jk}^2} \log\frac{x_{ji}}{2i}\right),
\end{align}
Above, $I_{\Delta,n}(x_i)$ is the integral corresponding to the Witten contact diagram of $n$ fields $\phi_\Delta$ of conformal dimension $\Delta$ inserted at positions $x_i$ defined in \ref{Eq: contact integral 1}. The normalisation $C_{\Delta}$ in \eqref{Eq: normalisation C_Delta} and the vertex coupling have been factored out. These expressions are in terms of the operators' positions and combine naturally into the cross-ratios obtained with the usual conformal transformations (see discussion in section \ref{subsec: CFT basics} around equation \eqref{Eq: cross-ratio} and Appendix \ref{App: list of correlators}).
\subsection{Contour Integration for Witten Diagrammatics}\label{subsec:massless}
			
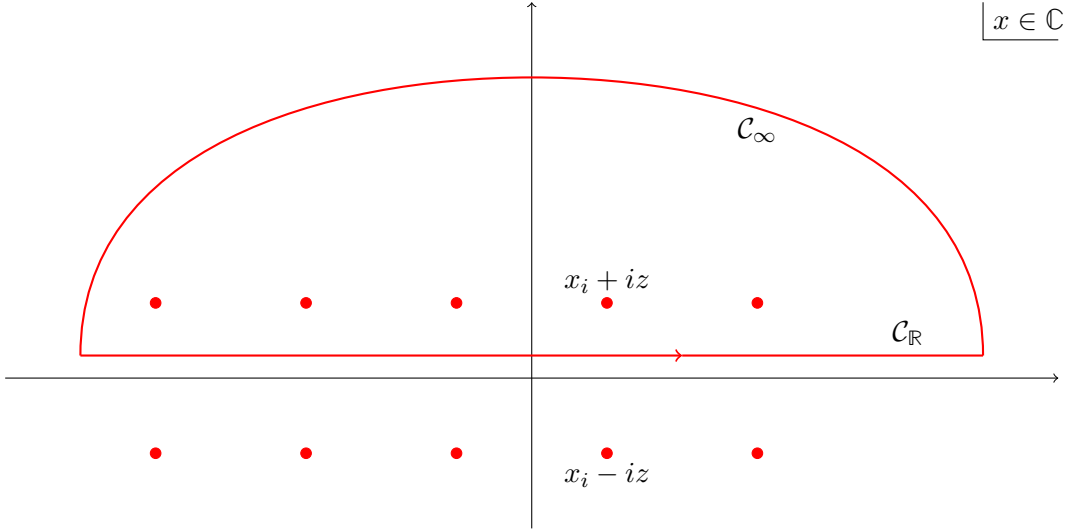
\begin{figure}
	\center
	\begin{tikzpicture}
		\draw[->] (0,-2) to (0,5);
		\draw[->] (-7,0) to (7,0);
		\draw [thick, red, ->] (-6,.3) to (2,.3);
		\draw [thick, red] (2,.3) to (6,.3);
		\filldraw[red] (1,1) circle (2pt);
		\node[anchor= south ] at (1,1)  {$x_i+i z$};
		\filldraw[red] (1,-1) circle (2pt);
		\node[anchor= north ] at (1,-1)  {$x_i-i z$};
		\filldraw[red] (-3,1) circle (2pt);
		\filldraw[red] (-3,-1) circle (2pt);
		\filldraw[red] (-5,1) circle (2pt);
		\filldraw[red] (-5,-1) circle (2pt);
		\filldraw[red] (-1,1) circle (2pt);
		\filldraw[red] (-1,-1) circle (2pt);
		\filldraw[red] (3,1) circle (2pt);
		\filldraw[red] (3,-1) circle (2pt);
		\draw [,thick,red] (-6,.3) to[out=90,in=180] (0,4) to[out=0,in=90](6,.3) ;
		\draw[] (6,5)to(6,4.5) to(7,4.5);
		\node[anchor = south west] at (6,4.5){$x\in \mathbb{C}$};
		\node[anchor= south] at (5,.3) {$\mathcal{C}_{\mathbb{R}}$};
		\node[anchor= south] at (3,3) {$\mathcal{C}_{\infty}$};
	\end{tikzpicture}
	\caption{Contour used for the integral over the variable $x$ parametrising the AdS$_2$ boundary. The contour can be chosen to close in the upper or lower complex  half-plane since the integrand is appropriately bounded at large $|x|$. }
	\label{Fig: contour integral}
\end{figure}

In the case of massless scalar fields, the integral corresponding to the contact Witten diagram with $n$ external points is
\begin{align}\label{Eq: massless contact integral}
	I_{\Delta=1}(x_1,...,x_n)= \int_0^\infty dz z^{n-2} \int_{-\infty}^{\infty} dx \frac{1}{\Pi_{i=1}^n(z^2+(x-x_i)^2)}.\vspace{.4cm}
\end{align}
The advantage of working in AdS$_2$ is that since the boundary only has one dimension, the integrated boundary coordinate $x$ can be analytically continued to the complex plane and the integral can be evaluated with the residue theorem. The contribution from the contour around infinity ($\mathcal{C}_\infty$ in Figure \ref{Fig: contour integral}) vanishes since the integrand is appropriately bounded at large $|x|$. 

The integrand in  \eqref{Eq: massless contact integral} has $2n$ poles at
\begin{align}
	x = x_j\pm i z
\end{align}
where $1\leq j\leq n$, with residues
\begin{align}
	\pm \frac{1}{2iz\Pi_{i\neq j}(x_{ij}^2+2izx_{ij})},
\end{align}
which are depicted in Figure \ref{Fig: contour integral}. Since, for $z>0$, the poles in the upper half-plane (UHP) come with a positive sign and those in the lower half-plane (LHP) come with a minus sign, the result is independent of the choice of closing the contour. However, when $z$ is real, these poles will have an additional factor $\textrm{sgn}(z)$. This is because the poles cross the $x\in \mathbb{R}$ axis when $z$ crosses $0$.\par \vspace{2mm}
We are thus left with the  integral
\begin{align}\label{Eq: z-integral}
	I(x_i) =\pi \int_0^\infty dz z^{n-3}\sum_{j=1}^{n}	\frac{1}{\Pi_{1=1,i\neq j}^{n}(x_{ij}^2+2izx_{ij})}.
\end{align}
The integrand of \eqref{Eq: z-integral} has a leading large $z$ behaviour 
\begin{align}
	\pi \sum_{j=1}^{n-1}\frac{z^{-1}}{(2i)^{n-2}}\frac{1}{\Pi_{i=1,i\neq j}^{n}x_{ij}}+O(z^{-2}),
\end{align}
which vanishes thanks to the identity
\begin{align}
	\sum_{j\in J}\frac{1}{\Pi_{i\in J, i\neq j}x_{ij}}=0,
\end{align}
so the integral is convergent for $n\geq 3$ as expected. Notice that the integrand of  \eqref{Eq: z-integral} has the same parity as the number of external fields. This leads to a simplification in computing the odd $n$-point functions since the $z$-integral can be extended to the range $\{-\infty, \infty\}$ and evaluated using contour integration once again (see Appendix \ref{App n odd}). The symmetry of the integrand dictates that all contact diagrams where $\sum \Delta$ is odd will lead to a rational function and not a logarithm.

For generic $n$, equation \eqref{Eq: z-integral} can be evaluated explicitly with the pole-matched, partial fraction decomposition of the integrand
\begin{align}
	\hspace{-.4cm} \sum_j \frac{z^{n-3}}{\Pi_{k \neq j}\left(2ix_{kj}(z-i\frac{x_{kj}}{2})\right) } 	&=\frac{1}{(2i)^{n-2}} \sum_{i\neq j}\frac{x_{ij}^{n-4}}{\Pi_{k\neq i\neq j}x_{ik}x_{kj}}\frac{-1}{(z+a_{ij})},
\end{align}
where
\begin{align}
	a_{ij} = \frac{x_{ij}}{2i}.
\end{align}
Using this decomposition, we obtain logarithm functions whose branch cut is chosen to be on the negative real axis. The choice of the branch of the logarithm is arbitrary since we do not cross any branch cut in the definite integration.\footnote{The author thanks Luke Corcoran for a discussion on this point.}
The convergent commuting of the sum and the integral is ensured by only taking the upper bound $\Lambda$ to infinity at the end of computations. This gives the result
\begin{align}
	I(x_i)&=\lim_{\Lambda\rightarrow\infty}\frac{-\pi}{(2i)^{n-2}}\sum_{i\neq j}\frac{x_{ij}^{n-4}}{\Pi_{k\neq i\neq j}x_{ik}x_{kj}}\left(\ln(a_{ij}+\Lambda)-\ln(a_{ij})\right),
\end{align}
which can be simplified by averaging over the permutation of the two indices $i$ and $j$. The first consequence is that the divergent term cancels in both cases. In the even-$n$ case
\begin{align}
	\log(\Lambda) \sum_{i\neq j}\frac{x_{ij}^{n-4}}{\Pi_{k\neq i\neq j}x_{ik}x_{jk}}=0.
\end{align}
In the odd-$n$ case, we have a vanishing leading term since
\begin{align}
	\ln(\Lambda-i\frac{x_{ij}}{2})-\ln(\Lambda+i\frac{x_{ij}}{2})\xrightarrow{\Lambda\rightarrow\infty}0. 
\end{align}
Thus, we can write the result as
\begin{align}\label{Eq: massless}
	I(x_i)&=\frac{\pi}{(2i)^{n-2}}\sum_{i\neq j}\frac{x_{ij}^{n-4}}{\Pi_{k\neq i\neq j}x_{ik}x_{kj}}\ln\left(\frac{x_{ij}}{2i}\right),
\end{align}
which is a real quantity for  both the even case
\begin{align}\label{Eq: result massless even}
	I_{even}(x_i)&=\frac{\pi}{2(2i)^{n-2}}\sum_{i\neq j}\frac{x_{ij}^{n-4}}{\Pi_{k\neq i\neq j}x_{ik}x_{jk}}\ln\left(x_{ij}^2\right) ,
\end{align}
\newpage
and the odd-$n$ case
\begin{align}\label{Eq: result massless odd}
	I_{odd}(x_i)&=\frac{\pi}{2(2i)^{n-2}}\sum_{i\neq j}\frac{x_{ij}^{n-4}}{\Pi_{k\neq i\neq j}x_{ik}x_{jk}}\left(\ln(a_{ij})-\ln(-a_{ij})\right)\nonumber \\
	&=	\frac{\pi}{2(2i)^{n-2}}\left(i\pi \sum_{i>j}\frac{x_{ij}^{n-4}}{\Pi_{k\neq i\neq j}x_{ik}x_{jk}}-i \pi \sum_{i<j}\frac{x_{ij}^{n-4}}{\Pi_{k\neq i\neq j}x_{ik}x_{jk}}\right)\nonumber \\
	&=	\frac{\pi^2 }{2(2i)^{n-3}}\sum_{i>j}\frac{x_{ij}^{n-4}}{\Pi_{k\neq i\neq j}x_{ik}x_{jk}}.
\end{align}
The correlator \eqref{Eq: solution massless all n} follows from \eqref{Eq: result massless odd} and \eqref{Eq: result massless even}.
This matches known literature for the case of the four-point functions, for example;
\begin{align}\label{Eq: four-point}
	I_{\Delta=1,n=4} =-\frac{\pi }{2}\left(  \frac{\log \left(\chi_1\right)}{ 1-\chi_1}+\frac{\log \left(1-\chi_1\right)}{ \chi_1}\right).
\end{align}
More cases are listed in Appendix \ref{App: list of correlators}.

\subsection{Massive Scalar Fields }
\label{subsec:massive }
The method used in subsection \ref{subsec:massless} is compelling in the generic $n$ case but quickly increases in complexity when $\Delta>1$. For $\Delta=2$, the result can still be computed with this method relatively efficiently.\footnote{Another method can be used to obtain the massive $n$-point functions from the massless cases, as seen in \ref{Sec: pinching} .} The integral 
\begin{align}\label{Eq: Integral Delta=2}
	I_{\Delta=2}(x_i) =\int dz z^{2n-2} \int dx\frac{ 1}{\Pi_{i=1}^{n}(z^2+(x-x_i)^2)^2}.
\end{align}
is evaluated by contour integration for the $x-$integral and partial fraction decomposition for the $z-$integral. Double poles lead to the less compact formula 
\begin{align}\label{Eq Delta=2 result}
	&I_{\Delta=2,n}=\sum_i\sum_{j\neq i}\frac{-\pi}{2(2i)^{2n-4}x_{ij}^2}\partial_{x_j}\left( \frac{x_{ji}^{2n-5}}{\prod_{k\neq j,k\neq i }x_{kj}^2x_{ki}^2}\ln\frac{x_{ji}}{2i}\right)\nonumber \\
	&+\sum_i \sum_{j\neq i}\partial_{x_i}\frac{-\pi}{(2i)^{2n-2}x_{ij}^2}\partial_{x_j}\left( \frac{x_{ji}^{2n-4}}{\prod_{k\neq j,k\neq i}x_{kj}^2x_{ik}^2} \ln\frac{x_{ji}}{2i}\right),
\end{align}\vspace{2mm}
which is derived in Appendix \ref{App: Delta=2 derivation}.
One expects a similar structure at higher $\Delta$, with a double sum over the external coordinates $x_{i,j}$ and $\partial^{2\Delta}$ derivatives and $\Delta$ terms. Some evidence of this is the pinching presented in subsection \ref{Sec: pinching} though subtleties in the order of limits prevent a general analysis. As such, the residue method loses its efficiency as we increase the dimension of the external operators.	
\newpage
\subsection{An Application: Topological Correlators}\label{sec:topological}
\label{sec: examples}
We now consider non-Abelian gauge theories in AdS$_2$ and an alternative construction to the Witten diagram computation in Appendix A.2 of \cite{Mezei:2017kmw}. For consistency with the notation in \cite{Mezei:2017kmw}, we denote the boundary coordinate by $t$ instead of $x$. We review the setting of \cite{Mezei:2017kmw} where the strong coupling action is that of Yang-Mills theory in AdS$_2$, completed with a regulating boundary term
\begin{align}
	S_{YM} &= \frac{1}{2g_{YM}^2}\int_{\text{AdS}_2} dx^2 \sqrt{-g}\textrm{Tr}\left(F_{\mu \nu}F^{\mu \nu}\right)\\
	S_{b^{y}}&=\frac{1}{g_{YM}^2}\int_{\partial \text{AdS}_2} dx \sqrt{-\gamma}\textrm{Tr}\left( A_i A^i-2A^iF_{\mu i}n^\mu \right),
\end{align}
where $\mu, \nu$ are the indices in the bulk coordinates of AdS$_2$, $i$ those of the boundary coordinates, and $n^\mu$ is a unit vector normal to the boundary of AdS$_2$. \par
\begin{figure}[H]
	\center
	\begin{tikzpicture}[scale=.8]
		\draw[->] (0,-2) to (0,5);
		\draw[->] (-7,0) to (7,0);
		\draw [thick, red, ->] (-6,.3) to (2,.3);
		\draw [thick, red] (2,.3) to (6,.3);
		\filldraw[red] (1,1) circle (2pt);
		\node[anchor= south ] at (2,1)  {$i (t_2-t_1)$};
		\filldraw[red] (1,-1) circle (2pt);
		\node[anchor= north ] at (2,-1)  {$-i(t_2-t_1)$};
		\draw [,thick,red] (-6,.3) to[out=90,in=180] (0,4) to[out=0,in=90](6,.3) ;
		\draw[] (6,5)to(6,4.5) to(7,4.5);
		\node[anchor = south west] at (6,4.5){$z\in \mathbb{C}$};
		\node[anchor= south] at (5,.3) {$\mathcal{C}_{\mathbb{R}}$};
		\node[anchor= south] at (3.5,3.5) {$\mathcal{C}_{\infty}$};
	\end{tikzpicture}
	
	\caption{Contour used for the $z-$integral in equation \eqref{Eq: z-integral topo}. The origin of the topological factor sgn$(t_1-t_2)$ is clear in this setup. The contour is closed in the UHP (the same analysis holds for the LHP closing). The pole contained within this contour depends on the sign of $(t_1-t_2)$ where, in this example, we have shown the case $t_1<t_2$.}
	\label{Fig: contour integral 3}
\end{figure}
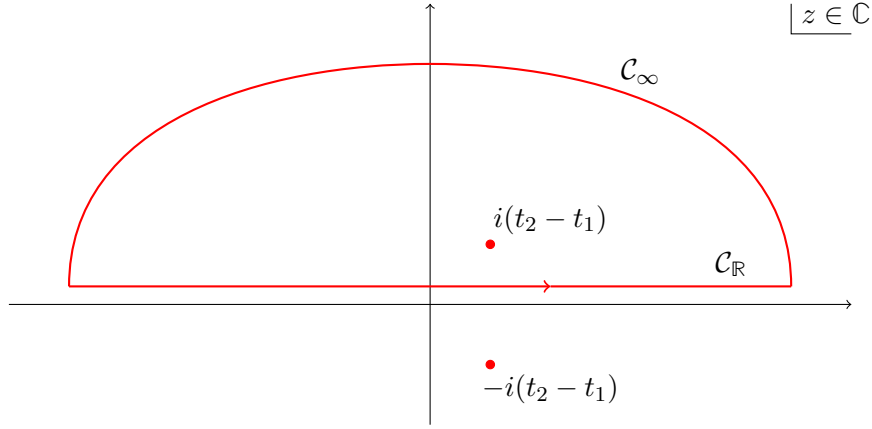

 
In radial coordinates, the equation of motion is solved by
\begin{align}
	F_{r \varphi} &= Q\sinh r & A_{\varphi} &= Q(\cosh r -1)&A_r&=0,
\end{align}
where $Q = Q_aT^a$ is an element of the Lie algebra of the theory, and in the following, indices $a,b,a_i$ are those of the gauge algebra. This gives the on-shell action
\begin{align}\label{Eq:on-shell action}
	\left(S_{tot}\right)_{\textrm{on-shell}} = -2\pi \frac{\textrm{Tr}(Q^2)}{g_{YM}^2}.
\end{align}
To relate the boundary fields to the bulk fields, the variation of the bulk action needs to be written in terms of the variation of the boundary field
\begin{align}
	\delta S_{tot} &= \frac{2}{g_{YM}^2} \int_{\partial B}dx \sqrt{-\gamma}\textrm{Tr}\left( A^i \delta a\right)	& a &= \lim_{x^\mu\rightarrow \partial B}\left( A_i -F_{\mu i}n^\mu\right),
\end{align}
where $a$ is thus the corresponding boundary field and $i$ is the index corresponding to the boundary coordinate ($t$).
The on-shell action \eqref{Eq:on-shell action} can be written in terms of the boundary fields $a$ through the equation
\begin{align}
	a(\varphi) &= -u Q u^{-1}+iu \partial_\varphi u^{-1}\\
	u_0Qu_0^{-1} &= \frac{i}{2\pi}\log \left( P\exp\left(i\int_0^{2\pi}d\varphi a(\varphi)\right)\right)\label{Eq : uQ u},
\end{align}
where the $\varphi-$dependant large gauge transformations at the boundary are parametrised by $u$ and $P \exp$ denotes the usual path ordered exponential. The expression for the on-shell action is then proportional to the trace of \eqref{Eq : uQ u} squared,
\begin{align}
	\Tr(Q^2) = 	\Tr((u_0 Q u_0^{-1})^2)  = -\frac{1}{4\pi^2}\Omega(a).
\end{align}
The expression for $\Omega(a)$ is a standard result in quantum mechanics and is solved by the Magnus expansion \cite{Wilcox:1967zz,2009PhR...470..151B}
\begin{align}
	\exp\left(\Omega\right)= P \exp\left(i\int d\varphi a(\varphi))\right).
\end{align}
This can be used to find the dual correlators through the holographic dictionary
\begin{align}
	<j^{a}(\varphi_1)j^{b}(\varphi_2)> &= \frac{\delta^{ab}}{4\pi g_{YM}^2}\\
	<j^{a}(\varphi_1)j^{b}(\varphi_2)j^{c}(\varphi_3)>& =-\frac{f^{abc}\textrm{sgn}{\varphi_{12}\varphi_{23}\varphi_{31}}}{4\pi g^2_{YM}}\label{eq: 3-point topological}\\
	<j^{a_1}(\varphi_1)j^{a_2}(\varphi_2)j^{a_3}(\varphi_3)j^{a_4}(\varphi_4)>& =-\frac{f^{a a_1 a_2}f^{a a_3 a_4}}{4\pi g^2_{YM}}\left(\textrm{sgn}{\varphi_{12}\varphi_{24}\varphi_{43}\varphi_{31}}-\textrm{sgn}{\varphi_{21}\varphi_{14}\varphi_{43}\varphi_{32}}\right)\nonumber \\
	&\quad +(2\leftrightarrow 3)+(2\leftrightarrow 4),
\end{align}
where the indices $a,b,c, a_i$ are those of the gauge algebra. Through Witten diagrams, these correlators of boundary terms can be computed explicitly using the contour integral method detailed above. 
The bulk-to-boundary propagators  in Poincaré coordinates for the gauge field $A_\mu$  \cite{DHoker:1999bve, Costa:2014kfa} are
\begin{align}
	G_{\mu}(z,t;t_i) &= \frac{z^2+(t-t_i)^2}{2\pi z}\partial_\mu \left(\frac{t-t_i}{z^2+(t-t_i)^2}\right),
\end{align}
or explicitly
\begin{align}
	G_{z}(z,t;t_i) &=\frac{t_i-t}{\pi  \left((t-t_i)^2+z^2\right)}&G_t(z,t,t_i)&=\frac{z^2-(t-t_i)^2}{2\pi z(t-t_i)^2+z^2}.\vspace{3mm}
\end{align}
\vspace{3mm}
The on-shell action is a pure boundary term
\begin{align}
	S_{on-shell} &= \frac{1}{2g_{YM}^2}\int_{\text{AdS}_2} dx^2\sqrt{-g}\textrm{Tr}\left(D_\mu A_\nu F^{\mu \nu}\right)+\frac{1}{g_{YM}^2}\int _{\partial \text{AdS}_2}dx\sqrt{-\gamma} \textrm{Tr}\left(A_i A^i-2A^i F_{\mu i}n^\mu\right)\nonumber \\
	&=\frac{1}{g_{YM}^2}\int _{\partial \text{AdS}_2}dx\sqrt{-\gamma} \textrm{Tr}\left(A_i A^i+A^i F_{i \mu }n^\mu\right).\vspace{3mm}
\end{align}
Explicitly, in the $(z,t)$ Poincaré coordinates, this gives\footnote{Note that the vector pointing out of the boundary goes in the $-z$ direction.}
\begin{align}
	S_{on-shell} &=-\frac{1}{g_{YM}^2}\int dt z \textrm{Tr}\left(A_t A_t-z A_tF_{t z }\right)|_{z=0}.
\end{align} 
\vspace{3mm}
The two-point correlators are given by Wick contractions acting on this term
\begin{align}
	<a^a(t_1)a^b(t_2)>&=\lim_{z\rightarrow 0}- \frac{1}{2g_{YM}^2}\int dt \delta^{ab} zG_t(z,t,t_2)\left(G_t(z,t;t_1)+z\partial_{[z}G_{t]}(z,t;t_1)\right) \\
	&=\lim_{z\rightarrow 0} \frac{(t_1-t_2)^2\delta^{ab}}{4 \pi g_{YM}^2 \left((t_1-t_2)^2+4 z^2\right)}\\
	&= \frac{\delta^{ab}}{4\pi g^2_{YM}}.
\end{align}
The three-point vertex is 
\begin{align}
	S_3 =- \frac{1}{g_{YM}^2}\int dt dz z^2 f_{abc}A^a_zA^b_t\partial_{[z}A^c_{t]},
\end{align}
which gives a correlator
\begin{align}
	\langle a^{a}(t_1)a^{b}(t_2)a^{c}(t_3)\rangle  = \frac{1}{ g_{YM}^2}\text{Perm}\left(f^{abc}I(t_1,t_2,t_3)\right),
\end{align}
where we define the single-Wick-contracted integral
\begin{align}
	I(t_1,t_2,t_3)	 = \int dt dz z^2 G_z(z,t;t_1) G_t(z,t;t_2)\partial_{[z}G_{t]}(z,t;t_3).
\end{align}
The anti-symmetrised derivative removes the $t_3$ dependence, and the parity of this integrand under $z\rightarrow -z$ is the same as that of the odd $n$ massless scalar case (see subsection \ref{subsec:massless}), so we can evaluate both the $z$ and $t$ integrals with a complex contour.\footnote{There is a convergence issue in $I({t_1,t_2,t_3})$, which is solved when considering the sum of the Wick contractions.  The leading term ($t^{-1}$) is always cancelled by the odd permutation of the indices $(1,2,3)$ and the next-to-leading term is convergent.}
Extending the $z$ variable to the entire real line, we have
\begin{align}
	I({t_1,t_2,t_3})&= \int_{-\infty}^{\infty}dt \int_0^{\infty} dz\frac{(t-t_1) \left((t-t_2)^2-z^2\right)}{4\pi ^3 z \left((t-t_1)^2+z^2\right) \left((t-t_2)^2+z^2\right)}\\
	&=\frac{1}{2} \int_{-\infty}^{\infty}dt \int_{-\infty}^{\infty} dz \, \, \textrm{sgn}(z)\frac{(t-t_1) \left((t-t_2)^2-z^2\right)}{4\pi ^3 z \left((t-t_1)^2+z^2\right) \left((t-t_2)^2+z^2\right)}.
\end{align}
This integral has a $t_i$-independent  contribution from the behaviour at $t\rightarrow \infty$. However, this is cancelled by the permutation and the antisymmetry of the structure constants. We will therefore ignore this contribution and evaluate the integral by contour integration. The $t-$integral evaluates to
\begin{align}
	I({t_1,t_2,t_3}) &=\int_{-\infty}^\infty\frac{dz}{z}\frac{2 z (t_1-t_2)+i (t_1-t_2)^2+4 i z^2}{8 \pi ^2  \left((t_1-t_2)^2+4 z^2\right)} \textrm{sgn}(z)^2\\
	&=\int_{-\infty}^\infty\frac{dz}{z}\frac{2 z (t_1-t_2)+i (t_1-t_2)^2+4 i z^2}{8 \pi ^2  \left((t_1-t_2)^2+4 z^2\right)}  \label{Eq: z-integral topo}.
\end{align}
The factors of $\textrm{sgn}(z)$ cancel and leave an analytic function in $z$. This integral also has a pole at 0 and at $\infty$, which can also be cancelled using the antisymmetry of the structure constants of the algebra by considering a combination of Wick contractions; for example, $	f^{a_1,a_2,a_3}\left(I_{t_1,t_2,t_3} -	I_{t_2,t_1,t_3}\right)  $. With this in mind, the integral can be evaluated using contour integration. The only remaining pole is at $z = \pm i \frac{t_1-t_2}{2}$ and therefore the integral will have a factor of $\textrm{sgn}(t_1-t_2)$, multiplying the residue at that point (see Figure \ref{Fig: contour integral 3}).\par 
The single-Wick-contracted integral is then
\begin{align}
	I(t_1,t_2,t_3) = \frac{1}{2\pi}\textrm{sgn}(t_1-t_2),
\end{align}
and therefore
\begin{align}
	\langle a^{a}(t_1)a^{b}(t_2)a^{c}(t_3)\rangle  = \frac{1}{8\pi g_{YM}^2}\left(\sum_{\sigma(\{1,2,3\})}\left(f^{a_1a_2a_3}\textrm{sgn}(t_1-t_2)\right)\right)|_{\{a_1,a_2,a_3\}\rightarrow \{a,b,c\}}.
\end{align}
Using the total antisymmetry of the structure constants, we have 
\begin{align}
	\sum_{\sigma(\{1,2,3\})}\left(f^{a_1a_2a_3}\textrm{sgn}(t_1-t_2)\right)|_{\{a_1,a_2,a_3\}\rightarrow \{a,b,c\}} = -2f^{abc}\, \textrm{sgn}(t_{12}t_{23}t_{31}),
\end{align}
which gives the final result
\begin{align}
	\langle a^{a}(t_1)a^{b}(t_1)a^{c}(t_1)\rangle  = -\frac{1}{4\pi g_{YM}^2} f^{abc}\, \textrm{sgn}(t_{12}t_{23}t_{31}).
\end{align}
This agrees with equation \eqref{eq: 3-point topological} through a simple change of coordinates.\footnote{For higher-point functions, there is the subtlety that the boundary field $a$ is not the boundary limit of the gauge field $A_\mu$, but rather has a dependence on both $A_\mu$ and $F_{\mu \nu}$. This implies that the bulk-to-boundary propagator receives corrections from multi-source terms. These questions are addressed in \cite{Mezei:2017kmw}.}

\subsection{A Basis of Interaction Terms}
This subsection considers deformations from generalized free field theory produced by effective interactions in a bulk AdS$_2$ field theory. In this holographic AdS$_2$/CFT$_1$ setup, the background AdS$_2$ metric is not dynamical, corresponding to the absence of a stress tensor in the boundary CFT$_1$. According to the usual dictionary, a massive free scalar field $\Phi$ in AdS$_2$ is dual to a boundary 1d generalised free field $\phi$. We deform this theory with quartic self-interactions with an arbitrary number $L$ of derivatives 
\be\label{phi4Lagrangian}
S=\int dx dz\,\sqrt{g}\,\big[ \,g^{\mu\nu}\,\partial_\mu\Phi\,\partial_\nu\Phi+m^2_{\Delta_\phi}\Phi^2+g_L\,(\partial^L\Phi)^4\,\big]\,,\qquad L=0,1,\dots
\ee
where we use the AdS$_2$ metric in Poincar\'e coordinates $ds^ 2=\frac{1}{z^2}(dx^ 2+dz^ 2)$. The mass $m^2_{\Delta_\phi}=\Delta_\phi(\Delta_\phi-1)$ is fixed in units of the AdS radius so that $\Delta_\phi$ is the dimension (independent of $g_L$) of the field $\Phi$ evaluated at the boundary, $\phi(x)$.\footnote{When we introduce an interaction, such as \eqref{phi4Lagrangian}, there will be Witten diagrams contributing to the mass renormalization of $\Phi$. We can always choose the bare mass so that the dictionary is preserved and $\Delta_{\phi}$ is not modified.}  We will limit our analysis to tree-level correlators and thus consider only contact diagrams, whose building blocks are the $D$-functions~\cite{Liu:1998ty,DHoker:1999kzh, Dolan:2003hv}  reviewed in Appendix~\ref{app:Dfunctions}. 
The writing $(\partial^L\Phi)^4$ above is symbolic, denoting a complete and independent set of quartic vertices with four fields and up to $4L$ derivatives.\footnote{The fact that a complete and independent basis of vertices is labelled by 1/4 the number of derivatives can be seen using integration by parts and the equations of motion, or by noticing that the counting of physically  distinct four-point interactions is equivalent to the counting of crossing-symmetric polynomial $S$-matrices in 2D Minkowski space (see discussion in~\cite{Mazac:2018ycv, Ferrero:2019luz}).} In the following, we will present a particularly convenient basis for these interactions, allowing us to derive a closed-form expression for the tree-level correlator in Mellin space. Consider the interaction Lagrangian
\begin{equation}\label{interactionlag}
	\mathcal{L}_{L}=g_{L}\left[\prod_{k=0}^{L-1}\left(\tfrac{1}{2}\partial_\mu \partial^\mu-(\Delta_{\phi}+k)(2(\Delta_{\phi}+k)-1)\right)\Phi^2 \right]^2 \, .
\end{equation}
This looks like a very complicated term, but it contains four fields $\Phi$ and $4L$ derivatives, so by the argument above, it must be effectively a linear combination of operators like $(\partial^{\ell}\Phi)^4$ for $\ell\leq L$. The advantage of this interaction is that the corresponding correlator computed via Witten diagrams reads 
\be\label{Ansatzcorrelator}
\hspace{-6mm}f^{(1)}_L(z) = \frac{4^{L-1}\pi^{-\frac32}\Gamma(2\Delta_{\phi}-\frac{1}{2}+2L)}{\Gamma(\Delta_{\phi}+\frac12)^4}  z^{2\Delta_\phi}\,(1+z^{2L}+(1-z)^{2L})\bar{D}_{\Delta_\phi+L, \Delta_\phi+L, \Delta_\phi+L, \Delta_\phi+L}(z)
\ee
where the $\bar{D}$-functions are listed in Appendix \ref{app:Dfunctions}. If one starts with some specific $4L$-derivative interaction, such as $(\partial^{L}\Phi)^4$, the explicit computation through Witten diagrams shows several other combinations of $D$-functions with different weights. Nevertheless, by the argument above, these results cannot be independent of those obtained using $\mathcal{L}_{L}$ and therefore, the result must be expressible as a linear combination $\sum_\ell a_{\ell}f^{(1)}_\ell(z)$. This requires a series of non-trivial identities among $\bar D$ functions, some of which are derived in \cite{Bianchi:2021piu}. 

Here we consider the interaction Lagrangian \eqref{interactionlag} and show that it leads to the correlator \eqref{Ansatzcorrelator} using Witten diagrams. The result of the Wick contractions is
\begin{multline}\label{wickforspecialint}
	\braket{\phi(x_1)\phi(x_2)\phi(x_3)\phi(x_4)}^{(1)} =\\ \sum_{\text{perms}} g_L \int \frac{dx dz}{z^2} \mathcal{D} \left(K_{\Delta_\phi}(x,z;x_1) K_{\Delta_{\phi}}(x,z;x_2)  \right) \mathcal{D} \left(K_{\Delta_\phi}(x,z;x_3) K_{\Delta_{\phi}}(x,z;x_4)  \right)\, ,
\end{multline}
where we defined
\begin{align}\label{Eq: D basis}
	\mathcal{D}= \prod_{k=0}^{L-1}\left(\frac{1}{2}\partial_{\mu}\partial^{\mu}-(\Delta_{\phi}+k)(2(\Delta_{\phi}+k)-1)\right) 
\end{align}
acting on the bulk point, and we used the bulk-to-boundary propagator with the conventions of \eqref{Eq: normalisation C_Delta}.
Using the identity recursively
\begin{align}
	-2x_{ij}^2\Delta^2 \tilde K_{\Delta+1}(x,z;x_i) \tilde K_{\Delta+1}(x,z;x_j) = (\tfrac{1}{2}\partial_{\mu}\partial^{\mu}-\Delta(2\Delta-1))(\tilde K_{\Delta}(x,z;x_i) \tilde K_{\Delta}(x,z;x_j)) \, ,
\end{align}
which can be derived from \eqref{identityder}, we obtain
\begin{align}
	\mathcal{D} \left(\tilde K_{\Delta_\phi}(x,z;x_1) \tilde K_{\Delta_{\phi}}(x,z;x_2)  \right) = (-2x_{12}^2)^L\left(\frac{\Gamma(\Delta_{\phi}+L)}{\Gamma(\Delta_{\phi})}\right)^2 \tilde K_{\Delta_\phi+L}(x,z;x_1) \tilde K_{\Delta_\phi+L}(x,z;x_2) \, .
\end{align}
Inserting this into equation \eqref{wickforspecialint}, summing over the permutations and remembering the definition of $\mathcal{C}_{\Delta}$ in \eqref{Eq: normalisation C_Delta}, we get 
\begin{multline}
	\braket{\phi(x_1)\phi(x_2)\phi(x_3)\phi(x_4)}^{(1)}_L = g_L [(x^2_{13}x^2_{24})^{L}+(x^2_{12}x^2_{34})^{L}+(x^2_{14}x^2_{23})^{L}] \times \\
	2^{2L-1} \pi^{-2}\left(\frac{\Gamma(\Delta_{\phi}+L)}{\Gamma(\Delta_{\phi}+\frac12)}\right)^4  
	D_{\Delta_{\phi}+L,\Delta_{\phi}+L,\Delta_{\phi}+L,\Delta_{\phi}+L}(x_1,x_2,x_3,x_4) \, .
\end{multline}
Using \eqref{Dbar} we immediately get
\begin{multline}\label{correlatorL}
	\braket{\phi(x_1)\phi(x_2)\phi(x_3)\phi(x_4)}^{(1)}_L=g_L (1+\chi^{2L}+(1-\chi)^{2L})\times\\ \frac{4^{L-1}\pi^{-\frac32}\,\Gamma(2\Delta_{\phi}-\frac{1}{2}+2L)}{\Gamma(\Delta_{\phi}+\frac12)^4\,(x^2_{13}\,x^2_{24})^{\Delta_{\phi}}}  \bar{D}_{\Delta_\phi+L, \Delta_\phi+L, \Delta_\phi+L, \Delta_\phi+L}(\chi)
\end{multline}
in perfect agreement with \eqref{Ansatzcorrelator}. 
\section{Mellin Amplitudes}\label{chapter: Mellin amplitudes}

In the higher-dimensional case, the Mellin representation of conformal correlators~\cite{Mack:2009mi,Penedones:2010ue} has proven to be an excellent tool, especially for the study of holographic CFTs~\cite{Fitzpatrick:2011ia,Paulos:2011ie,Rastelli:2017udc}.
The counting of independent cross-ratios for an $n$-point correlation function  of local operators in a $d$-dimensional CFT  is identical to that of independent variables scattering in $d+1$ dimensions. The Mellin representation, or Mellin amplitude, makes this correspondence manifest, expressing the correlators in a form that is the natural AdS counterpart of flat-space scattering amplitudes. 
%reminiscent of scattering amplitudes for dual resonance models. 
This construction has several nice features. First, the Mellin amplitude has simple poles located at the values of the twist of exchanged operators (there are, however, infinitely many accumulation points of such poles). Secondly, the crossing symmetry of the correlator maps to the amplitude crossing symmetry. Finally, the language of Mellin amplitudes is particularly suitable for large $N$ gauge theories, where perturbation theory is described in terms of Witten diagrams. An extensive introduction to higher dimensional Mellin amplitudes is presented in Appendix \ref{3 computations to understand Mellin formalism}. In the following,  we will use these properties as guiding principles for the definition of Mellin amplitudes for 1d CFTs.

\subsection{Nonperturbative Mellin Amplitude} \label{sec:nonpert}
Let $f(t)$ be a function describing a four-point correlator in a 1d CFT where the cross ratio $t$ is defined as
\begin{align}\label{t}
	t=\frac{\chi}{1-\chi}.
\end{align}
We define the one-parameter Mellin transform $\mc{M}_a[f(t)]$ as
\begin{align}\label{Mellin1para}
	\mc{M}_a[f(t)] =  =\int_0^\infty dt\, \left(\frac{t}{1+t}\right)^a f(t)\, t^{-1-s}\,,
\end{align} 
and Mellin amplitude $M_{a=0}(s)$ as
\begin{align}\label{Mellin}
M(s) =\frac{1}{\Gamma(s)\Gamma(2\Delta_\phi-s)}\int_0^\infty dt\, f(t)\, t^{-1-s}\,
\end{align}
which we multiply by an overall factor for future convenience.
In this case, the crossing relation becomes 
\begin{align}\label{crossing1def}%\label{cross-f-t}
	f(t)&=t^{2\Delta_\phi}f(\frac{1}{t}),&M(s)&=M(2\Delta_\phi -s)\, .
\end{align}
% becomes
%\begin{align}
%M(s)=M(2\Delta_\phi -s)\, .
%\end{align}
The goal of our discussion is to infer the analytic properties of the Mellin amplitude $M(s)$ from the physical requirements on the correlator $f(t)$. First, following \cite{Penedones:2019tng}, we recall a general result for the one-dimensional Mellin transform \eqref{Mellin}.

\subsubsection{A Theorem}\label{theorem}
Consider a function $F(t)$ in the vector space $\mathcal{F}_H^\Theta$ of complex valued functions that are holomorphic for $\textrm{arg}(t)\in \Theta$ and obey
\begin{equation}\label{boundnessF}
	|F(t)| \leq  \frac{C(h)}{|t|^h} \qquad h \in H \, ,
\end{equation}
where H is a subset of $\mathbb{R}$, typically of the form $H=(h_{min}, h_{max})$. Consider also the function $\hat{M}(s)$ in the vector space $\mathcal{M}_{H}^{\Theta}$ of complex valued functions that are holomorphic for $\text{Re}(s)\in H$ and exponentially suppressed in the limit $|\text{Im}(s)|\rightarrow  \infty$
\begin{align} \label{bound M}
	|\hat{M}(s)| \leq K(\textrm{Re}(s))e^{-|\text{Im}(s) \sup_{\Theta} \textrm{arg}(t)|} \qquad |\text{Im}(s)|\to \infty \, .
\end{align}
These two vector spaces exist independently but the following theorem holds\par
\rule{\textwidth}{0.4pt}
	\large
	\textbf{Theorem:} \par 
	Given a function $F(t)\in \mathcal{F}_H^\Theta$, its Mellin transform $\mathcal{M}[F](s)$ exists and $\mathcal{M}[F](s)\in \mathcal{M}_{H}^{\Theta}$. Furthermore, $\mathcal{M}^{-1}\mathcal{M}[F](t)=F(t)$ for any $\arg (t)\in \Theta$. Conversely, given $\hat{M}(s)\in \mathcal{M}_{H}^{\Theta}$ its inverse Mellin transform exists and $\mathcal{M}^{-1}[\hat{M}](s)\in \mathcal{F}_H^\Theta$.\\ Furthermore, $\mathcal{M}\mathcal{M}^{-1}[\hat{M}](s)=\hat M(s)$ for any $s\in H+i\mathbb{R}$.\par 
\rule{\textwidth}{0.4pt}\par
\vspace{3mm}
\normalsize
This is a classical result for the one-dimensional Mellin transform, so we will not prove it here. Instead, we will discuss how the physical 1d correlator violates the hypothesis of the theorem and how we can overcome this issue. The convergence of the $s$-channel OPE for $|\arg(t)|<\pi$ ensures that the function $f(t)$ is indeed analytic in a sectorial domain $\Theta$. Nevertheless, the condition \eqref{boundnessF} is violated in two ways :
\begin{itemize}
	\item When light operators ($\Delta<\Delta_\phi$) are exchanged in the OPE, the region $H$ is not well defined and the Mellin transform does not exist. This issue is analogous to the higher dimensional case of \cite{Penedones:2019tng} and we will solve it by implementing a finite number of subtractions in subsection \ref{convandsub}.
	\item The correlator $f(t)$ is not bounded for $t\to e^{i \pi}$ where it has a singularity controlled by the Regge limit \eqref{reggetheta}. This issue does not spoil the existence of the Mellin transform, but it gives a result that is not bounded by \eqref{bound M}. 
\end{itemize}
To understand this second point, let us present a simple example which will be useful to explain the issue. Consider the function $F(t)=(\frac{t}{1+t})^{2\Delta_{\phi}}$. This function is analytic for $|\arg(t)|<\pi$ and gives a convergent integral \eqref{Mellin} for $0<\text{Re}(s)<2\Delta_{\phi}$. However, although the bound \eqref{boundnessF} holds along the real axis, it is violated for $t\to e^{i \pi}$. The Mellin transform of this function reads
\begin{equation}\label{example}
	\int_{0}^{\infty} dt \left(\frac{t}{1+t}\right)^{2\Delta_{\phi}} t^{-1-s}=\frac{\Gamma(s)\Gamma(2\Delta_{\phi}-s)}{\Gamma(2\Delta_{\phi})} \,  .
\end{equation}
From this explicit expression we see immediately that for $|\text{Im}(s)|\to \infty$ the r.h.s.  is not bounded by $e^{-\pi |\text{Im}(s)|}$. It is bounded by
\begin{equation}
	\frac{\Gamma(s)\Gamma(2\Delta_{\phi}-s)}{\Gamma(2\Delta_{\phi})}\leq K(\textrm{Re}(s))  |\text{Im}(s)|^{2\Delta_{\phi}-1} e^{-\pi |\text{Im}(s)|} \qquad |\text{Im}(s)|\to \infty \, .
\end{equation}
The exponential decay is correctly predicted by the theorem, while the additional polynomial divergence can be related to the behaviour of the function $F(t)$ for $t\to e^{i\pi}$. 
In section \ref{boundedness}, we will show that this is a specific instance of a general relation between the large $s$ asymptotics of $M(s)$ to the Regge limit of $f(t)$.

\subsection{Convergence and Subtractions}\label{convandsub}
Let us now discuss the convergence of the integral \eqref{Mellin}. Let $f(t)$ be well-behaved for $t\in \mathbb{R}^+$; we do not want divergences in $t$ other than at $t=0$ and $t\rightarrow \infty$. This behaviour coincides with the CFT$_1$ correlators we are interested in. Consider the behaviour of $f(t)$  close to $t=0$. Using the conformal block expansion \eqref{Eq: OPE}, we find that the leading power is $f(t)\sim t^{\D_0}$ where $\Delta_0$ is the dimension of the lightest exchanged operator. 
Analogously, using the crossing symmetry relation~\eqref{crossing1def}, we find that the large $t$ behaviour of $f(t)$ is $f(t)\sim t^{2\Delta_\phi-\D_0}$. 
Therefore the integral converges in the strip
\be\label{convergence}
2\Delta_\phi-\D_0<\text{Re}(s)<\D_0\,, 
\ee
which is a well-defined interval \emph{only for $\D_0>\Delta_\phi$}.  
To give a nonperturbative definition of the Mellin transform, which allows for lighter operators to be exchanged,  we need to perform some subtractions along the lines of~\cite{Penedones:2019tng}.\footnote{See in particular Appendix B in~\cite{Penedones:2019tng} for the one-dimensional case.} One obvious example is GFF, where the identity operator is exchanged. We will consider this case explicitly in Appendix~\ref{sec:GFF}. For the moment, we consider the Mellin transform of the connected part of the correlator. Let us consider the following subtractions
\begin{align} \label{subtr0}
	f_0(t)&=f_{\text{conn}}(t)-\sum_{\D_{0}\leq \D \leq\Delta_\phi}\sum_{k=0}^{[\Delta_\phi-\D]} c_\D \frac{(-1)^{k}}{k!} \frac{\Gamma(\D+k)^2\Gamma(2\D)}{\Gamma(\D)^2\Gamma(2\D+k)} t^{\D+k}\,,\\
	f_{\infty}(t)&=f_{\text{conn}}(t)-\sum_{\D_{0}\leq \D \leq\Delta_\phi}\sum_{k=0}^{[\D_\phi-\D]} c_\D\frac{(-1)^k}{k!} \frac{\Gamma(\D+k)^2\Gamma(2\D)}{\Gamma(\D)^2\Gamma(2\D+k)} t^{2\Delta_\phi-\D-k}\,,\label{subtrinf}
\end{align}
where, for convenience, we write  $c_{\Delta_\phi \Delta_\phi \Delta}^2\equiv c_{\Delta}$.
For the function $f_0(t)$ we subtracted the $s$-channel contribution of all the operators (primaries and descendants) with scaling dimension below the threshold $\Delta=\Delta_\phi$, making use of the series expansion of the hypergeometric function in the conformal block. This improves the behaviour of the function at $t=0$. 
On the other hand, for $f_{\infty}(t)$ we subtracted all the $t$-channel operators below the threshold, thus improving the behaviour at $t=\infty$. The idea is to split the integral~\eqref{Mellin} into two parts, which are defined on (possibly non-overlapping) semi-infinite regions of the complex $s$ plane
\begin{align}\label{psi0def}
	\psi_0(s)&=\int_0^1 dt\, f_{\text{conn}}(t) \,t^{-1-s} & \text{Re}(s)&< \D_0 \,  , \\
	\psi_{\infty}(s)&=\int_1^{\infty} dt \,f_{\text{conn}}(t) \,t^{-1-s} & \text{Re}(s)&>2\Delta_\phi- \D_0 \label{psiinfdef}.
\end{align}
When the two regions do not overlap, we analytically continue $\psi_0(s)$ and $\psi_{\infty}(s)$ by considering the integrals of the functions  \eqref{subtr0} and \eqref{subtrinf} and adding a finite number of poles
\begin{align}
	\label{psi0ancont}
	\!\!\!\! \psi_0(s)&=\int_0^1 \!\!\!dt \, f_0(t)\, t^{-1-s}+\!\!\!\!\!\!\sum_{\D_{0}\leq \D \leq\Delta_\phi}\!\!\! \! \sum_{k=0}^{[\Delta_\phi-\D]} c_\D \tfrac{(-1)^{k}}{k!} \tfrac{\Gamma(\D+k)^2\Gamma(2\D)}{\Gamma(\D)^2\Gamma(2\D+k)} \tfrac{1}{s-\D-k}\,, \qquad \small{\text{Re}(s)< \tilde \D_0} \, , \\
	\!\!\!\! \psi_{\infty}(s)&=\int_1^{\infty} \!\!\!\!\!\!dt \, f_{\infty}(t)\, t^{-1-s}+\!\!\!\!\!\!\sum_{\D_{0}\leq \D \leq\Delta_\phi}\!\!\! \!\sum_{k=0}^{[\Delta_\phi-\D]} c_\D \tfrac{(-1)^{k}}{k!} \tfrac{\Gamma(\D+k)^2\Gamma(2\D)}{\Gamma(\D)^2\Gamma(2\D+k)}  \tfrac{1}{s-2\Delta_\phi+\D+k}\,,\small{\text{Re}(s)>2\Delta_\phi- \tilde \D_0} \, ,\label{psiinfancont}
\end{align}
where $\tilde \D_0>\Delta_\phi$ is the lightest exchanged operator above the threshold (notice that this operator could be either a primary or a descendant). Both these functions are now well-defined on the non-vanishing strip $2\Delta_\phi- \tilde \D_0<\text{Re(s)}<\tilde \D_0$ and therefore their sum yields a well-defined Mellin transform
\begin{align}\label{Mstrip}
	M(s)&=\frac{\psi_0(s)+\psi_{\infty}(s)}{\Gamma(s)\Gamma(2\Delta_\phi-s)} \, ,& 2\Delta_\phi- \tilde \D_0&<\text{Re(s)}<\tilde \D_0 \, .
\end{align}
The price to pay is a deformation of the integration contour in the inverse Mellin transform, which reads
\begin{align}\label{inverseMellin}
	f(t) = \int_\mathcal{C}\frac{ds}{2 \pi i}\,\Gamma(s)\Gamma(2\Delta_\phi-s)\,M(s)\,t^{s}\, .
\end{align}
To understand the form of the contour $\mathcal{C}$, we need to discuss the analytic structure of $M(s)$. One can follow the above-mentioned strategy to extend the definition \eqref{Mstrip} to the whole complex $s$ plane. To analytically continue $\psi_0(s)$ from the region $\text{Re}(s)<\D_0$ to the region $\text{Re}(s)<\tilde \D_0$ we subtracted a few exchanged operators in $f(t)$ and added a finite number of poles in \eqref{psi0ancont}. By adding more and more poles, we can further extend the area of analyticity. We then conclude that the Mellin block expansion defined by
\begin{equation}\label{Mellinpoles} 
	M(s)=\frac{\psi_0(s)+\psi_{\infty}(s)}{\Gamma(s)\Gamma(2\Delta_\phi-s)}
\end{equation}
with
\begin{align}\label{psi0sum}
	\psi_0(s)&=\sum_{\D}\sum_{k=0}^{\infty} c_\D \frac{(-1)^{k+1}\Gamma(\D+k)^2\Gamma(2\D)}{k!\Gamma(\D)^2\Gamma(2\D+k)}\frac{1}{s-\D-k} \, ,\\
	\psi_{\infty}(s)&=\sum_{\D}\sum_{k=0}^{\infty} c_\D \frac{(-1)^{k}\Gamma(\D+k)^2\Gamma(2\D)}{k!\Gamma(\D)^2\Gamma(2\Delta+k)}\frac{1}{s-2\Delta_\phi+\D+k} \label{psiinfsum}
\end{align}
provides a representation of $M(s)$, which is valid on the whole complex $s$ plane (excluding the point at infinity, which will be discussed in detail in subsection \ref{boundedness}). In particular, the expression~\eqref{Mellinpoles} immediately allows us to read off the position of the poles of $M(s)$.\footnote{In principle, there could be an additional singularity at $\infty$, but we postpone this discussion to section~\ref{sumrules}.} For any exchanged primary operator of dimension $\D$, two infinite sequences of poles run to the right of $s=\D$ and to the left of $s=2\Delta_\phi-\D$. Following the common nomenclature, we denote them as
\begin{align}
	\text{\emph{right} poles}: s_R&=\D+k\,, ~~\qquad\,\, \qquad k=0,1,2,\dots \label{leftpoles}\\
	\text{\emph{left} poles}: s_L&=2\Delta_\phi-\D-k\,, \qquad k=0,1,2,\dots \label{rightpoles}\\
	\text{Res}[M(s)]|_{s_L}&\equiv-\text{Res}[M(s)]|_{s_R}=\frac{(-1)^{k} \Gamma (2 \Delta) \Gamma (\D+k)}{k! \,\Gamma (\D)^2 \Gamma (2 \D+k) \Gamma (2   \Delta_\phi-\D-k )}\,. \label{residues}
\end{align}\par 
Notice that the precise identification of the sum over $k$ in \eqref{psi0sum} with the sum over descendants in the block expansion is a consequence of the choice $a=0$ in \eqref{Mellin1para}. The different choices of $a$ in \eqref{Mellin1para} would lead to a less transparent correspondence between poles and conformal descendants.

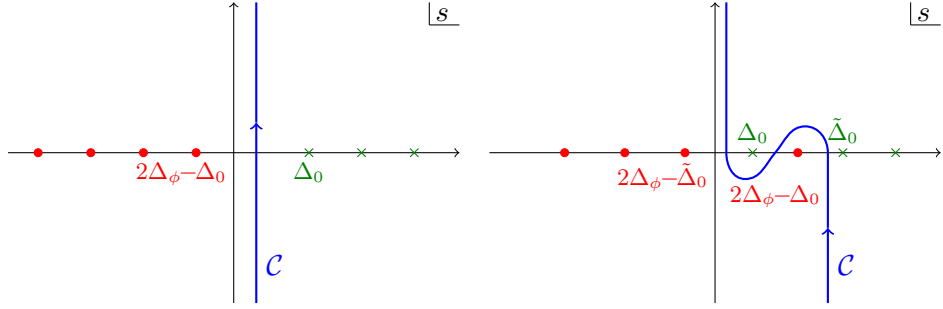
\begin{figure}[h]
	\center
	\begin{tikzpicture}
		\draw[->](-3,0)--(3,0);
		\draw[->](0,-2)--(0,2);
		\draw[] (2.6,2)--(2.6,1.7);
		\draw[] (3,1.7)--(2.6,1.7);
		\node[anchor = north east] at (3,2.05) {$s$};
		\draw[thick,->,blue] (0.3,-2)--(0.3,0.4);
		\draw[thick,blue] (0.3,0.4)-- (0.3,2);
		\node[blue, right] at (0.3,-1.5) {$\mathcal{C}$};
		\filldraw [red] (-0.5,0) circle (1.5pt);
		\filldraw [red] (-1.2,0) circle (1.5pt);
		\filldraw [red] (-1.9,0) circle (1.5pt);
		\filldraw [red] (-2.6,0) circle (1.5pt);
		\draw (1,0) node[cross=2pt,green!50!black] {};
		\draw (1.7,0) node[cross=2pt,green!50!black] {};
		\draw (2.4,0) node[cross=2pt,green!50!black] {};
		\node[below, green!50!black] at (1,0) {\footnotesize{$\D_0$}};
		\node[below, red] at (-0.7,0) {\footnotesize{$2\Delta_\phi\!\!-\!\! \D_0$}};
	\end{tikzpicture}\quad
	\begin{tikzpicture}
		\draw[->](-3,0)--(3,0);
		\draw[->](0,-2)--(0,2);
		\draw[] (2.6,2)--(2.6,1.7);
		\draw[] (3,1.7)--(2.6,1.7);
		\node[anchor = north east] at (3,2.05) {$s$};
		\node[blue, right] at (1.5,-1.5) {$\mathcal{C}$};
		\filldraw [red] (1.1,0) circle (1.5pt);
		\filldraw [red] (-0.4,0) circle (1.5pt);
		\filldraw [red] (-1.2,0) circle (1.5pt);
		\filldraw [red] (-2,0) circle (1.5pt);
		\draw (0.5,0) node[cross=2pt,green!50!black] {};
		\draw (1.7,0) node[cross=2pt,green!50!black] {};
		\draw (2.4,0) node[cross=2pt,green!50!black] {};
		\node[above, green!50!black] at (0.5,0) {\footnotesize{$\D_0$}};
		\node[below, red] at (.8,-0.3) {\footnotesize{$2\Delta_\phi\!\!-\!\! \D_0$}};
		\node[above, green!50!black] at (1.7,0) {\footnotesize{$\tilde \D_0$}};
		\node[below, red] at (-0.7,0) {\footnotesize{$2\Delta_\phi\!\!-\!\! \tilde \D_0$}};
		\draw [->,thick,blue] (1.5,-2) to[out=90,in=-90](1.5,-1) ;
		\draw [,thick,blue] (1.5,-1) to[out=90,in=-90](1.5,0) to[out=90,in=0](1.2,.35) to [out=180, in=50] (0.8,0)to [out=50-180, in =0](.4,-.35) to[out=180, in=-90] (.15,0)to[out=90, in = -90](.15,2);
	\end{tikzpicture}\quad
	\caption{\textbf{Left}: The contour for the inverse Mellin transform when $\D_0>\Delta_\phi$. Left poles are marked in \textcolor{red}{red} and right poles are \textcolor{green!50!black}{green}. \textbf{Right}: When $\D_0<\Delta_\phi$ left and right poles intersect, the contour must be deformed.}\label{fig:C}
\end{figure}

Given this structure of poles, we can now precisely define the contour $\mathcal{C}$ in \eqref{inverseMellin}. The contour $\mathcal{C}$ is chosen in such a way as to leave all the \emph{right} poles of $M(s)$ on its right and all the \emph{left} poles on its left. Suppose the lightest exchanged operator has dimension $\D_0>\Delta_\phi$. In that case, no analytic continuation is required in \eqref{psi0def} and \eqref{psiinfdef} (in other words, the set of left and right poles do not overlap) and any contour within the interval \eqref{convergence} will suffice, see the straight one on the left in Figure \ref{fig:C}. 

When lighter operators are exchanged, the contour needs to be deformed because the set of right poles intersects with the set of left poles. Figure~\ref{fig:C} shows an example with a single operator below the threshold. It is clear from the picture that a more complicated situation arises when a left and a right pole coincide. For instance, this happens in the GFF case, which we address in Appendix~\ref{sec:GFF}. More generally, this happens whenever there is an exchanged operator with dimension $\D=\Delta_\phi+\frac{\mathbb{Z}}{2}$. We do not expect this to be the case in a generic spectrum.\par 
\par 
We conclude this section by noticing that we can perform the sum over $k$ in~\eqref{Mellinpoles}, resumming all the conformal descendants in a crossing symmetric Mellin block expansion 
\begin{align} \label{hypergeom}
	M(s) &= \frac{1}{\Gamma(s)\Gamma(2\Delta_\phi-s)} \sum_\D\,c_\D [\mathcal{F}_\D(s)+\mathcal{F}_\D(2\Delta_\phi-s)]\,,\\
	\mathcal{F}_\D(s)&= \frac{{}_3F_2({\D, \D, \D - s}; {2 \D,1+ \D-s}; -1)}{\D-s} \, .
\end{align}
It is worth noting the importance of these finite subtractions in the context of generalized free field theory. This simple yet subtle example is shown in Appendix \ref{GFF Mellin amplitude}. 

\subsection{Regge Limit and Mellin Boundedness} \label{boundedness}
In this subsection, a bound on the large $s$ behaviour of the Mellin amplitude $M(s)$ will be derived using the Regge behaviour of the function $f(t)$.\footnote{In terms of the cross-ratio $t$, this corresponds to the limit $t\to e^{i\pi}$ described in \eqref{reggetheta}.}
The Regge limit is \footnote{This limit can be taken along any direction excluding the real line to avoid the branch cuts, but for definiteness we take it along the imaginary axis.}
\begin{align}
z\to \frac12+ i \infty.
\end{align}
This limit can be understood in terms of the higher-dimensional correlator in the diagonal limit, where it corresponds to the $u$-channel Regge limit.\footnote{In 1d there is no $u$-channel OPE expansion as it is impossible to bring $x_1$ close to $x_3$ without $x_2$ in-between. However, one can resort on the higher dimensional picture to understand that while the $u$-channel OPE would correspond to $z\to i\infty$ and $\bar z\to -i\infty$, the $u$-channel Regge limit is $z,\bar z\to i\infty$.} In particular,  four-point functions of a unitary CFT are bounded in the Regge limit \cite{Caron-Huot:2017vep,Maldacena:2015waa}, and we have \cite{Mazac:2018ycv}
\be\label{Reggebound}
\left|\tilde{g}\left(\textstyle{\frac{1}{2}}+iT\right)\right|\, \text{is bounded as } T\rightarrow \infty.
\ee
Translating into the $t$ cross-ratio \eqref{t}, the line parametrized by $z=\frac12+i \,\xi$ is mapped into the unit circle $t=e^{i \theta}$ for $\theta\in(-\pi,\pi)$ and the Regge limit occurs when $\theta\to \pi$. The Regge boundedness condition \eqref{Reggebound} for the function $f(t)$ then reads
\begin{equation}\label{reggetheta}
	f(e^{i\theta})=\mathcal{O}\left((\pi-\theta)^{-2\Delta_\phi}\right) \qquad \theta\to \pi .
\end{equation}
Looking at the direct definition of the Mellin transform \eqref{Mellin}, it may seem surprising that the large $s$ behaviour is controlled by a region ($t\sim -1$) which is far away from the integration contour. We must start by considering the inverse Mellin transform \eqref{inverseMellin} where the contour $\mathcal{C}$ is a straight line parametrized by $s=c+i\eta$ for some constant $2\D-\tilde \Delta_0<c<\tilde \Delta_0$ (the additional poles that are included in \eqref{psi0ancont} and \eqref{psiinfancont} for the analytic continuation will not affect this argument) and $\eta \in\mathbb{R}$. We take $t=e^{i\theta}$ and we integrate over~$\eta$  
\begin{equation}\label{inverseMellineta}
	f(e^{i \theta})=e^{i c \theta} \int_{-\infty}^{\infty} d\eta\, \Gamma(c+i \eta) \Gamma(2\Delta_{\phi}-c-i \eta)\, M(c+i\eta) \, e^{- \theta \eta} \, .
\end{equation}
We are interested in the behaviour of the integrand for $|\eta|\to \infty$. In this limit 
\begin{equation}\label{Gammaasyn}
	\Gamma(c+i \eta) \Gamma(2\Delta_{\phi}-c-i \eta) \sim e^{-\pi|\eta|}\, \eta^{2\Delta_{\phi}-1} \qquad |\eta|\to\infty \, .
\end{equation}
This means that the Gamma function prefactor accounts for the exponential behaviour \eqref{bound M} of $\hat{M}(s)$ for $|\text{Im}(s)|\to\infty$ predicted by the theorem in subsection \ref{theorem}. This essentially motivates our choice of prefactor in \eqref{Mellin}. In particular, the exponential in \eqref{Gammaasyn} combined with that in \eqref{inverseMellineta} shows that the regime $\theta\to \pm\pi$ is controlled by the region $\eta\sim \mp\infty$. Let us make this more precise by defining
\begin{align}\label{Hdef}
	H(\eta)&\equiv\Gamma(c+i \eta) \Gamma(2\Delta_{\phi}-c-i \eta)\, M(c+ i \eta) \,e^{\pi|\eta|} 
\end{align}
so that the integral \eqref{inverseMellineta} can be rewritten as
\begin{equation}\label{Laplacetransf}
	f(e^{i \theta})=e^{i c \theta} \int_{0}^{\infty} d\eta \,H(-\eta) \, e^{-\eta(\pi-\theta)}+ e^{i c \theta} \int_{0}^{\infty} d\eta \,H(\eta) \, e^{-\eta(\pi+\theta)} \,,
\end{equation}
where we recognize two Laplace transforms of the functions $H(\pm\eta)$. A singular behaviour for $\theta=\pi$ originates from the first term in the sum \eqref{Laplacetransf}, while the singularity at $\theta=-\pi$ arises from the second term. More specifically, Tauberian theorems for the Laplace transform imply that for a function $H(\eta)\sim k\, \eta^{ \a}$ as $\eta\to \infty$ then 
\begin{equation}
	\int_{0}^{\infty} d\eta\, H(\eta) \, e^{-\eta(\pi+\theta)}\sim k\, \Gamma(\a+1) (\theta+\pi)^{-\a-1}  \qquad \theta\to-\pi
\end{equation}
and similarly for the case $\theta\to\pi$. We are then led to the conclusion that the Regge behaviour \eqref{reggetheta} is reproduced by asking that
\begin{equation}
	H(\eta)\sim |\eta|^{2\Delta_{\phi}-1} \qquad |\eta|\to \infty \, .
\end{equation}
Combining this with \eqref{Hdef} and \eqref{Gammaasyn} we conclude that 
\begin{equation}
	M(c+i\eta)=\mathcal{O}(|\eta|^0) \qquad |\eta| \to \infty \, .
\end{equation}
Assuming that no Stokes phenomenon occurs for physical correlators, we can extend this behaviour for any $\text{arg}(s)$ such that
\begin{equation}\label{resultasymM}
	M(s)=\mathcal{O}(|s|^0) \qquad |s|\to\infty \, .
\end{equation}
The absence of Stokes phenomenon is an assumption for which we do not have proof. This assumption, however, is verified in all our examples and was also made in the higher-dimensional case \cite{Penedones:2019tng}. 
%In particular, notice that the function $M(s)$ has infinitely many poles that accumulate at $s=\infty$, where we would expect an essential singularity. Nevertheless, our prefactor in \eqref{Mellin} removes this singularity and leaves us with a bounded function \eqref{resultasymM}.

We conclude this subsection with an important remark about the perturbative regime, which we will consider in subsection \ref{sec:pert}. The result \eqref{resultasymM} is valid for the full non-perturbative Mellin amplitude. If the correlator contains a small parameter, it is often the case that, order-by-order in the perturbative expansion, the Regge behaviour is worse than in the full non-perturbative correlator.\footnote{A typical example of this phenomenon is the function $\frac{1}{1-gz}$, which is regular for $z\to\infty$ but its expansion at small $g$ is more and more divergent.} In Appendix~\ref{app:polesandseries}, we  illustrate this in detail in the context of the analytic sum rules discussed in the next section. Given this aspect, it is thereby useful to formulate our result in a more general form. Let us consider a correlator $\tilde f(z)$ with a Regge behaviour
\begin{equation}
	\tilde  f(z)=\mathcal{O}( z^{2\Delta_{\phi}+n}) \qquad z\to \frac12+i\infty
\end{equation}
for some positive integer $n$, then the associated Mellin amplitude will have a large $s$ asymptotics
\begin{equation}\label{Mlarges}
	M(s)=\mathcal{O}( |s|^n) \qquad |s|\to \infty \, .
\end{equation}

\subsection{Sum Rules} 
\label{sumrules}

A common way to express the well-known fact that an arbitrary set of CFT data does not necessarily lead to a consistent CFT is through a set of sum rules for the CFT data. In the following, we will start with our Mellin amplitude definition and derive an infinite set of sum rules. As we mentioned in the Introduction, these sum rules are not dispersive, according to the definition of \cite{Caron-Huot:2020adz}. This is related to the behaviour at infinity obtained using our one-dimensional definition. In subsection \ref{boundedness}, we described how the product of Gamma functions in our definition \eqref{Mellin} leads to a nice behaviour for the Mellin amplitude $M(s)$ at $s=\infty$. However, introducing that prefactor also leads to the appearance of spurious poles in the integral \eqref{inverseMellin}. In a generic CFT, it is not expected that operators with the exact dimension $s=2\Delta_\phi+n$ are present in the spectrum. Thus, the poles of the Gamma functions must be compensated by zeros in the Mellin amplitude. This strategy was used in \cite{Penedones:2019tng,Carmi:2020ekr} to derive dispersive sum rules for the higher dimensional case, where the Mellin amplitude needs to have double zeros. Here, we will use the same idea to derive a new set of sum rules characterised by single zeros of the Mellin amplitude. This makes these sum rules different and less powerful than the dispersive ones. Still, we believe that their derivation and the check of their validity on a set of known examples provide an important consistency check of our results.

One may be concerned because the presence of single or double zeroes for the Mellin amplitude seems to be related to the choice of the prefactor in \eqref{Mellin}. This is not the case. The choice to factor out a prefactor in \eqref{Mellin} is related to having a polynomial behaviour for the function $M(s)$ as $s\to\infty$. If we were to pick a different prefactor (for instance, using Gamma function squared, leading to double poles for the Mellin amplitude), the Mellin amplitude would contain an essential singularity at $s=\infty$, and this divergence would have to be compensated by the function $F_p(s)$, which we will use in \eqref{functional} to derive our sum rules. We can safely conclude that choosing a prefactor is a convenient trick, but it does not affect the resulting sum rules.

Finally, let us emphasize some important differences compared to the higher-dimensional strategy of the Mellin Polyakov bootstrap \cite{Polyakov:1974gs,Gopakumar:2016wkt,Gopakumar:2016cpb}. The derivation of the non-perturbative Polyakov consistency conditions used in \cite{Penedones:2019tng,Carmi:2020ekr} is quite subtle due to accumulation points in the twist spectrum of higher-dimensional CFTs. In our case, however, the situation is simpler. The twist accumulation points are related to the presence of a spin or, equivalently, to the need of introducing two Mandelstam variables. For us there is no spin and the only quantum number is the scaling dimension of the operators. Therefore, we do not expect any accumulation point in the spectrum and we will be able to impose the conditions \eqref{zeroes} without recurring to any analytic continuations. 

We start by summarizing the main properties of the Mellin amplitude $M(s)$ in~\eqref{Mellin}: 
\begin{itemize}
	\item $M$ is crossing symmetric
	\begin{equation}
		M(s) = M(2\Delta_\phi-s) \, .
	\end{equation}
	\item $M$ has poles at the location of the physical exchanged operators in the two channels, i.e. $s=\D+k$ and $s=2\Delta_\phi-\D-k$ for $k\in \mathbb{N}  \, .$
	\item Generically, $M$ has single zeros compensating the poles of the prefactor 
	\begin{align}\label{zeroes}
		M(2\Delta_\phi+k) =0 \quad\text{and} \quad M(-k)=0 \quad \text{for} \quad k \in \mathbb{N} \,.
	\end{align}
	Some of these zeros might be absent if the spectrum contains protected operators.
	\item $M$ is bounded for $|s|\to \infty$, see \eqref{resultasymM} .
	
	\item $M$ admits a crossing-symmetric Mellin block expansion
	\begin{align}
		M(s) = \sum_\D \, c_\D\, M_\D(s) 
	\end{align}
	with $M_\D(s)$ given by the comparison with \eqref{hypergeom}.
\end{itemize}

The properties above will allow us to define a set of sum rules along the lines of \cite{Penedones:2019tng, Carmi:2020ekr}. 
Let $\omega_p$ be the functional 
\begin{align}\label{functional}
	\omega_{p_i} = \oint_{\mathbb{C}|_{\infty}} \frac{ds}{2\pi i}M(s)F_{p_i}(s)\, ,
\end{align}
where the contour here is a very large circle around infinity. When $F_{p_i}(s)$ is a sufficiently suppressed function at $s\to \infty$, we can take the limit of infinite radius for the circle, and we get
\begin{align}
	\omega_{p_i} [M] =0	 \, .
\end{align}
For a nonperturbative Mellin amplitude characterised by the asymptotic behaviour \eqref{resultasymM} it is sufficient to ask that $F_{p_i}(s)\sim s^{-1-\e}$ for $\e>0$ as $|s|\to \infty$. As we mentioned at the end of subsection \ref{boundedness}, when considering a perturbative expansion around GFF, the Regge behaviour may worsen and a sufficiently suppressed function $F$ would be required as detailed in Appendix \ref{app:polesandseries}.\par \vspace{3mm}
The strategy to derive the sum rules consists of deforming the integration contour in \eqref{functional} to include all the poles of the integrand such that
\begin{align}\label{functionalwithM}
	\omega_{p_i} = \sum_{s^*} \text{Res}_{s=s^*}\left[M(s)\right]F_{p_i}(s^*)+\sum_{s^{**}} M(s^{**})\text{Res}_{s=s^{**}} \left[F_{p_i}(s)\right]=0 \, .
\end{align}
This equation already looks like a sum rule, but it depends on the value $M(s^{**})$ of the Mellin amplitude at the poles of $F_{p_i}(s)$. To avoid this issue, one can choose $F_{p_i}(s)$ to have simple poles at the position of the zeros of $M(s)$. Therefore, we need a function $F_{p_i}(s)$ with poles at $s=-k$ or at $s=2\Delta_{\phi} +k$. Furthermore, the function $F_{p_i}(s)$ must not be crossing symmetric. Indeed, using the position of the poles in \eqref{leftpoles} and \eqref{rightpoles} and crossing symmetry for the residues \eqref{residues} we get
\begin{align}\label{sumrulerightpoles}
	\omega_{p_i} = \sum_{s_R} \text{Res}_{s=s_R}(M(s))(F_{p_i}(s_R)-F_{p_i}(2\Delta_\phi-s_R)) \, ,
\end{align}
so that any crossing symmetric function $F$ would lead to a trivial vanishing of $\omega_{p_i}$. Using the explicit expression for the residues \eqref{residues}, we find the set of sum rules
\begin{align}\label{sumrulesgenericF}
	\sum_{\D,k} c_\D \frac{(-1)^{k+1}\Gamma(2\D)\Gamma(\D+k)}{\Gamma(\D)^2\Gamma(2\D+k)\Gamma(2\Delta_\phi-\D-k)\Gamma(k+1)}(F_{p_i}(\Delta+k)-F_{p_i}(2\Delta_\phi-\Delta-k))=0 \, .
\end{align}
A natural choice for the function $F$ is 
\begin{align}\label{Fp1p2}
	F_{p_1,p_2}(s)=\frac{1}{(s+p_1)(s+p_2)} \, ,\qquad p_1,p_2\in \mathbb{N} \, .
\end{align}
Notice that, despite the function $F_{p_1,p_2}(s)\sim \frac{1}{s^2}$ for $s\to \infty$, thanks to \eqref{sumrulerightpoles} only the crossing antisymmetric part of it matters, i.e. $F_{p_1,p_2}(s)-F_{p_1,p_2}(2\Delta-s)$, and one can easily check that this combination decays as $\frac{1}{s^3}$ for $s\to\infty$. Using this function, we can derive the nonperturbative sum rules
\begin{align}\label{sumrulesnonpert}
	\sum_{\D,k} c_\D \tfrac{(-1)^{k+1}\Gamma(2\D)\Gamma(\D+k)}{\Gamma(\D)^2\Gamma(2\D+k)\Gamma(2\Delta_\phi-\D-k)\Gamma(k+1)}\tfrac{2(\Delta+k-\Delta_{\phi})(p_1+p_2+2\Delta_{\phi})}{(\Delta+k+p_1)(\Delta+k+p_2)(2 \Delta_{\phi}-\Delta-k+p_1)(2 \Delta_{\phi}-\Delta-k+p_2)}=0 \, .
\end{align}
Performing the sum over $k$, one obtains sum rules of the form 
\begin{align}
	\sum_{\D} c_\D  \a_{\Delta}=0
\end{align} \label{formsumrule}
with 
\begin{align}\label{alphasumrule}
	\a_{\Delta}&=\frac{\Gamma(\Delta)}{\Gamma(2\Delta)\Gamma(2\Delta_{\phi}-\Delta)} \left(\mathcal{F}_{p_1,p_2}(\Delta)+\mathcal{F}_{-2\Delta_{\phi}-p_1,-2\Delta-p_2}(\Delta) \right) \, ,\\
	\mathcal{F}_{p_1,p_2}(\Delta)&=\tfrac{1}{p_1-p_2}\left(\tfrac{{}_3F_2(\Delta,p_1+\Delta,1+\Delta-2\Delta_{\phi};2\Delta,1+p_1+\Delta;1)}{(p_1+\Delta)}-\tfrac{{}_3F_2(\Delta,p_2+\Delta,1+\Delta-2\Delta_{\phi};2\Delta,1+p_2+\Delta;1)}{(p_2+\Delta)}\right)\, .
\end{align}
As already mentioned in the Introduction, these sum rules differ from those found in \cite{Mazac:2018ycv,Ferrero:2019luz,Penedones:2019tng,Carmi:2020ekr}, which are dispersive sum rules having double zeros at the dimension of double twist operators (twist in higher-$d$). Our functionals $\a_{\Delta}$ have single zeros at $\Delta=2\Delta_{\phi}+k$ for $k\in \mathbb{N}$ and $k\neq p_1,p_2$, implying that the functional changes sign at any of these zeros. The absence of any positivity property makes these sum rules less powerful and harder to use with the standard method of the modern conformal bootstrap. Recent developments in this direction have these positivity conditions \cite{Knop:2022viy} and have improved upon this method, confirming some of the results presented in \cite{Bianchi:2021piu}. 

\subsection{Perturbative Results}
\label{sec:pert}
 In this subsection, we consider the Lagrangian presented in \eqref{interactionlag} and use the Mellin transform to find the corresponding four-point correlators and anomalous dimension for generic $L$. 
Using \eqref{Ansatzcorrelator} as a basis for $4L$-derivative results, we can take its Mellin transform. The first step is to compute the Mellin transform of the function $\bar{D}_{\Delta_\phi\Delta_\phi\Delta_\phi\Delta_\phi}(t)$. In this section we consider the reduced Mellin amplitude $\hat{M}(s)\equiv M(s) \Gamma(s)\Gamma(2\Delta_{\phi}-s)$ and we need to compute
\be\label{MellinDbar}
\hat{M}_{\Delta_\phi}(s) = \int_0^\infty dt \,\bar D_{\Delta_\phi\Delta_\phi\Delta_\phi\Delta_\phi}(t)  \, \Big(\frac{t}{1+t}\Big)^{2\Delta_\phi}\,t^{-1-s}   \, .
\ee
A closed-form expression for the $\bar D$ functions is unavailable, and dealing with integral representations is quite hard. Therefore, we considered the case of integer $\Delta_{\phi}$, where simple explicit expressions for the $\bar D$ functions are known (see~\eqref{Dbar-explicit}-\eqref{Dbar-explicit-end}) and we inferred the general form
\begin{align}\label{MellinL0}
	\hat{M}_{\Delta_\phi}(s) &=\pi \csc(\pi s)\,\Big(  \pi \cot(\pi s)P_{\Delta_\phi}(s) -\sum_{k=1}^{2\Delta_\phi-1}\frac{P_{\Delta_\phi}(k)}{s-k} \Big)\, ,\\
	P_{\Delta_\phi}(s) &= 	2 \sum_{n=0}^{\Delta_{\phi}-1}   (-1)^{n} \frac{\Gamma(2 n + 1) \Gamma^4(\Delta_{\phi}) \Gamma(\Delta_{\phi}+n)}{\Gamma^4(n + 1)\Gamma(\Delta_{\phi}-n)\Gamma(2(\Delta_{\phi}+n))} (2\Delta_{\phi}-s)_n(s)_n\,,
\end{align}
The functions $P_{\Delta_\phi}(s)$ are effectively just polynomials of order $2\Delta_\phi-2$. Defining \be
Q_{\Delta_{\phi}} (s(s-2\Delta_{\phi}))\equiv P_{\Delta_{\phi}}(s) \, ,
\ee
we have, for the first few cases
\begin{align} \label{DfunctMellintable}
	{\renewcommand\arraystretch{1.3} 
		\begin{tabular}{ c | c }
			$\Delta_{\phi}$ &  Q$_{\Delta_{\phi}} (x)$ \\
			\hline 
			1  &  2 		\\																										
			2  & $\frac{1}{15} (5 + x) $ 	\\																					
			3  & $\frac{1}{315} (84 + 17 \,x + x^2) $		\\											
			4  & $\frac{1}{30030} (15444 + 2889 \,x + 206\, x^2 + 5\, x^3)$	\\
			5  & $\frac{1}{765765} (1400256 + 239640\, x + 17387\, x^2 + 570\, x^3 + 7\, x^4)$	\\
	\end{tabular}      }   
\end{align}

The functions $P_{\Delta_\phi}(s)$ can also be rewritten as
\begin{align}\label{PDelta}
	P_{\Delta_\phi}(s) &= 	2 \frac{\Gamma(\Delta_\phi)^4}{\Gamma(2\Delta_\phi)}	{}_4{F}_3(\{\textstyle\frac{1}{2},s,1-\Delta_\phi,2\Delta_\phi-s\};\{1,1,\Delta_\phi+\frac{1}{2}\};1)\, ,
\end{align}
Notice that in cross-ratio space a closed-form expression for the $\bar D$ functions is not known, while in Mellin space it looks reasonably simple, at least for integer $\Delta_{\phi}$. This is similar to what happens in the higher dimensional case, where this occurrence is even more striking as the reduced Mellin transform of the $\bar D$ functions is simply a product of Gamma functions.\footnote{It is often said that the Mellin transform of contact interactions is one, but this assumes that the correct product of Gamma function has been factored out~\cite{Mack:2009mi,Penedones:2010ue}.} In the one-dimensional case, we could not find such a simple representation for the contact interactions, but the fact we could write the result in a closed form is already a notable improvement compared to cross-ratio space, and as we will see, it will allow us to extract new CFT data successfully. 

Knowing the Mellin transform for the $\bar D$ functions, it is simple to compute the Mellin transform of \eqref{Ansatzcorrelator}
\begin{align}\label{ML}
	\hat{M}^{(1)}_{L}(s)&=\int_0^\infty dt \, f^{(1)}_L(t)\, \Big(\frac{t}{1+t}\Big)^{2\Delta_\phi}\,t^{-1-s}=  \,\sum_{k=0}^{2L}	c_{k,L} \, \hat M_{\Delta_\phi+L}(s+k)\, ,\\
	2 c_{k,L} &= \frac{\Gamma(2L+1)}{\Gamma(k+1)\Gamma(2L-k+1)}+\delta_{k,0}+\delta_{k,2L}\, .
\end{align}
Notice that the presence of double poles for integer values of $s$ in~\eqref{MellinL0} is not in contradiction with the general single-pole structure of the nonperturbative Mellin amplitude~\eqref{Mellinpoles}-\eqref{psiinfsum}, but just a consequence of the perturbative expansion of those single poles at this first order of perturbation theory as detailed in Appendix \ref{app:polesandseries}. Moreover, the structure of~\eqref{MellinL0}  is such that both single and double poles cancel (as they should) within the region of convergence~\eqref{convergence} of the integral~\eqref{MellinDbar}, which in this case ($\Delta_0=2\Delta_\phi$) is $0<\text{Re}(s)<2\Delta_\phi$.  The cancellation of the double poles is evident, given the poles of $\cot(\pi s)$ and the explicit poles in the sum.  The cancelling of the single poles stems  from a property of the polynomial $P_{\Delta_\phi}(s)$, which ensures the cancellation of the finite part in the expression multiplying $\csc(\pi s)$ when expanded  around integer values of~$s$, $0<s<2\Delta_\phi$. 
As we will see, this structure is consistent with the OPE expansion.

We stress that equation \eqref{ML} is a closed-form expression for the first-order perturbation around GFF generated by a quartic interaction with any number of derivatives. Under the assumption that the deformation from GFF described by these interactions only modifies two-particle  data, see~\eqref{pertdelta}-\eqref{pertOPE} below, one can extract these data. In particular, in section \ref{CFTdata}, we will show that the anomalous dimension of two-particle  operators receives the following correction
\begin{equation}\label{expansiondim}
	\Delta=2 \Delta_{\phi}+2 n+g_{L} \hat\gamma_{L,n}(\Delta_{\phi})
\end{equation}
with
\begin{align}
	\hat{\gamma}_{ L, n}^{(1)}(\Delta_{\phi})&=\frac{\Gamma(L+\Delta_{\phi})^4}{\Gamma(2L+2\Delta_{\phi})}\sum_{p=0}^{2n} \sum_{k=2L-p}^{2L}\sum_{l=0}^{k+p-2L}(-1)^k c_{k,L} \times\label{greatresult}\\   &\frac{(4\Delta_{\phi}+2n-1)_p (-2n)_p (2L-k-p)_l (1-\Delta_{\phi}-L)_l (2\Delta_{\phi}+k+p)_l (\tfrac12)_l}{(l!)^3 (2\Delta_{\phi})_p (\tfrac12+\Delta+L)_l} \nonumber  \, .
\end{align}
To compare these results with those computed with bootstrap methods in~\cite{Mazac:2018ycv, Ferrero:2019luz} for $L=0,1,2,3$, we have to change the basis in the space of couplings. Since the bootstrap approach is blind to the specific values of the couplings $g_L$ in \eqref{expansiondim}, one needs to establish a criterium to organize the set of independent data. The criterium that is used in~\cite{Mazac:2018ycv, Ferrero:2019luz} consists of setting
\begin{equation}\label{condvanandim}
	\gamma_{L,n}(\Delta_{\phi})= 0 \qquad n<L \, .
\end{equation}
In our approach, this is implemented by taking a linear combination
\begin{equation}\label{lincombandim}
	\gamma_{L,n} =\sum_{\ell=0}^{L} a_{\ell} \hat\gamma_{\ell,n}
\end{equation}
and fixing the $L+1$ $a_\ell$ coefficients in~\eqref{lincombandim}, using the $L$ conditions \eqref{condvanandim} and the normalization $\gamma_{L,L}(\Delta_{\phi})=1$. Following this strategy in section \ref{CFTdata}, we will reproduce the known results for $L\leq 3$ and present new results for $L\leq 8$ at any $\Delta$ and $n$. We stress, however, that equation \eqref{greatresult} is valid for any $L$, so, up to the algorithmic procedure of fixing the $a_{\ell}$ coefficients, one can easily extract the result for any given $L$.

%These results allow us to formulate the following recursive relation for first-order anomalous dimensions $\gamma^{\D_\phi}_{n,L}$ of exchanged operators when the external operators in four-point correlators have dimension $\Delta_\phi$ 
%\begin{align}\label{recursive}
%		\sum_{n=0}^{\lfloor{\frac{p}{2}\rfloor}}\gamma^{\Delta_\phi}_{n,L} G(n,\Delta_\phi,p) &= \sum_{l=0}^L a_l \sum_{k=0}^{2l} c_{k,l}(-1)^{k}\sum_{n=0}^{\lfloor{\frac{p}{2}\rfloor}}\gamma^{\Delta_\phi+\ell}_{n,0} G(n,\Delta+\ell,p-k) 
%\end{align}
%where $\gamma^{\Delta_\phi+\ell}_{n,0}$ is given in~\eqref{} and
%\be
%G(n,\Delta_\phi,2p) =\frac{2 (4 \Delta_\phi +4 n-1) \Gamma (2 n+4 \Delta_\phi -1) \Gamma (p+2 \Delta )^2}{\Gamma (2 \Delta_\phi )^2 \Gamma (2 n+1) \Gamma (-2 n+p+1) \Gamma (2 n+p+4 \Delta_\phi )}
%\ee
%
Below we will show how the interaction \eqref{interactionlag} leads to the correlator~\eqref{Ansatzcorrelator} through explicit Witten diagrammatics. We will also see, for the cases $L=0,1,2$, how other interaction terms lead to results that can be rearranged as linear combinations of the eigenfunctions \eqref{Ansatzcorrelator}. We will then proceed to  extract CFT data in the bootstrap normalization.

\subsection{CFT Data}\label{CFTdata}

Given the closed-form expression \eqref{ML} for the perturbative Mellin amplitude, we can use it to extract the CFT data
\begin{align}\label{pertdelta}
	\Delta&\equiv \D_{n,L}=2\D_\phi+2n+g_L\, \hat{\gamma}_{ L, n}^{(1)}(\Delta_{\phi})+\dots\, ,\\\label{pertOPE}
	c_{\D}&\equiv c_{n,L}=c_n^{(0)}(\Delta_{\phi})+g_L\,c_{L,n}^{(1)}(\Delta_\phi)+\dots \,,
\end{align}
where $n\in \mathbb{N}$ and $c_n^{(0)}$ is given in~\eqref{GFFOPE}. In~\eqref{pertdelta}-\eqref{pertOPE}, we assume that the AdS interaction only modifies the CFT data of the two-particle  operators exchanged in GFF. If we insert~\eqref{pertdelta}-\eqref{pertOPE} in the general Mellin OPE expansion~\eqref{Mellinpoles} at first order in $g_L$, one obtains double and single poles at $s=2\Delta_{\phi}+p$,  $p\in \mathbb{N}$. One can then compare the corresponding residues with the ones in the tree-level Mellin amplitude~\eqref{ML}, which amounts to solving the equations formally written as
\begin{align}\label{doublepoles}
	&\sum_{n}  c_n^{(0)}(\Delta_{\phi}) \hat{\gamma}_{ L, n}^{(1)}(\Delta_{\phi})\,{\textstyle{\frac{(-1)^p \Gamma(4\Delta_{\phi}+4n)\Gamma(2\Delta_{\phi}+p)^2}{\Gamma(2\Delta_{\phi}+2n)^2\Gamma(4\Delta_{\phi}+2n+p)\Gamma(p-2n+1)}}}\,
	=\lim_{s\rightarrow 2\Delta_{\phi}+p}(s-2\Delta_{\phi}-p)^2 \hat{M}^{(1)}_L(s) \,,
	\\\label{singlepoles}
	&\sum_{n}(c_{L,n}^{(1)}(\Delta_{\phi})\!+c_{n}^{(0)}(\Delta_{\phi}) \hat{\gamma}_{ L, n}^{(1)}(\Delta_{\phi})\partial_n)\,{\textstyle{\frac{(-1)^p \Gamma(4\Delta_{\phi}+4n)\Gamma(2\Delta_{\phi}+p)^2}{\Gamma(2\Delta_{\phi}+2n)^2\Gamma(4\Delta_{\phi}+2n+p)\Gamma(p-2n+1)}}}\,
	=\text{Res}_{s= 2\Delta_{\phi}+p} \hat{M}^{(1)}_L(s)
\end{align}
for the leading-order corrections $\hat{\gamma}_{ L, n}^{(1)}(\Delta_{\phi})$ and $c_{L,n}^{(1)}(\Delta_{\phi})$.  Since the function $\hat{M}^{(1)}_L(s)$ is known explicitly, it is possible to write down a linear system for the anomalous dimensions $\hat{\gamma}_{ L, n}^{(1)}(\Delta_{\phi})$. To do this, let us use the form \eqref{MellinL0} for $\hat{M}^{(1)}_{\Delta_{\phi}}(s)$ to rewrite \eqref{doublepoles}  as
\begin{align}\label{system}
	&\sum_{n}  c_n^{(0)}(\Delta_{\phi}) \hat{\gamma}_{ L, n}^{(1)}(\Delta_{\phi})\,{\textstyle{\frac{\Gamma(4\Delta_{\phi}+4n)\Gamma(2\Delta_{\phi}+p)^2}{\Gamma(2\Delta_{\phi}+2n)^2\Gamma(4\Delta_{\phi}+2n+p)\Gamma(p-2n+1)}}}\,
	=\sum_{k=0}^{2L}(-1)^{k} c_{k,L} P_{\Delta_{\phi}+L}(2\Delta_{\phi}+p+k) \, .
\end{align}
Notice that the sum on the left hand side of \eqref{system} is truncated by the $\Gamma(p-2n+1)$ in the denominator. This means that at a fixed value of $p$, equation \eqref{system} provides an invertible linear system which can be solved for $\gamma$. The system can also be inverted explicitly using the identity
\begin{align}
	\sum_{p=0}^{2n} \frac{(4 m+4\Delta_{\phi} -1)\, \Gamma(p+4\Delta_{\phi} +2m-1) (-1)^{p}}{\Gamma(2m-p+1)\,\Gamma(2\Delta+p)^2} \frac{\Gamma(2\Delta_{\phi}+p)^2}{\Gamma(4\Delta_{\phi}+2n+p)\Gamma(p-2n+1)}=\d_{m, n}
\end{align}
yielding \eqref{greatresult}:
\begin{align}
	\hat{\gamma}_{ L, n}^{(1)}(\Delta_{\phi})&=\frac{\Gamma(L+\Delta_{\phi})^4}{\Gamma(2L+2\Delta_{\phi})}\sum_{p=0}^{2n} \sum_{k=2L-p}^{2L}\sum_{l=0}^{k+p-2L}(-1)^k c_{k,L} \times\label{gammahatsums}\\   &\frac{(4\Delta_{\phi}+2n-1)_p (-2n)_p (2L-k-p)_l (1-\Delta_{\phi}-L)_l (2\Delta_{\phi}+k+p)_l (\tfrac12)_l}{(l!)^3 (2\Delta_{\phi})_p (\tfrac12+\Delta+L)_l} \nonumber  \,.
\end{align}
Equation \eqref{gammahatsums} has been derived under the assumption that $\Delta_{\phi}$ takes integer values. However, one can argue that it holds for any $\Delta_{\phi}$ by noticing that the result for $L=0$ agrees with that of \cite{Mazac:2018ycv}, which has been obtained without assuming integer $\Delta_{\phi}$. Furthermore, equation \eqref{ML} implies that the CFT data at $L=0$ fully determines that at higher $L$ or, in other words, equation \eqref{gammahatsums} could be rewritten as a combination of anomalous dimensions for $L=0$. This shows that \eqref{gammahatsums} holds for any $\Delta_{\phi}$.

Notice that all the sums in \eqref{gammahatsums} are finite, so for a given value of $L$ and $n$, it is straightforward to extract the value of the anomalous dimension $\hat{\gamma}_{ L, n}^{(1)}(\Delta_{\phi})$. It turns out expression \eqref{gammahatsums} can be rewritten as
\begin{align}\label{gammaLN2}
	\hat{\gamma}_{ L, n}^{(1)}(\Delta_{\phi})&=\hat{\mathcal{G}}_{L,n}(\Delta_{\phi}) \hat{\mathcal{P}}_{L,n}(\Delta_{\phi})\,,
\end{align}
where
\begin{align}\label{gammaLN3}
	\hat{\mathcal{G}}_{L,n}(\Delta_{\phi})=\tfrac{\sqrt{\pi }  4^{-2 \Delta -L+1} \Gamma (2 \Delta )^2\Gamma (L+\frac{1}{2}) \Gamma (L+\Delta )^4 \Gamma (L+2 \Delta -\frac{1}{2}) \Gamma
		(n+\Delta +\frac{1}{2}) \Gamma (L-n+\Delta )}{\Gamma (L+1) \Gamma (L+\Delta +\frac{1}{2})^2 \Gamma (L+2 \Delta ) \Gamma (n+\Delta )^3 \Gamma
		(2 n+2 \Delta -\frac{1}{2}) \Gamma (L+n+\Delta +\frac{1}{2})} \,,
\end{align}
while $\hat{\mathcal{P}}_{L,n}(\Delta_\phi)$ is a polynomial in $n$ and in $\Delta_{\phi}$ of degree $6L$. It is easy to extract these polynomials from \eqref{greatresult}, a short \texttt{Mathematica} notebook was attached to the  \cite{Bianchi:2021piu} arXiv submission where the function $\texttt{FindPolynomial[L,$\Delta$,n]}$ allows to extract $\hat{\mathcal{P}}_{L,n}$ for many values of $L$ (the function gets slower and slower at higher $L$, but in principle it works for any $L$).

To make contact with the bootstrap results, the coefficients $a_\ell$  in the definition~\eqref{ML} of  $M_L(s)$ should be chosen to have the  bootstrap normalization, namely $\gamma^{(1)}_{\ell,L}=0$ for $0\leq \ell< L$, and $\gamma^ {(1)}_{L,L} =1$.  In this case, we have
% we will obtain the leading-order corrections $\gamma_{n,L}^{(1)}$ and $c_{n,L}^{(1)}$ to dimensions and OPE coefficients of two-particle  operators~\eqref{doubletrace} 
\begin{align}\label{gammaLB2}
	\gamma_{ L, n}^{(1)}(\Delta_{\phi})&=\mathcal{G}_{L,n}(\Delta_{\phi}) \mathcal{P}_{L,n}(\Delta_{\phi}) \,,
\end{align}
where
\begin{align}\label{gammaLB3}
	\mathcal{G}_{L,n}(\Delta_{\phi})=\frac{4^{-L} \left(L+\frac{1}{2}\right)_{\Delta _{\phi }} \left(L+\Delta _{\phi }\right)_{\Delta _{\phi }} (-L+n+1)_{\Delta _{\phi }-1} \left(L+n+\Delta _{\phi }+\frac{1}{2}\right)_{\Delta _{\phi }-1}}{\Gamma \left(\Delta _{\phi }\right) \left(\Delta _{\phi }\right)_{3 L} \left(2 L+\Delta _{\phi }+\frac{1}{2}\right)_{\Delta _{\phi }-1} \left(L+2 \Delta _{\phi }-\frac{1}{2}\right)_{2 L} \left(n+\frac{1}{2}\right)_{\Delta _{\phi }} \left(n+\Delta _{\phi }\right)_{\Delta _{\phi }}} \,,
\end{align}
while $\mathcal{P}_{L,n}(\Delta_\phi)$ is a polynomial of degree $4L$ in $n$ and $5L$ in $\Delta_\phi$. The explicit polynomials for the first few values of $L$ are detailed in Appendix \ref{Ap: anomalous dimension} and, up to $L=3$, they perfectly agree with the result of \cite{Ferrero:2019luz}. The attached \texttt{Mathematica} notebook has values of $L$ ranging from $L=0$ to $L=8$ as well as a function~\texttt{FindBootstrapPolynomial[L,$\Delta$,n]} to compute $\mathcal{P}_{L,n}(\Delta_\phi)$ for arbitrary $L$.

\subsection{Bootstrapping the Anomalous Dimension}\label{Bootstrapping the anomalous dimension}
Since the interaction terms in \eqref{interactionlag} form a basis, we can use the results for the anomalous dimensions to directly bootstrap the result and find the resulting effective action in AdS$_2$. Consider an effective theory of massless scalar in $\text{AdS}_2$ with the Lagrangian based on the one in \eqref{phi4Lagrangian}

\be
S=\int dx dz\,\sqrt{g}\,\big[ \,g^{\mu\nu}\,\partial_\mu\Phi\,\partial_\nu\Phi+m^2_{\Delta_\phi}\Phi^2+\sum_L g_L\,(\partial^L\Phi)^4\,\big]\,,\qquad L=0,1,\dots
\ee

From the Mellin space analysis, the resulting anomalous dimension will be

\begin{align}
	\hat{\gamma}_n^{(1)}(\Delta_{\phi})&=\sum_L g_L\hat{\mathcal{G}}_{L,n}(\Delta_{\phi}) \hat{\mathcal{P}}_{L,n}(\Delta_{\phi})\,. 
\end{align}
 
 Explicitly for massless fields, we have
 \begin{align}
 	\gamma^{(1)}_h = \frac{2 g_0}{(h-1) h}+\frac{3g_1 \left((h-1)^2 h^2-4\right)}{35 (h-1) h}+\sum_{L\geq2} g_L\hat{\gamma}_L^{(1)}.
 \end{align}
For a given Regge bound $\gamma^{1} \sim_{\Delta>>1} \Delta^l$, this series truncates. Additionally, the requirement that the anomalous dimension vanishes for a certain protected operator fixes the ratio of the two coefficients, for example:
\begin{align}
	\gamma^{(1)}_{h=2}&=0 &\qquad \Rightarrow g_0 &= 0\\
	\gamma^{(1)}_{h=4}&=0 &\qquad \Rightarrow g_0 &= -6g_1
\end{align}
A particular choice stands out, which is for $\gamma^{(1)}_{h=\Delta=1}=0$. This gives a polynomial anomalous dimension proportional to the quadratic Casimir:
\begin{align}
	\gamma^{(1)}_{h, (\gamma_{h=1}=0)}&=\frac {g_0}2 h(h-1)
\end{align}
where $g_1=\frac{35}{6}g_0$. \par 
This analysis also points to one of the limits of the analytic conformal bootstrap. When the constraint from the growth of the anomalous dimension depends on the perturbative order, each order has additional contact terms which can potentially be added and the number of additional constraints to fix the full correlator increases. Up to this point, in the bootstrap of 1d dCFTs this has not yet been an issue (at third-order in ABJM and fourth-order in $\mc{N}=4$ SYM). However, this may indicate an upper limit for the order of perturbation one can bootstrap.\footnote{The author thanks Miguel Paulos for useful comments about this fact.} The next and final Chapter will reflect on this and other areas of the analytic bootstrap that can be improved. 

\chapter{Optimisation of the Bootstrap Procedure}\label{Chap: Optimisation and limits}
\begin{chapquote}{J.R.R. Tolkien, \textit{The Lord of the Rings}}
End? No, the journey doesn't end here. 
\end{chapquote}
As illustrated in the first Chapters of this thesis, the analytic conformal bootstrap is an extremely powerful tool to compute correlators in the context of conformal line defects in holographic theories. However, many improvements can still be made in a number of the steps essential to the bootstrap process. The first concerns the Ansatz. As this is the starting point of the analytic bootstrap process, its accuracy and reliability are crucial. Additionally, there are limits to the bootstrap procedure; these take the form of the increased number of constants to fix at each order of the bootstrap and the resolving the mixing problem. Finally, improvements can still be made for the physical input. All-order results may not be available depending on the setup, especially in non-supersymmetric cases. However, there is hope that QFT methods for non-perturbative computations might pave the way for these all-order results through lattice computations. 

\section{Ansatz and Mellin Bootstrap}
The first obvious choice of improvement is through the Ansatz: a rigorous treatment along the lines of the master integral's formalism for flat-space amplitudes is yet to be carried out. There has been work done in this direction in Mellin space \cite{Paulos:2011ie} and for contact integrals as seen above \cite{Bliard:2022xsm}, and the diagrams' structure is relatively well understood. However, in physical examples, the results are much simpler than the individual diagrams (usually at least one order of transcendentality less). This has already been observed for supergravity amplitudes in $\text{AdS}_5\times \text{S}^5$ and could be linked to the exchange diagram being expressed as a finite truncation of the originally infinite sum of $D-$functions\cite{DHoker:1999mqo}. There is some evidence that this could be linked to Yangian symmetry\cite{Rigatos:2022eos} given both ABJM and $\mc{N}=4$ are integrable theories. Such an explicit realisation of Yangian symmetry would be a powerful additional tool for the bootstrap \cite{Loebbert:2019vcj, Corcoran:2021gda}. \par 
\newpage
The requirements of having well-defined functions under crossing and braiding are already highly constraining. However, additional factors linked to the reasons above seem to favour powers of logarithms to polylogarithms in the results obtained in Chapters \ref{chapter: ABJM} and \ref{chapter N=4}. 
Additionally, the Ansatz will change drastically depending on the point around which one is expanding. This difference is visible between the free theories at weak and strong coupling, where the weak coupling sector seems to have a quicker increase in transcendentality \cite{Cavaglia:2022qpg, Barrat:2021tpn}. 
Finally, there might be simpler frameworks for the Ansatz and the results; for example in \cite{Rastelli:2016nze}, the Mellin amplitude was essential for an all-order result. In the perturbative results above, there is a drastic simplification between the generic $D-$functions and the result for correlators. For example, the $D_{1111}$ function, corresponding to a four-point contact diagram with external massless scalars, has the Mellin transform (as defined in section \ref{chapter: Mellin amplitudes})
\begin{align}\label{MellinL0}
	\hat{M}_{\Delta_\phi=1}(s) &=\pi \csc(\pi s)\,\Big(  2\pi \cot(\pi s) -\frac{2}{s-1} \Big)
\end{align}
Comparing this to the ABJM four-point functions, one immediately sees the simpler forms. The first-order solution $f^{(1)}(\chi)$ in equation \ref{Eq ABJM f1sol} has the Mellin transform 
\begin{align}
	\hat{M}[f^{(1)}(\chi)](s) &=6 \epsilon \Gamma (-s-2) \Gamma (s-1),
\end{align}
and the second order is simple enough to be written as
\begin{align}\label{f2 Mellin}
	\hat{M}[f^{(2)}(\chi)](s)&=6 (\epsilon^2  p_0(s)+1) \Gamma (-s+\epsilon^2  p_1(s)-2) \Gamma (s+\epsilon^2  p_2(s)-1)|_{\epsilon^2}\\
	p_0(s)&= \frac{30 \gamma  (s-2) ((s-1) s+2)-2 s (2 s (s ((s-5) s+17)-21)+69)+298}{3 (s-3) (s-2)^2}\\
	p_1(s)&=  \frac{(s+2) (s (s (s (2 s-9)+8)+34)+3)}{6 (s-3) (s-2)}\\
	p_2(s)&=-\frac{s \left(s \left(2 s^2+s-7\right)-31\right)+38}{6 (s-2)}.
\end{align}
Additionally, the differential operators relating $f(\chi)$ to the other correlators of elements of the displacement multiplet are linear in the cross-ratio. Therefore, the Mellin transform will correspond to a linear combination of shifted Mellin amplitudes and will therefore keep the `simple form'.
For example, the massless bosonic correlator is expressed in \ref{G1} in terms of two functions $G_1(\chi)$ and $G_2(\chi)$. These can be expressed in terms of a differential equation acting on $f(\chi)$ in equation \eqref{system-f} and can be written in terms of the Mellin amplitude
\begin{align}
	G_1(\chi)&=f(\chi)-\chi f'(\chi)+\chi^2f''(\chi) \\
	G_2(\chi)&=-\chi^2f'(\chi)-\chi^3 f''(\chi) \\
	M[G_1(\chi)](s)&=(s^2+1) M[f(\chi)](s)\\
	& 6 (s^2+1)\Gamma (-s-2) \Gamma (s-1)\\
	M[G_2(\chi)](s)&=-(s-3) (s-1) M[f(\chi)](s-1)\\
	&=-6 (s-3) (s-1) \Gamma (-s-1) \Gamma (s-2).
\end{align}
\newpage
This can also be done for the correlator of the mixed fluctuations detailed in \ref{4point-mixed}, for example:
\begin{align}
	M[h(\chi)]&= (s-2) (s-1) ((s-2) s+3)M[f(\chi)](s)-(-3 + s) (-2 + s) (-1 + s)^2M[f(\chi)](s-1)\\
	&=12 (1-3s+s^2) \Gamma (s+1) \Gamma (-s-2).
\end{align}
The Mellin transform from \cite{Bianchi:2021piu} seems promising at tree-level given the simplifications which occur when compared the generic $D-$functions. Even beyond this, there seems to be a clear perturbative structure which might be exploited (for example \ref{f2 Mellin}). Moreover, the different aspects of the bootstrap up to the normalisation can be implemented directly in Mellin space. 

\section{Physical Non-Perturbative Input}
One of the dangers of the ever-increasing bound on the large-$\Delta$ behaviour of the anomalous dimension is that one has an ever-increasing number of terms to fix with the bootstrap.\footnote{This is discussed in subsection \ref{Bootstrapping the anomalous dimension}} This implies that there might be a hard upper limit beyond which one cannot use the same algorithm to constrain the correlators via the analytic bootstrap. Even in the lucky cases of the 1/2-BPS defect theories in ABJM and $\mc{N}$=4 SYM, there is still a constant to fix at each order requiring input from other computations. In the latter, the topological sector is amenable to a localisation computation which gives an all-order result. In the former, the integrated correlator conditions \cite{Drukker:2022pxk} relate to another quantity, the Bremsstrahlung function, which can be computed non-perturbatively \cite{Bianchi:2017svd, Bianchi:2018scb}. In theory, another such integrated correlator condition should exist for the four-point function \cite{Cavaglia:2022qpg}, and one could also use the analogous quantities for higher-point functions. This is one of the difficulties when extending these methods to systems with fewer symmetries; they often lack non-perturbative results. \par 
One way to counter this is to use the non-perturbative framework developed for the standard model: Lattice Field Theory.  A well-defined discretised theory would provide non-perturbative data for analytic results and give insight into the whole range of the AdS/CFT correspondence. However, such a discretisation which preserves the symmetries, matches with continuum data and is renormalisable is hard to come by. This section presents the analysis carried out in \cite{Bliard:2022kne} of the cusped Wilson line in $\mc{N}=4$ SYM studied in \cite{Forini:2016gie,Forini:2017ene, Bianchi:2019ygz, Bianchi:2016cyv}, uncovering a discretisation which preserves more symmetry. Despite this, the lattice model requires fine-tuning to have finite quantities. 

\subsection{Green-Schwarz AdS$_5\times \text{S}^5$}
The Green-Schwarz AdS$_5\times \text{S}^5$ string is expected to be defined at the non-perturbative level. A valid question is whether the non-perturbative regime of the $\sigma$-model, which describes the AdS$_5\times~\text{S}^5$ string at tree-level in string perturbation theory, is accessible through a lattice discretization of the worldsheet (while target space remains continuous). This question is motivated by the success of the lattice as a UV non-perturbative regulator of Quantum Chromodynamics. This approach has been pioneered in~\cite{McKeown:2013vpa,Forini:2016gie,Bianchi:2019ygz, Forini:2021keh}, where a lattice-discretized version of $S_{\text{cusp}}$ has been introduced and also used to perform Monte Carlo simulations.\footnote{Other lattice approaches to AdS/CFT include~\cite{Catterall:2014vka,Schaich:2014pda,Catterall:2014vga,Catterall:2015ira, Schaich:2015ppr,Schaich:2015daa,Giedt:2016yfw,Catterall:2017lub,Jha:2017zad,Rinaldi:2017mjl,Hanada:2018qpf, Catterall:2020nmn,Watanabe:2020ufk,Gharibyan:2020bab,Bergner:2021ffz, Dhindsa:2021irw,Bergner:2021goh}, see also~\cite{Schaich:2018mmv} and references therein.} The goal of this system is to have a benchmark for a case where analytic results are still known to extend to systems where they are not. 

\subsection{$U(1)\times SU(4)$ Invariant Discretization}
\label{sec:discretization}

In the framework of the AdS/CFT correspondence~\cite{Maldacena:1998im,Rey:1998ik}, the expectation value of a light-like cusped Wilson loop in
$\mathcal{N}=4$ super Yang-Mills is equal to the partition function of an open
string propagating in AdS$_5\times \text{S}^5$ space and ending on the loop at the AdS
boundary. In practice, one writes
\begin{equation}
	\label{cusp_anom}
	\left\langle\mathcal{W}_{\text{cusp }}\right\rangle
	=
	\int \mathcal{D} Y \mathcal{D} \Psi \, e^{-S_{\text{cusp}}(X_\text{cl} + Y,\Psi)} \equiv e^{-\frac{f(g)}{8} V_{2}}
	\, .
\end{equation}
This gauge-fixed configuration in AdS$_5\times$S$^5$ has a superstring action describing quantum fluctuations around  this null-cusp background~\cite{Giombi:2010bj} 
\begin{eqnarray}
	\nonumber
	S^\text{cont}_{\rm cusp}=g \int dt ds&& \Bigg\{ 
	\left| \partial_t x + \tfrac{m}{2} x \right|^2
	+ \tfrac{1}{z^4} \left| \partial_s x - \tfrac{m}{2} x \right|^2
	+ \left( \partial_t z^M + \tfrac{m}{2} z^M + \tfrac{i}{z^2} z_N \eta_i \left(\rho^{MN}\right)_{\phantom{i}j}^{i} \eta^j \right)^2
	\\ \nonumber &&
	+ \tfrac{1}{z^4} \left( \partial_s z^M - \tfrac{m}{2} z^M \right)^2  %\\
	+i\, \left( \theta^i \partial_t \theta_i + \eta^i \partial_t \eta_i + \theta_i \partial_t \theta^i + \eta_i \partial_t \eta^i \right) - \tfrac{1}{z^2} \left( \eta^i \eta_i \right)^{2}  \\ \nonumber &&  
	+2i \, \Big[
	\tfrac{1}{z^3} z^M \eta^i \left( \rho^M \right)_{ij}
	\left( \partial_s \theta^j - \tfrac{m}{2} \theta^j - \tfrac{i}{z} \eta^j \left( \partial_s x -\tfrac{m}{2} x \right) \right)
	\\ && \qquad
	+ \tfrac{1}{z^3} z^M \eta_i \big( {\rho^M}^\dagger \big)^{ij}
	\left( \partial_s \theta_j - \tfrac{m}{2} \theta_j + \tfrac{i}{z} \eta_j \left( \partial_s x -\tfrac{m}{2} x \right)^* \right)
	\Big]
	\,\Bigg\}
	\, ,
	\label{S_cusp_cont}
\end{eqnarray}
where
\begin{itemize}
	\item  $x$ is a complex bosonic field whose real and imaginary parts
	parametrize the fluctuations of the string (in light-cone gauge) at the boundary of AdS$_5$
	\item $z^M$ are six real bosonic fields\footnote{$z=\sqrt{z^M
		z^M}$ is the radial coordinate of the AdS$_5$ space, while $u^M = z^M/z$
	identifies points on $S_5$.}
	\item the Gra\ss mann-odd fields $\theta^i = (\theta_i)^\dagger,~\eta^i =
	(\eta_i)^\dagger, \, i=1,2,3,4$ are complex anticommuting variables (no
	Lorentz spinor indices appear);
	\item the matrices $(\rho^{MN})_i^{\hphantom{i} j} = (\rho^{[M} \rho^{\dagger
		N]})_i^{\hphantom{i} j}$ are the $SO(6)$ generators.  $\rho^{M}_{ij}
	$\footnote{By convention, we will write the indices of $\rho$ as down and
		those of $\rho^\dag$ as up.}  are the (traceless) off-diagonal	blocks of
	$SO(6)$ Dirac matrices $\gamma^M$ in  chiral representation.
\end{itemize}
The massive parameter $m$ keeps track of the dimensionful light-cone momentum $P_+$, set to 1 in~\cite{Giombi:2009gd}. The action~\eqref{S_cusp_cont} is invariant under a $U(1)\times SU(4)$ global symmetry defined by
\begin{gather}
	z^M \to \text{Ad}(U)^{MN} 
	z^N \ , \quad \theta^i \to U^i_{\phantom{i}j} \theta^j \ , \quad \eta^i \to U^i_{\phantom{i}j} \eta^j \ , \\
	x \to e^{i \alpha} x \ , \quad \theta^i \to e^{i \alpha/2} \theta^i \ , \quad \eta^i \to e^{-i \alpha/2} \eta^j \ ,
\end{gather}
where $U$ is an element of $SU(4)$ and its representative in the adjoint, $\text{Ad}(U)$, is an element of $SO(6)$. While the original Green-Schwarz AdS$_5\times \text{S}^5$ string action is invariant under diffeomorphisms and $\kappa$-symmetry, these local symmetries have been fixed by the choice of light-cone gauge in equation~\eqref{S_cusp_cont}. Notice that the action is not invariant under worldsheet rotations, parity ($s \rightarrow -s$), or time reversal ($t\rightarrow -t$).\par 
To define the lattice-discretised theory, we need to provide a
discretised action and an explicit expression for the measure. We choose
to use a flat measure for the fields but we keep in mind that this choice is
quite arbitrary as it is not invariant under reparametrisation of the target
AdS$_5 \times S_5$ target space. Given a generic observable $A$, expectation values in the lattice discretised theory are defined by
\begin{equation}
	\label{path_int_latt}
	\langle A \rangle = \frac{1}{Z_{\text{cusp}}} \int dx dx^* d^6z d^4\theta d^4\theta^\dag d^4\eta d^4\eta^\dag
	\, e^{-S_{\text{cusp}}} A
	\, ,
\end{equation}
where $d f \equiv \prod_{s,t} d f(s,t)$ is the discretised measure. The partition function $Z_{\text{cusp}}$ is fixed by the requirement $\langle 1 \rangle = 1$,  and $S_{\text{cusp}}$ refers now to the discretised action that we choose to be 
\begin{eqnarray}\nonumber
	\!\!\!\!\!\!\!\!\!	
	S_{\rm cusp}=g
	\sum_{s, t} a^2&&\!\!\! \!\!\Bigg\{ \!\!
	\left| b_+ \hat\partial_t x + \tfrac{m}{2} x \right|^2
	\!\! + \tfrac{1}{z^4}  \left| b_- \hat\partial_s x - \tfrac{m}{2} x \right|^2
	\!\!+ \big( b_+ \hat\partial_t z^M + \tfrac{m}{2} z^M + \tfrac{i}{z^2} z^N \eta_i (\rho^{MN})_{\phantom{i}j}^i \eta^j \big)^2
	\\ \nonumber && 
	+ \tfrac{1}{z^4} \big( \hat\partial_s z^M \hat\partial_s z^M + \tfrac{m^2}{4} z^2 \big)
	+ 2 i\, \big( \theta^i \hat\partial_t \theta_i + \eta^i \hat\partial_t \eta_i \big)
	- \tfrac{1}{z^2} \left( \eta^i \eta_i \right)^2  \\\nonumber
	&& 
	+ 2i\, \Big[ \tfrac{1}{z^3} z^M \eta^i \left(\rho^M\right)_{ij}
	\big( b_+ \bar\partial_s \theta^j - \tfrac{m}{2} \theta^j - \tfrac{i}{z} \eta^j \big( b_- \hat\partial_s x - \tfrac{m}{2} x \big) \big)
	\\\label{S_cusp_latt} && \qquad
	+ \tfrac{1}{z^3} z^M \eta_i \big({\rho^M}^\dagger\big)^{ij} \big( b_+ \bar\partial_s \theta_j - \tfrac{m}{2} \theta_j + \tfrac{i}{z} \eta_j \big( b_- \hat\partial_s x^* -\tfrac{m}{2} x^* \big) \!\big)\!
	\Big]\Bigg\}
	\, .
	\label{Scusp}
\end{eqnarray}
The action is written in terms of the forward and backward discrete derivatives
\begin{equation}
	\hat{\partial}_\mu f(\sigma)\equiv\frac{f\left(\sigma +a e_\mu\right)-f\left(\sigma\right)}{a}
	\, , \qquad 
	\bar{\partial}_\mu f(\sigma)\equiv\frac{f\left(\sigma\right)-f\left(\sigma -a e_\mu\right)}{a}
	\,
	% &=\frac{1}{\mathcal{N}} \sum_{p_{i}} \frac{e^{i p(x+a)}-e^{i p x}}{a} f(p) \\ &=\frac{1}{\mathcal{N}} \sum_{p_{i}} i \hat{p} f(p) e^{i p x}e^{i \frac{p a}{2}}  \quad \hat{p}=\frac{2}{a} \sin \left(\frac{p a}{2}\right). \end{aligned}
\end{equation}
where $e_\mu$ is the unit vector in the direction $\mu=0,1$, and  $\sigma$ is a shorthand notation for $(s,t)$.\par \vspace{2mm}
Notice that the proposed discretized action~\eqref{S_cusp_latt} depends on four
parameters: $g$, $m$, and the auxiliary parameters $b_\pm$. The discretized action $S_{\rm cusp}$ reduces to the desired continuum action $S^\text{cont}_{\rm cusp}$ in the naïve $a \to 0$ limit
if $b_\pm \to 1$. However, as we will discuss in detail, the naïve choice $b_\pm
= 1$ produces undesired UV divergences at one loop. The values of $b_\pm$ need to be tuned so that these UV divergences cancel. This is a sign that the lattice regulator does not manage to reproduce the cancellation of UV divergences that occurs in dimensional regularisation.\par \vspace{2mm}
An important feature of the proposed discretised action and measure is that they
are invariant under the full $U(1)\times SU(4)$ internal symmetry group. This is
in contrast to the discretisation previously presented
in~\cite{Bianchi:2019ygz}. The key ingredient is the use of forward and backward
discrete derivatives for both the bosonic and fermionic parts of the action.
This is normally avoided for fields that satisfy first-order equations of motion
(usually fermions) since it breaks parity and time reversal.  This
is not an issue in our case because these symmetries are already broken in the continuum
action. In~\cite{Bianchi:2019ygz}, instead, the symmetric derivative was used
and, as in lattice QCD, a Wilson-like term was included to solve the resulting
doubling problem while breaking either the $U(1)$ or the $SU(4)$ symmetry. The perturbative series is obtained on the lattice as in the continuum by expanding the action around its minima. The $SU(4)$ symmetric point is a singularity for the action because of the terms proportional to inverse powers of the radial coordinate $z$.\footnote{All fields vanish in this point.} Consequently, the minimum of the action spontaneously breaks the internal symmetry. In the continuum, an absolute minimum of the action is given by $x=x^*=0$ and $z^M = \delta^{M6}$, and any other absolute minimum is obtained by acting with the $SU(4)$ symmetry. One can easily check that these minima are relative minima for the discretised action. We parametrise the fluctuations around the chosen minimum in the same way as it is done in the continuum~\cite{Giombi:2010bj}
\begin{equation}
	z=e^\phi\,,\qquad
	z^{a} = e^\phi \frac{y^a}{1+\frac{1}{4}y^2}
	\, , \qquad
	z^6 = e^\phi \frac{1-\frac{1}{4}y^2}{1+\frac{1}{4}y^2}
	\, , \qquad
	y^2 = \sum_{a=1}^5 (y^a)^2,
	\, 
\end{equation}
where $a=1,\dots,5$.
In terms of the new variables $\phi$ and $y^a$, the path-integral measure over
the $z^M$ fields reads
\begin{gather}\label{Jacobian discrete}
	\prod_{M=1}^6 dz^M = 
	e^{\sum_{s,t} \left\{ 6\phi+ 5\log\left( 1+\frac{y^2}{4} \right) \right\}}
	\,
	d\phi \prod_{a=1}^5 d y^a
	\, .
\end{gather}
The contribution of the Jacobian determinant above can be conveniently included
in the effective action
\begin{gather}
	S_{\text{eff}} = S_{\text{cusp}} - \sum_{s,t} \left\{ 6\phi+ 5\log\left( 1+\frac{y^2}{4} \right) \right\}
	\, ,
	\label{eq:Seff}
\end{gather}
in terms of which the expectation values of observables are
\begin{equation}
	\langle A \rangle = \frac{1}{Z_{\text{eff}}} \int dx dx^* d\phi d^5y d^4\theta d^4\theta^\dag d^4\eta d^4\eta^\dag
	\, e^{-S_{\text{eff}}} A
	\, .
\end{equation}
Notice that the sum in \eqref{Jacobian discrete} does not come with the corresponding $a^2$ factor, which means that in the naïve continuum limit it diverges like $a^{-2}$. This should not be
surprising: in the continuum, this term would be proportional to $\delta^2(0)$
which yields a quadratic divergence in a hard-cutoff regularization (but it is
set to zero in dimensional regularization).

The perturbative expansion, i.e. the expansion in powers of $g$, is
obtained by splitting the action $S_{\text{eff}} = S_0 + S_{\text{int}}$, where
$S_0$ contains all quadratic terms in the fields with a coefficient proportional
to $g^{-1}$, and $S_{\text{int}}$ contains all other terms. Notice that
$S_{\text{int}}$ also contains $g$-independent quadratic terms, which come from
expanding the Jacobian determinant. We focus here on the leading-order
quadratic action
\begin{eqnarray}
	\nonumber
	S_0 =
	g \,a^2
	\sum_{s, t}
	\Bigg\{ &&
	\left| b_+ \hat\partial_t x + \tfrac{m}{2} x\right|^2
	+ \left| b_- \hat\partial_s x - \tfrac{m}{2} x \right|^2
	\\ \nonumber &&
	+ b_+^2 (\hat\partial_t y^a)^2 + m b_+ y^a \hat\partial_t y^a
	+ (\hat\partial_s y^a)^2
	\\ \nonumber &&
	+ b_+^2 (\hat\partial_t \phi)^2 + m b_+ \phi \hat \partial_t \phi
	+ (\hat\partial_s \phi)^2 + m^2 \phi^2
	+2 i\left( \theta^i \hat{\partial}_t \theta_i +\eta^i \hat{\partial}_t \eta_i \right)
	\\ &&
	+2 i \eta^i (\rho^6)_{ij} \left( b_+ \bar{\partial}_s \theta^j - \tfrac{m}{2} \theta^j
	\right)
	+2 i \eta_i ({\rho^6}^\dagger)^{ij} \left( b_+ \bar{\partial}_s \theta_j - \tfrac{m}{2} \theta_j \right)
	\Bigg\}
	\, .
	\label{S0}
\end{eqnarray}

\subsection{One-Point Functions}\label{Subsec: 1pt function}

Let us turn to the one-point functions of the perturbative fields. Notice that
$\langle x \rangle = 0$ because of the $U(1)$ symmetry, and $\langle y^a \rangle
= 0$ because of the $SO(5) \subset SO(6) \simeq SU(4)$ which leaves the
perturbative vacuum invariant. $\phi$ is the only field with a non-vanishing
one-point function, which has been calculated in dimensional
regularisation~\cite{Giombi:2009gd,Giombi:2010bj,Giombi:2010fa}.  
This one-point function and any $n$-point function of bare
fields are not expected to be UV finite. It is known that $\langle \phi
\rangle$ is UV divergent in dimensional regularisation, and we will see that it
turns out to be UV divergent also in the lattice regularisation. The interest in
this one-point function lies in the fact that it appears as a sub-diagram in any
other $n$-point function, and ultimately its UV divergence contributes to any
physical observable. We will give an example of this mechanism in the next
subsection.

There are two classes of vertices contributing to the one-point function of
$\phi$: single-field vertices coming from the measure
\begin{gather}
	S_\phi 
	=
	-6 \sum_{s,t} \phi
	\ ,
\end{gather}
and three-field vertices coming from the action
\begin{eqnarray}
	S_{\phi \bullet \bullet}
	&=&
	g \sum_{s,t} a^2 \bigg\{
	-4 \phi \left| b_- \hat{\partial}_s x - \tfrac{m}{2} x \right|^2
	+ c_+ \hat{\partial}_t \phi \hat{\partial}_t (\phi^2)
	+ \hat{\partial}_s \phi \hat{\partial}_s \phi^2
	- 4 \phi (\hat{\partial}_s \phi)^2
	\nonumber \\ && \hspace{16mm}
	+ 2 c_+ \hat{\partial}_t y^a \hat{\partial}_t ( \phi y^a )
	- c_+ \hat{\partial}_t \phi \hat{\partial}_t (y^2)
	+ 2 \hat{\partial}_s y^a \hat{\partial}_s ( \phi y^a )
	- \hat{\partial}_s \phi \hat{\partial}_s (y^2)
	- 4 \phi (\hat{\partial}_s y^a)^2
	\nonumber \\ && \hspace{16mm}
	- 4 i \phi
	\left[
	\eta^i (\rho^6)_{ij} \left( b_+ \bar{\partial}_s \theta^j - \tfrac{m}{2} \theta^j \right)
	+ \eta_i ({\rho^6}^\dag)^{ij} \left( b_+ \bar{\partial}_s \theta_j - \tfrac{m}{2} \theta_j \right)
	\right]
	\bigg\}
	\ .
\end{eqnarray}

Notice that the insertion of $S_\phi$ produces a tree-level diagram, while the
insertion of $S_{\phi \bullet \bullet}$ produces a one-loop diagram. However,
because of the mismatch in the power of $g$ in $S_\phi$ and $S_{\phi \bullet
	\bullet}$, all these diagrams contribute to the same order in $g$, yielding
\begin{eqnarray}
	\langle \phi \rangle
	&=&
	\frac{3}{g m^2 a^2}
	+ \frac{2}{g m^2}
	\int_{-\pi/a}^{\pi/a} \frac{d^2q}{(2\pi)^2}
	\frac{
		c_- |\hat{q}_1|^2 + \tfrac{m^2}{4}
	}{
		c_+ |\hat{q}_0|^2 + c_- |\hat{q}_1|^2 + \frac{m^2}{2}
	}
	\nonumber \\ &&
	- \frac{1}{2gm^2}
	\int_{-\pi/a}^{\pi/a} \frac{d^2q}{(2\pi)^2}
	\frac{
		c_+ |\hat{q}_0|^2 - |\hat{q}_1|^2
	}{
		c_+ |\hat{q}_0|^2 + |\hat{q}_1|^2 + m^2
	}
	- \frac{5}{2gm^2}
	\int_{-\pi/a}^{\pi/a} \frac{d^2q}{(2\pi)^2}
	\frac{
		c_+ |\hat{q}_0|^2 - |\hat{q}_1|^2
	}{
		c_+ |\hat{q}_0|^2 + |\hat{q}_1|^2
	}
	\nonumber \\ &&
	- \frac{8}{gm^2}
	\int_{-\pi/a}^{\pi/a} \frac{d^2q}{(2\pi)^2}
	\frac{
		c_+ |\hat{q}_1|^2 + \frac{m^2}{4}
	}{
		|\hat{q}_0|^2 + c_+ |\hat{q}_1|^2 + \frac{m^2}{4}
	}
	+ O(g^{-2})
	\label{eq:phi-1}
	\ .
\end{eqnarray}

With the special choice $b_\pm=\bar{b}_\pm$, i.e. $c_\pm = 1$, one can use the
symmetry of the integrals under $p_0 \leftrightarrow p_1$ exchange to simplify
\begin{eqnarray}
	\langle \phi \rangle
	&=&
	- \frac{1}{g}
	\int_{-\pi/a}^{\pi/a} \frac{d^2q}{(2\pi)^2}
	\frac{
		1
	}{
		|\hat{q}|^2 + \frac{m^2}{4}
	} + O(g^{-2})
	\nonumber \\ &=&
	\frac{1}{g} \left\{
	\frac{1}{4\pi} \log \frac{(am)^2}{4}
	+ \frac{1}{4\pi}
	- I_0^{(0,0)} + O(a \log a)
	\right\} + O(g^{-2})
	\ ,
	\label{eq:phi-2}
\end{eqnarray}
which is logarithmically divergent. The definition of the
numerical constant $I_0^{(0,0)} \simeq 0.355$ is given in the 
appendix of \cite{Bliard:2022kne}. Notice that the measure, fermion-loop and $x$-loop
contributions are separately quadratically divergent, and the cancellation of these divergences is highly non-trivial.

In the general case $c_\pm = 1 + (am) \delta c_\pm$ where $\delta c_\pm =
O(a^0)$ and, after a lengthy calculation, one gets
\begin{align}
	\langle \phi \rangle=
	\frac{1}{g} \bigg\{\frac{- 8 \delta c_+ + \delta c_-}{\pi a}+ \frac{1}{4\pi} \log \frac{(am)^2}{4} + \frac{1}{4\pi}- I_0^{(0,0)} 	+ \frac{8 \delta c_+^2 - \delta c_-^2}{2\pi}+ O(a \log a) \bigg\} + O(g^{-2})
	\label{eq:phi-3}
	\ .
\end{align}
Notice that the naïve choice $b_\pm = 1$ corresponds to the choice $\delta c_\pm
= \mp 1/2$ which yields indeed a linear divergence for $\langle \phi \rangle$:
\begin{gather}
	\langle \phi \rangle
	=
	\frac{1}{g} \left\{ \frac{
		9
	}{2\pi a}
	+ O(\log a) \right\} + O(g^{-2})
	\label{eq:phi-4}
	\ .
\end{gather}

\subsection{Two-Point Function}

We turn now to the two-point function of the field $x$, which we calculate at
one loop. We will use the two-point function to extract the dispersion relation
of the $x$ particle propagating on the worldsheet. In dimensional regularisation
and at one loop \cite{Giombi:2010bj}, both the two-point function and the
dispersion relation are UV finite without the need for
renormalisation. We will also see that this is true at one loop in lattice
perturbation theory, provided that one has chosen $c_\pm = 1$. The naïve choice
$b_\pm = 1$ generates UV divergences in the dispersion relation. A valid question is whether these
divergences can be eliminated with a renormalisation procedure.

\begin{figure}
	\centering
	\includegraphics[scale=0.6]{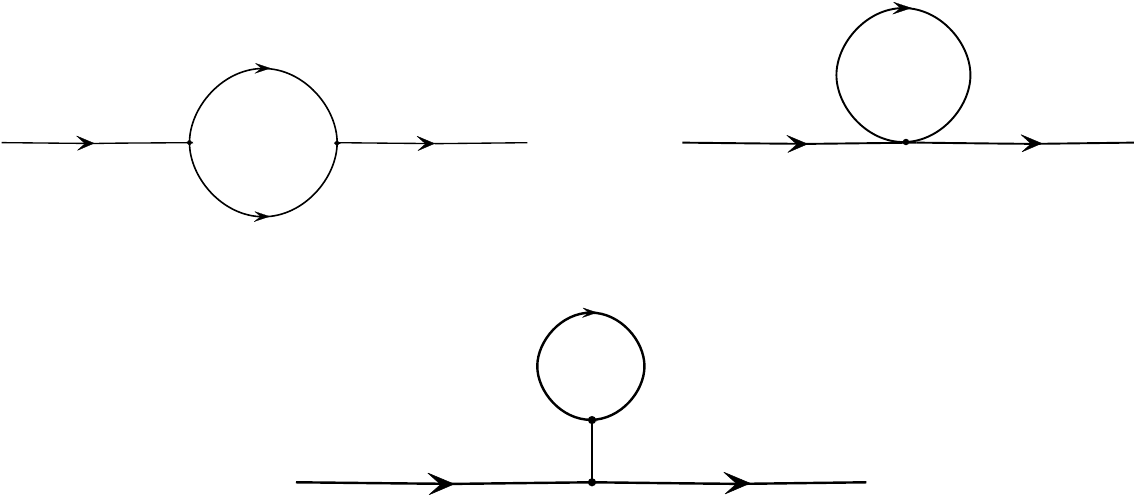}
	\caption{Topologies of diagrams contributing to the two-point function $\langle xx^* \rangle$ at one-loop in the discretized model in \eqref{Scusp} .}
	\label{fig:conn_two_p}
\end{figure}

There are two classes of vertices contributing to the two-point function of
$x$ at one loop: three-field vertices
\begin{eqnarray}
	S_{x x^* \bullet}
	&=&
	g \sum_{s,t} a^2 \bigg\{ -4 \phi \left| b_- \hat{\partial}_s x - \tfrac{m}{2} x \right|^2
	\nonumber \\ && \hspace{18mm}
	+ 2 \eta^i \rho^6_{ij} \eta^j \left( b_- \hat{\partial}_s x - \tfrac{m}{2} x \right)
	- 2 \eta_i ({\rho^6}^\dag)^{ij} \eta_j \left( b_- \hat{\partial}_s x^* - \tfrac{m}{2} x^* \right)
	\bigg\}
	\ ,
\end{eqnarray}
and four-field vertices
\begin{gather}
	S_{x x^* \bullet \bullet}
	=
	8 g \sum_{s,t} a^2 \phi^2 \left| b_- \hat{\partial}_s x - \tfrac{m}{2} x \right|^2
	\ ,
\end{gather}
combined to give Feynman diagrams with the three different topologies
illustrated in Figure~\ref{fig:conn_two_p}. Notice that the tadpole contribution
will be proportional to $\langle \phi \rangle$.

On general grounds, one sees that the two-point function has the following form
\begin{equation}\!\!\!\!\! 
	\langle \tilde{x}(p)x^*(0) \rangle
	=
	\frac{1}{g}
	\left\{
	c_+ |\hat{p}_0|^2 + c_- |\hat{p}_1|^2 + \frac{m^2}{2}
	+ \frac{1}{g} \left( c_- |\hat{p}_1|^2 + \frac{m^2}{4} \right) \Pi_a(p) + O(g^{-2})
	\right\}^{-1}
	\ . \label{eq:2pt-x}
\end{equation}
The factor $\left(c_- |\hat{p}_1|^2 + \tfrac{m^2}{4} \right)$ comes from the fact that, in all interaction vertices, $x$ always appears in the combination $\left(b_- \hat{\partial}_s x - \tfrac{m}{2} x \right)$ or its complex conjugate. The function $\Pi_a(p)$ has a representation in terms of amputated Feynman diagrams, and it is explicitly given by
\begin{eqnarray}
	\Pi_a(p)
	&=&
	- 4 g \langle \phi \rangle
	+ 4 \int_{-\pi/a}^{\pi/a} \frac{d^2q}{(2\pi)^2}
	\frac{1}{c_+ |\hat{q}_0|^2 + |\hat{q}_1|^2 + m^2}
	\nonumber \\ &&
	- 8 \int_{-\pi/a}^{\pi/a} \frac{d^2q}{(2\pi)^2}
	\frac{c_- |\hat{q}_1|^2 + \frac{m^2}{4}}{c_+ |\hat{q}_0|^2 + c_- |\hat{q}_1|^2 + \frac{m^2}{2}}
	\frac{1}{
		c_+ |\widehat{p+q}_0|^2 + |\widehat{p+q}_1|^2 + m^2
	}
	\nonumber \\ &&
	- 8 \int_{-\pi/a}^{\pi/a} \frac{d^2q}{(2\pi)^2}
	\frac{\hat{q}_0}{|\hat{q}_0|^2 + c_+ |\hat{q}_1|^2 + \frac{m^2}{4}}
	\frac{\widehat{p+q}_0^*}{
		|\widehat{p+q}_0|^2 + c_+ |\widehat{p+q}_1|^2 + \frac{m^2}{4}
	}
	\ .
\end{eqnarray}
All integrals in the above formula are logarithmically divergent, while the term proportional to $\langle \phi \rangle$ generally contains a linear divergence.
Up to terms that vanish in the $a \to 0$ limit, one can replace $c_\pm = 1$ in
the above integrals, obtaining the simpler expression
\begin{eqnarray}
	\Pi_a(p)
	&=&
	- 4 g \langle \phi \rangle
	+ 4 \int_{-\pi/a}^{\pi/a} \frac{d^2q}{(2\pi)^2}
	\frac{1}{|\hat{q}|^2 + m^2}
	- 8 \int_{-\pi/a}^{\pi/a} \frac{d^2q}{(2\pi)^2}
	\frac{|\hat{q}_1|^2 + \frac{m^2}{4}}{|\hat{q}|^2 + \frac{m^2}{2}}
	\frac{1}{
		|\widehat{p+q}|^2 + m^2
	}
	\nonumber \\ &&
	- 8 \int_{-\pi/a}^{\pi/a} \frac{d^2q}{(2\pi)^2}
	\frac{\hat{q}_0}{|\hat{q}|^2 + \frac{m^2}{4}}
	\frac{\widehat{p+q}_0^*}{
		|\widehat{p+q}|^2 + \frac{m^2}{4}
	}
	+ O(a \log a)
	\ .
\end{eqnarray}
As in the continuum, the leading divergence of the above integrals does not
depend on the external momentum, therefore the subtracted quantity $\Delta
\Pi_a(p) = \Pi_a(p) - \Pi_a(0)$ has a finite $a \to 0$ limit given by the corresponding continuum integrals, i.e.
\begin{eqnarray}
	\Delta \Pi_0(p)
	&=&
	- 8 \int_{-\infty}^\infty \frac{d^2q}{(2\pi)^2}
	\frac{q_1^2 + \frac{m^2}{4}}{q^2 + \frac{m^2}{2}}
	\left\{ \frac{1}{(p+q)^2 + m^2} - \frac{1}{q^2 + m^2} \right\}
	\nonumber \\ \label{Deltapi0} &&
	- 8 \int_{-\infty}^\infty \frac{d^2q}{(2\pi)^2}
	\frac{q_0}{|\hat{q}|^2 + \frac{m^2}{4}}
	\left\{
	\frac{p_0+q_0}{(p+q)^2 + \frac{m^2}{4}}
	- \frac{q_0}{q^2 + \frac{m^2}{4}}
	\right\}
	+ O(a \log a)
	\ ,
\end{eqnarray}
while all the divergences are contained in
\begin{gather}
	\Pi_a(0)
	=
	% - 4 g \langle \phi \rangle
	% - 4 \int_{-\pi/a}^{\pi/a} \frac{d^2q}{(2\pi)^2}
	% \frac{
		% |\hat{q}|^2
		% }{
		% \left( |\hat{q}|^2 + \frac{m^2}{4} \right)^2
		% }
	% + O(a \log a)
	% = \\ =
	% - 4 g \langle \phi \rangle
	% - 4 \int_{-\pi/a}^{\pi/a} \frac{d^2q}{(2\pi)^2}
	% \frac{
		% 1
		% }{
		% |\hat{q}|^2 + \frac{m^2}{4}
		% }
	% + m^2 \int_{-\infty}^\infty \frac{d^2q}{(2\pi)^2}
	% \frac{
		% 1
		% }{
		% \left( q^2 + \frac{m^2}{4} \right)^2
		% }
	% + O(a \log a)
	% = \\ =
	- 4 g \langle \phi \rangle
	- 4 \int_{-\pi/a}^{\pi/a} \frac{d^2q}{(2\pi)^2}
	\frac{
		1
	}{
		|\hat{q}|^2 + \frac{m^2}{4}
	}
	+ \frac{1}{\pi}
	+ O(a \log a)
	\ ,
\end{gather}
where we have used the symmetry of the integrals under $p_0 \leftrightarrow p_1$
exchange to simplify them.

With the choice $c_\pm = 1$, using equation~\eqref{eq:phi-2} one immediately sees that
all divergences cancel and $\Pi_0(0) = 1/\pi$. The two-point function is finite
in the continuum limit and 
\begin{gather}
	\lim_{a \to 0} \langle \tilde{x}(p)x^*(0) \rangle
	=
	\frac{1}{g}
	\left\{
	p^2 + \frac{m^2}{2}
	+ \frac{1}{g} \left( p_1^2 + \frac{m^2}{4} \right) \Pi_0(p) + O(g^{-2})
	\right\}^{-1}.
\end{gather}
The two-point function has poles at $p_0 = \pm i E(p_1)$ for every value of
$p_1$, where $E(p_1)$ is the energy of a single excitation with the quantum
numbers of the field $x$, propagating on the worldsheet with momentum $p_1$. 
In the continuum limit, this is found to be
\begin{eqnarray}
	E(p_1)^2 &=& p_1^2 + \frac{m^2}{2}
	+ \frac{1}{g} \left( p_1^2 + \frac{m^2}{4} \right) \Pi_0\left(\sqrt{p_1^2 + \frac{m^2}{2}},p_1\right) + O(g^{-2})
	\nonumber \\ &=&
	p_1^2 + \frac{m^2}{2}
	- \frac{1}{gm^2} \left( p_1^2 + \frac{m^2}{4} \right)^2
	+ O(g^{-2})
	\ , \label{eq:disprelation}
\end{eqnarray}
where we have used the on-shell value of $\Pi_0$. The obtained
dispersion relation coincides with the result in \cite{Giombi:2010bj}.\footnote{When comparing to~\cite{Giombi:2010bj}, notice that one has to redefine the worldsheet coordinates, resulting in square masses of the fluctuations rescaled by a factor of four.} 

However in the general case $c_\pm = 1 + (am) \delta c_\pm$ where $\delta c_\pm
= O(a^0)$, $\Pi_a(0)$ and $E(p_1)$ inherit the linear divergence from $\langle
\phi \rangle$. Using equation~\eqref{eq:phi-3} one obtains
\begin{gather}
	\Pi_a(0)
	=
	\frac{32 \delta c_+ - 4 \delta c_-}{\pi a}
	+ \frac{1 - 16 \delta c_+^2 + 2 \delta c_-^2}{\pi}
	+ O(a \log a)
	\ .
\end{gather}
For instance, for the naïve choice $b_\pm = 1$, which corresponds to $\delta c_\pm
= \mp 1/2$, one obtains for the dispersion relation
\begin{gather}
	E(p_1)^2 = p_1^2 + \frac{m^2}{2}
	+ \frac{1}{g} \left( p_1^2 + \frac{m^2}{4} \right)
	\left[ - \frac{18}{\pi a} + O(\log a) \right]
	+ O(g^{-2})
	\ .
\end{gather}
It is interesting to notice that once we have set $b_\pm = 1$, the divergence
in the dispersion relation cannot be eliminated by renormalising the remaining
available parameters, i.e. $g$ and $m$. In other words, the choice $b_\pm = 1$
is not stable under renormalisation. On the other hand, if one allows the
coefficients $b_\pm$ to be renormalised along with $m$ and $g$, then the
divergences in the dispersion relation are eliminated, e.g. by choosing
\begin{gather}
	b_+ = 1 + \frac{1}{g_R} \frac{ \frac{a\,m_R}{8} }{ 2 + \frac{a\,m_R}{2} } \left( \Pi_a(0) - \frac{1}{\pi} \right) 
	\ , \\
	b_- = 1 - \frac{1}{g_R} \frac{ 1 + \frac{5a\,m_R}{8} }{ 2 + \frac{a\,m_R}{2} } \left( \Pi_a(0) - \frac{1}{\pi} \right) 
	\ , \\
	m^2 = m_R^2 \left[ 1 + \frac{1}{2g_R} \left( \Pi_a(0) - \frac{1}{\pi} \right) \right]
	\ , \\
	g = g_R \left[ 1 + O(g^{-1}) \right]
	\ .
\end{gather}
This choice yields a dispersion relation in the continuum limit of the same form
as equation~\eqref{eq:disprelation}, except that the mass $m$ needs to be replaced by its renormalised counterpart $m_R$. One could also see that the one-loop renormalisation of the coupling constant can be chosen so that the cusp anomaly is finite. With this discussion, we do not want to imply that the chosen lattice theory is renormalisable (we do not know this). However, we conclude that if the lattice theory is renormalisable, then it is not sufficient to renormalize $m$ and $g$; one also needs to introduce additional coefficients in the action and either fine-tune their tree-level value or renormalise them.

\chapter{Conclusions and Outlook}\label{chap Outlook}
\vspace{-8mm}
In the first Chapters (\ref{chapter: Analytic Bootstrap}, \ref{chapter: ABJM}, \ref{chapter N=4}), the analytic conformal bootstrap was introduced and applied to two examples: the four-point correlator of operator insertions on the 1/2-BPS Wilson line in ABJM and the five-point correlator of operator insertions on the 1/2-BPS Wilson line in $\mc{N}=4$ SYM. In the first example, the superspace structure was developed in order to relate all the correlators of fields in the displacement multiplet to a single function $f(z)$. This function was then bootstrapped entirely up to third order in a strong coupling expansion and to all orders in the double-scaling limit. The mixing problem for the CFT data was solved at first order to obtain these results. The CFT data (generically mixed) was then extracted up to third order. \par 
In the second example, multipoint Ward identities and selection rules were used to compute the superblock expansion of the five-point correlator of 1/2-BPS operators inserted on the 1/2-BPS Wilson line in $\mc{N}=4$ SYM. This allows for the computation of the first-order correlator through the analytic conformal bootstrap. These higher-point correlators access a wider range of CFT data, which were then extracted. The analysis of the different channels confirms that the anomalous dimension does not contribute to mixing at this order.\par   \vspace{2mm}
 Extensions to this work fall into two parallel categories: bootstrapping correlators and  improving the constraints from the bootstrap. One can easily find more results to bootstrap; by increasing the order of perturbation, the number of insertions, or by considering different setups. Higher-order contributions increase in complexity due to the mixing of operators and somewhat because of the type of functions appearing in the ansatz. The higher-point correlators increase in complexity because of the wealth of possible operators exchanged in the many OPEs. Systems with lower supersymmetry will have fewer constraints from the corresponding selection rules. As such, these three extensions will have upper limits as to what can be fixed by the symmetry.\par 
When looking at improving the bootstrap, many constraints have not yet been explored. The first category are additional constraints from the symmetries of the theory. For example,  integrated correlator constraints relate the $n$-point results to functions known through integrability and should exist for most superconformal defects and for increasing number of insertions \cite{Cavaglia:2022yvv}. Integrability has also been extremely powerful in determining the spectrum of the 1/2-BPS setup in $\mc{N}=4$ SYM\cite{Grabner:2020nis}. This symmetry should also constraint the perturbative correlators through Yangian symmetry. 
The second category covers more general defects setups and can be explored in toy models. Three examples are the constraints from single-valuedness, from the flat-space limit and from non-perturbative sum rules. \par
\vspace{3mm}
Looking forwards, an immediate extension is the unmixing of the CFT data at higher orders, which would enable the computation of the correlators at the next orders in perturbation theory. One way to do this is to look at more general correlators, such as was done to solve the first-order mixing in the 1/2-BPS Wilson line in ABJM. A particular family of interest in this case are the 1/3-BPS operators, which are charged under R-symmetry. These form an infinite tower of protected operators which may have a topological limit. Additionally, these operators preserve the same symmetries as the 1/2-BPS operators on the 1/3-BPS Wilson line, whose parallel computation would be enlightening.   A direct extension of the 5-point bootstrap which is underway is the first connected, strong-coupling contribution to the 6-point function in $\mc{N}=4$ SYM. A more ambitious goal is to implement this algorithmic formalism to analyse $n-$point insertions of protected operators and finite deformations of the line. \par 
These two examples are but a small subset of holographic setups; a similar analysis holds in the 1/2-BPS defects in AdS$_3$/CFT$_2$ \cite{Correa:2021sky, Bliard-Correa} where there is a one-parameter family of 1/2-BPS defects. There is also a range of defects with lower supersymmetry related to each other through RG flow \cite{Beccaria:2022bcr} for which the corresponding deformation of the correlator may be solvable via analytic bootstrap. \par 
In this analysis, some features of the holographic setups stood out and point to some directions which could shed light on what makes these theories special. In order to really extend this analysis in the context of different holographic defects, one might consider the bootstrap of all protected multiplets inserted on superconformal lines such as described in \cite{Agmon:2020pde}. Additionally to constituting a dictionary of bootstrapped results, this would differentiate the cases of integrable and supersymmetric systems. One observation which is common to the two defects studied in this thesis is that the leading anomalous dimension does not mix and is proportional to the quadratic Casimir eigenvalue. This could be a consequence of either symmetries, or even of another aspect, such as a higher conformal symmetry \cite{Caron-Huot:2018kta}. The analysis of a diversity of setup will underline which aspects are crucial to the simplifications observed in this thesis.
\par
\vspace{2mm}
This diversity motivates the second category of results; generic constraints. 
For example, the constraint that the resulting functions should be single-valued is very constraining for the 4-point function and the determining of multivariable single-valued functions such as in \cite{Brown:2004ugm} would be essential to bootstrapping  six-point correlators to a non-trivial order without prior assumptions. Bootstrapping directly in Mellin is also a possibility where the simplicity of the correlators would be apparent as it was in this thesis. Additional constraints can also be found from integrability. Given that the examples here are integrable, one could use this symmetry by finding the explicit Yangian generators which in turn would bring some additional bootstrap constraints. In particular, multipoint correlators would be subject to strong restrictions in terms of an ansatz consistent with the (super-)conformal Yangian symmetry. This thesis also initiated the consequences from the flat-space space behaviour in terms of the large-weight spectrum of exchanged operators. In the future, this analysis of the flat-space limit could not only provide generic constraints for holographic dCFTs, but it might also provide some understanding of integrability on the line in terms of the flat-space integrbale theory.  Finally, the strong coupling description of a dCFT$_1$ in terms of an effective description in AdS$_2$ motivates the study of Witten diagrams described in the subsequent chapters in this Thesis. 

\vspace{6mm}
Chapters (\ref{Effective theories}, \ref{Chap: Optimisation and limits}) considered the analysis of defects through a different light: that of effective theories. This point of view allows for an analysis which does not depend on the details of the theory and finds features that can be widely applied to other conformal defects.\par \vspace{3mm}
This was first done through the analysis of Witten diagrams in AdS$_2$, whose structure for $n-$point contact diagrams can be found explicitly using the  elementary method of contour integration. This method was used to find all $n-$point contact diagrams for $\Delta=1$ and $\Delta=2$. Additionally, it was applied to the case of topological operators in Yang-Mills in AdS$_2$.  \par 
The second analysis was done through the definition of the Mellin amplitude for one-dimensional four-point conformal correlators. This formalism was then applied to an effective, high-derivative, interaction Lagrangian to solve all leading-order contributions and to find the associated CFT data.\par
The systematic analysis of scattering amplitudes and high-loop results in flat space has had drastic impact on computations in physical theories. The arsenal developed in this context is partially applicable to Witten diagrams.  Looking forwards, the analysis of master integrals and IBP relations, akin to those in scattering amplitudes, would provide a solid starting point for the Ansatz. Additionally, possible symmetries such as Yangian symmetry may explain the observed reduction of transcendental weight. An Ansatz informed by these symmetries would be a better starting point and could drastically improve the bootstrap process.  \par \vspace{1mm}

The Mellin formalism shows promise when it comes to the correlators computed in this thesis. Additionally, it has applicability beyond the context of defect CFTs. For example, it was recently used in the analysis of celestial conformal field theories \cite{Jiang:2022hho}. The correlators in this thesis have a Mellin transform that is simpler than the generic form for contact diagrams and may highlight features not yet considered. A clear direction for this Mellin formalism is to formulate a scattering limit akin to that in higher dimensions. \par \vspace{1mm}
The large-$\Delta$ behaviour of the anomalous dimension underlines an increase in the number of parameters needed to fix the bootstrap solution. This input of physical data, while accessible for supersymmetric theories, is a common problem in bootstrap (called polynomial ambiguities in the $\epsilon$-expansion bootstrap \cite{Dey:2017fab}). A possible solution to the need for additional physical input at each order could be in the interplay between analytic and numerical methods, provided by Lattice Field Theory, or by the numerical bootstrap. An example of the former was presented in Chapter \ref{Chap: Optimisation and limits} with the cusped Wilson line. The challenges posed by the discretisation of a model are clear: the lattice breaks most of the symmetry essential to these models. However, the success of lattice simulations provides ample motivation to develop formalisms or study setups which break less of these symmetries \cite{Costa:2022ezw}.\par \vspace{3mm}
More generally, the analytic bootstrap is an incredibly powerful tool in the context of dCFTs, which has produced results well beyond what is achievable by diagrammatic computations and is at the forefront of results along with the quantum spectral curve \cite{Cavaglia:2021bnz} and the numerical bootstrap \cite{El-Showk:2012cjh}. The basic ingredients of the conformal bootstrap are widely applicable to conformal systems without supersymmetry or integrability.  However, these methods can work in parallel to tackle the current roadblocks to all-order solutions. This thesis presents various methods parallel to the bootstrap, such as the integrated correlators, the Mellin amplitude, sum rules, and lattice descriptions. In the future, combining these with the power of the analytic bootstrap, the numerical bootstrap and integrability may hold the key to solving one-dimensional CFTs completely. \par \vspace{3mm}
The analysis of extended operators in quantum field theories is not only a natural question, but also an important one; with consequences in terms of quark radiation, generalised symmetries and more fundamentally the understanding of the full theory. The tools presented is this thesis are widely applicable to conformal defects in non-holographic setups and expansions different to the strong-coupling one. This $1d$ setup is also a perfect tool and starting point to understand the rich structure of higher dimensional theories. In general, conformal line defects can be constructed by tuning the parameters of the theory and of the line defect to reach the conformal point \cite{Cuomo:2021kfm}. As such, it is envisionable to find these objects not only in condensed matter systems, but at the heart of the standard model, such as at the conformal point of quantum chromodynamics (QCD).
\let\cleardoublepage\clearpage
	\appendix
\chapter{The Analytic Bootstrap }
	\section{Conformal Blocks}\label{Appendix Conformal blocks}
	Conformal blocks in a CFT are built from the Operator Product Expansion (OPE) which states that any two operators can be replaced by a weighted sum of exchanged operators. This principle, central to all conformal theories, is detailed in the next few sections with the examples of three-, four- and five-point scalar correlators. The four-point block expansion, which is by far the most famous, is derived in several ways. 
	\subsection*{OPE Operator (Three-Point)}
	The form of the OPE operator $\mathcal{F}(x_{12},\partial_{x_1})$ can easily be found by applying this OPE to the three-point function, which is fixed by conformal symmetry:
	\begin{align}
		\langle \phi_{\Delta_1}(x_1)\phi_{\Delta_2}(x_2)\phi_{\Delta_3}(x_3)\rangle & = \langle \sum_\Delta c_{\Delta_1 \Delta_2 \Delta}\mathcal{F}(x_{12};\partial_{x_1}) \mathcal{O}_\Delta(x_1)\phi_{\Delta_3}(x_3)\rangle\\
		&= \mathcal{F}(x_{12};\partial_{x_1}) \frac{c_{\Delta_1\Delta_2\Delta_3}}{x_{13}^{2\Delta_3}}\\
		&= \frac{c_{\Delta_1\Delta_2\Delta_3}}{x_{12}^a x_{13}^b x_{23}^c  } 
	\end{align}
where
\begin{align}
	a&=\Delta_1+\Delta_2-\Delta_3&b&=\Delta_1+\Delta_3-\Delta_2&c&=\Delta_2+\Delta_3-\Delta_1
\end{align}
	This gives an operator such that
	\begin{align}
		\mathcal{F}(x_{12};\partial_{x_1})x_{13}^{-2\Delta_3}
		&= x_{12}^{-a}x_{13}^{-b} x_{23}^{-c}.
	\end{align}
	By expanding 
	\begin{align}
		\partial_t^{(k)}\frac{1}{t^{2\Delta} }&= \frac{(-1)^k \Gamma(2\Delta+k)}{\Gamma(2\Delta)t^{2\Delta+k}}&\frac{1}{(p-q)^c } & = \sum_k \frac{\Gamma(k+c)q^k}{\Gamma(c)\Gamma(k+1)p^{c+k}}
	\end{align}
	we find
	\begin{align}
		\mathcal{F}_{\Delta_1,\Delta_2,\Delta}(x_{12};\partial_{x_1})x_{13}^{-2\Delta}
		&=\sum_k \frac{\Gamma(k+c)\Gamma(2\Delta)x_{12}^{k-a}}{\Gamma(c)\Gamma(k+1)\Gamma(2\Delta+k)} (-1)^k  \partial_{x_1}^{(k)}\frac{1}{x_{13}^{2\Delta_3}}\\
		&= \frac{1}{(x_{12})^a}{}_1F_1(c,2\Delta;-x_{12}\partial_{x_3})\frac{1}{x_{13}^{2\Delta}}
	\end{align}
	where the partial derivative has been expressed in terms of $x_3$ so that it can commute with the powers of $x_{12}$ in the series expansion. Alternatively, one can define the power expansion as `normal-ordered' such that 
	\begin{align}
		: (x\partial_x)^n: \, = x^n\partial_x^{(n)}
	\end{align}
	Therefore
	\begin{align}
		\mathcal{F}_{\Delta_1,\Delta_2,\Delta}(x_{12};\partial_{x_1})
		&=\sum_k \frac{\Gamma(k+c)\Gamma(2\Delta)x_{12}^{k-a}}{\Gamma(c)\Gamma(k+1)\Gamma(2\Delta+k)} (-1)^k  \partial_{x_1}^{(k)}\\
		&=\frac{1}{x_{12}^{\Delta_1+\Delta_2}}x_{12}^{\Delta}:{}_1F_1(\Delta_2-\Delta_1+\Delta,2\Delta;x_{12}\partial_{x_1}):
	\end{align}
	\subsection*{Four-Point Scalar Conformal Blocks}
	Applying the OPE to the first two scalars, one obtains
	\begin{align}
		\langle \phi_{\Delta_1}(x_1)\phi_{\Delta_2}(x_2)\phi_{\Delta_3}(x_3)\phi_{\Delta_4}(x_4)\rangle &=\sum_{\Delta}c_{\Delta_1\Delta_2\Delta}\langle \mathcal{F}_{\Delta_1, \Delta_2, \Delta}(x_{12},\partial_{x_1})\mathcal{O}_{\Delta}(x_1)\phi_{\Delta_3}(x_3)\phi_{\Delta_4}(x_4)\rangle\\
		&= \sum_{\Delta}c_{\Delta_1\Delta_2\Delta}c_{\Delta\Delta_3\Delta_4}\mathcal{F}_{\Delta_1, \Delta_2, \Delta}(x_{12},\partial_{x_1})f(x_1,x_3,x_4)
	\end{align}
	Notice that 
	\begin{align}
		f(x_1,x_3,x_4) = x_{13}^{-\Delta-\Delta_3+\Delta_4}x_{14}^{-\Delta-\Delta_4+\Delta_3}x_{34}^{-\Delta_3-\Delta_4+\Delta}
	\end{align}
	has the limit
	\begin{align}
		\lim_{x_4\rightarrow \infty}f(x_1,x_3,x_4) \sim x_4^{-2\Delta_4}
	\end{align}
	which is the same behaviour the correlator should have. Thus any derivative acting on the $x_{14}$ terms will vanish in this limit. We take this limit to find the following:
	\begin{align}
		\langle \phi_{\Delta_1}(x_1)\phi_{\Delta_2}(x_2)\phi_{\Delta_3}(x_3)\phi_{\Delta_4}(x_4)\rangle \sim_{x_4\rightarrow \infty} x_{4}^{-2\Delta_4}\sum_{\Delta}c_{\Delta}^2\mathcal{F}_{\Delta_1, \Delta_2, \Delta}(x_{12},\partial_{x_1})x_{13}^{-\Delta-\Delta_3+\Delta_4}
	\end{align}
	Using the representation above, we find
	\begin{align}
		\langle \phi_{\Delta_1}(0)\phi_{\Delta_2}(\chi)\phi_{\Delta_3}(1)\phi_{\Delta_4}(\infty)\rangle =\chi^{-\Delta_1-\Delta_2} \sum_{\Delta}c_{\Delta}^2 G_{\Delta}(\chi)
	\end{align}
	Where
	\begin{align}
		G_{\Delta}(\chi) &= \sum_k \frac{\Gamma(k+c)\Gamma(2\Delta)\chi^{k-a}}{\Gamma(c)\Gamma(k+1)\Gamma(2\Delta+k)} \frac{(-1)^{2k} \Gamma(b'+k)(1)^{-b'-k}}{\Gamma(b')}\\
		&=\chi^{\Delta-\Delta_1-\Delta_2} \sum_k \frac{\Gamma(k+c)\Gamma(2\Delta)\chi^{k}\Gamma(b'+k)}{\Gamma(c)\Gamma(k+1)\Gamma(2\Delta+k)\Gamma(b')}\\
		&=\chi^{\Delta-\Delta_1-\Delta_2} {}_2F_1(c,b',2\Delta,\chi)
	\end{align}
where
\begin{align}
	a &= \Delta_1+\Delta_2-\Delta& b'&=\Delta+\Delta_3-\Delta_4&c&=\Delta+\Delta_2-\Delta_1
\end{align}
In terms of the external dimensions, this gives
	\begin{align}
		\langle \phi_{\Delta_1}(x_1)\phi_{\Delta_2}(x_2)\phi_{\Delta_3}(x_3)\phi_{\Delta_4}(x_4)\rangle &=A(x_1,x_2,x_3,x_4)\sum_\Delta c_{12\Delta}c_{34\Delta}G_\Delta(\chi)
	\end{align}
	where
	\begin{align}
		A(x_1,x_2,x_3,x_4) &=x_{12}^{-\Delta _1-\Delta _2} x_{13}^{\Delta _4-\Delta _3} x_{14}^{-\Delta _1+\Delta _2+\Delta _3-\Delta _4} x_{24}^{\Delta _1-\Delta _2} x_{34}^{-\Delta _3-\Delta _4}\\
		G_{\Delta}(\chi) &= \chi^{\Delta} {}_2F_1(\Delta+\Delta_{2}-\Delta_1,\Delta+\Delta_{3}-\Delta_{4},2\Delta;\chi)
	\end{align}
\vspace{3mm}
	\subsection*{Using the Double Sum}
	Using the same principle, we apply the OPE twice to the four-point correlator of generic scalars:
	\begin{align}
		\langle \phi_{\Delta_1}(x_1)\phi_{\Delta_2}(x_2)\phi_{\Delta_3}(x_3)\phi_{\Delta_4}(x_4)\rangle &=\sum_{\Delta,\Delta'} c_{12\Delta}c_{34\Delta'}\mathcal{F}^{12}_{\Delta}(x_{12};\partial_{x_1})\mathcal{F}^{34}_{\Delta'}(x_{34};\partial_{x_3})\langle \mathcal{O}_\Delta(x_1)\mathcal{O}'_{\Delta'}(x_3) \rangle \\
		&=\sum_\Delta c_{12\Delta}c_{34\Delta}\mathcal{F}^{12}_{\Delta}(x_{12};\partial_{x_1})\mathcal{F}^{34}_{\Delta}(x_{34};\partial_{x_3})x_{13}^{-2\Delta}\\
		&=\sum_\Delta c_{12\Delta}c_{34\Delta}\tilde{G}_\Delta^{1234}(\chi)
	\end{align}
	which defines our conformal blocks (up to the choice in decomposing prefactor and blocks).
	So \begin{align}
		\tilde{G}_\Delta^{1234}(\chi)&=\mathcal{F}^{12}_{\Delta}(x_{12};\partial_{x_1})\mathcal{F}^{34}_{\Delta}(x_{34};\partial_{x_3})x_{13}^{-2\Delta}
	\end{align}
	The conformal transformations bring
	\begin{align}
		\langle \phi_{\Delta_1}(x_1)\phi_{\Delta_2}(x_2)\phi_{\Delta_3}(x_3)\phi_{\Delta_4}(x_4)\rangle = \mathcal{A}(x_i) \langle \phi_{\Delta_1}(0)\phi_{\Delta_2}(\chi)\phi_{\Delta_3}(1)\tilde{\phi}_{\Delta_4}(\infty)\rangle 
	\end{align}
	where 
	\begin{align}
		\tilde{\phi}_\Delta(\infty) = \lim_{u\rightarrow\infty} u^{2\Delta}\phi_{\Delta}(u).
	\end{align}
	This means we want the finite part of the limit 
	\begin{align}
		x_4^{2\Delta_4}G^{1234}.
	\end{align}
	Setting the values
	\begin{align}
		a &= \Delta_1+\Delta_2-\Delta&a'&=\Delta_3+\Delta_4-\Delta\\
		c&= \Delta+\Delta_2-\Delta_1&c'&=\Delta+\Delta_4-\Delta_3
	\end{align}We can identify this term in the series
	\begin{align}
		G_{\Delta}^{1234} = \frac{1}{x_{12}^ax_{34}^{a'}}\sum_{k,k'} \frac{\Gamma(k+c)\Gamma(k'+c')\Gamma(2\Delta)^2 :(x_{12}\partial_{x_1})^k(x_{34}\partial_{x_3})^{k'}:}{\Gamma(c)\Gamma(c')\Gamma(k+1)\Gamma(k'+1)\Gamma(2\Delta+k)\Gamma(2\Delta+k')} x_{13}^{-2\Delta}
	\end{align}
	as the one solving 
	\begin{align}
		x^{k'-a'+2\Delta_4}\simeq 1 \Leftrightarrow k' = a'-2\Delta_4.
	\end{align}
	Though $ a'-2\Delta_4=\Delta_3-\Delta_4-\Delta=-c'$ is not a positive integer for arbitrary values of each dimension, we add this as an assumption and see where we get. Choosing this term in the double series and setting the coordinates to the ones of the conformal cross-ratio, we obtain
	\begin{align}
		&\frac{1}{\chi^a}\sum_k \frac{\Gamma(k+c)(-1)^{c'}\Gamma(c')\Gamma(2\Delta)^2 \chi^k \partial_{x_1}^{(k)}(-\partial_{x_1})^{c'} x_{13}^{-2\Delta}}{\Gamma(c)\Gamma(c')\Gamma(2\Delta-c')\Gamma(k+1)\Gamma(2\Delta+k)}|_{x_1\rightarrow0,x_3\rightarrow1} \\
		&= \sum_{k}\chi^{-a}\frac{\Gamma(k+c)\Gamma(2\Delta)^2\Gamma(2\Delta+k-c')}{\Gamma(c)\Gamma(c')\Gamma(2\Delta)\Gamma(k+1)\Gamma(2\Delta+k)}\chi^k \\
		&= \chi^{-a}{}_2F_1(c,2\Delta-c',2\Delta,\chi).
	\end{align}
	This gives 
	\begin{align}
		G_{\Delta}^{1234} = \chi^{-\Delta_1-\Delta_2}\chi^{\Delta}{}_2F_1(\Delta-\Delta_{12},\Delta+\Delta_{34},2\Delta;\chi)
	\end{align}
	The factor of $\chi$ can be reabsorbed into the prefactor giving
	\begin{align}
		\langle \phi_{\Delta_1}(x_1)\phi_{\Delta_2}(x_2)\phi_{\Delta_3}(x_3)\phi_{\Delta_4}(x_4)\rangle &=A(x_1,t_2,t_3,t_4)\sum_\Delta c_{12\Delta}c_{34\Delta}G_\Delta(\chi)
	\end{align}
	where
	\begin{align}
		A(t_1,t_2,t_3,t_4) &=t_{12}^{-\Delta _1-\Delta _2} t_{13}^{\Delta _4-\Delta _3} t_{14}^{-\Delta _1+\Delta _2+\Delta _3-\Delta _4} t_{24}^{\Delta _1-\Delta _2} t_{34}^{-\Delta _3-\Delta _4}\\
		G_{\Delta}(\chi) &= \chi^\Delta  {}_2F_1(\Delta+\Delta_{2}-\Delta_1,\Delta+\Delta_{3}-\Delta_{4},2\Delta;\chi)
	\end{align}\begin{center}
		
	\end{center}
	\subsection*{Using the Conformal Casimir}
	We can write a four-point function in the operator formalism as \cite{Dolan:2000ut}
	\begin{align}
		\langle 0| \phi_{\Delta_4}(x_4)\phi_{\Delta_3}(x_3)\phi_{\Delta_2}(x_2)\phi_{\Delta_1}(x_1)|0\rangle
	\end{align}
	We then insert a representation of the identity as the sum over the quadratic Casimir eigenstates satisfying $\hat{C}\mathcal{O}_h = h(h-1)\mathcal{O}_h$.
	\begin{align}
		&\langle 0| \phi_{\Delta_4}(x_4)\phi_{\Delta_3}(x_3)\sum_h|\mathcal{O}_h\rangle \langle \mathcal{O}_h|\phi_{\Delta_2}(x_2)\phi_{\Delta_1}(x_1)|0\rangle\\
		&=\sum_h\langle\phi_{\Delta_1}(x_1)\phi_{\Delta_2}(x_2)\mathcal{O}_h\rangle\langle\mathcal{O}_h\phi_{\Delta_3}(x_3)\phi_{\Delta_4}(x_4)\rangle
	\end{align}
	Applying a generic conformal transformation to all points in a three-point function leaves it invariant
	\begin{align}
		(\hat{L}_1+\hat{L}_2+\hat{L})\langle\phi_{\Delta_1}(x_1)\phi_{\Delta_2}(x_2)\mathcal{O}_h\rangle&=0\\
		C^{(2)}_{12}\langle\phi_{\Delta_1}(x_1)\phi_{\Delta_2}(x_2)\mathcal{O}_h\rangle&=\hat{C}^{(2)}_{\mathcal{O}}\langle\phi_{\Delta_1}(x_1)\phi_{\Delta_2}(x_2)\mathcal{O}_h\rangle\\
		&=h(h-1)\langle\phi_{\Delta_1}(x_1)\phi_{\Delta_2}(x_2)\mathcal{O}_h\rangle
	\end{align} 
	Therefore, the conformal block solves the homogeneous differential equation
	\begin{align}
		\left(C^{(2)}_{12}-h(h-1)\right)A(x_1,...,x_4)G_{\Delta}(\chi) =0
	\end{align}
	This translates as
	\begin{align}
		\chi  \left(\left(\Delta _{12}-\Delta _{34}-1\right) \chi  G_{\Delta }'(\chi )-(\chi -1) \chi  G_{\Delta }''(\chi )+\Delta _{12}\Delta _{34} G_{\Delta }(\chi )\right) = h(h-1)G_{\Delta }(\chi )
	\end{align}
	where $\Delta_{ij}=\Delta_i-\Delta_j$.\par 
This is solved by 
	\begin{align}
		G_{\Delta}(\chi) &= \chi^\Delta  {}_2F_1(\Delta-\Delta_{12},\Delta+\Delta_{34},2\Delta;\chi)
	\end{align}
	as expected.
	\subsection*{Shadow Block Formalism}
	Similarly to the conformal Casimir above, one can insert the identity explicitly in the four-point function. This is the shadow block formalism \cite{Simmons-Duffin:2012juh}. In this context, one explicitly inserts the identity as a complete set of states. This operator takes the form
	\begin{align}
		\mathbb{1} = \int dx \, \sum_h  \mathcal{O}_h(x)\ket{0}\bra{0}\tilde{\mathcal{O}}_h(x)
	\end{align}
	Where $\tilde{\mathcal{O}}_h$ is an operator of conformal dimension $1-h$ by dimension counting and is called the \textit{shadow operator}. In \cite{Simmons-Duffin:2012juh}, the explicit realisation is given as
	\begin{align}
		`\tilde{\mathcal{O}}_h(X) = \int d^dY \frac{\mathcal{O}_h(Y)}{(X-Y)^{2h}}
	\end{align}
	Inserting this operator explicitly in a four-point function gives
	\begin{align}
		\langle \phi_{\Delta_1}(x_1) \phi_{\Delta_2}(x_2) \phi_{\Delta_3}(x_3) \phi_{\Delta_4}(x_4)\rangle &= \int dx_5  \sum_h \bra{0} \phi_{\Delta_4}(x_4) \phi_{\Delta_3}(x_3) \mathcal{O}_h(x_5)\ket{0}\bra{0}\tilde{\mathcal{O}}_h(x_5) \phi_{\Delta_2}(x_2) \phi_{\Delta_1}(x_1)\ket{0}\\
		&= \int dx_5 \sum_h \langle\phi_{\Delta_1}(x_1)\phi_{\Delta_2}(x_2) \tilde{\mathcal{O}}_h(x_5) \rangle \langle\mathcal{O}_h(x_5) \phi_{\Delta_3}(x_3) \phi_{\Delta_4}(x_4)\rangle
	\end{align}
	Even though in one dimension, the shadow operator will not satisfy unitarity for $h>1$; we consider the shadow operator an effective local operator with conformal dimension $\tilde{h} = 1-h$. Performing the $x_5$ integral of this product of three-point functions, we obtain:
	\begin{align}
		&\int dx_5 \sum_h c_{\Delta_1\Delta_2 h}c_{\Delta_3\Delta_4h} (x_{12}^{\tilde{h}-\Delta_1-\Delta_2}x_{15}^{\Delta_2-\Delta_1-\tilde{h}}x_{25}^{\Delta_1-\Delta_2-\tilde{h}}) (x_{53}^{\Delta_4-\Delta_3-h}x_{54}^{\Delta_3-h-\Delta_4}x_{34}^{h-\Delta_3-\Delta_4}) \\
		&= A(x_1,...,x_4)\int dx_5 \sum_h c_h^2 (\chi^{\tilde{h}-\Delta_1-\Delta_2}x_{5}^{\Delta_2-\Delta_1-\tilde{h}}(x_{5}-\chi)^{\Delta_1-\Delta_2-\tilde{h}}) (1-x_{5})^{\Delta_4-\Delta_3-h}\\
		&=\sum_h c^2_h (a \, G_h(\chi) +b \, \tilde{G}_{h}(\chi))
	\end{align}
	Where the form of the blocks is
	\begin{align}
		G_h(\chi) &= \chi ^{h-\Delta_1-\Delta_2} \, {}_2F_1(h-\Delta_{12},h+\Delta_{34},2h;\chi )\\
		\tilde{G}_{h}(\chi)&= \chi ^{\tilde{h}-\Delta_1-\Delta_2} \, {}_2F_1(\tilde{h}-\Delta_{12},\tilde{h}+\Delta_{34},2\tilde{h};\chi )
	\end{align}
	These correspond to the two solutions of the differential equation above where we can choose the one with the correct boundary terms for unitary operators.\footnote{The exact expressions of a and b depend on the range of integration. However, since we subtract the shadow contribution anyway, one is free to rescale the remaining term as a basis.} Here, we must subtract this solution to find the correct conformal blocks. The subtraction procedure is detailed in \cite{Simmons-Duffin:2012juh} in terms of the monodromy $x_{12}\rightarrow e^{2\pi i}x_{12}$. This will not be detailed as the form of the conformal blocks has already been derived several times above.
	
	\subsection*{Five-point Scalar Blocks}\label{App:five-point scalar bloc}
	We do the same thing for the five-point function illustrated in Figure \ref{diagram: 5pt block} where we use the blocks corresponding to the OPE of points 1 and 2 and points 3 and 4.	
	This setup makes explicit the fact that the objects of weight $h_1$ are those exchanged in the $(\chi_1,\chi_2)\rightarrow(0,1)$. These are used as a basis to find the superconformal blocks in the symmetric channel in section \ref{Section: superblocks}. The blocks in the antisymmetric channel can be found using the principles of crossing and braiding. 
	Using the notation established above, we have
	\begin{align}
		\langle \prod_{i=1}^{5} \phi_{\Delta_i}(x_i)  \rangle =\sum_{h_1,h_2} c_{12h_1}c_{34h_2}c_{h_1h_2\Delta_5}\mathcal{F}_{12h_1}(x_{12},\partial_{x_1})\mathcal{F}_{34h_2}(x_{34},\partial_{x_3}) \langle \mathcal{O}_{h_1}(x_1)\mathcal{O}_{h_2}(x_3) \phi_{\Delta_5}(x_5)\rangle
	\end{align}
	As before, in the limit $x_5\rightarrow\infty$, any derivative acting on $x_{15}$ or $x_{35}$ leads to a vanishing term, so we can safely ignore these terms in this limit. This gives
	\begin{align}
		\sum_{h_1,h_2} c^3_{h_1,h_2} \sum_{k,l} \frac{(h_1-\Delta_{12})_k (h_2-\Delta_{34})_k (-1)^{k+l}}{(2h_1)_{k} (2h_2)_l k! l!}x_{12}^{k+h_1-\Delta_1-\Delta_2}x_{34}^{l+h_2-\Delta_3-\Delta_4}\partial_{x_1}^{(k)}\partial_{x_3}^{(l)}x_{13}^{h_1+h_2-\Delta_5}
	\end{align} 
	Taking the other limits to obtain the cross-ratios, we have
	\begin{align}
		& \chi_1^{-\Delta_1-\Delta_2} (1-\chi_2)^{-\Delta_3-\Delta_4}\chi_2^{\Delta_5} \sum_{h_1,h_2} c^3_{h_1,h_2}\left(\frac{\chi_1}{\chi_2}\right)^{h_1} \left(\frac{1-\chi_2}{\chi_2}\right)^{h_2} \times \\
		&F_2\left(
		h_1+h_2-\Delta_5,h_1-\Delta_{12},h_2-\Delta_{34};2h_1,2h_2\,|\,\frac{\chi_1}{\chi_2},\frac{\chi_2-1}{\chi_2}\right)
	\end{align} 
	Using the identity for the Appell function
	\begin{align}
		F_2(A,B_1,B_2,C_1,C_2;z,w) = (1-w)^{-A}F_2(A,B_1,C_2-B_2,C_1,C_2;\frac{z}{1-w},\frac{w}{w-1})
	\end{align}
	This gives the final result
	\begin{align}
		\langle \prod_{i=1}^{5} \phi_{\Delta_i}(\chi_i)  \rangle =& \chi_1^{-\Delta_1-\Delta_2} (1-\chi_2)^{-\Delta_3-\Delta_4}\sum_{h_1,h_2} c^3_{h_1,h_2} \chi_1^{h_1}(1-\chi_2)^{h_2}\times \\
		& F_2\left(
		h_1+h_2-\Delta_5,h_1-\Delta_{12},h_2+\Delta_{34};2h_1,2h_2\,|\,\chi_1,1-\chi_2\right)
	\end{align}
	Here, the cross-ratios and the prefactor are those found through the conformal transformations bringing the points 
	\begin{align}
		\{x_i\}&\longrightarrow \{0,\chi_i,,1,\infty\}&
		\chi_i &= \frac{x_{1i}x_{n,n-1}}{x_{1,n-1}x_{in}}
	\end{align}
	The conformal prefactor for this transformation is
	\begin{align}
		\left(\frac{x_{45}x_{15}}{x_{14}}\right)^{\Sigma-2\Delta_5}x_{15}^{-2\Delta_1}x_{25}^{-2\Delta_2}x_{35}^{-2\Delta_3}x_{45}^{-2\Delta_4}
	\end{align}
	This gives the final result for external weights $(11112)$
	\begin{align}
		\frac{1}{\left( x_{14}x_{25}x_{35}\right)^2}& \chi_1^{-2} (1-\chi)^{-2}\sum_{h_1,h_2} c^3_{h_1,h_2} \chi_1^{h_1}(1-\chi_2)^{h_2}\times \\
		& F_2\left(
		h_1+h_2-2,h_1,h_2;2h_1,2h_2\,|\,\chi_1,1-\chi_2\right)
	\end{align}
	Where $F_2$ is the Appell Function defined as
	\begin{align}\label{Eq: Appell Function}
		F_2\left(
		\alpha,\beta_1,\beta_2;\gamma_1,\gamma_2\,|\,x,y\right) = \sum_{k,l} \frac{(\beta_1)_k(\beta_2)_l(\alpha)_{k+l}}{(\gamma_1)_k(\gamma_2)_l(1)_k(1)_l}x^ky^l
	\end{align}
\section{Polyakov Blocks}\label{App: Polyakov}
Polyakov blocks mentioned in section \ref{Intro Conformal} are an alternative basis in which to decompose the OPE expansion, they are defined to be crossing-symmetric, Regge-bounded and to provide the same OPE expansion as the conformal blocks. 
\subsubsection{Perturbative Polyakov Blocks}

It will be useful to illustrate how Polyakov blocks and conformal blocks operate perturbatively. 
The four-point correlator of a scalar of conformal dimension $\Delta $ is
\begin{align}
	\langle \phi_\Delta(x_1)\phi_\Delta(x_1)\phi_\Delta(x_1)\phi_\Delta(x_1)\rangle  = \frac{(C_\Delta)^4}{(x_{13}x_{24})^{2\Delta}}f(\chi)\\
	f(\chi) = \sum_h c^2_{\Delta\Delta h }\chi^{-2\Delta}G_h(\chi) = \sum_h c^2_{\Delta\Delta h }\chi^{-2\Delta}P_h(\chi).
\end{align}
where $G_h$ are the conformal blocks and $P_h(\chi)$ are the Polyakov blocks. The four-point conformal blocks in $d=1$ are the eigenfunctions of the quadratic Casimir
\begin{align}\label{Eq: conformal blocks}
	G_h(\chi) = \chi^h {}_2F_1(h,h,2h;\chi).
\end{align}
The Polyakov blocks are not eigenvalues of the quadratic Casimir and depend non-trivially on the exchanged and external dimension. Though the Polyakov blocks are not known in closed form in position space, their double discontinuity is equal to that of the conformal blocks in the $t$-channel
\begin{align}
	dDisc[\chi^{-2\Delta}P^{(t)}_h(\chi)] &= dDisc[\chi^{-2\Delta}G^{(t)}_{h}(\chi)] \nonumber \\
	&= 2\sin^2\left(\frac{\pi}{2}(h-2\Delta)\right)(1-\chi)^{-2\Delta}G_h(1-\chi).
\end{align}
The double discontinuity of the Polyakov block in the $t$-channel is given by the replacement $\chi\rightarrow 1-\chi$ since the Polyakov block is crossing-symmetric.\footnote{We choose a prefactor $\frac{1}{(x_{13}x_{24})^{2\Delta}}$ to have a crossing-symmetric function $f(\chi)$. However, we keep the normalisation of the blocks in equation \eqref{Eq: conformal blocks} to be consistent with the literature and use the combination $\chi^{-2\Delta}P(\chi)$ to work with a truly crossing-symmetric quantity, without the need for a prefactor. }
\begin{align}
	dDisc[\chi^{-2\Delta}P^{(s)}_h(\chi)] = 2\sin^2\left(\frac{\pi}{2}(h-2\Delta)\right)\chi^{-2\Delta}G_h(\chi).
\end{align}
If we expand the four-point correlator in a small strong coupling parameter $\epsilon$
\begin{align}
	f(\chi) = f^{0}(\chi)+\epsilon f^{1}(\chi)+O(\epsilon^2),
\end{align}
one can look at the structure and properties of these two expansions. The first order is GFF where the spectrum is $\{0,h=2\Delta+2n\}$, and the correlator is  obtained with the pairwise Wick contractions between fields
\begin{align}
	f^{0}(\chi) =1+  \chi^{2\Delta}+\left(\frac{\chi}{1-\chi}\right)^{2\Delta}.
\end{align}
The conformal decomposition 
\begin{align}
	f^0(\chi)&=1+\sum_{n}c_{2\Delta+2n,\Delta,\Delta}^2G_{2\Delta+2n}(\chi),
\end{align}
gives the OPE coefficients.
Whereas the Polyakov blocks vanish at the position of the double trace operators
\begin{align}
	f^0(\chi) = P_{\Delta,0}(\chi)+\sum_{n}c_n^{(0)}P_{2\Delta+2n}(\chi),
\end{align}
giving the identity contribution in all channels
\begin{align}
	P_{\Delta,0}(\chi) = 1+  \chi^{2\Delta}+\left(\frac{\chi}{1-\chi}\right)^{2\Delta}.
\end{align}
At first order in a perturbative expansion, assuming that there are no new exchanged operators, the spectrum is $h=\{0,2\Delta+2n+\epsilon \gamma^{(1)}_n\}$, where the identity operator receives no corrections.\footnote{At strong coupling, this means that the first-order Witten diagrams are contact diagrams and not exchange diagrams.} The first order OPE expansion is then
\begin{align}
	f^{1}(\chi) & = \sum_n c^{(1)}_nG_n(\chi)+c_n^{(0)}\left(\frac{g\gamma_n}{2}\right)\partial_nG_n(\chi)\\
	&= \sum_n c_n^{(0)}\left(\frac{g\gamma_n}{2}\right)\partial_nP_n(\chi).
\end{align}
Only one term in the Polyakov block expansion is non-vanishing in this case
\begin{align}
	f^{1}(\chi) = c_0^{(0)}\left(\frac{g\gamma_0}{2}\right)(\partial_nP_n)|_{n=0}(\chi).\label{Eq: contact Witten diagram}
\end{align}
If there is a new exchanged operator $\Delta_E\neq 2\Delta+2n$ at this order ($c_{\Delta_E\Delta\Delta}=O(\sqrt{\epsilon})$), the expansion is changed by a factor:
\begin{align}\label{Eq: Polyakov pert exp}
	f^{1}(\chi) =c_0^{(0)}\left(\frac{g\gamma_0}{2}\right)(\partial_nP_n)|_{n=0}(\chi)+c_{\Delta_E\Delta\Delta}^2P_{\Delta_E}(\chi)
\end{align}
In the strong coupling language, this corresponds to having an exchange Witten diagram with exchange dimension $\Delta_E$. Hence, the Polyakov blocks are given by exchange Witten diagrams, up to the contact diagram contribution from equation \eqref{Eq: contact Witten diagram}.

\chapter{ABJM Four-Point Bootstrap}
The bootstrap of the four-point correlator for the displacement multiplet presented in Chapter \ref{chapter: ABJM} requires constraints from the symmetry preserved by the line. Details on the preserved algebra and its embedding in the algebra of ABJM are presented in section \ref{ABJM symmetry} along with the representation theory needed to consider the displacement multiplet, as well as the exchanged operators in the OPE expansion. In the next section \ref{App: AdS correlators}, additional perturbative correlators are presented which were not necessary in the main text in \ref{sec:sigmamodel}. Then, the weak-coupling description of the displacement multiplet is presented in section \ref{susyonfields}. Finally, the details of the solving of the mixing are presented in section \ref{ABJM solving mixing}.
	\section{ABJM Symmetry and Representations}\label{ABJM symmetry}
	\subsection*{$osp(6|4)$ Algebra and its Subalgebra $su(1,1|3)$}\label{algebra} 
	\label{app:algebra-superconf}
	We now list the commutation relations for the $osp(6|4)$ superalgebra. Let us start with the three-dimensional conformal algebra
	\begin{align}
		[P^{\mu},K^{\nu}]&=-2\d^{\mu\nu} D-2 M^{\mu\nu} & [D,P^{\mu}]&= P^{\mu} & [D,K^{\mu}]&=-K^{\mu}\\
		[M^{\mu\nu}, M^{\rho\sigma}]&=\d^{\sigma[\mu} M^{\nu]\rho}+\d^{\rho[\nu} M^{\mu]\sigma} & [P^{\mu},M^{\nu\rho}]&=\delta^{\mu[\nu}P^{\rho]} & [K^{\mu},M^{\nu\rho}]&=\delta^{\mu[\nu}K^{\rho]} 
	\end{align}
	Then we have the SU$(4)$ generators
	\begin{align}
		[{J_I}^J,{J_K}^L]=\d_I^L {J_K}^J-\d^J_{K} {J_I}^L
	\end{align}
	
	Fermionic generators $Q^{IJ}_\a$ and $S^{IJ}_\a$ respect the reality condition $\bar Q_{IJ\a}=\frac12 \e_{IJKL} Q^{KL}_{\a}$ and similarly for $S$. Anticommutation relations are
	\begin{align}\label{Qosp}
		\{Q^{IJ}_\a,Q^{KL\b}\}&=2i\,\e^{IJKL} {(\gamma^\mu)_{\a}}^{\b} P_\mu \qquad  \{S^{IJ}_\a,S^{KL\b}\}=2i\,\e^{IJKL} {(\gamma^\mu)_{\a}}^{\b} K_\mu \\  \{Q^{IJ}_\a,S^{KL\b}\}&=\e^{IJKL} ({(\gamma^{\mu\nu})_{\a}}^{\b} M_{\mu\nu}+2\d_{\a}^{\b} D)+2\d_{\a}^{\b}\e^{KLMN}(\d_M^J {J_N}^I-\d_M^I {J_{N}}^J)
	\end{align}
	Finally, mixed commutators are
	\begin{align}
		[D,Q^{IJ}_{\a}]&=\frac12 Q^{IJ}_{\a} & [D,S^{IJ}_{\a}]&=-\frac12 S^{IJ}_{\a} \\
		[M^{\mu\nu},Q^{IJ}_{\a}]&=-\frac12 {(\gamma^{\mu\nu})_{\a}}^{\b} Q^{IJ}_{\b} & [M^{\mu\nu},S^{IJ}_{\a}]&=-\frac12 {(\gamma^{\mu\nu})_{\a}}^{\b} S^{IJ}_{\b}  \\
		[K^{\mu},Q_{\a}^{IJ}]&=-i\,{(\g^{\mu})_{\a}}^{\b} S^{IJ}_{\b} & [P^{\mu},S^{IJ}_{\a}]&=-i\,{(\g^{\mu})_{\a}}^{\b} Q^{IJ}_{\b}\\
		[{J_I}^J,Q^{KL}_{\a}]&=\d_I^K Q^{JL}_\a+\d_{I}^L Q^{KJ}_\a-\frac12\d_I^J Q^{KL}_{\a} &  [{J_I}^J,S^{KL}_{\a}]&=\d_I^K S^{JL}_\a+\d_{I}^L S^{KJ}_\a-\frac12\d_I^J S^{KL}_{\a}
	\end{align}
	
	%\section{The subalgebra $su(1,1|3)$}\label{subalgebra}
	
	Inside the $osp(6|4)$ it is possible to identify the $su(2|3)$ (or, more precisely $su(1,1|3)$) subalgebra preserved by the 1/2-BPS Wilson line. The $su(1,1)$ generators are those of the one-dimensional conformal group, i.e. $\{D,P\equiv P_1,K\equiv K_1\}$,
	%Since for building irreducible representations it will be important to choose the correct real section, compared to the previous section we make the transformations $P_1\to i P_1$ and $K_1\to i K_1$ in order to obtain the correct $su(1,1)$ commutation relations
	satisfying
	\begin{align}
		[P,K]&=-2 D & [D,P]&=P & [D,K]&=-K
	\end{align}
	The SU$(3)$ generators ${R_a}^b$ are traceless, i.e. ${R_a}^a=0$ and they are given in terms of the original $su(4)$ ones by
	\begin{align}
		{R_{a}}^{b}&=\begin{pmatrix}
			{J_2}^2+\frac13 {J_1}^1 & {J_2}^3 & {J_2}^4\\
			{J_3}^2 & {J_3}^3+\frac13 {J_1}^1 & {J_3}^4\\
			{J_4}^2 & {J_4}^3  & -{J_3}^3-{J_2}^2-\frac23 {J_1}^1\\
		\end{pmatrix}             
	\end{align}
	Their commutation relations are
	\begin{align}
		[{R_{a}}^{b},{R_{c}}^{d}]&=\d_{a}^{d} {R_{c}}^{b} -\d^{b}_{c} {R_{a}}^{d}
	\end{align}
	The last bosonic symmetry is the $u(1)$ algebra generated by
	\begin{align}
		J_0=3i M_{23}-2 {J_1}^1
	\end{align}
	and commuting with the other bosonic generators.
	
	The fermionic generators are given by a reorganization of the preserved supercharges\newline $\{Q^{12}_+,Q^{13}_+,Q^{14}_+,Q^{23}_-,Q^{24}_-,Q^{34}_-\}$, together with the corresponding superconformal charges. Our notation is
	\begin{equation}
		Q^a=Q^{1a}_{+} \qquad S^a=i\, S^{1a}_{+}  \qquad  \bar Q_a=i\,\frac12 \epsilon_{abc} Q_{-}^{bc} \qquad  \bar S_a=\frac12 \epsilon_{abc} S_{-}^{bc}
	\end{equation}
	The $i$ factors are chosen to compensate those in the algebra \eqref{Qosp}
	%transformations on $P_1$ and $K_1$ 
	so that %anticommutators read
	\begin{align}
		\{Q^a,\bar Q_b\}&=2 \d^a_b P &  \{S^a,\bar S_b\}&=2\d^a_b K\\  \{Q^a,\bar S_b\}&= 2\d_b^a ( D+\tfrac13 J_0)-2 {R_b}^a & \{\bar Q_a, S^b\}&= 2\d_b^a(D-\tfrac13 J_0)+2 {R_a}^b 
	\end{align}
	Finally, non-vanishing mixed commutators are \small
	\begin{align}
		[D,Q^a]&=\frac12 Q^a & [D,\bar Q_a]&=\frac12 \bar Q_a &  [K,Q^a]&=S^a & [K,\bar Q_a]&= \bar S_a\\
		[D,S^a]&=-\frac12 S^a & [D,\bar S_a]&=-\frac12 \bar S_a &  [P,S^a]&=-Q^a & [P,\bar S_a]&=- \bar Q_a\\
		[{R_a}^b,Q^c]&=\d_a^c Q^b-\tfrac13 \d_a^b Q^c & [{R_a}^b,\bar Q_c]&=-\d_c^b \bar Q_a+\tfrac13 \d_a^b \bar Q_c & [J_0,Q^a]&=\tfrac12 Q^a & [J_0,\bar Q^a]&=-\tfrac12 \bar Q^a\\
		[{R_a}^b,S^c]&=\d_a^c S^b-\tfrac13 \d_a^b S^c & [{R_a}^b,\bar S_c]&=-\d_c^b \bar S_a+\tfrac13 \d_a^b \bar S_c & [J_0,S^a]&=\tfrac12 S^a & [J_0,\bar S^a]&=-\tfrac12 \bar S^a
	\end{align}\normalsize
	
	\subsection*{Representations of $su(1,1|3)$}\label{representations}
	\label{app:reprs}
	Here we present a summary of the representation theory of the $su(1,1|3)$ algebra. A detailed analysis can be found in \cite{Bianchi:2017ozk}. The algebra is characterised by four Dynkin labels $[\D,j_0,j_1,j_2]$ associated to the Cartan generators of the bosonic subalgebra $su(1,1)\oplus u(1)\oplus su(3)$. The two SU$(3)$ Cartan generators are defined as
	\begin{equation}
		J_1={R_1}^1-{R_2}^2 \qquad J_2={R_1}^1+2{R_2}^2
	\end{equation}
	The highest weight state is characterised by
	\begin{align}
		S^a\ket{\D,j_0,j_1,j_2}^{\text{hw}}&=0 & \bar S_a\ket{\D,j_0,j_1,j_2}^{\text{hw}}&=0 & E^+_a\ket{\D,j_0,j_1,j_2}^{\text{hw}}&=0
	\end{align}
	where $E^+_a$ are raising generators of SU$(3)$ in the Weyl Cartan basis (see \cite{Bianchi:2017ozk}). The long multiplet is built by acting with supercharges, momentum and SU$(3)$ lowering generators on the highest weight state. The dimension of the long multiplet is 
	\begin{equation}
		\text{dim} \mathcal{A}^{\D}_{j_0;j_1,j_2}=27 (j_1+1)(j_2+1)(j_1+j_2+2)
	\end{equation}
	and unitarity requires
	\begin{equation}\label{unitaritybounds}
		\D\geq\left\{\begin{array}{l}
			\frac13(2j_1+j_2-j_0) \quad j_0\leq \frac{j_1-j_2}{2}\\
			\frac13(j_1+2j_2+j_0) \quad j_0 > \frac{j_1-j_2}{2}
		\end{array}\right.
	\end{equation}
	
	There are several shortening conditions one can impose. The multiplets $\mathcal{B}_{j_0;j_1,j_2}$ are obtained by imposing
	\begin{equation}
		Q^a\ket{\D,j_0,j_1,j_2}^{\text{hw}}=0
	\end{equation}
	for the three cases  
	\begin{align}
		a&=1 & \D&=\frac13(2 j_1+ j_2-j_0) & &\mathcal{B}_{j_0,j_1,j_2}\\
		a&=1,2 & \D&=\frac13(j_2-j_0)  \quad j_1=0 & &\mathcal{B}_{j_0,j_2}\\
		a&=1,2,3 & \D&=-\frac13 j_0 \qquad \quad \, j_1=j_2=0 & &\mathcal{B}_{j_0}
	\end{align}
	where compared to \cite{Bianchi:2017ozk}, we simplified notation leaving the number of indices to indicate the fraction of supercharges annihilating each multiplet. The conjugate ones are given by
	\begin{equation}
		\bar Q_a\ket{\D,j_0,j_1,j_2}^{\text{hw}}=0
	\end{equation}
	for the three cases  
	\begin{align}
		a&=3 & \D&=\frac13 ( j_1+2j_2+j_0) & &\bar{\mathcal{B}}_{j_0,j_1,j_2}\\
		a&=2,3 & \D&=\frac13 ( j_1+j_0) \quad j_2=0 & &\bar{\mathcal{B}}_{j_0,j_1}\\
		a&=1,2,3 & \D&=\frac13j_0 \qquad \qquad  j_1=j_2=0 & &\bar{\mathcal{B}}_{j_0}
	\end{align}
	The remaining multiplets are listed for completeness, but they are not relevant to our setup
	\begin{align}
		&\hat{\mathcal{B}}_{j_0,j_1,j_2} & \D&=\frac{j_1+j_2}{2} & j_0&=\frac{j_1-j_2}{2}\\
		&\hat{\mathcal{B}}_{j_0,0,j_2} & \D&=\frac{j_2}{2} & j_0&=\frac{-j_2}{2} & &j_1=0\\
		&\hat{\mathcal{B}}_{j_0,j_1,0} & \D&=\frac{j_1}{2} & j_0&=\frac{j_1}{2} & &j_2=0
	\end{align}
	
	We also list the recombination of long multiplets at the unitarity bound. For $j_0< \frac{j_1-j_2}{2}$ the unitarity bound is for $\D=\frac13(2j_1+j_2-j_0)$ and we have
	\begin{equation}
		\mathcal{A}^{-\frac13 j_0+\frac23 j_1+\frac13 j_2}_{j_0,j_1,j_2}=\mathcal{B}_{j_0,j_1,j_2} \oplus \mathcal{B}_{j_0+\frac12,j_1+1,j_2}
	\end{equation}
	Similarly, for $j_0> \frac{j_1-j_2}{2}$ one has
	\begin{equation}
		\mathcal{A}^{\frac13 j_0+\frac13 j_1+\frac23 j_2}_{j_0,j_1,j_2}=\bar{\mathcal{B}}_{j_0,j_1,j_2} \oplus \bar{\mathcal{B}}_{j_0-\frac12,j_1,j_2+1}
	\end{equation}
	For $j_0=\frac{j_1-j_2}{2}$ we have
	\begin{equation}
		\mathcal{A}^{j_1+j_2}_{\frac{j_1-j_2}{2},j_1,j_2}=\hat{\mathcal{B}}_{\frac{j_1-j_2}{2},j_1,j_2} \oplus \hat{\mathcal{B}}_{\frac{j_1-j_2}{2}+\frac12,j_1+1,j_2}\oplus \hat{\mathcal{B}}_{\frac{j_1-j_2}{2}-\frac12,j_1+1,j_2+1}\oplus \hat{\mathcal{B}}_{\frac{j_1-j_2}{2},j_1+1,j_2+1}
	\end{equation}
	
	For vanishing Dynkin labels, the decomposition is different. We first list all short multiplets with vanishing labels as
	\begin{align}
		&\{\bar{\mathcal{B}}_{j_0,0,j_2},\mathcal{B}_{j_0,j_2},\hat{\mathcal{B}}_{j_0,0,j_2}\}  & j_1&=0 & j_2&>0 \\
		&\{\mathcal{B}_{j_0,j_1,0},\bar{\mathcal{B}}_{j_0,j_1},\hat{\mathcal{B}}_{j_0,j_1,0}\}  & j_1&>0 & j_2&=0 \\
		&\{\mathcal{B}_{j_0},\bar{\mathcal{B}}_{j_0}\}  & j_1&=0 & j_2&=0 
	\end{align}
	The decompositions of long multiplet at the unitarity bound for these cases are shown in Table~\ref{longmultdec}.
	
	\begin{table}[htbp]
		\begin{center}
			\def\arraystretch{1.5}
			\begin{tabular}{|c|c|c|}
				\hline 
				& $j_0<-\frac{j_2}{2}$  & $\mathcal{A}^{\frac13 (j_2-j_0)}_{j_0,0,j_2}=\mathcal{B}_{j_0,j_2}\oplus \mathcal{B}_{j_0+\frac12,1,j_2}$  \\
				$j_1=0$		& $j_0>-\frac{j_2}{2}$ & $\mathcal{A}^{\frac13 (2j_2+j_0)}_{j_0,0,j_2}=\bar{\mathcal{B}}_{j_0,0,j_2}\oplus \bar{\mathcal{B}}_{j_0-\frac12,0,j_2+1}$\\
				& $j_0=-\frac{j_2}{2}$ & $\mathcal{A}^{\frac{j_2}{2}}_{-\frac{j_2}{2},0,j_2}=\hat{\mathcal{B}}_{-\frac{j_2}{2},0,j_2}\oplus \hat{\mathcal{B}}_{-\frac{j_2+1}{2},0,j_2+1}\oplus \hat{\mathcal{B}}_{\frac{1-j_2}{2},1,j_2}\oplus \hat{\mathcal{B}}_{-\frac{j_2}{2},1,j_2+1}$\\[1ex]
				\hline 
				\To   	& $j_0<\frac{j_1}{2}$  & $\mathcal{A}^{\frac13 (2j_1-j_0)}_{j_0,j_1,0}=\mathcal{B}_{j_0,j_1,0}\oplus \mathcal{B}_{j_0+\frac12,j_1+1,0}$\\
				$j_2=0$		& $j_0>\frac{j_1}{2}$ & $\mathcal{A}^{\frac13 (j_1+j_0)}_{j_0,j_1,0}=\bar{\mathcal{B}}_{j_0,j_1}\oplus \bar{\mathcal{B}}_{j_0-\frac12,j_1,1}$\\
				& $j_0=\frac{j_1}{2}$ & $\mathcal{A}^{\frac{j_1}{2}}_{\frac{j_1}{2},j_1,0}=\hat{\mathcal{B}}_{\frac{j_1}{2},j_1,0}\oplus \hat{\mathcal{B}}_{\frac{j_1+1}{2},j_1+1,0}\oplus \hat{\mathcal{B}}_{\frac{j_1-1}{2},j_1,1}\oplus \hat{\mathcal{B}}_{\frac{j_1}{2},j_1+1,1}$\\[1ex]
				\hline 
				\multirow{ 2}{*}{$j_1=j_2=0$}		&$j_0<0$  & $\mathcal{A}^{-\frac{j_0}{3}}_{j_0,0,0}=\mathcal{B}_{j_0}\oplus \mathcal{B}_{j_0+\frac12,1,0}$\\
				&$j_0>0$  & $\mathcal{A}^{\frac{j_0}{3}}_{j_0,0,0}=\bar{\mathcal{B}}_{j_0}\oplus \bar{\mathcal{B}}_{j_0-\frac12,0,1}$\\[1ex]
				\hline
			\end{tabular}
		\end{center}
		\caption{Decomposition of long multiplets into short ones for the case of some vanishing Dynkin labels.}\label{longmultdec}
	\end{table}

\section{AdS$_4\times \mathbb{C}P^3$ Strong Coupling Correlators}\label{App: AdS correlators}
	
	\subsection*{Four-Point Function of Massive Bosonic Fluctuations}
	
	The  complex field  $X$ appearing in the AdS$_2$  action in \eqref{lagr}-\eqref{lagrend} is the AdS/CFT dual  of the displacement operator insertion  $\mathbb{D}$, which has  protected dimension $\Delta=2$.    
	Due to conformal invariance, the four-point correlator reads then
	\begin{equation}
		\langle X(t_1)\,\Xb(t_2)\,X(t_3)\,\Xb(t_4)\rangle
		=  \frac{\big[C_{X}
			(\lambda)\big]^2}{t_{12}^4 t_{34}^4}\,G(\chi) \ , 
		\label{4point-X}
	\end{equation}
	where  we have used
	\begin{equation}\la{4.2}
		\langle X(t_1)\,\Xb(t_2)\rangle=\frac{C_{{X}}(\lambda)}{t_{12}^4}\,.
	\end{equation}
	Again, the normalisation factor $C_X(\lambda)$ may be chosen to be in direct correspondence with the Bremsstrahlung function~\eqref{Bremss} to realise a direct identification of $X$ with the displacement operator $\mathbb{D}$.   In view of~\eqref{2-p-insertions}, this would mean $C_X(\lambda)\equiv 24 B_{1/2}(\lambda)$.  
	
	The   disconnected contribution to~\eqref{4point-X} reads 
	\begin{equation}\label{X-disconn}
		\cor{X(t_1)\,\Xb (t_2)\,X(t_3)\,\Xb(t_4)}_\text{disconn.} =\frac{\big[C_{{X}}(\lambda)\big]^2}{t_{12}^4 t_{34}^4}\,\Big[1+\frac{\chi^4}{(1-\chi)^4}\,\Big]\,.
	\end{equation}
	The subleading, connected contribution  is obtained by evaluating Witten diagrams from the four-point interaction vertices $L_{4X}$ in \eqref{lagr}-\eqref{lagrend} and reads
	\begin{eqnarray}
		&&\langle X(t_1)\,\Xb (t_2)\,X(t_3)\,\Xb(t_4)\rangle_\text{conn} =\frac{1}{T}\,\big(\mathcal{C}_{\Delta=2}\big)^4 \, \mathcal{Q}\,, \qquad \mathcal{C}_{\Delta=2}=\frac{3}{2\pi}\\\nonumber
		&&\qquad\mathcal{Q}=8 (D_{2 2 2 2} + t_{12}^2 D_{33 2 2}+ t_{34}^2 D_{2 2 3 3} +  t_{23}^2 D_{2 3 3 2} + t_{14}^2 D_{3 2 2 3}\\
		&& \qquad\qquad~~ - 8 t_{24}^2 D_{23 2 3}- 
		8 t_{13}^2 D_{3 2 3 2}  +    16 t_{13}^2 t_{24}^2 D_{3 3 3 3})  \,.
	\end{eqnarray}
	Explicitly and adopting the standard normalization, one gets
	\begin{eqnarray}\nonumber
		\!\!\!   \cor{X\Xb X\Xb}_\text{conn.} \! \! \! &=&\epsilon\,\frac{\big[\mathcal{C}_{\Delta=2}\big]^2}{ t_{12}^4 t_{34}^4}\,
		\Big[\frac{115 \chi ^5-345 \chi ^4+543 \chi ^3-511 \chi ^2+246 \chi -48}{3 (1-\chi )^5}
		\\\label{X-conn}
		&&\qquad\qquad\qquad + \frac{2 \chi ^4 (5 \chi +3)}{(\chi -1)^5}\log\chi+2\big(\textstyle{5-\frac{8}{\chi }}\big) \log (1-\chi )
		\Big]\,.
	\end{eqnarray}
	Writing everything in terms of  the invariant $z$, and confronting the superspace prediction \eqref{corrD} with \eqref{X-disconn} and \eqref{X-conn} we obtain 
	\begin{eqnarray}\nonumber
		&&\!\!\!\!\!\!\!
		\frac{1}{36}\Big[36 f(z)-36 (z^4+z) f'(z) +18 z^2 (-14 z^3+3 z^2+1) f''(z) -6 z^3 \left(55 z^3-39 z^2+3 z+1\right) f^{(3)}(z)\\\nonumber
		&&~~. -3 z^4 \left(46 z^3-63 z^2+18 z-1\right) f^{(4)}(z) -3 (z-1)^2 z^5 (7 z-1) f^{(5)}(z) -(z-1)^3 z^6 f^{(6)}(z) \,\Big]\\\nonumber
		&&\!\!\!\!=1+z^4+\epsilon\,\Big[\textstyle-16 -2 z-16 z^4-2 z^3-\frac{7 z^2}{3}
		+2 (8 z-3) z^4 \log (-z)\\
		&&\qquad\qquad\qquad\qquad \textstyle+\big( 6-\frac{16}{z}+6 z^4-16 z^5\big) \log (1-z)\Big]\,.
	\end{eqnarray}
	 The function $f(z)$ in~\eqref{fstring} solves this non-trivial sixth-order differential equation.
	
	Again, one may repeat the analysis for the correlator $\cor{X(t_1)\,\bar X(t_2)\,\bar X(t_3)\,X(t_4)}$. The result coincides with~\eqref{X-conn}
	after the transformation $\chi\to \chi/(\chi-1)$, and neglecting the imaginary part of the logarithm.

	\subsection*{Four-Point Function of Mixed Fluctuations}
The four-point correlator mixing two AdS $X$ fluctuations and two $\cp^3$ $w$ fluctuations reads
	\begin{equation}
		\langle X(t_1)\,\Xb (t_2)\,w^{a_3}(t_3)\,\wb_{a_4}(t_4)\rangle
		%=\langle \langle  \mathbb{D}(t_1)  \bar{\mathbb{D}}(t_2)  \mathbb{O}^{a_3}(t_3)  \bar{\mathbb{O}}_{a_4}(t_4)\rangle\rangle
		=  \frac{C_{{X}}(\lambda)\,C_{{w}}(\lambda)}{t_{12}^4 t_{34}^2}\,\delta^{a_3}_{a_4}\,G(\chi) \ .
		\label{4point-mixed}
	\end{equation}
	The disconnected contribution is
	\begin{equation}\label{mixed-disconn}
		\cor{X(t_1)\,\Xb (t_2)\,w^{a_3}(t_3)\,\wb_{a_4}(t_4)}_\text{disconn.} =\frac{C_{X}(\lambda) C_{w}(\lambda) }{t_{12}^4 t_{34}^2}\,\delta^{a_3}_{a_4}\,.
	\end{equation}
	The connected contribution  is obtained by evaluating Witten diagrams from the four-point interaction vertices $L_{4X}$ in \eqref{lagr}-\eqref{lagrend}, and reads
	\begin{eqnarray}
		&&\langle X(t_1)\,\Xb (t_2)\,w^{a_3}(t_3)\,\wb_{a_4}(t_4)\rangle_\text{conn} =\frac{1}{T}\,\big(\mathcal{C}_{\Delta=2}\big)^2 \big(\mathcal{C}_{\Delta=1}\big)^2 \, \delta^{a_3}_{a_4}\,\mathcal{Q}_{2X\,2w} \,, \\\nonumber
		&&\qquad\mathcal{Q}_{2X\,2w}=4 \Big[\,D_{22 1 1} + 2 t_{12}^2 D_{33 1 1} + 2 t_{34}^2 D_{2 2 2 2} -   2 t_{24}^2 D_{2 3 1 2} - 2 t_{23} D_{2 3 2 1} \\
		&& \qquad\qquad\qquad  \qquad - 2 t_{14}^2 D_{3 2 1 2} - 2 t_{13}^2 D_{3 2 2 1} +    4 (t_{14}^2 t_{23}^2 + t_{13}^2 t_{24}^2 - t_{12}^2 t_{34}^2) D_{3 3 2 2}\,\Big]\,,
	\end{eqnarray}
	explicitly 
	\begin{eqnarray} 
		\!\!\!   \cor{X(t_1)\,\Xb (t_2)\,w^{a_3}(t_3)\,\wb_{a_4}(t_4)}_\text{conn.} \! \! \! &=&\epsilon\,\frac{\mathcal{C}_{\Delta=2}\,\mathcal{C}_{\Delta=1}\, }{ t_{12}^4 t_{34}^2}\,\delta^{a_3}_{a_4}\,\Big[\frac{4 (\chi -2) \log (1-\chi )}{\chi }-8\Big]\,.
	\end{eqnarray}
	Once again, writing everything in terms of the invariant $z$ and equating to the superspace prediction, the differential equation obtained
	\be\nonumber
	(\!1\!-z)\, z^4 f^{(4)}\!(z)-(\!1\!
	-3 z) \,z^3 f^{(3)}\!(z)+3 z^2\, f''(z)+6 z f'(z)+6 f(z)
	=1+\epsilon \big[-8+\textstyle{\frac{4 (z-2) }{z}}\,\log (1-z)\big]\\
	\ee
	is  solved by the function $f(z)$ in~\eqref{fstring}.
\section{Weak Coupling Field Representation}\label{susyonfields}
	\subsection*{Supersymmetry Transformation of the Fields}
\label{app:susytransf}

The supersymmetry transformations of the scalar fields under the preserved supercharges read
\begin{align}
	Q^a Z&=\bar \chi^a_+     &  \bar Q_a Z&=0 &  Q^a \bar Z&=0  &\bar Q_a \bar Z&=i\chi^+_a  \\
	Q^a Y_b&=-\d^a_b \bar \psi_+ &  \bar Q_a Y_b&=i\e_{abc} \bar \chi^c_- & Q^a \bar Y^b&=-\e^{abc}\chi^-_c & \bar Q_a \bar Y^b&=-i\d^b_a\psi^+
\end{align}
and similarly for fermions
\begin{align}
	\bar Q_a \psi^+&=0    & Q^a \psi^+&=-2iD_1 \bar Y^a-\frac{4\pi i}{k}[ \bar Y^a l_B-\hat{l}_B \bar Y^a]  \\
	Q^a \psi^-&=-2D\bar Y^a  &\bar Q_a \psi^-&=-\frac{8\pi }{k}\e_{abc}\bar Y^b Z \bar Y^c  \\
	\bar Q_a \chi^+_b&=2i\e_{abc}\bar D \bar Y^c & Q^a \chi^+_b&=2i\d^a_b D_1 \bar Z+\frac{8\pi i}{k} [\bar Z\Lambda^a_b-\hat\Lambda^a_b\bar Z]     \\
	Q^a \chi^-_b&=2\d^a_b D \bar Z &\bar Q_a \chi^-_b&=-2\e_{abc}D_1 \bar Y^c-\frac{4\pi }{k}\e_{acd}[\bar Y^c  \Theta^d_b-\hat\Theta^d_b \bar Y^c]  \\
	Q^a \bar \psi_+&=0    & \bar Q_a \bar \psi_+&=2 D_1  Y_a+\frac{4\pi }{k}[ Y_a \hat{l}_B-l_B  Y_a]  \\
	\bar Q_a \bar \psi_-&=2 i\bar D Y_a  & Q^a \bar \psi_-&=-\frac{8\pi i}{k}\e^{abc} Y_b \bar Z  Y_c  \\
	Q^a \bar \chi_+^b&=2\e^{abc} D  Y_c & \bar Q_a \bar \chi_+^b&=-2\d^b_a D_1 Z-\frac{8\pi }{k} [Z \hat\Lambda_a^b-\Lambda_a^b Z]     \\
	\bar Q_a \bar \chi_-^b&=-2i\d^b_a \bar D Z & Q^a \bar \chi_-^b&=-2i\e^{abc}D_1 Y_c-\frac{4\pi i}{k}\e^{acd}[  Y_c \hat\Theta_d^b-\Theta_d^b Y_c]
\end{align}
where we used the definitions
\begin{align}
	D&=D_2-i D_3 & \bar D&=D_2+i D_3
\end{align}
and the entries of the supermatrices
\begin{align}
	\begin{pmatrix} \L_a^b  & 0\\
		0 & \hat \L_a^b\end{pmatrix}&=\begin{pmatrix}Y_a\bar Y^b+\frac12\d^b_a l_B   & 0\\
		0 & \bar Y^b Y_a+\frac12\d^b_a \hat{l}_B \end{pmatrix} \\
	\begin{pmatrix} \Theta_a^b  & 0\\
		0 & \hat \Theta_a^b\end{pmatrix}&=\begin{pmatrix} Y_a\bar Y^b-\d^b_a (Y_c\bar Y^c+Z \bar Z)  & 0\\
		0 & \bar Y^b Y_a-\d^b_a (\bar Y^c Y_c+\bar Z Z)\end{pmatrix}\\
	\begin{pmatrix} l_B  & 0\\
		0 & \hat{l}_B\end{pmatrix}&=\begin{pmatrix} (Z \bar Z-Y_a \bar Y^a)  & 0\\
		0 &  (\bar Z Z -\bar Y^a Y_a )\end{pmatrix}
\end{align}
Notice that, due to the last identity, the bosonic part of the superconnection reads
\begin{align}
	\mathcal{L}_B=\frac{2\pi i}{k} \begin{pmatrix} l_B  & 0\\
		0 & \hat{l}_B\end{pmatrix}
\end{align}
Finally, we can list the transformation properties of the gauge fields
\begin{align}
	Q^a A_1&=\frac{2\pi i}{k} ( \bar \psi_+ \bar Y^a-\bar \chi^a_+ \bar Z- \e^{abc} Y_b \chi_c^-) & \bar Q_a A_1&=\frac{2\pi }{k} (Z\chi_a^+-Y_a\psi^+ + \e_{abc} \bar \chi^b_- \bar Y^c) \\
	Q^a A&=0 &  \bar Q_a A&=-\frac{4\pi i }{k} ( Y_a \psi^- - Z \chi_a^- + \e_{abc} \bar \chi^b_+ \bar Y^c)\\
	Q^a \bar A&=\frac{4\pi }{k} ( \bar \psi_- \bar Y^a-\bar \chi^a_- \bar Z+\e^{abc} Y_b \chi_c^+) &  \bar Q_a \bar A&=0\\
	Q^a \hat A_1&=\frac{2\pi i}{k} ( \bar Y^a \bar \psi_+ - \bar Z \bar \chi^a_+ - \e^{abc} \chi_c^- Y_b ) & \bar Q_a \hat A_1&=\frac{2\pi }{k} (\chi_a^+ Z-\psi^+  Y_a + \e_{abc}  \bar Y^c \bar \chi^b_-) \\
	Q^a \hat A&=0 &  \bar Q_a \hat A&=-\frac{4\pi i }{k} (  \psi^- Y_a- \chi_a^- Z+\e_{abc}  \bar Y^c \bar \chi^b_+)\\
	Q^a \bar {\hat A}&=\frac{4\pi }{k} (  \bar Y^a \bar \psi_- - \bar Z \bar \chi^a_-  +  \e^{abc}  \chi_c^+ Y_b) &  \bar Q_a \bar{\hat A}&=0
\end{align}

To check the closure of these transformations and use them on local operators, it is important to keep in mind the equations of motion.
For the gauge field we are interested in the components  $\mathcal{F}= \mathcal{F}_{21} -i  \mathcal{F}_{31}$ and   $\bar{\mathcal{F}}=  \mathcal{F}_{21} +i  \mathcal{F}_{31}$ of the field strength. In particular, we focus on the first one, which respects the equation
\begin{equation}\label{Feqmot}
	\mathcal{F} =\frac{2\pi i}{k} \begin{pmatrix} Z\overleftrightarrow{D}\bar Z+ Y_a \overleftrightarrow{D}\bar Y^a+\bar\psi_+\psi^- +\bar\chi_+^a\chi^-_a & 0\\
		0& -\bar Z\overleftrightarrow{D}Z- \bar Y^a \overleftrightarrow{D}Y_a -\psi^-\bar\psi_+ -\chi^-_a\bar\chi_+^a
	\end{pmatrix} 
\end{equation}
where the operator $\overleftrightarrow{D}$ has the usual definition $Z\overleftrightarrow{D}\bar Z\equiv ZD\bar Z-DZ\bar Z$.
For the fermions, we need the equation
\begin{equation}
	\slashed{D}\psi_J=\frac{2\pi}{k} \left(\bar C^I C_I\psi_J-\psi_J C_I \bar C^I +2 \psi_I C_J \bar C^I-2 \bar C^I C_J \psi_I + 2\e_{ILKJ} \bar C^I \bar \psi^L \bar C^K\right)
\end{equation}
whose projection yields (listing only the components needed for our computations)
\begin{align}\label{fermeq}
	D\psi^+ = i D_1 \psi^- &+ \frac{2\pi i}{k}\left( \hat l_B \psi^- - \psi^- l_B +2\bar Y^a Z\chi^-_a -2 \chi^-_a Z \bar Y^a-2\bar Y^a \bar \chi_+^b \bar Y^c \e_{abc}\right)\\
	D \chi _a^+=i D_1\chi ^-_a &+\frac{2 \pi i}{k}  \left(\chi ^-_b \O_a^b- \hat \O_a^b \chi ^-_b-2 \bar{Z} Y_a \psi ^-+2 \psi ^-
	Y_a \bar{Z}+\right)\\
	&+\frac{4 \pi i}{k}\epsilon _{a c d} \left(\bar{Y}^c \bar{\psi }_+ \bar{Y}^d+\bar{Y}^d \bar{\chi }_+^c \bar{Z}-\bar{Z} \bar{\chi }_+^c \bar{Y}^d\right)
\end{align}
with 
\begin{equation}
	\O_a^b= \Theta_a^b+ \Lambda_a^b-\frac12 \d_a^b l_B.
\end{equation}

\subsection*{Weak Coupling Representation}

In the following,  we will also find  it convenient to accommodate the original scalar and fermionic fields of ABJM theory according to  the new R-symmetry pattern
\begin{align}
	C_I&=(Z, Y_a) &  \bar C^I&=(\bar Z, \bar Y^a)\\
	\psi_I^\pm&=(\psi^{\pm},\chi_a^{\pm}) &\bar \psi^I_\pm&=(\bar\psi_{\pm},\bar\chi^a_{\pm})\label{fermions}
\end{align}
where $Y^a$ ($\bar Y_a$) and $\chi^{a\pm}$ ($\bar\chi_{a\pm}$) change in the $\mathbf{3}$ ($\mathbf{\bar 3}$) of SU$(3)$, whereas $Z$ and $\psi^{\pm}$ $(\bar \psi_{\pm})$ are singlet.  Moreover,   we have expressed them in the basis of eigenvectors of $\gamma_1=\sigma_1$, e.g. \begin{equation}
	\psi_+=(\psi_1+\psi_2) \qquad \psi_-=(\psi_1-\psi_2)
\end{equation}
with the rules $\psi^-=-\psi_+$ and $\psi^+=\psi_-$.
The two gauge fields and, consequently, the covariant derivative can be instead split according to the new spacetime symmetry pattern, namely
\begin{equation}\label{gaugecomps}
	A_{\mu}=(A_1,A=A_2-i A_3,\bar A=A_2+i A_3)\qquad \hat A_{\mu}=(\hat A_1,\hat A=\hat A_2-i \hat A_3,\hat{ \bar{ A}}=\hat A_2+i \hat A_3)
\end{equation}
and $D_\mu=(D_1,D=D_2-i D_3,\bar D=D_2+i D_3) .$  In terms of  these new fields, the superconnection \eqref{superconnection} for the case of the straight line takes the following form 
\begin{align}\label{straightlineconna}
	\mathcal{L}(t)&=\begin{pmatrix}  A_1  & 0\\
		0 & \hat A_1 
	\end{pmatrix}+\frac{2\pi i}{k}\begin{pmatrix}  Z \bar Z-Y_a \bar Y^a  & 0\\
		0 & \bar Z Z -\bar Y^a Y_a 
	\end{pmatrix} + \sqrt{\frac{2\pi}{ k}} \begin{pmatrix} 0  & - i\bar \psi_+ \\
		\psi^+ & 0
	\end{pmatrix}.
\end{align}

This subsection aims to provide an explicit realization of the ABJM displacement supermultiplet  in terms of the fundamental fields of the underlying theory. Since these fields live along the Wilson lines and, in most cases, are obtained by varying the  superconnection \eqref{superconnection}, they naturally possess the structure of a supermatrix. 
The lowest component of the multiplet $\mathbb{F}$ has quantum numbers $[\frac{1}{2},\frac{3}{2},0,0]$ and thus we expect it to be a merely fermionic object. On the other hand,  the only field or combination of elementary fields  with these quantum numbers, which  can appear in the
entries of the supermatrix, is the bosonic complex scalar $Z$. Therefore we shall write the following  Ansatz for $\mathbb{F}$ 
\be\label{Fmatrix}
\mathbb{F}=i\sqrt{\frac{2\pi}{k}}\,\tilde\epsilon\begin{pmatrix} 0 & Z\\0 & 0\end{pmatrix}\,,
\ee
where $\tilde\epsilon$ is a fermionic parameter endowing $\mathbb{F}$ with its anticommuting nature.\footnote{We remark that $\tilde\epsilon$ is just a bookkeeping device to keep the memory of the Gra\ss mann nature of the operator insertions when constructing their explicit representation in terms of super matrices, and it appears in the definitions \eqref{Fmatrix}, \eqref{Omatrix}, \eqref{Lambdamatrix}, \eqref{Dmatrix} of $\mathbb{F}, \mathbb{O}, \mathbb{\Lambda}, \mathbb{D}$  below. In any explicit gauge theory computation of four-point functions, the presence of such fermionic parameters (each of the fields in the correlator should have a different one) will result in an overall factor, which should be dropped when comparing the final result with any alternative calculation. In particular, there is no $\tilde\epsilon$ dependence  in the correlators of the quantum fluctuations of the fundamental dual string evaluated in section~\ref{sec:sigmamodel}. The fermionic/bosonic nature of these fluctuations is indeed already explicit in the corresponding fluctuation Lagrangian.} To construct the next element of the multiplet, we have to act with $\theta_a\,Q^a$ on the operator inserted in the Wilson line, where $\theta_a$ is a Gra\ss mann-odd parameter that we can identify with one of the fermionic coordinates of  the superspace constructed in subsection \ref{Sec: ABJM Displacement}. In the four-dimensional case, the only contribution would come from the action of $\theta_a\,Q^a$ directly on the operator  
since the connection and the (open) Wilson line are invariant.  This is not the case for operators inserted in  the fermionic Wilson loop of ABJM  theory. There is an additional contribution coming from the variation of the Wilson line.  However, this amounts to ``covariantize''  the usual action of the supersymmetry on the operator as follows
\begin{align}
	\label{deltacov}
	\delta_{\text{cov}}\, \sbullet[0.75] &= \theta_a \,Q^a  \sbullet[0.75] +2 \, \sqrt{\frac{2\pi}{k}}\,\left[\begin{pmatrix}0&0\\ \,\theta_a\,\bar Y^a&0
	\end{pmatrix}\,, \sbullet[0.75]  \right]\,, \\
	\bar\delta_{\text{cov}} \, \sbullet[0.75]&=\bar \theta^a \,\bar{Q}_a \sbullet[0.75]+ 
	2 \, \sqrt{\frac{2\pi}{k}}\,\left[\begin{pmatrix}0&\bar\theta^a Y_a\\ 0&0
	\end{pmatrix}\,,\sbullet[0.75]\right]\,,\label{deltacovbar}
\end{align}
where $\theta_a$ and $\bar \theta^a$ are Gra\ss mann odd parameters associated to supertranslations. The commutators in \eqref{deltacov} and \eqref{deltacovbar} compensates the super-gauge transformation of the Wilson line induced by the supersymmetry transformation \eqref{susyvarL} with the matrix $\mathcal{G}$ given by
\begin{align}
	\mathcal{G}=2 \, \sqrt{\frac{2\pi}{k}}\begin{pmatrix}0&\bar\theta^a Y_a\\ \,\theta_a\,\bar Y^a&0
	\end{pmatrix} \, .
\end{align}

The action of $\delta_{\text{cov}}$ on   $\mathbb{F}$ defined in \eqref{Fmatrix} is quite straightforward to evaluate once   we use the transformations in Appendix~\ref{app:susytransf}. The final result is
\be\label{Omatrix}
\delta_{\text{cov}}\mathbb{F}=\, i\sqrt{\frac{2\pi}{k}} \theta_a\,\tilde\epsilon
\begin{pmatrix}
	2\sqrt{\frac{2\pi}{k}}\,Z\,\bar Y^a & - \bar\chi^a_+\\ 
	0&2\sqrt{\frac{2\pi}{k}}\,\bar Y^a\,Z 
\end{pmatrix}\equiv 
\theta_a\,\mathbb{O}^a\,.
\ee
Namely, we have obtained the second component of our supermultiplet~\eqref{displacement}, the one associated to the $R$-symmetry breaking. An identical expression can be obtained by exploiting the transformation under the action of $\mathsf{J}^a$ appearing in the superconnection \eqref{straightlineconna}
\begin{equation}
	{\delta_{\mathsf{J}^a}(Z,{Y_b})=(0, i \delta^a_b Z),\qquad
		{\delta_{\mathsf{J}^a}(\bar Z,\bar {Y}^b})=(-i\bar {Y}^a},0),\qquad {\delta_{\mathsf{J}^a}}\psi^+=0,\qquad  {\delta_{\mathsf{J}^a}}\bar\psi_+=-i\bar\chi^a_+,
\end{equation}
Using \eqref{O}, one can apply the action of these broken generators on the superconnection \eqref{straightlineconna}, recovering the same result for $\mathbb{O}^a$.
Similarly, the form of $\bar{\mathbb{O}}_a(t)$ can be obtained  by looking at the explicit action of $\bar{\mathsf{J}}_a$ on the fields. 
Applying once more $\delta_{\text{cov}}$, we reach the third component  
\begin{align} \nonumber
	\delta_{\text{cov}}\,\big(\,\theta_a\,\mathbb{O}^a\,\big)&=2i\sqrt{\frac{2\pi}{k}}\, \epsilon^{nrs}\,\theta_r \theta_s\,\tilde\epsilon
	\begin{pmatrix}
		\sqrt{\frac{2\pi}{k}}\,(\epsilon_{abn} \,\bar\chi^a_+ \,\bar Y^b - Z\,\chi^-_n)
		& - D\, Y_n \\ 
		\frac{4\pi}{k}\, \epsilon_{abn} \bar Y^a \,Z \bar Y^b&\sqrt{\frac{2\pi}{k}} \,(\epsilon_{abn} \,\bar Y^b\,\bar\chi^a_+ -\chi^-_n \,Z\,)
	\end{pmatrix}
	\\ \label{Lambdamatrix}
	& \equiv-
	\epsilon^{nrs}\,\theta_r \,\theta_s\,\mathbb{\Lambda}_n\,.
\end{align}
Also, in this case, the expression \eqref{Lambdamatrix} for $\mathbb{\Lambda}_n$ perfectly agrees with the one obtained by acting with the broken generator on the superconnection \eqref{straightlineconna} according to \eqref{La}. 

Finally,  applying again $\delta_{\text{cov}}$ to \eqref{Lambdamatrix} we obtain the top component $\mathbb{D}$ of the supermultiplet \eqref{displacement}, namely the displacement operator
\begin{align} \nonumber
	\!\!\!\!\!
	\delta_{\text{cov}} \big( - \epsilon^{nrs}& \theta_r  \theta_s  \mathbb{\Lambda}_n \big)
	\!=\! 2   \sqrt{\frac{2\pi}{k}} \epsilon^{nrs}\theta_n \theta_r  \theta_s\,\tilde\epsilon\! ~~\times\\ &\times\begin{pmatrix}\!
		\sqrt{\frac{2\pi}{k}}\, i(2Z\, D\bar Z- 2D Y_a\bar Y^a +\bar\chi^a_+\chi^-_a) & -i \,D\bar\psi_+
		\\ \nonumber
		D\psi_+ -i\,\widetilde{\mathcal{D}}_t\psi^-
		& 
		\sqrt{\frac{2\pi}{k}}\, i(2 D\bar Z\,Z -2 \bar Y^a\, D Y_a -\chi^-_a\,\bar\chi^a_+) \!
	\end{pmatrix}\equiv
	\\ \label{Dmatrix}
	& \equiv 2
	\epsilon^{nrs}\,\theta_n \,\theta_r  \,\theta_s \,\mathbb{D}\,.
\end{align}
where $\widetilde{\mathcal{D}}_t$ is the covariant derivative constructed only with the bosonic part of the superconnection.
The $\mathbb{D}$ insertion can also be constructed out of its abstract definition~\eqref{D} by acting with a broken translation on the superconnection, as was done in equation (5.18) of~\cite{Bianchi:2017ozk}. In this case, the two expressions are not identical, and the difference between the two definitions is proportional to the matrix
\be
\begin{pmatrix}
	\sqrt{\frac{2\pi}{k}}\bar\psi_+ \,\psi^- &  0
	\\ 
	\,\widetilde{\mathcal{D}}_t \psi^-  & - \sqrt{\frac{2\pi}{k}}\psi^-\,\bar\psi_+
\end{pmatrix}
{  \propto}\, \mathcal{D}_t
\begin{pmatrix}
	0 &  0
	\\ 
	\psi^-  &0
\end{pmatrix}\,
\ee
This difference is compatible with the construction~\eqref{D}, which is blind to total derivatives.
For the case of Wilson lines insertions, total derivatives are implemented~\cite{Bianchi:2017ozk} as the following modification for an operator $\mathcal{O}(t)$ inserted into the loop
\begin{align}
	\int dt ~\mathcal{W}
	[\mathcal{O}(t)]=&\int dt ~[\mathcal{W}
	[\mathcal{O}(t)]+\partial_t(\mathcal{W}
	[ \Sigma(t) ] )]=\nonumber\\=&\int dt ~(\mathcal{W}
	[\mathcal{O}(t)]+\mathcal{W}
	[\mathcal{D}_t \Sigma(t) ] )=
	\int dt ~\mathcal{W}
	[\mathcal{O}(t)+\mathcal{D}_t \Sigma(t) ] \,,
\end{align}
where $\mathcal{D}_t$ is the covariant derivative defined in~\eqref{susyvarL}. 
We conclude that our supermatrix construction perfectly agrees with the abstract structure outlined in subsection \ref{Sec: ABJM Displacement}.
\newpage 
\section{Solving the Leading Mixing in ABJM}\label{ABJM solving mixing}
The following section presents the details of the solving of the mixing for the first-order anomalous dimension. This is done through the consideration of another correlator that exchanges four-particle operators in GFF. This leads to the conclusion that the degeneracy of the dimension is not lifted at this order and allows for an easier bootstrap in section \ref{Subsec: ABJM bootstrap}. 
Mixing cannot occur between operators of different $U(1)$-charge so we focus on the uncharged operators.\footnote{The mixing between operators with different $U(1)$ charges does not occur. This can be seen from the vanishing of the relevant three-point functions and the symmetry under the $U(1)$ reflection $\mathbb{F}\rightarrow \bar{\mathbb{F}}$.} 
\subsection*{Length-2 Operators}
The only length-2 operators are
\begin{align}
	&F\partial^{(n)}\bar{F} &&F\partial^{(n)}F &&\bar{F}\partial^{(n)}\bar{F} &
\end{align}
where $n$ is an integer (odd in the second and third case). For even $n$, there is only one operator at each dimension, so no mixing occurs. However, for odd $n$, there is generically mixing. We divide the weight matrix into these three sectors, which we denote $\{0,-1,1\}$ for the operator $\mathcal{O}_0$, $\mathcal{O}_{-1}$, and $\mathcal{O}_1$. 
We look at the set of vanishing three-point functions
\begin{align}
	<FF\mathcal{O}_1> &= <FF\mathcal{O}_0> = 0\\
	<F\bar{F}\mathcal{O}_1> &= <F\bar{F}\mathcal{O}_{-1}> =0
\end{align}
These appear in the OPE expansion of the vanishing four-point function, which at first order gives
\begin{align}
	<FFF\bar{F}>^{(1)}|_{\log(\chi)}=\frac{1}{x_{12}x_{34}}\sum_{\Delta} c_{FF\mathcal{O}}\gamma_{\mathcal{O}\mathcal{O}'}c_{\mathcal{O}'F\bar{F}} G_{\Delta}(\chi).
\end{align}
From the analysis of the three-point functions, we know that the only non-vanishing terms are
\begin{align}
	<FFF\bar{F}>^{(1)}|_{\log(\chi)}=\frac{1}{x_{12}x_{34}}\sum_{\Delta} c_{FF\mathcal{O}_{-1}}\gamma^{(1)}_{\mathcal{O}_{-1}\mathcal{O}'_0}c_{\mathcal{O}'_0 F\bar{F}} G_{\Delta}(\chi).
\end{align}
Since only one operator of these types appears at each weight, the vanishing of the correlator implies the vanishing of the anomalous dimension. Therefore, there is no mixing at this order between the operators $\mathcal{O}_0$ and $\mathcal{O}_1$. Likewise, with the $U(1)$ conjugate. We are left between the mixing of  $\mathcal{O}_{\pm1}$. \par
I would expect the $U(1)$ symmetry to identify the anomalous dimension of the operators $\mathcal{O}_1$ and $\mathcal{O}_{-1}$. Therefore we have a structure
\begin{align}
	\gamma = \begin{pmatrix} 0& \gamma_{1} \\ \gamma_{1}&0\end{pmatrix}
\end{align}
Therefore the square has $\gamma_{\mathcal{O}_1\mathcal{O}_{-1}}^2$ in the diagonal components, so we can safely ignore the mixing from length-2 operators at this order.
\subsection*{Length-4 Operators}
The anomalous dimension $\gamma^{(1)}_\Delta$ of exchanged operators computed at first order, such as in \cite{Bianchi:2020hsz} is a weighted average of anomalous dimensions from operators which have the same weight in free theory, such as 
\begin{align}
	\mathbb{F}\partial^3 \bar{\mathbb{F}}&&\mathbb{F}\partial \mathbb{F} \bar{\mathbb{F}}\partial \bar{\mathbb{F}}
\end{align}
which have conformal weight $\Delta=4$. This is especially relevant in the bootstrap  process, where these contributions must be disentangled to use the data from the previous order, such as in \cite{Ferrero:2021bsb}.
\begin{align}
	\sum_\Delta \left(\gamma^{(1)}_\Delta\right)^2 c^{(0)}_\Delta G_\Delta(\chi)
\end{align}
At first order, since the relevant vertices at strong coupling have four legs, the solving of the mixing only needs to include two-particle and four-particle operators. In particular, the protected operator $\mathcal{O}_2 = \mathbb{F}\partial \mathbb{F} $ exchanges both  two-particle and four-particle operators in free theory. 
The strong coupling free theory result is obtained by Wick contracting the elementary fields.\footnote{In this case, the elementary field is $\mathbb{F}$. This differs from the generalised free field result since $\mathcal{O}_2$ is a composite operator, not an elementary excitation. In the free theory, both length-two and length-four operators will be exchanged.}
\begin{align}
	\langle \mathcal{O}_2 \bar{\mathcal{O}}_2 \bar{\mathcal{O}}_2\mathcal{O}_2 \rangle  &= \lim_{\{t_2,t_4,t_6,t_8\}\rightarrow\{t_1,t_3,t_5,t_7\}}\partial_{t_2}\partial_{t_4}\partial_{t_6}\partial_{t_8}\langle \mathbb{F}\mathbb{F}\bar{\mathbb{F}}\bar{\mathbb{F}}\bar{\mathbb{F}}\bar{\mathbb{F}} \mathbb{F}\mathbb{F}\rangle\\
	&=\frac{1}{(t_{13}t_{24})^4} \frac{(\chi-1)^4}{ \chi^4}
\end{align}

This gives
\begin{align}
	f(\chi)&= \frac{1}{(1-\chi)^4}&	\hat{f}(\chi)&= \frac{1}{(1-\chi)^4\chi^4}\\
	h(\chi)&=(1-\chi)^4 &	\hat{h}(\chi) &= \frac{(1-\chi)^8}{\chi^4}
\end{align}

In particular, notice that even though the unitarity bound is $h>4$ in the charged channel, the lightest operator exchanged in the free theory has dimension $\Delta_L = 8$ and that this weight is the same as the operator $\mathbb{F}\partial\mathbb{F}\partial^2\mathbb{F}\partial^3\mathbb{F}$.\par
In the $s$-channel of $f$, the free spectrum is $h^{(0)}=1+n$ and the OPE coefficients are
\begin{align}
	\tilde{c}_h& =
	\frac{\sqrt{\pi } 2^{-2h -3} (h+1)^2 (h+2) (h  (h +2)+9) \Gamma (h+3)}{9 \Gamma \left(h +\frac{3}{2}\right)}
\end{align}\par 
In the $s$-channel of $h$, the free spectrum is the same ($h^{(0)}=1+n$) and the   OPE coefficients are
\begin{align}
	c_{h^{(0)}} = (-1)^{h^{(0)}}\tilde{c}_{h^{(0)}}
\end{align}
as expected from arguments of S-parity argued in \cite{Billo:2013jda,Bianchi:2020hsz}\par 
In the $t$-channel of $h$, then free spectrum is $h^{(0)}=2+2n$ with OPE coefficients:
\begin{align}
	\mathbb{c}_h& = \frac{\sqrt{\pi } 4^{-h } \left((-1)^{h }+1\right) (h -6) (h -4) (h -2) (h +1) (h +3) (h +5) \Gamma (h )}{9 \Gamma \left(h -\frac{1}{2}\right)}
\end{align}
Notice that the difference between the free solution and the GFF solution will be from the exchange of two-particle  operators since the exchanged fields in GFF for $\mathcal{O}_2\sim \mathbb{F}\partial\mathbb{F}$ will be of the form $\mathbb{F}\partial\mathbb{F}\partial^{(n)}\mathbb{F}\partial\mathbb{F} $. 
Since the GFF for chiral fields of dimension $\Delta_\phi$ is given by
\begin{align}
	\frac{\sqrt{\pi } 2^{-2 \Delta -1} \Gamma (\Delta ) \Gamma (\Delta +2 \Delta \phi +2)}{\Delta  \Gamma \left(\Delta +\frac{3}{2}\right) \Gamma (2 \Delta \phi )^2 \Gamma (\Delta -2 \Delta \phi +1)}
\end{align}
One can dissociate the two-particle  and the four-particle contributions.

\begin{align}
	\langle \mathcal{O}_2 \bar{\mathcal{O}}_2 \bar{\mathcal{O}}_2\mathcal{O}_2 \rangle^{(0)}  &= \lim_{\{t_2,t_4,t_6,t_8\}\rightarrow\{t_1,t_3,t_5,t_7\}}\partial_{t_2}\partial_{t_4}\partial_{t_6}\partial_{t_8}\langle \mathbb{F}\mathbb{F}\bar{\mathbb{F}}\bar{\mathbb{F}}\bar{\mathbb{F}}\bar{\mathbb{F}} \mathbb{F}\mathbb{F}\rangle^{(0)}\\
	&=\frac{1}{(t_{13}t_{24})^4} \frac{(\chi-1)^4}{ \chi^4}
\end{align}
\subsection*{Bootstrap of $\mathbb{F}\partial\mathbb{F}$ }
Using the same constraints as for $\mathbb{F}$, the specific ones for the chiral ring detailed in \ref{Section: chiral ring}, and the CFT data for the exchanged operator $\mathbb{F}\bar{\mathbb{F}}$ the first-order perturbation can be fixed entirely. This gives
\begin{align}	
	f^{(1)}_{\mathcal{O}_2}(\chi)=4\epsilon\frac{(\chi-4) (\chi-1)^2 \log (1-\chi)-\chi (-4 \chi+(\chi+3) \chi \log (\chi)+4)}{ (\chi-1)^5 \chi}
\end{align}
where 
\begin{align}
	\epsilon = -\frac{1}{2\pi \sqrt{2\lambda}}
\end{align}
is the strong-coupling expansion parameter and $\lambda$ the 't Hooft coupling for ABJM. 
This gives the following CFT data:
\begin{align}
	\gamma_s^{(1)} &= \epsilon \Delta(\Delta+2)\\
	\gamma_t^{(1)} &= \epsilon (\Delta(\Delta-1)-20)
\end{align}
In particular, this would mean that both the two-particle and the four-particle operators with vanishing $U(1)$ charge have anomalous dimensions proportional to the quadratic Casimir and we have 
\begin{align}
	\langle(\gamma^{(1)})^n\rangle =\langle\gamma^{(1)}\rangle ^n
\end{align}
\subsection*{Chiral Ring on the Line }\label{Section: chiral ring}
The OPE expansion of two protected chiral operators reveals a tower of such operators organised by their charge under the $U(1)$ symmetry and the shortening condition of the associated superprimary. This is similar to the case of $\mathcal{N}=4$ SYM, where one can form higher dimension protected operators by considering the totally antisymmetric states $\phi^{[a_1}..\phi^{a_n]}$ built from combinations of the displacement multiplet, these form a chiral ring.

\subsection*{The $\phi \times \phi$ Chiral OPE}
It is useful to review the chiral OPE presented in \cite{Bianchi:2020hsz} which uses similar arguments to \cite{Dolan:2000uw,Howe:1996rb,Poland:2010wg} before generalising this to chiral operators with larger charge.
The Displacement multiplet is a protected chiral operator satisfying 
\begin{align}
	[\bar{Q},\phi] = [\bar{S},\phi] = 0.
\end{align}
Combining this with the commutation relation $[\bar{Q},P]=\bar{S}$, this implies that $\bar{Q}$ and $\bar{S}$ also annihilate all the operators in the the OPE 
\begin{align}
	\phi_{j_0} \times \phi_{j_0} &\sim \sum \mathcal{O}_{2j_0} &\qquad [\bar{Q},\mathcal{O}_{2j_0}]&=[\bar{S},\mathcal{O}_{2j_0}]=0
\end{align}
The chiral condition implies that the exchanged operator is either a chiral superprimary or a chiral superdescendent. We separate these cases as 
\begin{align}
	\phi_{j_0} \times \phi_{j_0} &\sim \mathcal{O}^1_{2j_0,0,0} + \bar{Q}(\mathcal{O}^2_{2j_0+\frac{1}{2},1,0})+\bar{Q}^2(\mathcal{O}^3_{2j_0+1,0,1})+\bar{Q}^3(\mathcal{L}_{2j_0+\frac{3}{2},0,0})
\end{align}
where the first three operators are short multiplets and satisfy the shortening conditions 
\begin{align}
	\bar{Q}\mathcal{O}^1_{2j_0,0,0} &= 0 &\bar{Q}^2(\mathcal{O}^2_{2j_0+\frac{1}{2},1,0})&=0&\bar{Q}^3(\mathcal{O}^3_{2j_0+1,0,1})&=0
\end{align}
The highest weight super descendent $\bar{Q}^3(\mathcal{L}_{2j_0+\frac{3}{2},0,0})$ does not have any shortening conditions and is, therefore, generically a long operator. Additionally, the long operators satisfy the unitarity bound
\begin{align}
	\Delta_{\mathcal{L}} > \frac{2j_0}{3}+\frac{1}{2}
\end{align}
where the equality was ruled out thanks to the recombination of the long operator at the equality. This corresponds to the following bound for the exchanged operator
\begin{align}
	\Delta_{\bar{Q}^3\mathcal{L}}>  \frac{2j_0}{3}+2
\end{align}
\subsection*{Chiral $O_{j_r}\times O_{j_s}$ OPE and Chiral Ring}
The three operators exchanged in the $\phi\times \phi$ OPE are protected chiral operators satisfying 
\begin{align}
	[\bar{Q},\mathcal{O}]&=[\bar{S},\mathcal{O}]=0
\end{align}
despite not all being superprimary operators. This is sufficient to generalise use the argument above in the OPE of two such chiral fields:
\begin{align}
	\mathcal{O}_{j_r}\times \mathcal{O}_{j_s}\sim   \mathcal{O}^1_{j_r+j_s,0,0} + \bar{Q}(\mathcal{O}^2_{j_r+j_s+\frac{1}{2},1,0})+\bar{Q}^2(\mathcal{O}^3_{j_r+j_s+1,0,1})+\bar{Q}^3(\mathcal{L}_{j_r+j_s+\frac{3}{2},0,0})
\end{align}
with a following unitarity bound for the first long exchanged operator
\begin{align}
	\Delta_{\bar{Q}^3\mathcal{L}}>\frac{j_r+j_s}{3}+2,
\end{align}
where the equality was again excluded thanks to the recombination rule at the bound. Additionally, since only one super descendent will contribute at each weight, the block expansion becomes the usual 1dconformal block expansion. \par 
By iterating this analysis, we have a family of operators with integer multiples of $j_0$ $U(1)$ charge. 
\begin{align}
	&\{\bar{\mathcal{B}}_{nj_0}\} = \{\bar{\mathcal{B}}_{n_0},\bar{\mathcal{B}}_{2j_0},..\}\\
	&\{\bar{\mathcal{B}}_{nj_0+\frac{1}{2},1}\} = \{\bar{\mathcal{B}}_{j_0+\frac{1}{2},1},\bar{\mathcal{B}}_{2j_0+\frac{1}{2},1},..\}\\
	&\{\bar{\mathcal{B}}_{nj_0+1,0,1}\} = \{\bar{\mathcal{B}}_{j_0+1,0,1},\bar{\mathcal{B}}_{2j_0+1,0,1},..\}
\end{align}
These operators with increasing $U(1)$ charge have protected dimensions and form the chiral ring generated by the chiral field $\phi$. Since they were obtained by iterating the chiral OPE, they will be considered as operators of length $n$. For example 
\begin{align}
	\bar{Q}\bar{\mathcal{B}}_{2j_0+\frac{1}{2},1} \sim \Phi \partial \Phi 
\end{align}
is an operator of length two.
And the correlation functions will be of the same type, with
\begin{align}
	\frac{\langle \mathcal{O}_\Delta \bar{\mathcal{O}}_\Delta\mathcal{O}_\Delta \bar{\mathcal{O}}_\Delta\rangle}{\langle \mathcal{O}_\Delta \bar{\mathcal{O}}_\Delta\rangle^2}& = f(\chi) = \chi^{2\Delta}\hat{f}(\chi)\\
	\frac{\langle \mathcal{O}_\Delta \bar{\mathcal{O}}_\Delta \bar{\mathcal{O}}_\Delta \mathcal{O}_\Delta \rangle}{\langle \mathcal{O}_\Delta \bar{\mathcal{O}}_\Delta \rangle \langle\bar{\mathcal{O}}_\Delta \mathcal{O}_\Delta \rangle}& = h(\chi) =\left( \frac{ \chi}{1-\chi}\right)^{2\Delta}\hat{h}(\chi)
\end{align}
And the OPE expansions are
\begin{align}
	f(\chi) &=_{\chi\rightarrow 0} 1+\sum_h \tilde{c}_h  \chi^h {}_2F_1(h,h+3,2h+3;\chi)\\
	\hat{f}(\chi) &= \hat{f}(1-\chi)\\
	h(\chi) &=_{\chi\rightarrow 0} 1+ \sum_h c_h  \chi^h {}_2F_1(h,h,2h+3;\chi)\\
	\hat{h}(\chi)&=_{\chi\rightarrow 1} \sum_{h_L>2+\frac{2 j_R}{3}} \mathbb{c}_h (1-\chi)^h {}_2F_1(h,h,2h;1-\chi)
\end{align}
	
\chapter{$\mc{N}=4$ SYM Multipoint Superconformal Blocks}\label{app: N=4 multipoint}
This section contains information that was not essential to the understanding of the structure and computation of correlators in Chapter \ref{chapter N=4}. Section \ref{App N=4 multiplets} details some information about the 1/2-BPS multiplets. Sections \ref{App: superblocks symmetric} and \ref{App: superblocks asymmetric} list the superconformal blocks corresponding to the OPE expansion in the symmetric and asymmetric channels respectively. These are then used to extract the CFT data in section \ref{N=4 free theory data} and to constrain the correlator in section \ref{N=4 conformal bootstrap}.

\section{Multiplets}\label{App N=4 multiplets}
The 1/2-BPS multiplets have a straightforward structure since they can only be acted upon with one family of supercharges
\begin{align}
	\mc{D}_k :\qquad  [0,k]_{s=0}^{h=k} \rightarrow [1,k-1]_{s=1}^{k+\frac 12}\rightarrow \begin{matrix}
		[0,k-1]_{s=2}^{h=k+1}\\
		\\
		[2,k-2]_{s=0}^{h=k+2}
	\end{matrix} \rightarrow [1,k-2]_{s=1}^{h=k+\frac 32}\rightarrow [0,k-2]_{s=0}^{h=k+2}
\end{align}

For the blocks of protected operators, the normalisation corresponding to having a unit topological sector gives free OPE coefficients consistent with the literature \cite{Giombi:2018qox} where 
\begin{align}
	c_{i,j,k} = \frac{\sqrt{i! j!k!}}{(ij,k)!(ik,j)!(jk,i)!}
\end{align}
where we used the shorthand $(ij,k) = \frac{i+j-k}{2}$.\par 
The generic long operators have a full family of descendants from all the supercharges acting on the superconformal primary which can be found in the long version of  \cite{Ferrero:2021bsb}.

%	\section{Multiplet structure of operators}\label{App: Multiplets}
\newpage 
\section{Symmetric Superconformal Blocks }\label{App: superblocks symmetric}
This section consists of a list of the superconformal blocks arising from the OPE expansions between points 1 and 2 and points 3 and 4. Since these two channels \textit{are} related by the crossing $\chi_1\rightarrow 1-\chi_2$, we refer to this as the \textit{symmetric} blocks. The Figure \ref{Figure: OPE diagram symmetric} illustrates this OPE where the operators $\mc{I}, \mc{D}_2$ and $\mc{L}^{h}_{0,[0,0]}$ are exchanged in both channels.These correspond to the identity $\mc{I}$, the 1/2-BPS operator of weight $2$, $\mc{D}_2$ and the uncharged long operator $\mc{L}^h_{0,[0,0]}$. 
It is useful to use variables that vanish in the OPE limit. Therefore, we define
\begin{align}
	\nu_2=1-\chi_2.
\end{align}
The superconformal blocks $\mc{G}_{A,B}$ will correspond to the one with operator $A$ being exchanged in the $\chi_1\rightarrow0$ limit and $B$ being exchanged in the $\nu_2 \rightarrow 0$ limit. These can be written as a sum of five-point scalar blocks $g_{a,b}$, weighted by the appropriate R-symmetry block $h_{[a_1,a_2],[b_1,b_2]}$. The R-symmetry block is found by applying the $SU(4)$ quadratic Casimir to points 12 and 34 and uniquely determines the form of $h_{[a_1,a_2],[b_1,b_2]}$ from the representation of $A$ and $B$. The full $(r_i,s_i,t,\chi_i)-$ dependant superblock can then be decomposed in the same way as the correlator, in the Ward identity channels described in \ref{definition N=4 f-functions}. The corresponding blocks decomposition will be denoted $f^i_{A,B}$ to avoid confusion with the asymmetric super blocks. 
\subsection*{$\bullet  \quad \scaleto{\mathbf{\mathcal{G}_{\mathcal{I},\mathcal{D}_2}}}{14pt}$}
	The selection rules allow for only one term
	\begin{align}
		&\mathcal{G}_{\mathcal{I},\mathcal{D}_2} = h_{[0,0][0,2]}g_{0,2}=\frac{r_1}{\chi_1^2}\\
		&\mathcal{G}_{\mathcal{D}_2,\mathcal{I}} = h_{[0,2][0,0]}g_{2,0}=\frac{s_2}{\nu_2^2}
	\end{align}
	where the crossing symmetry above is clear. In terms of the functions $f_i$ described above, this gives
	\begin{align}
		f^0_{\mc{I},\mc{D}_2}& = 1&f^1_{\mc{I},\mc{D}_2}& =- \frac{1}{\chi_1}&f^2_{\mc{I},\mc{D}_2}& = 0&f^3_{\mc{I},\mc{D}_2}& = 0
	\end{align}
	
	\begin{align}
		f^0_{\mc{D}_2,\mc{I}}& = 1&f^1_{\mc{D}_2,\mc{I}}& =0 &f^2_{\mc{D}_2,\mc{I}}& = \frac{1}{\nu_2}&f^3_{\mc{D}_2,\mc{I}}& = 0
	\end{align}
	\subsection*{$\bullet  \quad \scaleto{\mathbf{\mathcal{G}_{\mathcal{D}_2,\mathcal{D}_2}}}{14pt}$}
	The selection rules restrict the superconformal block to the following expansion (where we have already eliminated the vanishing contributions)
	\begin{align}
		\mathcal{G}_{\mathcal{D}_2,\mathcal{D}_2} = \beta_1 h_{[0,0][0,2]}g_{4,2}+\beta_2 h_{[0,2][0,0]}g_{4,2}+\beta_3 h_{[0,2][2,0]}g_{2,3}\nonumber \\
		+\beta_4 h_{[2,0][0,2]}g_{3,2}+\beta_5 h_{[0,2][0,2]}g_{2,2}+\beta_6 h_{[2,0][2,0]}g_{3,3}
	\end{align}
	The Ward identities constrain the coefficients up to an overall constant which is normalised to  $\beta_5=\frac{-1}{5}$ for the lowest conformal weight $g_{2,2}$ to match the normalisation conventions
	\begin{align}
		\beta_1 &= \frac{3}{70}\beta_5&\beta_2&= -\frac{3}{70}\beta_5&\beta_3&= -\frac{1}{4}\beta_5&\beta_4&= \frac{1}{4}\beta_5&\beta_6&=\frac{3}{40}\beta_5
	\end{align}
	In terms of the Ward identity channels, this gives
	\begin{align}
		f^0_{\mc{D}_2,\mc{D}_2}& = 1\\
		f^1_{\mc{D}_2,\mc{D}_2}& =\sum_{k,l}\frac{2 (k+1) (l+1) (2 (k+1) l+3 k+2)}{\nu_2^2 \chi_1^2 (k+l+1)}G_{\mc{D}_2,\mc{D}_2} ^{k,l}\\
		f^2_{\mc{D}_2,\mc{D}_2}& =-\sum_{k,l}\frac{ (k+1) (l+1) (2 k (l+1)+3 l+2)}{\nu_2^2 \chi_1^2 (k+l+1)}G_{\mc{D}_2,\mc{D}_2} ^{k,l}\\
		f^3_{\mc{D}_2,\mc{D}_2}&= -\sum_{k,l}\frac{(k+1) (k+2) (l+1) (l+2)}{2\nu_2^2\chi_1^2}	G_{\mc{D}_2,\mc{D}_2} ^{k,l}
	\end{align}
	One can define the block
	\begin{align}
		G_{\mc{D}_2,\mc{D}_2} ^{k,l}= \frac{ 3 (2)_{k+l}}{2(5)_k (5)_l  }\chi_1^{k+2} \nu_2^{l+2}\\
		\sum_{k,l}	G_{\mc{D}_2,\mc{D}_2} ^{k,l} = 3 \chi_1^{2} \nu_2^{2}F_2\left(
		2,1,1;5,5\,|\,\chi_1,\nu_2\right)
	\end{align}
	and the function $	f^3_{\mc{D}_2,\mc{D}_2}$ is then the derivative of this Appell function. 
	\subsection*{$\bullet  \quad \scaleto{\mathbf{\mathcal{G}_{\mathcal{D}_2,\mathcal{L}_{h_2}}}}{18pt}$}
	The selection rules restrict the superconformal block to the following expansion (where we have already eliminated the vanishing contributions)
	\begin{align}
		\mathcal{G}_{\mathcal{D}_2,\mathcal{L}_{h_2}} = &\beta_1 h_{[0,2][0,0]}g_{2,h_2}+\beta_2 h_{[0,2][0,0]}g_{2,h_2+2}+\beta_3 h_{[0,2][0,0]}g_{2,h_2+4}\nonumber \\
		&+\beta_4 h_{[0,2][2,0]}g_{2,h_2+1}+\beta_5 h_{[0,2][2,0]}g_{2,h_2+3}+\beta_6 h_{[0,2][0,2]}g_{2,h_2+2}\nonumber\\
		&+\beta_7 h_{[2,0][2,0]}g_{3,h_2+1}+\beta_8 h_{[2,0][2,0]}g_{3,h_2+3}+\beta_9 h_{[2,0][0,2]}g_{3,h_2+2}\nonumber \\
		&+\beta_{10} h_{[0,0][0,2]}g_{4,h_2+2}
	\end{align}
	The Ward identities constrain the constants (normalised to $g_{2,h_2}$)
	\begin{align}
		\beta_1 &= 1&\beta_2 &= \frac{3 h_2  (h_2 +1) (h_2 +3)}{10 (h_2 -1) (2 h_2 +1) (2 h_2 +5)}\nonumber \\
		\beta_3 &= \frac{(h_2 +1) (h_2 +2) (h_2 +3)^2 (h_2 +4)}{16 (h_2 -1) (2 h_2 +3) (2 h_2 +5)^2 (2 h_2 +7)}&
		\beta_4 &= \frac{h_2 }{2 (h_2 -1)}\nonumber \\
		\beta_5 &= \frac{(h_2 +1) (h_2 +2) (h_2 +3)}{8 (h_2 -1) (2 h_2 +3) (2 h_2 +5)}&\beta_6 &= \frac{h_2 +1}{5(1-h_2) }\nonumber \\
		\beta_7&= -\frac{h_2  (h_2 +1)}{20 (h_2 -1)}&\beta_8 &= -\frac{(h_2 +1) (h_2 +2) (h_2 +3)^2}{80 (h_2 -1) (2 h_2 +3) (2 h_2 +5)}\nonumber \\
		\beta_9 &= -\frac{(h_2 +1) (h_2 +2)}{40 (h_2 -1)}\nonumber \\
		\beta_{10} &= \frac{(h_2 +1) (h_2 +2) (h_2 +3)}{700 (h_2 -1)}
	\end{align}
	\begin{align}
		f^0_{\mc{D}_2,\mc{L}_{\Delta_2}}& = 0\\
		f^1_{\mc{D}_2,\mc{L}_{\Delta_2}}&= -\sum _{k,l} \frac{(k+1) (\Delta_2+l) \left(-2 \Delta_2 (\Delta_2+1)+k \left(-\Delta_2^2+\Delta_2+4 l+2\right)+4 l\right)}{\nu_2 \chi_1^2 (\Delta_2+l+1) (\Delta_2+k+l)}G_{\mc{D}_2,\mc{L}_{\Delta_2}} \\
		f^2_{\mc{D}_2,\mc{L}_{\Delta_2}} &=\sum _{k,l}  \frac{(k+1)}{\nu_2 \chi_1^2 (\Delta_2+l+1) (\Delta_2+k+l)} P_{\mc{D}_2,\mc{L}_{\Delta_2}}G_{\mc{D}_2,\mc{L}_{\Delta_2}} \\
		f^3_{\mc{D}_2,\mc{L}_{\Delta_2}}&=\sum _{k,l} \frac{(k+1) (k+2) (\Delta_2+l)}{\nu_2 \chi_1^2} G_{\mc{D}_2,\mc{L}_{\Delta_2}}
	\end{align}
	Where 
	\begin{align}
		P_{\mc{D}_2,\mc{L}_{\Delta_2}}^{k,l}&=-\Delta_2 (\Delta_2+1) (\Delta_2 (k+2)-3 k-4)+(4 k+6) l^2+l (\Delta_2 (-\Delta_2 (k+2)+5 k+6)-2)\\
		G_{\mc{D}_2,\mc{L}_{\Delta_2}}^{k,l} &=-\frac{\chi_1^2 \nu_2^{\Delta_2}}{4 (\Delta_2-1)} \frac{(1)_k \chi_1^k \nu_2^l (\Delta_2+2)_l (\Delta_2+1)_{k+l}}{(5)_k \Gamma (k+1) \Gamma (l+1) (2\Delta_2+4)_l}
	\end{align}
	and the sum is 
	\begin{align}
		\sum_{k,l} 	G_{\mc{D}_2,\mc{L}_{\Delta_2}}^{k,l} = -\frac{\chi_1^2 \nu_2^{\Delta_2}}{4 (\Delta_2-1)} F_2\left(
		1+\Delta_2,1,\Delta_2+2;5,2(\Delta_1+2)\,|\,\chi_1,\nu_2\right)
	\end{align}
	and $	f^3_{\mc{D}_2,\mc{L}_{\Delta_2}}$ is related to the direct derivative of this Appell function.
	\subsection*{$\bullet  \quad \scaleto{\mathbf{\mathcal{G}_{\mathcal{L}_{h_1},\mathcal{D}_2}}}{18pt }$}
	This is given by the crossing symmetric of $\mathcal{G}_{\mathcal{D}_2,\mathcal{L}_{h_2}}$
	
	\subsection*{$\bullet  \quad \scaleto{\mathbf{\mathcal{G}_{\mathcal{L}_{h_1},\mathcal{L}_{h_2}}}}{18pt}$}
	The selection rules restrict the terms to the following:
	\begin{align}
		\mathcal{G}_{\mathcal{L}_{h_1},\mathcal{L}_{h_2}} &= \beta _1 h_{[0,0][0,2]} g_{\Delta_1,\Delta_2+2}+\beta _2 h_{[0,0][0,2]} g_{\Delta_1+2,\Delta_2+2}+\beta _3 h_{[0,0][0,2]} g_{\Delta_1+4,\Delta_2+2}\nonumber \\
		&+\beta _4 h_{[2,0][2,0]} g_{\Delta_1+1,\Delta_2+1}+\beta _5 h_{[2,0][2,0]} g_{\Delta_1+1,\Delta_2+3}+\beta _6 h_{[2,0][2,0]} g_{\Delta_1+3,\Delta_2+1}\nonumber \\
		&+\beta _7 h_{[2,0][2,0]} g_{\Delta_1+3,\Delta_2+3}+\beta _8 h_{[2,0][0,2]} g_{\Delta_1+1,\Delta_2+2}+\beta _9 h_{[2,0][0,2]} g_{\Delta_1+3,\Delta_2+2}\nonumber \\
		&+\beta _{10} h_{[0,2][0,0]} g_{\Delta_1+2,\Delta_2}+\beta _{11} h_{[0,2][0,0]} g_{\Delta_1+2,\Delta_2+2}+\beta _{12} h_{[0,2][0,0]} g_{\Delta_1+2,\Delta_2+4}\nonumber \\
		&+\beta _{13} h_{[0,2][2,0]} g_{\Delta_1+2,\Delta_2+1}+\beta _{14} h_{[0,2][2,0]} g_{\Delta_1+2,\Delta_2+3}+\beta _{15} h_{[0,2][0,2]} g_{\Delta_1+2,\Delta_2+2}\nonumber \\
		&+\beta _{16} h_{[0,1][0,1]} g_{\Delta_1+2,\Delta_2+2}
	\end{align}
	The Ward identities fix all the coefficients (except for $\beta_{16}$ which multiplies $h_{[0,1][0,1]}=0$) up to a normalisation factor. We set $\beta_4=1$ since it is the lowest weight term which is crossing symmetric
	\[
	\{\Delta_1,\chi_1,h_{[a][b]}\}\leftrightarrow\{\Delta_2,1-\chi_2,h_{[b][a]}\}
	\]
	This  gives the coefficients consistent with crossing symmetry
	\begin{align}
		\beta_4&=1&\beta_1&=\tilde{\beta}_{10}&\beta_2&=\tilde{\beta}_{11}&\beta_3&=\tilde{\beta}_{12}&\beta_5&=\tilde{\beta}_6&\beta_8&=\tilde{\beta}_{13}&\beta_9&=-\tilde{\beta}_{14}&\beta_{16}&=0
	\end{align}
	Where the tilde indicates the replacement $\Delta_1\leftrightarrow\Delta_2$ and 
	\begin{align}
		\beta _1&=-\frac{\Delta _1-1}{\Delta _2} \\
		\beta _5&=\frac{\left(\Delta _2+3\right) \left(\Delta _1+\Delta _2\right) \left(\Delta _1+\Delta _2+1\right)}{4 \Delta _2 \left(2 \Delta _2+3\right) \left(2 \Delta _2+5\right)}&\\
		\beta _2&=-\frac{3 \left(\Delta _1+3\right) \left(\Delta _1+\Delta _2\right) \left(\Delta _1+\Delta _2+1\right)}{10 \left(2 \Delta _1+1\right) \left(2 \Delta _1+5\right) \Delta _2}\\
		\beta _3&=-\frac{\left(\Delta _1+3\right) \left(\Delta _1+4\right) \left(\Delta _1+\Delta _2\right) \left(\Delta _1+\Delta _2+1\right) \left(\Delta _1+\Delta _2+2\right) \left(\Delta _1+\Delta _2+3\right)}{16 \Delta _1 \left(2 \Delta _1+3\right) \left(2 \Delta _1+5\right){}^2 \left(2 \Delta _1+7\right) \Delta _2}\\
		\beta _8&=\frac{1}{2}\left(\frac{\Delta _1}{\Delta _2}+1\right)\\
		\beta _9&=\frac{\left(\Delta _1+3\right) \left(\Delta _1+\Delta _2\right) \left(\Delta _1+\Delta _2+1\right) \left(\Delta _1+\Delta _2+2\right)}{8 \Delta _1 \left(2 \Delta _1+3\right) \left(2 \Delta _1+5\right) \Delta _2}\\
	\end{align}
	The corresponding blocks in the Ward identity channels are then:
	\begin{align}
		f^0_{\mc{L}_{\Delta_1},\mc{L}_{\Delta_2}}&=0\non \\
		f^1_{\mc{L}_{\Delta_1},\mc{L}_{\Delta_2}}&=\sum _{k,l}\frac{\Delta_2+l}{\nu_2 \chi_1 (\Delta_1+k+1) (\Delta_2+l+1) (\Delta_1+\Delta_2+k+l-1)} (P^1)^{k,l}_{\mc{L}_{\Delta_1},\mc{L}_{\Delta_2}}G^{k, l}_{\mc{L}_{\Delta_1},\mc{L}_{\Delta_2}}\non \\
		f^2_{\mc{L}_{\Delta_1},\mc{L}_{\Delta_2}}&=\sum _{k,l}\frac{\Delta_1+k}{\nu_2 \chi_1 (\Delta_1+k+1) (\Delta_2+l+1) (\Delta_1+\Delta_2+k+l-1)} (P^2)^{k,l}_{\mc{L}_{\Delta_1},\mc{L}_{\Delta_2}}G^{k, l}_{\mc{L}_{\Delta_1},\mc{L}_{\Delta_2}} \non \\
		f^3_{\mc{L}_{\Delta_1},\mc{L}_{\Delta_2}}&=\sum _{k,l} \frac{(\Delta_1+k) (\Delta_2+l)}{\nu_2 \chi_1} G^{k, l}_{\mc{L}_{\Delta_1},\mc{L}_{\Delta_2}}
	\end{align}
	Where 
	\begin{align}
		G^{k, l}_{\mc{L}_{\Delta_1},\mc{L}_{\Delta_2}}&= \frac{ (\Delta_1+2)_k (\Delta_2+2)_l (\Delta_1+\Delta_2)_{k+l}}{\Delta_1 \Delta_2 \Gamma (k+1) \Gamma (l+1) (2 \Delta_1+4)_k (2 \Delta_2+4)_l} \chi_1^{k+\Delta_1} \nu_2^{l+\Delta_2} \\
		(P^1)^{k,l}_{\mc{L}_{\Delta_1},\mc{L}_{\Delta_2}}&=-k^2 \left(-\Delta_2^2+\Delta_2+4 l+2\right)\non \\
		&\quad+k \left((\Delta_2+1) (\Delta_2+2)+\Delta_1^2 (\Delta_2+l+1)+\Delta_1 \left(2\Delta_2^2+\Delta_2-5 l-1\right)\right)\non \\
		&\quad +\Delta_1 (\Delta_1+1) (\Delta_2 (\Delta_1+\Delta_2)+\Delta_1+(\Delta_1-3) l-1)\\
		(P^2)^{k,l}_{\mc{L}_{\Delta_1},\mc{L}_{\Delta_2}}&=-\Delta_2 (\Delta_2+1) ((\Delta_1+1) (\Delta_1+\Delta_2-1)+(\Delta_2-3) k)\non \\
		&\quad +l^2 \left(-\Delta_1^2+\Delta_1+4 k+2\right)\non \\
		&\quad -l \left(\Delta_1^2 (2\Delta_2+1)+\Delta_1 \left(\Delta_2^2+\Delta_2+3\right)+\Delta_2 (\Delta_2+(\Delta_2-5) k-1)+2\right)
	\end{align}
	Notice that 
	\begin{align}
		\sum_{k,l}  G^{k, l}_{\mc{L}_{\Delta_1},\mc{L}_{\Delta_2}} =  \frac{\chi_1^{\Delta_1} \nu_2^{\Delta_2}}{\Delta_1\Delta_2 }F_2\left(
		\Delta_1+\Delta_2,\Delta_1+2,\Delta_2+2;2(\Delta_1+2),2(\Delta_2+2)\,|\,\chi_1,\nu_2\right)
	\end{align}
	So we have a situation very similar to the four-point case where the superblocks and scalar blocks are not that dissimilar. Note that $f^3_{\mc{L}_{\Delta_1},\mc{L}_{\Delta_2}}$ is just a derivative of the Appell function.
\section{Blocks in the Asymmetric Channel}\label{App: superblocks asymmetric}
This section consists of a list of the superconformal blocks arising from the OPE expansions between points 1 and 2 and points 4 and 5. Since these two channels are \textit{not} related by the crossing $\chi_1\rightarrow 1-\chi_2$, we refer to this as the \textit{asymmetric} blocks. The Figure \ref{Figure: OPE diagram asymmetric} illustrates this OPE where the operators $\mc{I}, \mc{D}_1, \mc{D}_2,\mc{D}_3,\mc{L}^{h}_{0,[0,0]} $ and $\mc{L}^{h}_{0,[0,1]}$ are exchanged. These correspond to the identity $\mc{I}$, the 1/2-BPS operator of weight $k$, $\mc{D}_k$, the uncharged long operator $\mc{L}^h_{0,[0,0]}$, and the charged long Operator $\mc{L}^h_{0,[0,1]}$. 
It is useful to use variables that vanish in the OPE limit. Therefore, we define
\begin{align}
	\nu_1&=\frac{\chi_1-\chi_2}{1-\chi_2} &\nu_2&=1-\chi_2
\end{align}
The superconformal blocks $\mc{G}_{A,B}$ will correspond to the one with operator $A$ being exchanged in the $\nu_1\rightarrow0$ limit and $B$ being exchanged in the $\nu_2 \rightarrow 0$ limit. These can be written as a sum of five-point scalar blocks $g_{a,b}$, weighted by the appropriate R-symmetry block $h_{[a_1,a_2],[b_1,b_2]}$. The R-symmetry block is found by applying the $SU(4)$ quadratic Casimir to points 12 and 45 and uniquely determines the form of $h_{[a_1,a_2],[b_1,b_2]}$ from the representation of $A$ and $B$. The full $(r_i,s_i,t,\chi_i)-$ dependant superblock can then be decomposed in the same way as the correlator, in the Ward identity channels described in \ref{definition N=4 f-functions}. The corresponding blocks decomposition will be denoted $g^i_{A,B}$ to avoid confusion with the symmetric super blocks. 
	\subsection*{$\bullet  \quad \scaleto{\mathbf{\mathcal{G}_{\mathcal{I},\mathcal{D}_1}}}{14pt}$}
	\begin{align}
		\mc{G}_{\mc{D}_1,\mc{I}}=\frac{r_1 s_2}{\chi_1^2 (1-\chi_2)^2}
	\end{align}
	\begin{align}
		g^0_{\mc{D}_1,\mc{I}} &=1&g^1_{\mc{D}_1,\mc{I}} &=0&g^2_{\mc{D}_1,\mc{I}} &=0&g^3_{\mc{D}_1,\mc{I}} &=-\frac{1}{(\nu_1 \nu_2)^2}
	\end{align}
	\subsection*{$\bullet  \quad \scaleto{\mathbf{\mathcal{G}_{\mathcal{D}_2,\mathcal{D}_1}}}{14pt}$}
	\begin{align}
		g^0_{\mc{D}_2,\mc{D}_1} &=1\\
		g^1_{\mc{D}_2,\mc{D}_1} &=\frac{1}{\nu_2}\sum _{k_2} -\frac{6 (k_2-2) \nu_1^{k_2}}{(k_2+2) (k_2+3) (k_2+4)}\\
		g^2_{\mc{D}_2,\mc{D}_1} &=\frac{1}{\nu_2}\sum _{k_2} \frac{12 (1+k_2) \nu_1^{k_2}}{(k_2+2) (k_2+3) (k_2+4)}\\
		g^3_{\mc{D}_2,\mc{D}_1} &=\frac{1}{\nu_2^2}\sum _{k_2} \frac{12 (k_2-2) \nu_1^{k_2}}{(k_2+3) (k_2+4) (k_2+5)}
	\end{align}
	Note that the normalisation was chosen so that the OPE coefficient of the protected operator $c_{112}$ corresponds to the localisation result. 
	\subsection*{$\bullet  \quad \scaleto{\mathbf{\mc{G}_{\mc{D}_2,\mc{D}_3}}}{14pt}$}
	\begin{align}
		g^0_{\mc{D}_2,\mc{D}_3} &=-\frac{21}{2}\\
		g^1_{\mc{D}_2,\mc{D}_3} &=-\sum _{k_1, k_2} \frac{(k_1-2) (k_1+1)}{(k_2+5)} G_{\mc{D}_2,\mc{D}_3}^{k_1,k_2}\\
		g^2_{\mc{D}_2,\mc{D}_3} &=\sum_{k_1,k_2} \frac{2 (k_1+1)^2}{(k_2+5) (k_2+6)}G_{\mc{D}_2,\mc{D}_3}^{k_1,k_2}\\
		g^3_{\mc{D}_2,\mc{D}_3} &=-\sum_{k_1,k_2} \frac{2 (k_1+1) (k_1+2) (k_2+1)}{(k_1+5) (k_2+6) (k_2+7)}G_{\mc{D}_2,\mc{D}_3}^{k_1,k_2}
	\end{align}
	where
	\begin{align}
		G_{\mc{D}_2,\mc{D}_3}^{k_1,k_2}&=\frac{\Gamma (8) \Gamma (k_2+1)}{4 \Gamma (k_1+5) \Gamma (-k_1+k_2+1)}(-\nu_1)^{k_1} \nu_2^{k_2}
	\end{align}
	\subsection*{$\bullet  \quad \scaleto{\mathbf{\mc{G}_{\mc{L}^h_{0,[0,0]},\mc{D}_1}}}{18pt}$}
	\begin{align}
		g^0_{\mc{L}^h_{0,[0,0]},\mc{D}_1} &=0\\
		g^1_{\mc{L}^h_{0,[0,0]},\mc{D}_1} &=\sum_{k_1} h^3+h^2 (k_1+1)+h k_1-(k_1-3) k_1G_{\mc{L}^h_{0,[0,0]},\mc{D}_1}^{k_1,k_2}\\
		g^2_{\mc{L}^h_{0,[0,0]},\mc{D}_1} &=\sum _{k_1}-((h + h^2 - 2 k1) (h + k1))G_{\mc{L}^h_{0,[0,0]},\mc{D}_1}^{k_1,k_2} \\
		g^3_{\mc{L}^h_{0,[0,0]},\mc{D}_1} &= \sum _{k_1}-\frac{\left(h^2+h-2 k_1\right) (h+k_1-1)}{\nu_1 \nu_2}G_{\mc{L}^h_{0,[0,0]},\mc{D}_1}^{k_1,k_2}
	\end{align}
	where
	\begin{align}
		G_{\mc{L}^h_{0,[0,0]},\mc{D}_1}^{k_1,k_2}&=\frac{(-1)^{k_1+1} \Gamma (2 h+4) \Gamma (h+k_1) \Gamma (h+k_1+1)}{\left(h^2-1\right) \Gamma (h+1)^2 \Gamma (k_1+1) \Gamma (2 h+k_1+4)}(-\nu_1)^{h+k_1-1}  \nu_2^{-1}
	\end{align}
	\subsection*{$\bullet  \quad \scaleto{\mathbf{\mc{G}_{\mc{L}^h_{0,[0,0]},\mc{D}_3}}}{18pt}$}
	This block vanishes from selection rules since the uncharged long does not appear in the $\mc{D}_1\times \mc{D}_3$ OPE \cite{Liendo:2018ukf}. 
	\newpage 
	\subsection*{$\bullet  \quad \scaleto{\mathbf{\mc{G}_{\mc{D}_2,\mc{L}^h_{0,[0,1]}}}}{18pt}$}
	\begin{align}
		g^0_{\mc{D}_2,\mc{L}^h_{0,[0,1]}} &=0\\
		g^1_{\mc{D}_2,\mc{L}^h_{0,[0,1]}} &=\sum_{k_1,k_2}(k_1-2)G_{\mc{D}_2,\mc{L}^h_{0,[0,1]}}^{k_1,k_2}\\
		g^2_{\mc{D}_2,\mc{L}^h_{0,[0,1]}} &=\sum_{k_1,k_2} (-2 (k_1+1) )G_{\mc{D}_2,\mc{L}^h_{0,[0,1]}}^{k_1,k_2}\\
		g^3_{\mc{D}_2,\mc{L}^h_{0,[0,1]}} 	&=\sum_{k_1,k_2}\frac{2 (k_1+2) (h-k_1+k_2-2)}{(k_1+5) \nu_2}G_{\mc{D}_2,\mc{L}^h_{0,[0,1]}}^{k_1,k_2}\\
	\end{align}
	where
	\begin{align}
		G_{\mc{D}_2,\mc{L}^h_{0,[0,1]}}^{k_1,k_2}&=\frac{(k_1+1) (h+k_2-1)  (h-1)_{k_2} (h+3)_{k_2}}{2 (h-2) (h-1) (5)_{k_1} (1)_{k_2} (2 h+4)_{k_2} (h+k_2-1)_{-k_1}}(-\nu_1)^{k_1} \nu_2^{h+k_2-2}
	\end{align}
	\subsection*{$\bullet  \quad \scaleto{\mathbf{\mc{G}_{\mc{L}^h_{0,[0,0]},\mc{L}^h_{0,[0,1]}}}}{18pt}$}
	\begin{align}
		g^0_{\mc{L}^{h_1}_{0,[0,0]},\mc{L}^{h_2}_{0,[0,1]}}&=0\\
		g^1_{\mc{L}^{h_1}_{0,[0,0]},\mc{L}^{h_2}_{0,[0,1]}}&=\sum_{k_1,k_2}\left(-\left(h_2^2+h_2+3\right)k_1-h_2^2 (h_2+1)+k_1^2\right)G_{\mc{L}^h_{0,[0,0]},\mc{L}^h_{0,[0,1]}}^{k_1,k_2}\\
		g^2_{\mc{L}^{h_1}_{0,[0,0]},\mc{L}^{h_2}_{0,[0,1]}}&=\sum_{k_1,k_2}\left((h_2^2+h_2-2k_1) (h_2+k_1)\right)G_{\mc{L}^{h_1}_{0,[0,0]},\mc{L}^{h_2}_{0,[0,1]}}^{k_1,k_2}\\
		g^3_{\mc{L}^{h_1}_{0,[0,0]},\mc{L}^{h_2}_{0,[0,1]}}&=\sum_{k_1,k_2}\frac{\left(h_2^2+h_2-2k_1\right) (h_2+k_1-1)}{\nu_1 \nu_2}G_{\mc{L}^{h_1}_{0,[0,0]},\mc{L}^{h_2}_{0,[0,1]}}^{k_1,k_2}
	\end{align}
	where
	\begin{align}
		G_{\mc{L}^{h_1}_{0,[0,0]},\mc{L}^{h_2}_{0,[0,1]}}^{k_1,k_2}&=\frac{\left((h_1-1)_{k_2} (h_2+1)_{k_1} (h_1+h_2+3)_{k_2}\right)}{(h_2-1) h_2 (h_2+1)(2 h_1+4)_{k_2} (2 h_2+4)_{k_1} (h_1-h_2+k_2)_{-k_1}}\frac{\nu_2^{h_1+k_2-2} (-\nu_1)^{h_2+k_1-1} }{ k_1! k_2!  }
	\end{align}

\chapter{Effective Theories in AdS$_2$}
This Appendix groups the Witten diagram computations in the first section of Chapter \ref{Effective theories}. The first section (\ref{Section: contact diagrams}) details the derivations for the odd-massless case and the $\Delta=2$ case which were not included in \ref{Section AdS2 Witten diagrams}. Then, a list of results corresponding to contact diagrams is included. In section \ref{App exchange}, the case of the exchange diagram, presented in \ref{Eq: Exchange solution Delta=1} and in \ref{App: Polyakov} is gone through in detail. 
\section{Contact Diagrams}\label{Section: contact diagrams}
\subsection*{$n$-odd Massless Scalars}\label{App n odd}
\begin{figure}
	\center
	\begin{tikzpicture}[scale=.5]
		\draw[->] (0,-2) to (0,5);
		\draw[->] (-7,0) to (7,0);
		\draw [thick, red, ->] (-6,.3) to (2,.3);
		\draw [thick, red] (2,.3) to (6,.3);
		\filldraw[red] (1,1) circle (2pt);
		\node[anchor= south ] at (1.5,1)  {$x-i z$};
		\filldraw[red] (1,-1) circle (2pt);
		\node[anchor= north ] at (1.5,-1)  {$x+i z$};
		\filldraw[red] (-1,1) circle (2pt);
		\filldraw[red] (-1,-1) circle (2pt);
		\draw [,thick,red] (-6,.3) to[out=90,in=180] (0,4) to[out=0,in=90](6,.3) ;
		\draw[] (6,5)to(6,4.5) to(7,4.5);
		\node[anchor = south west] at (6,4.5){$x\in \mathbb{C}$};
		\node[anchor= south] at (5,.3) {$\mathcal{C}_{\mathbb{R}}$};
		\node[anchor= south] at (3.5,3.5) {$\mathcal{C}_{\infty}$};
	\end{tikzpicture}
	\begin{tikzpicture}[scale=.5]
		\draw[->] (0,-2) to (0,5);
		\draw[->] (-7,0) to (7,0);
		\draw [thick, red, ->] (-6,.3) to (2,.3);
		\draw [thick, red] (2,.3) to (6,.3);
		\filldraw[red] (1,1) circle (2pt);
		\node[anchor= south ] at (1.5,1)  {$x+i z$};
		\filldraw[red] (1,-1) circle (2pt);
		\node[anchor= north ] at (1.5,-1)  {$x-i z$};
		\filldraw[red] (-1,1) circle (2pt);
		\filldraw[red] (-1,-1) circle (2pt);
		\draw [,thick,red] (-6,.3) to[out=90,in=180] (0,4) to[out=0,in=90](6,.3) ;
		\draw[] (6,5)to(6,4.5) to(7,4.5);
		\node[anchor = south west] at (6,4.5){$x\in \mathbb{C}$};
		\node[anchor= south] at (5,.3) {$\mathcal{C}_{\mathbb{R}}$};
		\node[anchor= south] at (3.5,3.5) {$\mathcal{C}_{\infty}$};
	\end{tikzpicture}
	\caption{The contour here is closed in the UHP (the same analysis holds for the LHP closing). However, since $z$ is now defined on the entire real line, the pole enclosed in the given contour depends on the sign of $z$. The case where $z<0$ (Left) will have a residue of the opposite sign when compared to that where $z>0$ (Right).}
	\label{Fig: contour integral 2}
\end{figure}
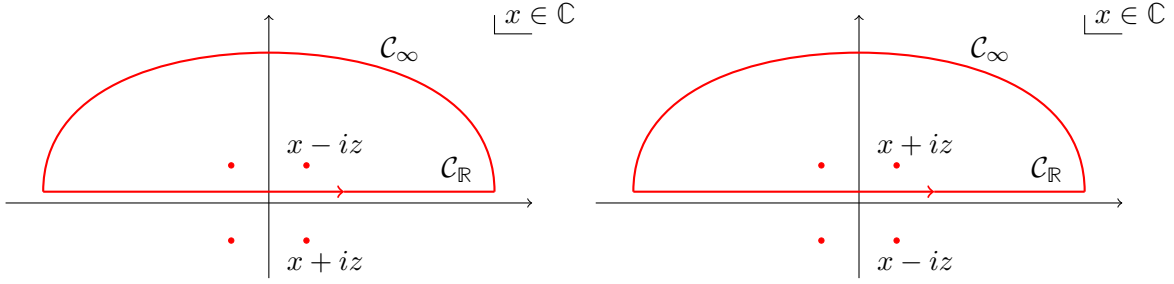

The massless odd-$n$ case can be solved  with contour integration for the $x$ and the $\chi$ coordinates.\footnote{This is true for any convergent integral with odd $\sum_i \Delta_i$. The resulting correlator will thus be a rational polynomial in the cross-ratios, though it may not have a simple form.} Since the integrand and the residue of the pole in the $x$ coordinates are antisymmetric under $z\rightarrow -z$, we can extend the region of integration of $z$ to the entire real line. This is best seen in the trivial example of the conformal massless three-point function
\begin{align}
	\lim_{\Lambda\rightarrow \infty} \Lambda^{2}I(0,1,\Lambda) = \int_0^\infty \frac{dz}{z^2}\int_{-\infty}^{\infty}dx \frac{z^3}{(z^2+x^2)(z^2+(x-1)^2)}.
\end{align}
Since the integrand is antisymmetric under $z\rightarrow -z$, we need to compensate for the sign change when extending the range of the integral over $z$
\begin{align}\label{Eq 3point sign z}
	\frac{1}{2}\int_{-\infty}^\infty \frac{dz\textrm{sgn}(z)}{z^2}\int_{-\infty}^{\infty}dx \frac{z^3}{(z^2+x^2)(z^2+(x-1)^2)}.
\end{align}
When considering the $x$-contour integral \eqref{Eq 3point sign z}, we are now faced with two situations for the contour integral; the first ($z<0$) is depicted on the left of Figure \ref{Fig: contour integral 2}, the second ($z>0$) is depicted on the right.
The sign of the pole included in the contour cancels the sign contribution from equation \eqref{Eq 3point sign z}. The range of $z$ can be extended after having done the $x$-integral to obtain the same conclusion. In so doing, one obtains
\begin{align}
	\pi \int_{-\infty}^\infty dz \frac{1 }{4 z^2+1} = \frac{\pi^2}{2}.
\end{align}
This reasoning holds for a general $n$-point function whose integrand is antisymmetric under $z\rightarrow -z$.  In one of the examples in subsection \ref{sec:topological}, we will see the case of topological operators where the polynomial dependence on the external coordinates is just a constant (up to a sign-dependent factor).\par
We have an analytic integral over the entire real $z$ line.
\begin{align}
	I(x_i)&= \frac{2\pi i }{2} \int_{-\infty}^{+\infty} dz z^{n-2}  \sum_{j=1}^{n-1}	\frac{1}{2 i z\Pi_{i\neq j}(x_{ij}^2+2izx_{ij})}\\
	&=\frac{\pi  }{2} \int_{-\infty}^{+\infty} dz z^{n-3}  \sum_{j=1}^{n-1}	\frac{1}{\Pi_{i\neq j}(x_{ij}^2+2izx_{ij})}.
\end{align}
The integrand has a good large-$|z|$ behaviour, and we can analytically continue $z$ and evaluate this integral by contour integration, neglecting the vanishing contribution from the contour at $\infty$. There are poles at positions 
\begin{align}
	z^*=-\frac{x_{ij}}{2i},
\end{align} 
with residues
\begin{align}
	\hspace{-.2cm} \! Res_{z=z^*}\left(\frac{z^{n-3} }{\Pi_{k\neq j}(x_{kj}^2+2izx_{kj})}\right)
	%&=  	\frac{(\frac{x_{ji}}{2i})^{n-3} }{\Pi_{k\neq j} (2i)x_{kj}\Pi_{k\neq j}(\frac{x_{ij}}{2i}-\frac{x_{kj}}{2i})}\nonumber \\
	=	\frac{x_{ij}^{n-4} }{(2i)^{n-4}\Pi_{k\neq j\neq i}x_{kj}x_{ki}}.
\end{align}
We close the contour in the UHP, where only the poles where $x_i-x_j>0$ contribute.
This gives the final result for odd-$n$ and ordered $x_i$
\begin{align}\label{Eq:massless odd result}
	I_{n\,odd}(x_i)	&=\frac{\pi^2  }{ 2(2i)^{n-3}} \sum_{i>j}^{n}\frac{( x_i-x_j)^{n-4}}{\Pi_{k\neq j\neq i}x_{kj}x_{ki}}.
\end{align}
This formula agrees with the canonical case of $n=3$, and the explicit results for $n=5$ and $n=7$ are given in Appendix \ref{App: list of correlators}.
\subsection*{Derivation of 	\texorpdfstring{$\Delta=2$}{Lg}}\label{App: Delta=2 derivation}
One may, of course, be interested in higher $\Delta$. This is where this method loses some of its power. While it is mighty in the generic $n$ regime, it quickly increases complexity when $\Delta$ increases. However, the complexity will only be combinatorial and not intrinsic.
The integrand of
\begin{align}
	I(x_i) =\int dz z^{2n-2} \int dx\frac{ 1}{\Pi_{i=1}^{N-1}(z^2+(x-x_i)^2)^2}
\end{align}
only has double poles (and single poles from the expansion around these poles) at the position $x = x_i+i z$ with residue
\begin{align}
	&\underset{x = x_i+iz}{Res} \left(\! \frac{ 1}{\Pi(z^2+(x-x_i)^2)^2}\right) \!=\!\partial_x\left( \frac{1}{(x-(x_i-iz))^2}	\frac{ 1}{{\displaystyle \prod_{j,j\neq i}}(z^2+(x-x_j)^2)^2}\right)\Biggr|_{x=x_i+iz} \\
	&= -\sum_i	\frac{1}{4z^2}(\frac{1}{iz}+\partial_{x_i})\left( \frac{1}{\Pi_{j\neq i}x_{ij}^2(2i z +x_{ij})^2}\right),\nonumber
	%+\frac{1}{4iz^3}\left( \frac{1}{\Pi_{j\neq i}x_{ij}^2(2i z +x_{ij})^2}\right)
\end{align}
where the residue was massaged into a more usable form. As in the massless case, these can be integrated using partial fraction decomposition. By comparison of simple and double poles, we have
\begin{align}
	\sum_i \frac{z^{2n-n_0}}{\prod_{k}(z+a_{ik})^2} &= \sum_{i\neq j}\frac{c_{ij}}{(z+a_{ij})^2}-\sum_{i\neq j}\frac{\partial_{a_{ij}}c_{ij}}{(z+a_{ij})}\\
	&= -\sum_{i\neq j}\partial_{a_{ij}}\left(\frac{c_{ij}}{ (z+a_{ij})}\right),\\
	c_{ij}&= 	\frac{(-a_{ij})^{2n-n_0}}{\prod_{k\neq j}(a_{ki}-a_{ji})^2}. 
\end{align}\bigskip
as long as the $n_0>-1$. This is integrated by sight
\begin{align}
	\int_0^\Lambda dz \left( \sum_{i\neq j}\frac{c_{ij}}{(z+a_{ij})^2}-\sum_{i}\frac{\partial_{a_{ij}}c_{ij}}{(z+a_{ij})}\right) =\sum_{i\neq j}\partial_{a_{ij}}\left(c_{ij}\ln(a_{ij})\right)+(\partial_{a_i}c_{ij})\ln(\Lambda).
\end{align}
\vspace{2mm} Explicitly, we have
\begin{align}
	&\int  dz z^{2n-2}\frac{-1}{4z^2}\sum_i \partial_{x_i}\left( \frac{1}{\prod_{j\neq i}x_{ij}^2(2i z +x_{ij})^2}\right)\\
	&=\sum_i \sum_{j\neq i}\partial_{x_i}\frac{1}{4(2i)^{2n-3}x_{ij}^2}\partial_{x_j}\left( \frac{x_{ji}^{2n-4}}{\prod_{k\neq j,k\neq i}x_{kj}^2x_{ik}^2}\ln\frac{x_{ji}}{2i}\right),
\end{align}
and 
\begin{align}
	&\int dz \frac{z^{2n-5}}{4i}\left( \frac{1}{\prod_{j\neq i}x_{ij}^2(2i z +x_{ij})^2}\right)\nonumber\\
	&=\sum_i\sum_{j\neq i}\frac{1}{4i(2i)^{2n-4}x_{ij}^2}\partial_{x_j}\left( \frac{x_{ji}^{2n-5}}{\prod_{k\neq j,k\neq i }x_{kj}^2x_{ki}^2}\ln\frac{x_{ji}}{2i}\right).
\end{align}
As in the massless case, the divergent terms cancel, and the logarithm is a well-defined function with a branch cut on the negative real axis.
This gives the result
\begin{align}
	I_{\Delta=2,n}(x_i)&=\sum_i\sum_{j\neq i}\frac{-\pi}{2(2i)^{2n-4}x_{ij}^2}\partial_{x_j}\left( \frac{x_{ji}^{2n-5}}{\prod_{k\neq j,k\neq i }x_{kj}^2x_{ki}^2}\ln\frac{x_{ji}}{2i}\right)\nonumber \\
	+\sum_i &\sum_{j\neq i}\partial_{x_i}\frac{-\pi}{(2i)^{2n-2}x_{ij}^2}\partial_{x_j}\left( \frac{x_{ji}^{2n-4}}{\prod_{k\neq j,k\neq i}x_{kj}^2x_{ik}^2}\ln\frac{x_{ji}}{2i}\right).
\end{align}

\subsection*{Library of Contact Correlators}\label{App: list of correlators}
%In the main body, results are naturally written in terms of the cross-ratios $\chi_i$ defined in equation \eqref{Eq: cross-ratio}. However, they also hold for the  external coordinates; for example $\{x_1,...,x_4\}$ combine naturally to form the cross-ratio $u_1$ in \eqref{Eq: four-point} in the case of the four-point function 
%\begin{align}
%	I_{\Delta=1,n=4}(x_1,..,x_4)
	%&=-\frac{\pi}{2}\Big( \frac{\log x_{12}}{x_{23} x_{13}x_{24}x_{14}}+\frac{\log x_{13}}{x_{12}x_{23}x_{34}x_{14}}+\frac{\log x_{23}}{x_{12}x_{13}x_{34}x_{24}}\nonumber \\
	%&+\frac{\log x_{34}}{x_{13}x_{23}x_{14}x_{24}}+\frac{\log x_{24}}{x_{12}x_{23}x_{14}x_{34}}+\frac{\log x_{14}}{x_{12}x_{13}x_{24}x_{34}}\nonumber \Big)\\
%	&= -\frac{\pi}{2(x_{13}x_{24})^2}\left(\frac{x_{13}x_{24}}{x_{14}x_{23}}\log\left(\frac{x_{12}x_{34}}{x_{13}x_{24}} \right)  +\frac{x_{13}x_{24}}{x_{12}x_{34}}\log\left(\frac{x_{14}x_{23}}{x_{13}x_{24}} \right) \right).
%\end{align} 
Below, we include a few examples $I(0,u_1,...,u_{n-3},1,\infty)$ of the contact integral \eqref{Eq: contact integral} evaluated in the cross-ratios defined in \eqref{Eq: cross-ratio}, where we use the notation $I_{\Delta,n}$ for equal dimension operators and $I_{[\Delta_1,...,\Delta_n]}$ to include external operators of different dimensions.

\begin{align}
	&I_{1,3}  = \frac{\pi ^2}{2}&&&&&\qquad \qquad&\\
	&I_{1,4} =-\frac{\pi }{2}\left(  \frac{\log \left(\chi_1\right)}{ 1-\chi_1}+\frac{\log \left(1-\chi_1\right)}{ \chi_1}\right)\\
	&I_{1,5} = \frac{\pi ^2}{4 \chi_2\left(1-\chi_1\right)}\\
	&I_{1,6}= \frac{\pi}{8}\left(\frac{\left(\chi_1-1\right){}^2 \log \left(1-\chi_1\right)}{\chi_1 \left(\chi_1-\chi_2\right) \left(\chi_2-1\right) \left(\chi_1-\chi_3\right) \left(\chi_3-1\right)}\right. \nonumber \\
	&\left.+\frac{\left(\chi_1-\chi_2\right){}^2 \log \left(\chi_2-\chi_1\right)}{\left(\chi_1-1\right) \chi_1 \left(\chi_2-1\right) \chi_2 \left(\chi_1-\chi_3\right) \left(\chi_2-\chi_3\right)}-\frac{\left(\chi_2-1\right){}^2 \log \left(1-\chi_2\right)}{\left(\chi_1-1\right) \left(\chi_1-\chi_2\right) \chi_2 \left(\chi_2-\chi_3\right) \left(\chi_3-1\right)}\right.\nonumber \\
	&\left. +\frac{\chi_2^2 \log \left(\chi_2\right)}{\chi_1 \left(\chi_1-\chi_2\right) \left(\chi_2-1\right) \left(\chi_2-\chi_3\right) \chi_3}+\frac{\left(\chi_3-1\right){}^2 \log \left(1-\chi_3\right)}{\left(\chi_1-1\right) \left(\chi_2-1\right) \left(\chi_1-\chi_3\right) \left(\chi_2-\chi_3\right) \chi_3}\right.\nonumber \\
	&\left.+\frac{\left(\chi_2-\chi_3\right){}^2 \log \left(\chi_3-\chi_2\right)}{\left(\chi_1-\chi_2\right) \left(\chi_2-1\right) \chi_2 \left(\chi_1-\chi_3\right) \left(\chi_3-1\right) \chi_3}-\frac{\left(\chi_1-\chi_3\right){}^2 \log \left(\chi_3-\chi_1\right)}{\left(\chi_1-1\right) \chi_1 \left(\chi_1-\chi_2\right) \left(\chi_2-\chi_3\right) \left(\chi_3-1\right) \chi_3}\right.\nonumber\\
	&\left.-\frac{\chi_3^2 \log \left(\chi_3\right)}{\chi_1 \chi_2 \left(\chi_1-\chi_3\right) \left(\chi_2-\chi_3\right) \left(\chi_3-1\right)}-\frac{\chi_1^2 \log \left(\chi_1\right)}{\left(\chi_1-1\right) \left(\chi_1-\chi_2\right) \chi_2 \left(\chi_1-\chi_3\right) \chi_3}\right)\\
	&I_{1,7} =\frac{\pi ^2}{16 \left(\chi_1-1\right) \left(\chi_2-1\right) \chi_2 \left(\chi_1-\chi_3\right) \left(\chi_3-1\right) \chi_3 \left(\chi_1-\chi_4\right) \left(\chi_2-\chi_4\right) \chi_4} \nonumber \\
	&\Big(\left(\chi_2-1\right) \left(\chi_3-1\right) \left(\chi_2-\chi_4\right) \chi_1^2\nonumber \\
	&+\chi_2 \left(\chi_3^2+\chi_4 \left(\chi_3+\chi_4-2\right) \chi_3-\chi_4-\chi_2 \left(\chi_3-1\right) \left(\chi_3+\chi_4+1\right)\right) \chi_1 \nonumber \\
	&+\chi_2 \left(\chi_2 \left(\chi_3-1\right) \left(\chi_3+\left(\chi_3+1\right) \chi_4\right)-\chi_3 \left(\chi_3 \left(\chi_4^2+\chi_4+1\right)-3 \chi_4\right)\right) \Big)
\end{align}
For the massive cases, we find agreement between the result of pinching and that of the formula \eqref{Eq Delta=2 result} which gives for the first few cases:
\begin{align}
	&I_{2,3}= \frac{3 \pi }{8}\\
	&I_{2,4}=-\frac{\pi  ((\chi -1) \chi +1)}{8 (\chi -1)^2 \chi ^2}-\frac{\pi  \left(2 \chi ^2+\chi +2\right) \log (1-\chi )}{16 \chi ^3}+\frac{\pi  (\chi  (2 \chi -5)+5) \log (\chi )}{16 (\chi -1)^3}\\
	&I_{2,5} =\frac{\pi}{32} \big(\left. \right.\nonumber \\
	&+\frac{(\chi_1-v)^2 \left(\chi_1^3 (2 \chi_2-1)+\chi_1^2 (3 (\chi_2-2) \chi_2+2)+\chi_1 \chi_2 (2 (\chi_2-3) \chi_2+3)-(\chi_2-2) \chi_2^2\right) \log (\chi_2-\chi_1)}{(\chi_1-1)^3 \chi_1^3 (1-\chi_2)^3 \chi_2^3}\nonumber \\
	&+\frac{(\chi_2-1)^2 \left(\chi_1^2 (-(\chi_2 (2 \chi_2+3)+2))+\chi_1 (\chi_2+1) (\chi_2 (\chi_2+5)+1)-\chi_2 (\chi_2 (2 \chi_2+3)+2)\right) \log (1-\chi_2)}{(\chi_1-1)^3 \chi_2^3 (\chi_1-\chi_2)^3} \nonumber \\
	&+\frac{(\chi_1-1)^2 \left((\chi_1 (2 \chi_1+3)+2) \chi_2^2-(\chi_1+1) (\chi_1 (\chi_1+5)+1) \chi_2+\chi_1 (\chi_1 (2 \chi_1+3)+2)\right) \log (1-\chi_1)}{\chi_1^3 (\chi_2-1)^3 (\chi_1-\chi_2)^3}\nonumber \\
	&+\frac{\chi_2^2 \left(7 \chi_1^2-(\chi_1+1) \chi_2^3+(\chi_1+2) (2 \chi_1+1) \chi_2^2-7 \chi_1 (\chi_1+1) \chi_2\right) \log (\chi_2)}{\chi_1^3 (\chi_2-1)^3 (\chi_1-\chi_2)^3}\nonumber \\
	&+\frac{\chi_1^2 \left(\chi_1^3 (\chi_2+1)-\chi_1^2 (\chi_2+2) (2 \chi_2+1)+7 \chi_1 \chi_2 (\chi_2+1)-7 \chi_2^2\right) \log (\chi_1)}{(\chi_1-1)^3 \chi_2^3 (\chi_1-\chi_2)^3}\nonumber \\
	&+\frac{-2 \chi_1^4 ((\chi_2-1) \chi_2+1)+\chi_1^3 \left(2 \chi_2^3+\chi_2^2+\chi_2+2\right)-2 \chi_2^2 ((\chi_2-1) \chi_2+1)}{2 (\chi_1-1)^2 \chi_1^2 (\chi_2-1)^2 \chi_2^2 (\chi_1-\chi_2)^2} \nonumber \\
	&+\frac{\chi_1 \left(-2 \chi_2^4+\chi_2^3-6 \chi_2^2+\chi_2-2\right)+\chi_2 \left(2 \chi_2^3+\chi_2^2+\chi_2+2\right)}{2 (\chi_1-1)^2 \chi_1 (\chi_2-1)^2 \chi_2^2 (\chi_1-\chi_2)^2} \big)
\end{align}
\vspace{3mm}\par 
From pinching, we define the integral with the pinched weights at positions $x_i$, e.g. $I_{[1,2,2,1]}$
\begin{align}
%	I_{[2,2,2]}&= \frac{3 \pi }{8}\\
%	I_{[1,1,1,2]}&= \frac{\pi ^2}{4 x_{13} x_{14} x_{24}^2 x_{34}}\\
%	I_{[2,2,1,1]}&= -\frac{\pi  (\chi +2) \log (1-\chi )}{8 \chi ^3}+\frac{\pi }{8 \chi ^2(1-\chi)}+\frac{\pi  \log (\chi )}{8 (\chi -1)^2}\\
%	I_{[2,1,2,1]}&=-\frac{(2 \pi  \chi +\pi ) \log (1-\chi )}{8 \chi ^2}+\frac{\pi }{8 (\chi -1) \chi }+\frac{\pi  (2 \chi -3) \log (\chi )}{8 (\chi -1)^2}\\
%	I_{[1,2,2,1]}&=\frac{\pi  \log (1-\chi )}{8 \chi ^2}+\frac{\pi }{8 (\chi -1)^2 \chi }-\frac{\pi  (\chi -3) \log (\chi )}{8 (\chi -1)^3}\\
%	I_{[2,2,2,1]}&= \frac{\pi ^2}{8 \chi (1-\chi)}\\
%	I_{[3,1,2,1]}&= \frac{3 \pi ^2}{16 \chi }\\
	I_{[1,2,1,1,1]}&= -\frac{\pi  \chi_2^2 \log (\chi_2)}{8 \chi_1^2 (\chi_2-1) (\chi_1-\chi_2)^2}+\frac{\pi  (\chi_1 (\chi_1-2 \chi_2+2)-\chi_2) \log (1-\chi_1)}{8 \chi_1^2 (\chi_2-1) (\chi_1-\chi_2)^2}\nonumber \\
	&+\frac{\pi  (\chi_2-\chi_1 (\chi_1+2 \chi_2-2)) \log (\chi_2-\chi_1)}{8 (\chi_1-1)^2 \chi_1^2 (\chi_2-1) \chi_2}-\frac{\pi  \left(\chi_1^2-2 \chi_1 (\chi_2+1)+3 \chi_2\right) \log (\chi_1)}{8 (\chi_1-1)^2 \chi_2 (\chi_1-\chi_2)^2}\nonumber \\
	&+\frac{\pi  (\chi_2-1)^2 \log (1-\chi_2)}{8 (\chi_1-1)^2 \chi_2 (\chi_1-\chi_2)^2}\\
	I_{[2,1,1,2,1]}&= -\frac{\pi ^2 (2 \chi_1 \chi_2+\chi_1-3 \chi_2)}{16 (\chi_1-1)^2 \chi_2^2}\\
	I_{[1,1,1,1,2,1]}&= \frac{\pi ^2 \left(\chi_1^2 \left(\chi_2-1\right){}^2+\chi_2 \left(\chi_2+\left(2 \chi_2-3\right) \chi_3\right)+\chi_1 \chi_2 \left(2 \chi_3-\chi_2 \left(\chi_3+2\right)+1\right)\right)}{16 \left(\chi_1-1\right){}^2 \left(\chi_2-1\right){}^2 \chi_2 \left(\chi_1-\chi_3\right) \chi_3}
\end{align}
There are also some divergent cases
\begin{align}
	I_{[2,1,1]}&= \frac{\pi}{2} (1- \log (\epsilon ))\\
	I_{[1,3,1,1]}&= \frac{5 \pi }{16 (\chi -1)^2 \chi ^2}-\frac{\pi  ((\chi -3) \chi +3) \log (\chi )}{8 (\chi -1)^3 \chi ^2}+\frac{\pi  \left(\chi ^2+\chi +1\right) \log (1-\chi )}{8 (\chi -1)^2 \chi ^3}\nonumber \\
	&\qquad-\frac{3 \pi  \log (\epsilon )}{8 (\chi -1)^2 \chi ^2}
\end{align}
We also find agreement between the pinching of the $2n$-point function of massless correlators and the $n$-point function of $\Delta=2$ correlators up to $n=8$. Still, these were omitted from the text since they are bulky and not elucidating.
Notice that the prefactor \eqref{Eq: prefactor} has no neighbouring terms of  type $x_{i,i+1}$ except for the $x_{n-1,n}$ term and the $n=3$ cases so that no single pinching will lead to divergences in the prefactor. In general, one expects the divergences appearing in the pinching to be physical divergences that need to be regularised, not artefacts of this method.

\subsection*{$D$-functions}
\label{app:Dfunctions}
The quartic contact diagrams with external conformal dimensions $\D_i$ are expressed in terms of $D$-functions \cite{Liu:1998ty,DHoker:1999kzh, Dolan:2003hv},  defined  for the general case of $\text{AdS}_{d+1}$ as
\be \label{D-function}
\!\!\!\!\!\!\!\!
D_{\Delta_{1}\Delta_{2}\Delta_{3}\Delta_{4}}(x_1,x_2,x_3,x_4) =\! \!\int \!\!\frac{dz d^dx}{z^{d+1}} 
\tilde{K}_{\Delta_{1}}\!(z,x;x_1) \tilde{K}_{\Delta_{2}}\!(z,x;x_2) \tilde{K}_{\Delta_{3}}\!(z,x;x_3) \tilde{K}_{\Delta_{4}}\!(z,x;x_4)\,
\ee
For vertices with derivatives, the following identity is useful 
\begin{align}
	&g^{\m\n}\partial_\m \tilde{K}_{\Delta_1}(z,x;x_1)\ \partial_\n\tilde{K}_{\Delta_2}(z,x;x_2) 
	\\   & \qquad = 
	\Delta_1 \Delta_2
	\left[\tilde{K}_{\Delta_1}(z,x;x_1)\tilde{K}_{\Delta_2}(z,x;x_2)-2x_{12}^2 \tilde{K}_{\Delta_1+1}(z,x;x_1)\tilde{K}_{\Delta_2+1}(z,x;x_2)\right]\ \,,
	\label{identityder}
\end{align}
where $g^{\m\n}= {z^2}\delta^{\m\n}$ and $\del_\m=(\del_z,\del_r)$,  $r=0,1, 2, ...,\text{d}-1$.  \par \vspace{2mm}
To make explicit the covariant form of the correlator, it is useful to introduce the ``reduced" functions $\bar D$~\cite{Dolan:2003hv},  defined as  ($\Sigma \equiv \frac{1}{2}\sum_i \Delta_i$) 
\be \label{Dbar}
\!\!\!\!
D_{\Delta_{1}\Delta_{2}\Delta_{3}\Delta_{4}}= 
\frac{\pi^{d\ov 2}\Gamma\left(\Sigma-{d\ov2}\right)}{2\, \Gamma\left(\Delta_1\right)\Gamma\left(\Delta_2\right)\Gamma\left(\Delta_3\right)\Gamma\left(\Delta_4\right)}
\frac{x_{14}^{2(\Sigma-\Delta_1-\Delta_4)} x_{34}^{2(\Sigma-\Delta_3-\Delta_4)}}
{x_{13}^{2(\Sigma-\Delta_4)} x_{24}^{2\Delta_2}}\bar{D}_{\Delta_1\Delta_2\Delta_3\Delta_4}(u,v)
\ee
which depends only on the cross-ratios $u=\frac{x_{12}x_{34}}{x_{13}x_{24}}\,,  v=\frac{x_{14}x_{23}}{x_{13}x_{24}}$.  Their explicit expression in terms Feynman parameters integral is
\begin{equation}
	\bar{D}_{\Delta_1\Delta_2\Delta_3\Delta_4}(u,v)=
	\int d\alpha d\beta d\gamma\  \delta(\alpha+\beta+\gamma-1)\ 
	\alpha ^{\Delta _1-1} \beta ^{\Delta _2-1} \gamma ^{\Delta _3-1} 
	\frac{\Gamma \left(\Sigma-\Delta _4\right) \Gamma\left(\Delta_4\right)}
	{\big(\alpha  \gamma + \alpha  \beta\, u  + \beta  \gamma\, v\big)^{\Sigma-\Delta_4}}\,.
	\label{Dbar-integral}
\end{equation}
In  d=1 as usual they only depend on the single variable  $\chi$ 
\begin{align}
	u=\chi^2, \quad v=(1-\chi)^2
\end{align}
\be\label{Dbar1d}
\bar D_{\Delta\Delta\Delta\Delta}(\chi)=\frac{\Gamma(\Delta)^4}{\Gamma(2\Delta)} (1-\chi)^{-2\Delta}\!\int_{-\infty}^{+\infty}\!d\tau\,e^{-\tau} {}_2F_1\big(\Delta,\Delta,2\Delta,\textstyle-\frac{4\chi}{(1-\chi^2)}\cosh^2\frac{\tau}{2}\big)\, .
\ee

Some explicit expression for $\bar{D}$-functions  read
%, which may  be evaluated directly via \eqref{Dbar-integral} or 
%derived  in terms of derivatives of the first one $\bar{D}_{1111}(u,v)$ using the identities contained in \cite{Dolan:2003hv}, are
\begin{eqnarray}\label{Dbar-explicit}
	\bar D_{1111}&=&-\frac{2 \log (1-\chi )}{\chi }-\frac{2 \log (\chi )}{1-\chi }\, ,\\
	%\bar{D}_{2,2,1,1} &=&-\frac{(\chi +2) \log (1-\chi )}{3 \chi ^3}+\frac{1}{3 (1-\chi ) \chi ^2}+\frac{\log (\chi )}{3 (1-\chi )^2}\\
	%\bar D_{1,2,2,1}&=&   \frac{\log (1-\chi )}{3 \chi ^2}+\frac{1}{3 (1-\chi )^2 \chi }-\frac{(\chi -3) \log (\chi
		%   )}{3 (\chi -1)^3}\\
	%\bar{D}_{1,2,1,2} &=& -\frac{(2 \chi +1) \log (1-\chi )}{3 \chi ^2}-\frac{1}{3 (1-\chi ) \chi }+\frac{(2 \chi -3) \log (\chi )}{3 (1-\chi )^2}\\
	\bar{D}_{2222} &=&-\frac{2 \left(\chi ^2-\chi +1\right)}{15 (1-\chi )^2 \chi ^2}+\frac{\left(2 \chi ^2-5 \chi +5\right) \log (\chi )}{15 (\chi -1)^3}-\frac{\left(2 \chi ^2+\chi +2\right) \log (1-\chi )}{15 \chi ^3}\, ,\\
	%\bar{D}_{2,3,1,2}&=&-\frac{(\chi  (3 \chi +4)+3) \log (1-\chi )}{15 \chi ^4}-\frac{\chi  (3 \chi -8)+3}{15 (\chi -1)^2 \chi ^3}+\frac{(3 \chi -5) \log (\chi )}{15 (\chi -1)^3}\\
	%\bar{D}_{2,3,2,1} &=&\frac{(2 \chi +3) \log (1-\chi )}{15 \chi ^4}+\frac{2 (\chi -1) \chi -3}{15 (\chi -1)^3 \chi ^3}+\frac{(5-2 \chi ) \log (\chi )}{15 (\chi -1)^4}\\
	%\bar{D}_{3,3,1,1} &=& -\frac{2 (\chi  (\chi +3)+6) \log (1-\chi )}{15 \chi ^5}+\frac{3-\chi  (2 \chi +3)}{15 (\chi -1)^2 \chi ^4}+\frac{2 \log (\chi )}{15 (\chi -1)^3}\\
	%\bar{D}_{2,3,2,3}&=&\frac{\left(12 \chi ^3-42 \chi ^2+56 \chi -35\right) \log (\chi )}{105 (\chi -1)^4}+\frac{-24 \chi ^4+48 \chi ^3+5 \chi ^2-29 \chi +18}{210 (\chi -1)^3 \chi ^3}\nonumber\\
	%&&+\frac{\left(-12 \chi ^3-6 \chi ^2-8 \chi -9\right) \log (1-\chi )}{105 \chi ^4}\\\nonumber
	%\bar{D}_{2,3,3,2}&=&\frac{\left(9 \chi ^2+10 \chi +9\right) \log (1-\chi )}{105 \chi ^4}+\frac{\left(-9 \chi ^3+35 \chi ^2-49 \chi +35\right) \log (\chi )}{105 (\chi -1)^5}\nonumber\\
	%&&+\frac{18 \chi ^4-43 \chi ^3+26 \chi ^2-43 \chi +18}{210 (\chi -1)^4 \chi ^3}\\
	%\bar{D}_{3,3,2,2}&=&\frac{\left(9 \chi ^2-28 \chi +28\right) \log (\chi )}{105 (\chi -1)^4}+\frac{\left(-9 \chi ^3-8 \chi ^2-6 \chi -12\right) \log (1-\chi )}{105 \chi ^5}\nonumber\\
	%&&+\frac{-18 \chi ^4+29 \chi ^3-5 \chi ^2-48 \chi +24}{210 (\chi -1)^3 \chi ^4}\\
	\bar{D}_{3333}&=&\frac{\left(8 \chi ^4-36 \chi ^3+64 \chi ^2-56 \chi +28\right) \log(\chi)}{105 (\chi -1)^5}+\frac{\left(-8 \chi ^4-4 \chi ^3-4 \chi ^2-4 \chi -8\right) \log (1-\chi )}{105 \chi ^5}\nonumber\\
	&&+\frac{-24 \chi ^6+72 \chi ^5-74 \chi ^4+28 \chi ^3-74 \chi ^2+72 \chi -24}{315 (\chi -1)^4 \chi ^4}\, .\label{Dbar-explicit-end}
\end{eqnarray}

Further expressions can be found through the identities in~\cite{Dolan:2003hv}.  Useful relations between $\bar D$-function of consequent weight are

\begin{align}\label{Dbarconseq}
	\Delta\,\bar{D}_{\Delta \Delta \Delta \Delta }&= \bar{D}_{\Delta \Delta \Delta+1 \Delta+1}+\bar{D}_{\Delta \Delta+1 \Delta \Delta +1}+\bar{D}_{\Delta+1 \Delta \Delta \Delta +1} \, ,\\
	\label{Dfunctioncrossing1}
	(\Delta_2+\Delta_4-\Sigma) \,\bar{D}_{\Delta_1 \Delta_2 \Delta_3 \Delta_4} &= \bar{D}_{\Delta_1 \Delta_2+1 \Delta_3 \Delta_4+1} - \bar{D}_{\Delta_1+1 \Delta_2 \Delta_3+1 \Delta_4}\, ,\\	
	(\Delta_1+\Delta_4-\Sigma) \,\bar{D}_{\Delta_1 \Delta_2 \Delta_3 \Delta_4}& =\bar{D}_{\Delta_1+1 \Delta_2 \Delta_3 \Delta_4+1}-(1-\chi)^2 \bar{D}_{\Delta_1 \Delta_2+1 \Delta_3+1 \Delta_4}\, ,\\
	(\Delta_3+\Delta_4-\Sigma)\, \bar{D}_{\Delta_1 \Delta_2 \Delta_3 \Delta_4}&= \bar{D}_{\Delta_1 \Delta_2 \Delta_3+1 \Delta_4+1}-\chi^2\bar{D}_{\Delta_1+1 \Delta_2+1 \Delta_3 \Delta_4}\, ,\\
	\label{Dfunctioncrossing2}
	\bar{D}_{\Delta_1 \Delta_2 \Delta_3 \Delta_4}&=(1-\chi)^{2(\Delta_1+\Delta_4-\Sigma)}\bar{D}_{\Delta_2 \Delta_1 \Delta_4 \Delta_3}\, ,\\
	&=\bar{D}_{\Sigma-\Delta_3 \Sigma- \Delta_4 \Sigma-\Delta_1 \Sigma-\Delta_2}\, ,\\
	&=\chi^{2(\Delta_3+\Delta_4-\Sigma)}\bar{D}_{\Delta_4 \Delta_3 \Delta_2 \Delta_1}\, .
\end{align}
\section{Exchange Diagrams}\label{App exchange}
This section contains complementary material relating to the derivation and interpretation of the exchange diagrams. First, details on the conformal Casimir and its relation to the equation of motion in AdS are given.  Then the derivation of the relationship between exchange and contact diagrams from \cite{Rastelli:2017udc,Zhou:2020ptb} is reviewed. This is then applied to the toy model of $\phi^3$ interaction in AdS$_2$, which gives insight into Polyakov blocks whose perturbative strong coupling structure is shown explicitly.
\subsection*{Quadratic Casimir }\label{App: quad Casimir}
The $n$-Casimir of the boundary conformal group is given by
\begin{align}
	C^{(n)}_{i_1..i_n} = \frac{1}{2}\{\sum_{k=1}^n L_{i_k}^{(0)},\sum_{k=1}^n L_{i_k}^{(0)}\}-\frac{1}{2} [\sum_{k=1}^n L_{i_k}^{-\alpha},\sum_{k=1}^n L_{i_k}^{\alpha}],
\end{align}
where $L^{(0)}$ are elements of the Cartan and the others are the simple roots. Explicitly for a d=1 conformal boundary, the differential expression of the operators is:
\begin{align}
	D &= L_0 = \Delta+x\partial_x& P &= L_{-1} = -\partial_x&	K& = L_{+1} = -2\Delta x-x^2\partial_x
\end{align}
This leads to a linear Casimir
\begin{align}
	C^{(1)}_a &=  \Delta(\Delta-1),
\end{align}
which is the mass-squared of the bulk operator and a quadratic Casimir:
\begin{align}
	C^{(2)}_{x,y} &=2 (x-y) (- \Delta_1 \partial_y + \Delta_2 \partial_x )-(y-x)^2 \partial_x\partial_y+(\Delta_1+\Delta_2-1) (\Delta_1+\Delta_2).
\end{align}
The quadratic Casimir of the AdS$_2$ isometries is the Laplacian of AdS,
this is best seen in flat embedding coordinates where the generators are given by 
\begin{align}
	J_{AB} = -i(X_A\partial_B-X_B\partial_A),
\end{align}
where $X_A X^A=1$. The Quadratic Casimir of the AdS isometries in embedding coordinates is then
\begin{align}
	-\frac{1}{2}\mathcal{L}_a \mathcal{L}^a &=- \frac{1}{2}J_{AB}J^{AB}\\
	&= \frac{1}{2}(X_A\partial_B-X_B\partial_A)(X^A\partial^B-X^B\partial^A)\\
	&=X_AX^A\partial_B\partial^B +(1-d)X_B\partial^B\\
	&=\partial_A\partial^A,
\end{align}
which is the coordinate-independent Laplacian. 

A case of interest here is when we have $\Delta_1=\Delta_2$. The conformal quadratic Casimir then simplifies further to
\begin{align}
	C^{(2)}_{z} = 2 \Delta  (2 \Delta -1) f(z)-z \left((z-1) z f''(z)+(2 \Delta  (z-2)+z) f'(z)\right),
\end{align}
where we have made a change of variable 
\begin{align}
	z = 1-\frac{x}{y}. 
\end{align}
Such changes of variables can be made to reduce this differential equation into a single variable differential equation for each of the $(s,t,u)$ exchange channels.

\subsubsection{Relating the exchange and contact diagrams}\label{Sec: relating exchange to contact diagrams}
We review the analysis from \cite{1999NuPhB.562..395D,Zhou:2018sfz}, which goes through a detailed computation of the exchange diagram and the $z$ integral. However, they specialise in the case where the resulting exchange correlator can be written in terms of a finite sum of $ D$ functions. The toy model does not satisfy the conditions needed for such a simplification, but it is still useful to illustrate the Polyakov blocks.

\begin{figure}[h]
	\center
	\begin{tikzpicture}
		\def\x{0.8}
		\filldraw[blue!10!white] (0,0) circle (4*\x cm);
		\draw[] (0,0) circle (4*\x cm);
		
		\draw[] (-2.8*\x,-2.8*\x) to (-1.5*\x,0);
		\draw[] (-2.8*\x,2.8*\x) to (-1.5*\x,0);
		\draw[] (2.8*\x,2.8*\x) to (1.5*\x,0);
		\draw[] (2.8*\x,-2.8*\x) to (1.5*\x,0);
		\draw[] (-1.5*\x,0) to (1.5*\x,0);
		
		\node[anchor =west] at (1.5*\x,0) {$-\lambda$};
		\node[anchor =east] at (-1.5*\x,0) {$-\lambda$};
		\node[anchor=south] at (0,0) {$\Delta_E$};
		
		\node[anchor=south east] at (-2.8*\x,2.8*\x) {$x_1$};
		\node[anchor=north east] at (-2.8*\x,-2.8*\x) {$x_4$};
		\node[anchor=north west] at (2.8*\x,-2.8*\x) {$x_3$};
		\node[anchor=south west] at (2.8*\x,2.8*\x) {$x_2$};
	\end{tikzpicture}
	\caption{Exchange Witten diagram for four external insertions of identical scalars of weight $\Delta_\phi$ exchanging a scalar of weight $\Delta_E$ in the $t$ channel.}
	\label{Witten exchange}
\end{figure}
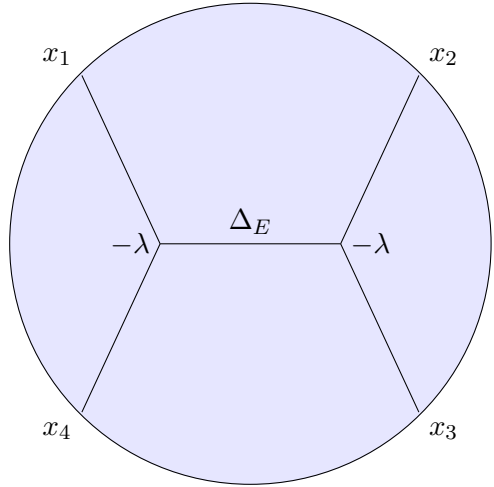

The integral we are interested in, corresponding to the Witten diagram \eqref{Witten exchange}, is
\begin{align}
	J^{(t)}(x_1,...,x_4) &= \int \frac{d^2w}{w_0^2}I(w,x_2,x_3)K_{\Delta_1}(x_1,w)K_{\Delta_4}(x_4,w),\\
	I(w,x_2,x_3) &= \int \frac{d^2z}{z_0^2}G_{\Delta_E}(w,z)K_{\Delta_2}(x_2,z)K_{\Delta_3}(x_3,z).\label{Eq: I three-point}
\end{align}
The integral $I(z_a,x_a;x_2,x_3)$ is a boundary-boundary-to-bulk three-point function and has conformal symmetry. 

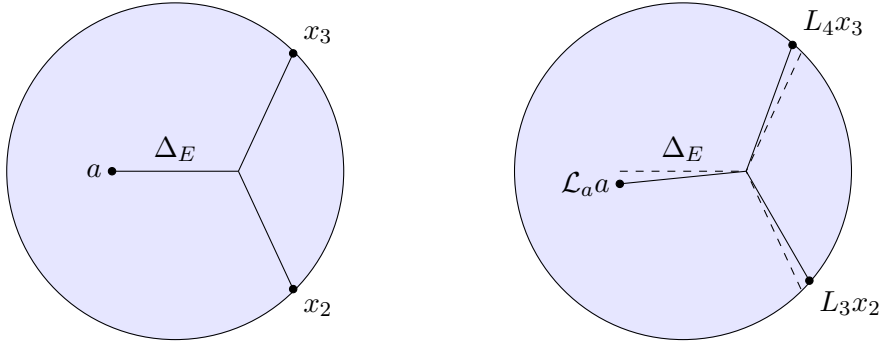
\begin{figure}[h!]
	\center
	\begin{tikzpicture}[scale=.7]
		\def\x{0.8}
		\filldraw[blue!10!white] (0,0) circle (4*\x cm);
		\draw[] (0,0) circle (4*\x cm);

		\draw[] (2.8*\x,2.8*\x) to (1.5*\x,0);
		\draw[] (2.8*\x,-2.8*\x) to (1.5*\x,0);
		\draw[] (-1.5*\x,0) to (1.5*\x,0);
		\filldraw[] (-1.5*\x,0) circle (2pt);
		\filldraw[] (2.8*\x,-2.8*\x)  circle (2pt);
		\filldraw[] (2.8*\x,2.8*\x)  circle (2pt);
		
		\node[anchor =east] at (-1.5*\x,0) {$a$};
		\node[anchor=south] at (0,0) {$\Delta_E$};
		
		\node[anchor=north west] at (2.8*\x,-2.8*\x) {$x_2$};
		\node[anchor=south west] at (2.8*\x,2.8*\x) {$x_3$};
	\end{tikzpicture}
	\hspace{2cm}
	\begin{tikzpicture}[scale=.7]
		\def\x{0.8}
		\filldraw[blue!10!white] (0,0) circle (4*\x cm);
		\draw[] (0,0) circle (4*\x cm);

		\draw[] (2.6*\x,3*\x) to (1.5*\x,0);
		\draw[] (3*\x,-2.6*\x) to (1.5*\x,0);
		\draw[] (-1.5*\x,-.3*\x) to (1.5*\x,0);
		\filldraw[] (-1.5*\x,-.3*\x) circle (2pt);
		\filldraw[] (3*\x,-2.6*\x)  circle (2pt);
		\filldraw[] (2.6*\x,3*\x)  circle (2pt);
		
		\node[anchor =east] at (-1.5*\x,-.3*\x) {$\mathcal{L}_a a$};
		\node[anchor=south] at (0,0) {$\Delta_E$};
		
		\node[anchor=north west] at (3*\x,-2.6*\x) {$L_3 x_2$};
		\node[anchor=south west] at (2.6*\x,3*\x) {$L_4 x_3$};

		\draw[dashed] (2.8*\x,2.8*\x) to (1.5*\x,0);
		\draw[dashed] (2.8*\x,-2.8*\x) to (1.5*\x,0);
		\draw[dashed] (-1.5*\x,0) to (1.5*\x,0);
	\end{tikzpicture}
	\caption{The system is invariant under an infinitesimal transformation \eqref{Eq: Infinitesimal transformation}.}
\end{figure}

\par 
$I(w,x_2,x_3)$ in \eqref{Eq: I three-point} is invariant under global transformations generated by $\mathbf{\mathcal{L}}_a+\vec{L}_2+\vec{L}_3$, where the first term generates the isometries of $\text{AdS}_2$ and the other two generate the conformal transformation of the boundary.\footnote{We write the transformations under the conformal group as vectors $\vec{L}_2 = (L_{-1},L_0,L_1)$ acting on point 2.}  As such, we can write
\begin{align}
	(\mathbf{\mathcal{L}}_a+\vec{L}_2+\vec{L}_3)I(z_a,x_a;x_2,x_3) = 0, \label{Eq: Infinitesimal transformation}
\end{align}
and can therefore relate the Casimirs of the generators
\begin{align}
	-\frac{1}{2}\mathbf{\mathcal{L}}_a^2 I (z_a,x_a;x_2,x_3) &=-\frac{1}{2} (\vec{L}_2+\vec{L}_3)^2I(z_a,x_a;x_2,x_3)\\
	&=C^{(23)}_2 I(z_a,x_a;x_2,x_3) .
\end{align}
In one dimension, the quadratic Casimir of the $\text{AdS}_2$ isometries is the Laplacian of the bulk \ref{App: quad Casimir}. This Laplacian will allow us to eliminate the bulk-to-bulk propagator through the equation of motion \eqref{Eq: Equation of motion ads2}. Linking the previous elements together, we obtain
\begin{align}
	(C_{(23)}^{(2)}-m^2_E)I(z_a,x_a;x_2,x_3) &= (\nabla_{a}^2-m_E^2)I(z_a,x_a;x_2,x_3)\\
	%&=\int \frac{dz_b dx_b}{z_b^2}	(\nabla_{AdS}^2-m_E^2) G_{BB}(a,b) K_{\Delta_\phi}(z_b,x_b;x_3) K_{\Delta_\phi}(z_b,x_b;x_4)\\
	%&=\int dz_b dx_b  \delta^{(2)}(a-b)  K_{\Delta_\phi}(z_b,x_b;x_3) K_{\Delta_\phi}(z_b,x_b;x_4)\\
	&= K_{\Delta_\phi}(z_a,x_a;x_2) K_{\Delta_\phi}(z_a,x_a;x_3).
\end{align}
The quadratic Casimir acting on points 3 and 4 commutes with the other coordinates so that we can write a differential equation relating the full exchange diagram to the contact term
\begin{align}
	(C_{(23)}^{(2)}-m^2_E)	J(x_1,x_2,x_2,x_3) &=\int \frac{dz_a dx_a}{z_a^2} \Pi_{i=1}^4K_{\Delta_\phi}(z_a,x_a;x_i)\\
	&= \frac{A}{(x_{13}x_{24})^{2\Delta_\phi}} \bar{D}_{\Delta_\phi \Delta_\phi \Delta_\phi \Delta_\phi}(\chi).
\end{align}
The same analysis holds for any two legs attached to a bulk-to-bulk propagator, though the final differential equation might depend on many variables.

\subsubsection*{Polyakov Blocks and Exchange Witten Diagrams}\label{App: Polyakov}
The computation of the exchange Witten diagram in the main text \eqref{Eq: Exchange solution Delta=1} is the sum of solutions to second-order differential equations. The three-point function and the symmetry of the correlator provide the integration constants. The example given in the main text, corresponding to the Polyakov block of external dimension $\Delta=1$ and exchanged dimension $\Delta_E=1$, can be computed with the toy model of a massless scalar theory in AdS$_2$ with a $\phi^3$ interaction. This corresponds to the action
\begin{align}
	S_{\phi^3} = \int \frac{dtdz}{z^2}\left( \partial_\mu \phi \partial^\mu \phi-\frac{\lambda}{3!} \phi^3\right).
\end{align}
GFF gives the leading order contribution. The NLO ($O(\lambda)$) correlators come from the constant vertex, giving the three-point function 
\begin{align}
	<\phi(x_1)\phi(x_2)\phi(x_3)> & =\frac{\lambda}{\pi^3} \frac{\pi^2}{2x_{12}x_{23}x_{13}},
\end{align}
and the OPE coefficient
\begin{align}
	c_{111}&= \frac{\lambda}{2\pi}.
\end{align}
The first sub-leading term ($O(\lambda^2)$) in the four-point function is generated by the exchange diagram, which contributes
\begin{align}
	<\phi(x_1)\phi(x_2)\phi(x_3)\phi(x_4)> &=\sum_{s,t,u} \lambda^2J_{\Delta_E=1,\Delta=1}(x_1,...,x_4)\\
	&=\frac{\lambda^2}{\pi^5}\frac{\pi}{4}\frac{1}{(x_{13}x_{24})^2}P^{(0)}_{(1,1)}(\chi),
\end{align}
where $P^{(0)}_{(1,1)}(\chi)$ is written explicitly in equation \eqref{Eq: Exchange solution Delta=1} and has a small $\chi$ expansion
\begin{align}\label{OPE Polyakov 2}
	<\phi(x_1)\phi(x_2)\phi(x_3)\phi(x_4)>&=\frac{1}{(x_{13}x_{24})^2} \frac{\lambda^2}{4\pi^4}\left(\frac{c_1-6 \zeta (3)}{\chi^2}+\frac{2 \left(\pi ^2-3 c_2 \right)}{3 \chi}\right)+O\left(\chi^0 \right),
\end{align}
where we have kept the integration constants from the differential equation. 
The first term in the expansion is set to zero since it corresponds to a correction to the weight of the identity operator. The second term corresponds, in the conformal s-channel OPE, to
\begin{align}\label{OPE Polyakov 1}
	\frac{1}{(x_{13}x_{24})^2}\sum_h c_{11h}^2 \chi^{h-2} {}_2F_1(h,h,2h,\chi) & = \frac{1}{(x_{13}x_{24})^2}\left(\frac{c_{111}^2}{\chi}+O\left(\chi^0 \right) \right),
\end{align}
where by equating the expansions in the conformal and Polyakov blocks, this corresponds to $c_{\Delta_E \Delta\Delta}$ in equation \eqref{Eq: Polyakov pert exp}. 
Equating equations \ref{OPE Polyakov 1} and \ref{OPE Polyakov 2} fixes the integration constants
\begin{align}
	c_1 = 6\zeta(3)\\
	c_2= -\frac{\pi^2}{6}. 
\end{align}
This provides the correct numerical factor for the Polyakov block computed in \ref{Eq: Exchange solution Delta=1} and the maximally symmetric and Regge-bounded function.

\chapter{Mellin Transform}
The Mellin transform is a topic in itself so the following Appendix starts with an introduction in section \ref{3 computations to understand Mellin formalism}, before using the 1d Mellin transform defined in the main text in subsection \ref{sec:nonpert} to consider the case of GFF in section \ref{GFF Mellin amplitude}. Section \ref{sum rules appendix} then tests the sum rules presented in \ref{sumrules} and addresses some of the issues linked to perturbations and poles. The final section \ref{Ap: anomalous dimension} lists the expressions for the anomalous dimensions obtained in the model described in \ref{sec:pert}. 

\section{3 Computations to Understand Mellin Formalism}\label{3 computations to understand Mellin formalism}
The simple Mellin transform is the following transform, closely related to the Laplace transform
\begin{align}
	M[g](s) &= \int_{0}^\infty \frac{dz}{z} g(z)z^s \\
	&=\int_{0}^\infty du \, g(e^u) \, e^{s u},
\end{align}
where the anti-Mellin transform is done through a contour integral parallel to the imaginary axis.
As the Fourier and Laplace transform, the Mellin transform reveals and highlights specific theory elements, making it sometimes simpler to spot patterns and solve correlators. From its inception, it has revealed itself to be very useful in hypergeometric integrals and their generalisations. Given the omnipresence of hypergeometric functions in the context of CFTs, one could expect there to be applications of the Mellin transform to CFT. The founding work in \cite{Mack:2009gy} revealed the power of Mellin transforms for dimension-independent formulations of conformal correlators. Since then, it has been a major tool and a very active area of research whose focus has been perturbative and non-perturbative computations.
Further work \cite{Penedones:2010ue,Rastelli:2016nze,Paulos:2016fap} developed the tools for efficient strong-coupling computations in this setting, as well as motivation to consider the Mellin transform of correlators as amplitudes by linking them to flat-space scattering amplitudes. The literature over the past decade and more is quite extensive and detailed, so the goal of this section is to be specific rather than extensive to give a feel for the Mellin transform.\footnote{I refer the interested reader to \cite{Penedones:2016voo, Penedones:2010ue, Penedones:2019tng} for a more exhaustive treatment of the subject.}\par 
\newpage 
 In this line, I will present three key computations in the Mellin formalism, these are:
\begin{itemize}
	\item The Mellin amplitude of contact Witten diagram
	\item From Mellin to position space
	\item Scattering amplitudes
\end{itemize}
From this, one can take away the key idea that Mellin formalism is useful because the quantities' simplicity underlines the theory's important aspects. At strong coupling, contact interactions in AdS give constant amplitudes, and exchanges operators give poles, which allow for a `Feynman Rules' approach. The main obstacle to the development of this formalism is that going from position to Mellin space and vice versa is relatively hard, even for the simplest quantities. This motivates the quest to understand how to compute quantities directly in Mellin space and relate them to physical observables, such as flat-space scattering amplitudes. 

\subsection*{Mellin Amplitudes of Contact Witten Diagrams}
We will approach this from the point of view of boundary correlators of operators propagating in a bulk theory on AdS$_{d+1}$.
We introduce a redundant coordinate system called embedding lightcone coordinates ($X_\mu X^\mu =-R^2$)
\begin{align}
	X^\mu &= R\left(\frac{1-x^2-z^2}{2z},\frac{x^\mu}{z},\frac{1+x^2+z^2}{2z}\right)&P^\mu = \lim_{z\rightarrow0} zX^\mu&
\end{align}
which map AdS$_{d+1}$ to flat space in d+2 dimensions. Note that $P^\mu$ parametrises the coordinates of the boundary CFT. These lead to easier integrals and simpler propagators. In these coordinates, for example, the bulk-to-boundary propagator is simply
\begin{align}
	G_{\Delta}(X,P) = \frac{C_{\Delta} }{\left(-2P\cdot X \right)^\Delta}
\end{align}
The $n-$point contact integral in this context is 
\begin{align}
	I_{\Delta_1,...,\Delta_n}(P_1^\mu,...,P_m^\mu) = \int_{\text{AdS}_{d+1}}dX \prod_{i=1}^{n} \frac{C_{\Delta_i}}{\left(-2P\cdot X\right)^{\Delta_i}}
\end{align}
which we can write as 
\begin{align}
	I_{\Delta_1,...,\Delta_n}(P_1^\mu,...,P_m^\mu) =  \prod_{i=1}^{n}  \left(\frac{C_{\Delta_i}}{\Gamma(\Delta_i)} \right)\int_0^\infty [dt]\prod_{i}^{n}t_i^{\Delta_i-1}\int dX e^{2Q\cdot X}
\end{align}
where
\begin{align}
	[dt] = \prod_{i=0}^{n} dt_i
	Q^\mu = \sum_{i=0}^n t_i P_i^\mu
\end{align}
and we use the Schwinger parametrisation 
\begin{align}
	\frac{1}{A^{p_i}}=\frac{1}{\Gamma(p_i)} \int_0^\infty  dt_i t_i^{p_i-1}e^{-t_i A}.
\end{align}
The boundary integral of AdS$_{d+1}$ is now easy:
\begin{align}
	\int dX e^{2Q\cdot X} = \pi^{\frac{d}{2}}\int \frac{dz}{z}z^{-\frac{d}{2}}e^{-z+\tfrac{Q^2}{z}}
\end{align}
where
\begin{align}
	Q^2 = (\sum_iP_i t_i)^2 = \sum_{i>j}-(-2P_i\cdot P_j)t_i t_j .
\end{align}
The total integral is now
\begin{align}
	I_{\Delta_1,...,\Delta_n}(P_1^\mu,...,P_m^\mu) =  \pi^{\tfrac d 2}\prod_{i=1}^{n}  \left(\frac{C_{\Delta_i}}{\Gamma(\Delta_i)} \right)\int_0^\infty [dt]\prod_{i}^{n}t_i^{\Delta_i-1}\int \frac{dz}{z}z^{-\frac{d}{2}}e^{-z-\sum_{i>j}P_{ij}\tfrac{t_i t_j}{z}}
\end{align}
where $P_{ij} = -2P_i \cdot P_j = |x_i-x_j|^2$. By rescaling the variables $s_i = \frac{t_i}{\sqrt{z}}$, the $z-$integral is simply a gamma function.
\begin{align}
	I_{\Delta_1,...,\Delta_n}(P_1^\mu,...,P_m^\mu) &=  \pi^{\tfrac d 2}\prod_{i=1}^{n}  \left(\frac{C_{\Delta_i}}{\Gamma(\Delta_i)} \right)\int_0^\infty [ds]\prod_{i}^{n}s_i^{\Delta_i-1}e^{-\sum_{i>j}P_{ij}s_i s_j }\int \frac{dz}{z}z^{\frac{\sum \Delta_i-d}{2}}e^{-z}\\
	&=A \int_0^\infty [ds]\prod_{i}^{n}s_i^{\Delta_i-1}e^{-\sum_{i>j}P_{ij}s_i s_j }
\end{align}
where 
\begin{align}
	A = \pi^{\tfrac d 2} \Gamma\left(\tfrac{\sum \Delta_i-d}{2}\right)\prod_{i=1}^{n}  \left(\frac{C_{\Delta_i}}{\Gamma(\Delta_i)} \right)
\end{align}
Since the conformal symmetry fixes the integral up to $n=3$, we separate the cases $n=1,2,3$ from the others to make some symmetry explicit. This is also where the Mellin transform comes into play; we introduce the Mellin transform of the exponential function
\begin{align}
	\exp^{-P_{ij}s_i s_j} = \int\frac{d\gamma_{ij}}{2\pi i} \frac{\Gamma(\gamma_{ij})}{(P_{ij}s_i s_j)^{\gamma_{ij}}}
\end{align}
to write
\begin{align}
	I &=A \int_0^\infty [ds]\prod_{i}^{n}s_i^{\Delta_i-1}e^{-s_i \sum_{j>1}P_{1j}s_j+s_2s_3P_{23} }e^{-\sum_{j>3,i\geq 2}P_{ij}s_i s_j }\\
	&=A\int_0^\infty \prod_{i=1}^{n}\left( \frac{ds_i}{s_i}s_i^{\Delta_i}\right) \exp^{-\sum_{j>1}s_1 s_j P_{1j}-s_2s_3P_{23}} \prod_{i\geq 2, j>3}^{n}\int \frac{d\gamma_{ij}}{2\pi i} \frac{\Gamma(\gamma_{ij})}{\left( P_{ij}s_is_j\right)^{\gamma_{ij}}}
\end{align}
Collecting the $s_i$-dependent factor for $j\geq4$, we can integrate them
\begin{align}
	\int_0^\infty \tfrac{ds_i}{s_i} s_i^{\Delta_i}\exp^{-s_1s_i P_{1i}} s_i^{-\sum_{j=2}^{i-1}\gamma_{j,i}-\sum_{j=i+1}^{n}\gamma_{i,j}} &= \int_0^\infty \tfrac{ds_i}{s_i} \exp^{-s_1s_i P_{1i}}s_i^{\gamma_{1i}}\\
	&=\frac{\Gamma(\gamma_{1i})}{(s_1P_{1i})^{\gamma_{1i}}}
\end{align}
where we have defined redundant variables through the following
\newpage
\begin{align}\label{Eq: Mellin variables}
	&\gamma_{ii} = -\Delta_i\\
	&\gamma_{ij} = \gamma_{ji}\\
	&\sum_{j=1}^n \gamma_{ij}= 0 .
\end{align}
Therefore, integrating $s_j$ ($j\geq4$), we obtain
\begin{align}
	I &=A \int \prod_{j>i>3}^{n} \frac{d\gamma_{ij}\Gamma(\gamma_{ij})}{\left(P_{ij}\right)^{\gamma_{ij}}} \prod_{ j=4}^{n} \frac{\Gamma(\gamma_{1j})}{\left(P_{1j}\right)^{\gamma_{1j}}}\frac{\Gamma(\gamma_{2j})}{\left(P_{2j}\right)^{\gamma_{2j}}}\frac{\Gamma(\gamma_{3j})}{\left(P_{3j}\right)^{\gamma_{3j}}}I_s \\
	I_s &= \int \frac{ds_1ds_2ds_3}{s_1s_2s_3} s_1^{\gamma_{12}+\gamma_{13}}s_2^{\gamma_{12}+\gamma_{23}}s_3^{\gamma_{13}+\gamma_{23}}\exp^{-s_1s_2P_{12}-s_1s_3P_{13}-s_2s_3P_{23}} \\
	&=  \int \frac{dt_1dt_2dt_3}{t_1t_2t_3} t_1^{\gamma_{23}}t_2^{\gamma_{13}}t_3^{\gamma_{12}}\exp^{-t_3P_{12}-t_2P_{13}-t_1P_{23}} \\
	&=\frac{\Gamma(\gamma_{12})}{\left(P_{12}\right)^{\gamma_{12}}}\frac{\Gamma(\gamma_{13})}{\left(P_{13}\right)^{\gamma_{13}}}\frac{\Gamma(\gamma_{23})}{\left(P_{23}\right)^{\gamma_{23}}}
\end{align}
This gives us the final result 
\begin{align}
	I =  \pi^{\tfrac d 2} \Gamma\left(\tfrac{\sum \Delta_i-d}{2}\right)\prod_{i=1}^{n}  \left(\frac{C_{\Delta_i}}{\Gamma(\Delta_i)} \right) \int\prod_{j>i}^{n} \frac{[d\gamma_{ij}]}{2\pi i}\frac{\Gamma(\gamma_{ij})}{\left(P_{ij}\right)^{\gamma_{ij}}} .
\end{align}
Here, the integral is over the non-redundant variables $\gamma_{ij}$ satisfying the relations in \ref{Eq: Mellin variables}. If one defines a truncated Mellin transform or Mellin amplitude where 
\begin{align}
	f(P_{ij}) =\int\prod_{j>i}^{n} \frac{[d\gamma_{ij}]}{2\pi i}\frac{\Gamma(\gamma_{ij})}{\left(P_{ij}\right)^{\gamma_{ij}}} M(\gamma_{ij})
\end{align}
this is the claim that contact Witten diagrams give a constant Mellin amplitude. 

\subsection*{From Mellin to Position Space}
As expected, doing the Mellin integral (or anti-Mellin transform) is trivial for the cases $n=1,2,3$:
\begin{itemize}
	\item $n=1$ \par
	\begin{align}
		I_{n=1,\Delta} = \pi^{\tfrac d 2} \Gamma\left(\tfrac{\Delta-d}{2}\right) \left(\frac{C_{\Delta}}{\Gamma(\Delta)} \right) \delta_{\Delta,0}
	\end{align}
	where $\sum_{i}\gamma_{1i} = -\Delta =0$ was used
	\item $n=2$ \par
	\begin{align}
		I_{n=2}(\Delta_1,\Delta_2,P_{12}=|\vec{x}_1-\vec{x}_2|^2) =\delta_{\Delta_1,\Delta_2}	\pi^{\tfrac d 2} \Gamma\left(\tfrac{2\Delta-d}{2}\right)\left(\frac{C_{\Delta}}{\Gamma(\Delta)} \right)^2  \frac{\Gamma{(\Delta)}}{\left(P_{12}\right)^{\Delta}} 
	\end{align}
	where $\gamma_{12}=\Delta_1=\Delta_2$
	\item $n=3$ \par
	\begin{align}
		I =  \pi^{\tfrac d 2} \Gamma\left(\tfrac{\sum \Delta_i-d}{2}\right)\prod_{i=1}^{n}  \left(\frac{C_{\Delta_i}}{\Gamma(\Delta_i)} \right) \frac{\Gamma(\gamma_{12})\Gamma(\gamma_{13})\Gamma(\gamma_{23})}{\left(P_{12}\right)^{\gamma_{12}}\left(P_{13}\right)^{\gamma_{13}}\left(P_{23}\right)^{\gamma_{23}}} 
	\end{align}
	where 
	\begin{align}
		\gamma_{12} &= \Delta_1+\Delta_2-\Delta_3\\
		\gamma_{13} &= \Delta_1+\Delta_3-\Delta_2\\
		\gamma_{23} &= \Delta_2+\Delta_3-\Delta_1
	\end{align}
\end{itemize}

For $n=4$, we obtain our first non-trivial integrals since solving the redundancy equations \ref{Eq: Mellin variables} leaves two variables free
\begin{align}
	\gamma_{ij} = \begin{pmatrix}-\Delta_1 &\gamma_{12}&\Delta_1-\gamma_{12}-\gamma_{14}&\gamma_{14}\\ &-\Delta_2 &\gamma_{14}+\tfrac 12 \Delta_{23,14}&\tfrac 12 \Delta_{124,3}-\gamma_{12}-\gamma_{14} \\ &&-\Delta_3&\gamma_{12}-\frac 12\Delta_{12,34}\\ &&&-\Delta_4
	\end{pmatrix}
\end{align}
where 
\be
\Delta_{a_1 a_2 .. ,b_1 b_2..} =\sum_i \Delta_{a_i}-\sum_j \Delta_{b_j}
\ee
\begin{align}
	I_{n=4,\Delta_i}(P_{ij}) &= \pi^{\tfrac d 2} \Gamma\left(\tfrac{\sum \Delta_i-d}{2}\right)\prod_{i=1}^{n}  \left(\frac{C_{\Delta_i}}{\Gamma(\Delta_i)} \right) 
	P_{13}^{-\Delta_1} P_{23}^{\frac{1}{2} \Delta_{14,23}} P_{24}^{-\frac{1}{2} \Delta_{124,3}} P_{34}^{\frac{1}{2} \Delta_{12,34}} \int \frac{d\gamma_{12}d\gamma_{14}}{(2\pi i)^2}M(\gamma_{12},\gamma_{14})\\
	M(\gamma_{12},\gamma_{14})&=\left(\frac{P_{14} P_{23}}{P_{13} P_{24}}\right)^{-\gamma _{14}} \left(\frac{P_{12} P_{34}}{P_{13} P_{24}}\right)^{-\gamma _{12}} \prod_{j>i=1}^{4}\Gamma\left(\gamma_{ij}\right)
\end{align}
where the conformal cross-ratios 
\begin{align}
	u &= \frac{P_{12} P_{34}}{P_{13} P_{24}}\\
	v&=\frac{P_{14} P_{23}}{P_{13} P_{24}}
\end{align}
and conformal prefactors appear naturally in this formalism. Additionally, the crossing of operators is related to the exchange of the Mellin variables $\gamma_{ij}$. Additionally, each cross-ratio is in 1-1 correspondence with a non-redundant Mellin variable. \par For simplicity, we will keep to the case where $\Delta_i = \Delta$, so we have 
\begin{align}
	I_{n=4,\Delta_i}(P_{ij}) &= \pi^{\tfrac d 2} \Gamma\left(\tfrac{4\Delta-d}{2}\right) \left(\frac{C_{\Delta_i}}{\Gamma(\Delta_i)} \right)^4 
	P_{13}^{-\Delta} P_{24}^{- \Delta} \int \frac{d\gamma_{12}d\gamma_{14}}{(2\pi i)^2}M(\gamma_{12},\gamma_{14})\\
	M(\gamma_{12},\gamma_{14})&=v^{-\gamma _{14}} u^{-\gamma _{12}}  \Gamma\left(\gamma_{12}\right)^2  \Gamma\left(\gamma_{14}\right)^2 \Gamma\left(\Delta-\gamma_{12}-\gamma_{14}\right)^2
\end{align}
To solve this integral, one can relate the Mellin integral to an Euler-type integral through Hypergeometric functions (see  \cite{Gradshteyn:1943cpj}):
\begin{align}
	{}_2F_1(a,b,c,z) & = \frac{\Gamma(c)}{\Gamma(a)\Gamma(b)} \int \frac{ds}{2\pi i} \frac{\Gamma(s)\Gamma(1-s)\Gamma(b-s)}{\Gamma(c-s)} (-z)^{-s}\\
	&=\frac{\Gamma(c)}{\Gamma(a)\Gamma(b)} \int \frac{ds}{2\pi i} \frac{\Gamma(s)\Gamma(a-s)\Gamma(b-s)\Gamma(c-a-b+s)}{\Gamma(c-a)\Gamma(c-b)} (1-z)^{-s}\\
	&=	\frac{\Gamma(c)}{\Gamma(b)\Gamma(c-b)}\int_0^1 \frac{dt}{t(1-t)}t^b(1-t)^{c-b}(1-zt)^{-a}.
\end{align}
We can therefore rewrite the integral
\begin{align}
	\int \frac{d\gamma_{12}d\gamma_{14}}{(2\pi i)^2}&M_{n=4,\Delta}(\gamma_{12},\gamma_{14})\non \\
	& =  \int \frac{d\gamma_{12}d\gamma_{14}}{(2\pi i)^2} \Gamma\left(\gamma_{12}\right)^2\Gamma\left(\gamma_{14}\right)^2\Gamma \left(\Delta-\gamma_{12}-\gamma_{14}\right)^2 u^{-\gamma_{12}}v^{-\gamma_{14}}\\
	&=\int \frac{\gamma_{14}}{2\pi i} \Gamma\left(\gamma_{14}\right)^2\Gamma \left(\Delta-\gamma_{14}\right)^2\int_0^1 \frac{dt}{t(1-t)}\left(\frac{t(1-t)}{1-(1-u)t}\right)^{\Delta-\gamma_{14}}\\
	&=\frac{\Gamma(\Delta)^4}{\Gamma(2\Delta)v^\Delta} \int_0^1 \frac{dt}{t(1-t)}\left(\frac{vt(1-t)}{1-(1-u)t}\right)^{\Delta}{}_2F_1(\Delta,\Delta,2\Delta,1-\frac{v t(1-t)}{1-(1-u)t})\\
	&=\frac{\Gamma(\Delta)^4}{\Gamma(2\Delta)v^\Delta} \int_0^1 \frac{dt}{t(1-t)}X^{\Delta}{}_2F_1(\Delta,\Delta,2\Delta,1-X).
\end{align}
In the familiar case $\Delta=1$ this is 
\begin{align}
	\int \frac{d\gamma_{12}d\gamma_{14}}{(2\pi i)^2}M_{n=4,\Delta}(\gamma_{12},\gamma_{14})& =   \int_0^1 dt\left(\frac{1}{1-(1-u)t-vt(1-t)}\right)\log\left( \frac{v t(1-t)}{1-(1-u)t} \right)\\
	& = \int_0^1 dt\left(\frac{1}{(1-t(1-\chi))(1-t(1-\bar{\chi}))}\right)\log\left( \frac{(1-t)t(1-\chi)(1-\bar{\chi})}{1-t(1-\chi \bar{\chi})}\right)
\end{align}
where
\begin{align}
	u&=\chi\bar{\chi}&v&=(1-\chi)(1-\bar{\chi})
\end{align} and the remaining integral can be evaluated easily using partial fractions to give to usual box integral result
\begin{align}
	\int \frac{d\gamma_{12}d\gamma_{14}}{(2\pi i)^2}M_{n=4,\Delta=1}(\gamma_{12},\gamma_{14}) = \frac{1}{2(\chi-\bar{\chi})} \left( \log(\chi\bar{\chi})\log(\frac{1-\chi}{1-\bar{\chi}})+2\Li(\chi)-2\Li(\bar{\chi})\right)
\end{align}
Explicitely computing the explicit Mellin and Anti-Mellin transform is often impractical, especially when considering more complex cases. Instead, using Mellin amplitudes directly to find quantities such as CFT data and scattering amplitudes is more convenient. 
\subsection*{Scattering Amplitudes}
The conditions on the Mellin variables in \eqref{Eq: Mellin variables} can be written as the on-shell conserved momenta of  $n$ particles of mass $m_i^2=\Delta_i$:
\begin{align}
	&\gamma_{ij} = p_i\cdot p_j\\
	&p_j \cdot \sum p_i = 0\\
	&p_i^2 = -m^2
\end{align}
The Mellin amplitudes can then be expressed in terms of Mandelstam variables
\begin{align}
	&-(p_1+p_2)^2 = \Delta_1+\Delta_2-2\gamma_{12} = s\\
	&-(p_1+p_3)^2 = \Delta_1+\Delta_3-2\gamma_{13}  =t\\
	&-(p_1+p_4)^2 = \Delta_1+\Delta_4-2\gamma_{14} = u\\
	&s+t+u = \sum_i \Delta_i
\end{align}
In \cite{Penedones:2010ue, Fitzpatrick:2011hu}, the Mellin amplitudes for Witten diagrams were related to flat-space amplitudes in a certain limit:
\begin{align}\label{Eq: Penedones flat space}
	\frac{\mc{T}_n(l_s,k_i)}{l_s^{n\tfrac{d-1}{2}-d-1}} = \lim_{\theta\rightarrow \infty} \frac{1}{\mc{N}} \int \frac{d\alpha}{2\pi i} \alpha^{\frac{d-\sum \Delta_i}{2}}\text{e}^{\alpha}M_n(\gamma_{ij}=\frac{\theta^2}{2\alpha} l_s^2 k_i \cdot k_j) \theta^{n\frac{d-1}{2}-d-1}
\end{align}
where $l_s$ is a scattering length intrinsic to the theory, $k_i$ are the external momenta, $\mc{N}$ is a normalisation factor to match with flat-space amplitudes.  The limit in $\theta=\frac{R}{l_s}$ controls the flat-space limit and suppresses the disconnected diagram contribution naturally, finally the integral in $\alpha$, like the anti-Mellin transform is also on a contour parallel to the imaginary axis.
We will illustrate this formula in the following examples.
\subsubsection*{Disconnected Contributions}
For disconnected contributions, the Mellin amplitude has no integration variables, so the replacement conditions for the Mellin variables in equations \ref{Eq: Mellin variables} give the momentum conservation condition expected from flat-space scattering.
\subsubsection*{Contact Interactions}
The Mellin transform of a four-point function with a contact interaction $-\lambda_4\phi_\Delta^4$is 
\begin{align}
	M_{n=4,\Delta} =  \lambda_4\pi^{\tfrac d 2} \Gamma\left(\tfrac{4\Delta-d}{2}\right) \left(\frac{C_{\Delta_i}}{\Gamma(\Delta_i)} \right)^4 
	P_{13}^{-\Delta} P_{24}^{- \Delta} \Gamma\left(\gamma_{12}\right)^2  \Gamma\left(\gamma_{14}\right)^2 \Gamma\left(\Delta-\gamma_{12}-\gamma_{14}\right)^2
\end{align}
The associated `truncated' Mellin amplitude is
\begin{align}
	\hat{M}_{n=4,\Delta} &= \lambda_4\pi^{\tfrac d 2} \Gamma\left(\tfrac{\sum \Delta-d}{2}\right) \left(\frac{C_{\Delta_i}}{\Gamma(\Delta_i)} \right)^4 \\
	\hat{M}_{n,\Delta} &= \lambda_n \mc{N} \Gamma\left(\tfrac{\sum \Delta-d}{2}\right)
\end{align}
However, the interaction constant $\lambda_4$ is a dimensionful parameter. To make it dimensionless, we define a new interaction constant through dimensional analysis by using the only dimensionful parameter of our theory: the radius $R$ of AdS:
\begin{align}
	\tilde{\lambda_n} &= \lambda_n R^{d+1+n\frac{1-d}{2}}\\
	&=\lambda_n \left(\theta l_s \right)^{d+1+n\frac{1-d}{2}}
\end{align}
The scattering amplitude formula \eqref{Eq: Penedones flat space} then gives
\begin{align}
	\frac{\mc{T}_n(l_s,k_i)}{l_s^{n\tfrac{d-1}{2}-d-1}} &= \lim_{\theta\rightarrow \infty} \frac{1}{\mc{N}} \int \frac{d\alpha}{2\pi i} \alpha^{\frac{d-\sum \Delta_i}{2}}\text{e}^{\alpha}\tilde{\lambda}_n \mc{N}\Gamma\left(\tfrac{\sum \Delta-d}{2}\right)\left(\theta l_s \right)^{d+1+n\frac{1-d}{2}} \theta^{n\frac{d-1}{2}-d-1}\\
	\mc{T}_n(l_s,k_i) &= \tilde{\lambda}_n  \int \frac{d\alpha}{2\pi i}\alpha^{\frac{d-\sum \Delta_i}{2}}\text{e}^{\alpha}\Gamma\left(\tfrac{\sum \Delta-d}{2}\right)\\
	&= \tilde{\lambda}_n
\end{align}

\subsubsection*{Derivative Interactions}
Though the preceding example may seem slightly artificial, derivative interactions are where this formalism starts showing some of its power (as well as exchange diagrams). Using the identity
\begin{align}
	\partial_\mu K_{\Delta_1} \partial^\mu K_{\Delta_2} = \Delta_1\Delta_2 (K_{\Delta_1}K_{\Delta_2}-2P_{12}K_{\Delta_1+1}K_{\Delta_2+1}),
\end{align}
the diagram corresponding to the Wick contraction such that the derivatives of an interaction such as $\lambda_{4,2}(\partial \phi_\Delta)^2\phi_\Delta^2$ are acting on the legs $1$ and $2$ gives the conformal integral 
\begin{align}
	\lambda_{4} \Delta_1 \Delta_2 (D_{\Delta_1\Delta_2\Delta_3\Delta_4}-2P_{12}D_{\Delta_1\Delta_2\Delta_3+1\Delta_4+1}). 
\end{align}
Which gives the Mellin integral
\begin{align}
	\lambda_{4,2} \Delta_1 \Delta_2 \int [d\gamma_{ij}] \prod_{i<j} \frac{\Gamma(\gamma{ij})}{(P_{ij})^{\gamma_{ij}}}\left(1-2P_{12}\frac{\Gamma(\gamma_{12}+1)\Gamma\left(\tfrac{\sum \Delta+2-d}{2}\right)\Gamma(\Delta_1)\Gamma(\Delta_2))}{\Gamma(\gamma_{12})P_{12}\Gamma\left(\tfrac{\sum \Delta-d}{2}\right)\Gamma(\Delta_1+1)\Gamma(\Delta_2+1)}\right)
\end{align}
and the truncated Mellin amplitude
\begin{align}
	M_{\partial^2,n=4}&=\tilde{\lambda_{4,2}} \Delta_1 \Delta_2 \left(\theta l_s \right)^{d+1+n\frac{1-d}{2}-2} \left(1-2\frac{\gamma_{12}}{\Delta_1\Delta_2}\tfrac{\sum \Delta+2-d}{2}\right)\mc{N} \Gamma\left(\tfrac{\sum \Delta-d}{2}\right)
\end{align}
Notice that the effective interaction term $\tilde{\lambda}_{4,2}$ is additionally suppressed due to the derivative interactions, so the constant term above will vanish in the $\theta\rightarrow \infty$ limit. The other term gives 

\begin{align}\label{Eq: Penedones flat space}
	\frac{\mc{T}_{\partial^2,n=4}(l_s,k_i)}{l_s^{n\tfrac{d-1}{2}-d-1}} &= -2k_1 \cdot k_2\tilde{\lambda_{4,2}}\Gamma\left(\tfrac{\sum \Delta-d}{2}+1\right) l_s^{d+1+n\frac{1-d}{2}}   \int \frac{d\alpha}{2\pi i} \alpha^{\frac{d-\sum \Delta_i}{2}}\text{e}^{\alpha}\frac{1}{2\alpha} \\
	\mc{T}_{\partial^2,n=4}(l_s,k_i)&= -\tilde{\lambda_{4,2}} k_1 \cdot k_2
\end{align}
Which corresponds to the contribution from the equivalent Wick contractions on a Feynman diagram with the same interaction term. It is easy to see that the $\theta$-suppression means that for  a generic interaction term, only the term with the highest power of $\gamma_{ij}$ survives so that for a product of fields $\phi_i$ pairwise contracted with a derivative power $\alpha_{i,j}$, the resulting flat-space amplitude will be 
\begin{align}
	\mc{T}_n = \lambda_n \prod_{i,j}\left(k_i\cdot k_j \right)^{\alpha_{i,j}}
\end{align}
\section{Generalized Free Field Theory}\label{GFF Mellin amplitude}
\label{sec:GFF}
\begin{figure}[h]
	\center
	\begin{tikzpicture}
		\draw[->](-3,0)--(3,0);
		\draw[->](0,-2)--(0,2);
		\draw[] (2.6,2)--(2.6,1.7);
		\draw[] (3,1.7)--(2.6,1.7);
		\node[anchor = north east] at (3,2.05) {$s$};
		\filldraw [red] (1,0) circle (1.5pt);
		\filldraw [red] (0,0) circle (1.5pt);
		\filldraw [red] (-0.7,0) circle (1.5pt);
		\filldraw [red] (-1.4,0) circle (1.5pt);
		\filldraw [red] (-2.1,0) circle (1.5pt);
		\draw (0,0) node[cross=2pt,green!50!black] {};
		\draw (1,0) node[cross=2pt,green!50!black] {};
		\draw (1.7,0) node[cross=2pt,green!50!black] {};
		\draw (2.4,0) node[cross=2pt,green!50!black] {};
		\node[below] at (1,0) {\footnotesize{$2\Delta_\phi$}};
	\end{tikzpicture}\quad
	\begin{tikzpicture}
		\draw[->](-3,0)--(3,0);
		\draw[->](0,-2)--(0,2);
		\draw[] (2.6,2)--(2.6,1.7);
		\draw[] (3,1.7)--(2.6,1.7);
		\node[anchor = north east] at (3,2.05) {$s$};
		\node[blue, right] at (1.1,-1.5) {$\mathcal{C}$};
		\filldraw [red] (1,0) circle (1.5pt);
		\filldraw [red] (0,0) circle (1.5pt);
		\filldraw [red] (-0.7,0) circle (1.5pt);
		\filldraw [red] (-1.4,0) circle (1.5pt);
		\filldraw [red] (-2.1,0) circle (1.5pt);
		\draw (0.2,0) node[cross=2pt,green!50!black] {};
		\draw (1.2,0) node[cross=2pt,green!50!black] {};
		\draw (1.9,0) node[cross=2pt,green!50!black] {};
		\draw (2.6,0) node[cross=2pt,green!50!black] {};
		\node[above, green!50!black] at (0.3,0) {\footnotesize{$\e$}};
		\node[below, red] at (0.8,0) {\footnotesize{$2\Delta_\phi$}};
		\node[above, green!50!black] at (1.5,0) {\footnotesize{$2\Delta_\phi\!\!+\!\!\e$}};
		\draw [->,thick,blue] (1.1,-2) to[out=90,in=-90](1.1,-1) ;
		\draw [,thick,blue] (1.1,-1) to[out=90,in=-90](1.1,0) to[out=90,in=0](.9,.35) to [out=180, in=60] (0.6,0)to [out=60-180, in =0](.3,-.35) to[out=180, in=-90] (.1,0)to[out=90, in = -90](.1,2);
	\end{tikzpicture}\quad
	\caption{\textbf{Left}: The configuration of the poles for GFF. The left and right poles coincide in $s=0$ and $s=2 \Delta_\phi$. \textbf{Right}: the right poles are slightly shifted to the right so that the contour $\mathcal{C}$ can run in the middle.}\label{fig:GFFpoles}
\end{figure}
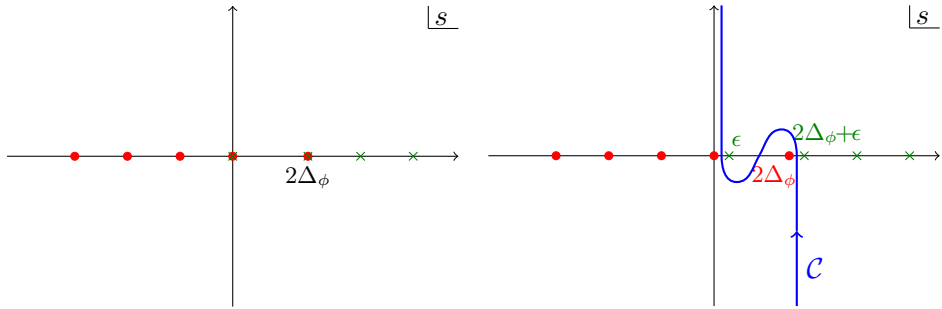

Let us consider the simplest possible example of 1d CFT. The GFF correlator for four identical scalars of dimension $\Delta_\phi$ reads
\begin{equation}\label{fGFF}
	f^{\text{GFF}}(t)=1+ t^{2\Delta_\phi}+\left(\frac{t}{1+t}\right)^{2\Delta_\phi}.
\end{equation}
Of course, in this case, we know very well the spectrum of exchanged operators, which includes the identity,  ``two-particle'' operators of the schematic form 
\be\label{doubletrace}
[\phi\phi]_n\sim\phi\, \square^n \phi,
\ee 
and their conformal descendants. The conformal primary operators have dimension $\D_n=2\Delta_\phi+2n$ for $n\geq 0$. As usual, the fact that the dimensions of these operators are separated by integers creates huge degeneracies. In particular, we should be worried by the coincidence of left and right poles in \eqref{leftpoles} and \eqref{rightpoles}. This happens for $s=0$, where the first right pole (associated to the exchange of the identity in the $s$-channel) coincides with the second left pole (associated to the exchange of $\phi^2$ in the $t$-channel) and at $s=2\Delta_\phi$ for the crossing symmetric case. The situation is depicted in Figure \ref{fig:GFFpoles}.\par 

Coincident poles generate a problem in the realization of the contour $\mathcal{C}$, which was defined precisely in such a way as to separate the left and right poles. To avoid this issue, one can slightly split the poles by shifting the right (or left) poles by a small amount, as shown in Figure \ref{fig:GFFpoles}. The result of the procedure outlined in subsection \ref{sec:nonpert} leads to a Mellin transform which depends on the shift $\e$
\begin{equation}\label{resultGFFepsilon}
	M^{\text{GFF}}_{\e}(s)=\frac{1}{\Gamma(s)\Gamma(2\Delta_\phi-s)}\left(\frac{\Gamma(s)\Gamma(2\Delta_\phi+\e-s)}{\Gamma(2\Delta_\phi)}+\frac{1}{s-\e}+\frac{1}{s-2\Delta_\phi-\e}-\frac{1}{s}-\frac{1}{s-2\Delta_\phi}\right) \, .
\end{equation}
Taking the inverse Mellin transform \eqref{inverseMellin} with the contour $\mathcal{C}$ depicted in Figure \ref{fig:GFFpoles} and then taking the limit $\e\to0$, one recovers \eqref{fGFF}. Taking the limit $\e\to0$ before integrating leads to the straightforward result $M_0(s)=\frac{1}{\Gamma(2\Delta_\phi)}$; however, the contour $\mathcal{C}$ is not well-defined and one cannot consistently take the inverse Mellin transform.\footnote{However, the correction to this is of order $M_{\epsilon}(s)-M_0(s) =\frac{\epsilon}{s^2}+\frac{\epsilon}{(s-2\Delta_\phi)^2}$ which is only non-negligible in the $s\rightarrow 0$ and $s\rightarrow2\Delta_\phi$ limit. Therefore, for all other purposes (for example, for comparisons at perturbative level or sum rule results), we can effectively take the $M^{GFF}_0(s)$ result.} Notice that this procedure is simply the limit in the weak sense normally considered in distribution theory. In particular, the identity 
\begin{equation}
	\lim_{\e\to0}\int_{p-i \infty}^{p+i\infty}\frac{ds}{2\pi i } \left(\frac{1}{s-p-\e}-\frac{1}{s-p+\e}\right) t^{s}=t^p
\end{equation}
tells us that the function $\frac{1}{s-p-\e}-\frac{1}{s-p+\e}$ in the limit $\e\to 0$ defines a delta function $\delta(s-p)$ for Mellin integration.\footnote{Attentive readers will probably have noticed that the individual terms in $\lim_{\e\to0}\int_{p-i \infty}^{p+i\infty}\frac{ds}{2\pi i } (\frac{1}{s-p-\e})$ diverge and one cannot use the residue theorem. However, the combination does not have this problem since $\left(\frac{1}{s-p-\e}-\frac{1}{s-p+\e}\right) = \frac{2\epsilon}{(s-p)^2-\e^2}$ which ensures convergence at large $s$.} In this sense, we can say that the Mellin transform for GFF is a distribution rather than a function.
This issue is caused by the exchange of the identity operator and by the degeneracy typical of GFF. 
An alternative route for obtaining the Mellin transform of the GFF correlator would be to use equations \eqref{psi0sum} and \eqref{psiinfsum} with the well-known expressions of the GFF OPE coefficients
\begin{equation}\label{GFFOPE}
	c^{(0)}_n =\frac{2 \Gamma (2 n+2 \Delta_\phi )^2 \Gamma (2 n+4 \Delta_\phi -1)}{\Gamma (2 \Delta_\phi )^2 \Gamma (2 n+1) \Gamma (4 n+4 \Delta_\phi -1)}
\end{equation}
and summing over $\D_n=2\Delta_\phi+2n$. Doing this, one finds the result $M(s)=\frac{1}{\Gamma(2\Delta_\phi)}$ which is consistent with the fact that equations \eqref{psi0sum} and \eqref{psiinfsum} were obtained under the assumption of non-coincident poles. Also, in this case, one could slightly shift the position of the poles in $\psi_{0}(s)$ (or $\psi_{\infty}(s)$) and recover an expression which coincides with $M_{\e}^{GFF}$ in the weak limit $\e\to 0$.

\section{Sum Rules for the 1d Mellin Transform}\label{sum rules appendix}

Testing the sum rules \eqref{sumrulesnonpert} or  \eqref{sumrulesgenericF} on a fully nonperturbative spectrum is a task which is momentarily out of reach. Therefore, we start by testing it on the simplest possible 1d CFT, i.e. generalised free field theory and perturbations thereof. 

\subsection*{Generalized free theories} 
\label{subsubsec:GFFsumrules}
One may immediately raise several objections to our attempt to apply the sum rules \eqref{sumrulesgenericF} to GFF theories. First of all, as discussed in section \ref{sec:GFF}, the definition of the Mellin amplitude for GFF requires the inclusion of a cut-off to regulate the exchange of the identity operator. Furthermore, a crucial assumption in deriving the sum rules was that $M(s)$ must have zeros for $s=-k$ and $s=2\Delta_{\phi}+k$. Still, GFF is precisely the example where two-particle  operators with dimension $2\Delta_{\phi}+2n+k$ are exchanged; therefore, those zeros are absent. We will see that these issues can be avoided by considering the function
\begin{align} \label{GFFkernel}
	F_p(s)=\frac{1}{(2\Delta_{\phi}+2p-s)(2\Delta_{\phi}+2p-s+1)} \, .
\end{align}
Since this function has two poles and no residue at infinity, we have the property
\begin{align}
	\text{Res}_{s=2\Delta_{\phi}+2p} (F_p(s))+ \text{Res}_{s=2\Delta_{\phi}+2p+1} (F_p(s))=0 \, .
\end{align}
Furthermore, one can easily check from \eqref{resultGFFepsilon} that $M_{\epsilon}^{\text{GFF}}(2\Delta_{\phi}+2p)=M_{\epsilon}^{\text{GFF}}(2\Delta_{\phi}+2p+1)$ for $p\in \mathbb{N}$. Combining these two properties, it is clear that the last term in \eqref{functionalwithM} vanishes even though $M(s)$ has no zeros at the positions of the poles of $F_p(s)$. We are then left with a sum over the residues of $M(s)$. For GFF, the position of the poles in principle depends on the regulator $\epsilon$, but the regulator's role in \eqref{resultGFFepsilon} is to separate the left and right poles. This is precisely what we have done to go from \eqref{functionalwithM} to \eqref{sumrulerightpoles}. Therefore, equation \eqref{sumrulesnonpert} can be used with $\e\to 0$ and, inserting the GFF spectrum $\Delta=2\Delta_{\phi}+2n$ we end up with the following sum rule
\begin{align}
	\sum_{n,k} c^{(0)}_n \tfrac{(-1)^{k+1}\Gamma(4(\Delta_{\phi}+n))\Gamma(2\Delta_{\phi}+2n+k)}{\Gamma(2\Delta_{\phi}+2n)^2\Gamma(4(\Delta_{\phi}+n)+k)\Gamma(-2n-k)\Gamma(k+1)}\tfrac{2 (2 \Delta_{\phi} +4 p+1) (\Delta_{\phi} +k+2 n)}{(k+2 n-2 p-1) (k+2 n-2
		p) (2 \Delta_{\phi} +k+2 n+2 p+1) (k+2 (\Delta_{\phi} +n+p))}=0 \,  .
\end{align}
Notice that the $\Gamma(-2n-k)$ factor in the denominator kills all the terms in this sum that are not compensated by a pole in the second ratio. The sum over $k$ then receives only two contributions at $k=2p-2n$ and $k=2p-2n-1$, which are present only for $p\geq n$. We are then left with a finite sum over $n$
\begin{align} \label{bosonicGFFsumrule}
	\sum_{n=0}^{p}c_n^{(0)}\frac{ \Gamma (2 p+1) \Gamma (4 (n+\Delta_{\phi} )) \Gamma (2 (p+\Delta_{\phi} )) \left(\Delta_{\phi} -2 n^2-4 \Delta_{\phi}  n+n+2 \Delta_{\phi}  p\right)}{\Gamma (2 (n+\Delta_{\phi} ))^2 \Gamma (2(p-n+1)) \Gamma (2 (n+p+2 \Delta_{\phi}+1 ))}=0 \, .
\end{align}
Each value of $p$ leads to an equation for $c_p^{(1)}$ in terms of all the other OPE coefficients with a lower index. In other words, this sum rule can be solved recursively for $c^{(0)}_n$ determining everything in terms of $c^{(0)}_0$, which sets the overall normalization. Doing so, one easily finds that the unique solution to this sum rule is provided by \eqref{GFFOPE}. The same strategy can be used to determine the OPE coefficients for a fermionic GFF and a GFF with $O(N)$ symmetry.  For these theories, sum rules are obtained inserting~\eqref{GFFkernel} into~\eqref{functional} and using the GFF spectrum, and read
\begin{align}
	\!\!\!\! \sum_{\D,k} c_\D \frac{(-1)^{k+1}\Gamma(2\D)\Gamma(\D+k)}{\Gamma(\D)^2\Gamma(2\D+k)\Gamma(2\Delta_\phi-\D-k)\Gamma(k+1)}(\frac{1}{\Delta+k+p}-\frac{1}{2\Delta_\phi-\Delta-k+p})=0\,.
\end{align}
The GFF spectrum $\Delta = 2\Delta_\phi+n$ for exchanged operators has the effect of truncating the sum above because of the factor of $\Gamma(2\Delta_\phi-\Delta-k)$ in the denominator. The example of bosonic GFF in section \ref{subsubsec:GFFsumrules} illustrates the case of even $n$, namely $\Delta=2\Delta_\phi+2n$.  Here we extend this to  odd integers, namely  $\Delta=2\Delta_\phi+2n+1$,  in covering the free fermionic model. We also consider the free bosonic and fermionic models with O($N$) symmetry.

\bigskip

In the case of a free fermion theory, the spectrum $\Delta=2\Delta_\phi+2n+1$ of exchanged operators leads to the sum rule
\begin{align} \label{Sum rule GFF fermions}
	\sum_{n=0}^p c_{n}\frac{2 \Gamma (2 p+1) \Gamma (2 (p+\Delta_\phi )) \left(\Delta_\phi -2 \Delta_\phi  p+2 n^2+4 \Delta_\phi  n+n\right) \Gamma (4 n+4 \Delta_\phi +2)}{\Gamma (2 p-2 n+1) \Gamma (2 n+2 \Delta_\phi +1)^2 \Gamma (2 (p+n+2 \Delta_\phi +1))}=0.
\end{align}
Just like in the bosonic case, see~\eqref{bosonicGFFsumrule}, this is a recursive relation for the OPE coefficients, whose general solution is
\begin{align}\label{OPE fermion}
	c_n^{(0)}=\frac{2 \Gamma (2 n+2 \Delta_\phi +1)^2 \Gamma (2 n+4 \Delta_\phi )}{\Gamma (2 \Delta_\phi ) \Gamma (2 \Delta_\phi +1) \Gamma (2 n+2) \Gamma (4 n+4 \Delta_\phi +1)}\,.
\end{align}
This is confirmed by the vanishing of \eqref{Sum rule GFF fermions}. As usual, this is true up to an overall scaling, and the choice in \eqref{OPE fermion} is set by requiring $c_0^{(0)}=1$.

\bigskip

Consider a bosonic four-point function with $O(N)$ symmetry. We write the Mellin amplitude as  a sum of the singlet, antisymmetric and traceless symmetric contributions
\begin{align}
	\hat{\mathcal{M}}^{1234}(s)  = 	\hat{\mathcal{M}}^S\delta^{12}\delta^{34}+\hat{\mathcal{M}}^A(\delta^{13}\delta^{24}-\delta^{14}\delta^{23})+\hat{\mathcal{M}}^T(\frac{\delta^{13}\delta^{24}+\delta^{14}\delta^{23}}{2}-\frac{\delta^{12}\delta^{34}}{N}),
\end{align}
with scalar coefficient functions $\hat{\mathcal{M}}^{S}(s)$, $\hat{\mathcal{M}}^{A}(s)$ and $\hat{\mathcal{M}}^{T}(s)$.  
In these channels, exchanged operators will be of the form $\phi_i\partial^{2n}_x\phi^i$,$\phi^{[i}\partial^{2n+1}_x\phi^{j]}$,$\phi^{(i}\partial^{2n}_x\phi^{j)}$ respectively, with same spectra of exchanged operators previously seen ($\Delta=2\Delta_\phi+2n$,  $\Delta=2\Delta_\phi+2n+1$,   $\Delta=2\Delta_\phi+2n$ respectively).   Therefore we get the corresponding OPE coefficients 
\begin{align}
	c_n^{S} & = 	c^S_0 \frac{2 \Gamma (2 n+2 \Delta_\phi )^2 \Gamma (2 n+4 \Delta_\phi -1)}{\Gamma (2 \Delta_\phi )^2 \Gamma (2 n+1) \Gamma (4 n+4 \Delta_\phi -1)}\,,\\
	c_n^{A} & = 	c^A_0 \frac{2 \Gamma (2 n+2 \Delta_\phi +1)^2 \Gamma (2 n+4 \Delta_\phi )}{\Gamma (2 \Delta_\phi )\Gamma (2 \Delta_\phi+1 )\Gamma (2 n+2) \Gamma (4 n+4 \Delta_\phi +1)}\,,\\
	c_n^{T} & = 	c^T_0 \frac{2 \Gamma (2 n+2 \Delta_\phi )^2 \Gamma (2 n+4 \Delta_\phi -1)}{\Gamma (2 \Delta_\phi )^2 \Gamma (2 n+1) \Gamma (4 n+4 \Delta_\phi -1)}\,,
\end{align}
up to a normalisation factor which can be easily found by looking at the first identity contribution in the different channels. This gives the known result (see~\cite{Ferrero:2019luz})
\begin{align}
	c_n^{S} & = 	 \frac{1}{N} \frac{2 \Gamma (2 n+2 \Delta_\phi )^2 \Gamma (2 n+4 \Delta_\phi -1)}{\Gamma (2 \Delta_\phi )^2 \Gamma (2 n+1) \Gamma (4 n+4 \Delta_\phi -1)}\,,\\
	c_n^{T} & = 	 \frac{2 \Gamma (2 n+2 \Delta_\phi )^2 \Gamma (2 n+4 \Delta_\phi -1)}{\Gamma (2 \Delta_\phi )^2 \Gamma (2 n+1) \Gamma (4 n+4 \Delta_\phi -1)}\,,\\ 
	c_n^{A} & = 	- \frac{2 \Gamma (2 n+2 \Delta_\phi +1)^2 \Gamma (2 n+4 \Delta_\phi )}{\Gamma (2 \Delta_\phi )^2\Gamma (2 n+2) \Gamma (4 n+4 \Delta_\phi +1)}\,.
\end{align}
The same procedure for free fermions with O($N$) symmetry leads to  
\begin{align}
	c_n^{S} & = \frac{1}{N}	\frac{2 \Gamma (2 n+2 \Delta_\phi +1)^2 \Gamma (2 n+4 \Delta_\phi )}{\Gamma (2 \Delta_\phi ) \Gamma (2 \Delta_\phi +1) \Gamma (2 n+2) \Gamma (4 n+4 \Delta_\phi +1)}\,, \\
	c_n^{T} & = \frac{2 \Gamma (2 n+2 \Delta_\phi +1)^2 \Gamma (2 n+4 \Delta_\phi )}{\Gamma (2 \Delta_\phi ) \Gamma (2 \Delta_\phi +1) \Gamma (2 n+2) \Gamma (4 n+4 \Delta_\phi +1)} \,, \\ 
	c_n^{A} & = 	- \frac{2\Delta_\phi \Gamma (2 n+2 \Delta_\phi )^2 \Gamma (2 n+4 \Delta_\phi -1)}{\Gamma (2 \Delta_\phi )^2 \Gamma (2 n+1) \Gamma (4 n+4 \Delta_\phi -1)}\,.
\end{align}

\subsection*{Perturbative Sum Rules}
In this section we further test our sum rules by using known CFT data for a class of perturbations around GFF. These perturbations, which will be treated in great details in subsection \ref{sec:pert}, are constructed by introducing an effective field theory in AdS$_2$ background and considering the 1d boundary conformal field theory through the usual holographic dictionary. In particular, we will be interested in quartic contact interaction with derivatives. A classification of these independent contact interactions is given e.g. in~\cite{Mazac:2018ycv}, where the authors find there is a one-parameter family labelled by~$L$, where~$4L$ is the number of derivatives in the schematic interaction~$(\partial^L \Phi)^4$. As we mentioned at the end of subsection \ref{boundedness}, single terms in the perturbative expansion of the correlator may have a worse Regge behaviour than the general bound \eqref{Reggebound}. In particular, let us consider a perturbed correlator 
\begin{equation}\label{pertcorr}
	f(t)=f^{\text{GFF}}(t)+g_{L} f_L^{(1)}(t)+O(g_L^2),
\end{equation}
where $L$ labels the maximum number of derivatives in the quartic interaction (i.e. the interaction term may involve a combination of terms with $\ell\leq L$ derivatives) and $g_L$ is the associated coupling. The Regge behaviour of this correlator is determined by the term with the maximum number of derivatives, and it reads \cite{Mazac:2018ycv}
\begin{equation}
	f_L^{(1)}(z)\sim z^{2\Delta+2L-1} \qquad z \to \frac12+ i \infty.
\end{equation}
According to our discussion in subsection \ref{sumrules}, the associated Mellin amplitude will behave as
\begin{equation}
	M^{(1)}_L(s)\sim |s|^{2L-1} \qquad |s|\to \infty,
\end{equation}
and we need to choose a function $F_p(s)$ which vanishes at infinity faster than $|s|^{-2L}$. Here we will derive and check the sum rules for the cases $L=0$ and $L=1$. The strategy is the following. We use equation \eqref{sumrulesgenericF} to write down nonperturbative sum rules with a specific function $F_p(s)$, which will be chosen to decay sufficiently fast at $|s|\to \infty$ at a given value of $L$. We then expand the CFT data as
\begin{align}\label{expansionDelta}
	\Delta&=2 \Delta_{\phi}+ 2n+g_L \gamma_{L,n}^{(1)}+\mathcal{O}(g_L^2) \, ,\\
	c_{\Delta}&= c^{(0)}_n+ g_L c^{(1)}_{L,n} +\mathcal{O}(g_L^2)\label{expansionc}
\end{align}
and derive perturbative sum rules for $\gamma^{(1)}_{L,n}$ and $c^{(1)}_{L,n}$. We then check that these sum rules are satisfied by the $L=0,1$ results obtained in \cite{Mazac:2018ycv}, which read
\begin{align}\label{andimL0}
	\gamma_{0,n}^{(1)} &= \frac{\left(\frac{1}{2}\right)_n \left((\Delta_{\phi} )_n\right){}^2 \left(2 \Delta_{\phi} -\frac{1}{2}\right)_n}{(1)_n \,(2 \Delta_{\phi} )_n \left(\left(\Delta_{\phi} +\frac{1}{2}\right)_n\right){}^2}\, ,\\ \label{andimL1}
	\gamma_{1,n}^{(1)}&=A_{\Delta_{\phi}}^{-1} \gamma_{0,n}^{(1)}  \frac{2 n (4 \Delta_{\phi} +2 n-1)}{(\Delta_{\phi} +n-1) (2 \Delta_{\phi} +2 n+1)}( 16 \Delta_{\phi} ^5-13 \Delta_{\phi} ^3-3 \Delta_{\phi} ^2+16 \Delta_{\phi}  n^4+8 n^4+64 \Delta_{\phi} ^2 n^3\nonumber \\
	&+16 \Delta_{\phi}  n^3-8 n^3+96 \Delta_{\phi} ^3 n^2+8 \Delta_{\phi} ^2 n^2-24 \Delta_{\phi}  n^2-2 n^2+64 \Delta_{\phi} ^4 n-28 \Delta_{\phi} ^2 n-2 \Delta_{\phi}  n+2 n)
\end{align}
where the result for $\gamma^{(1)}_{1,n}$ differs from \cite{Mazac:2018ycv} by an overall factor 
\begin{equation}
	A_{\Delta_{\phi}} \equiv \frac{\Delta_{\phi}  (\Delta_{\phi} +1) (\Delta_{\phi} +2) (4 \Delta_{\phi}-1) (4 \Delta_{\phi} +1)^2 (4 \Delta_{\phi} +3)}{(2 \Delta_{\phi} +1)^2 (2 \Delta_{\phi} +3)}  \,  ,
\end{equation}
which we introduced to normalize the anomalous dimension as $\gamma_{1,1}^{(1)}=1$. Notice that $\gamma_{1,0}^{(1)}=0$. This is equivalent to a choice of basis for the set of independent interactions that can be built with up to one derivative. We discuss this issue in detail in subsection \ref{sec:pert}. The OPE coefficients $c^{(1)}_n$ are given by the relation
\begin{equation}\label{OPEcoeff1}
	c^{(1)}_{L,n}=\partial_n(\gamma^{(1)}_{L,n} c^{(0)}_{n}) \, .
\end{equation}

\begin{figure}[htbp]
	\center
	\includegraphics[width=.9\linewidth]{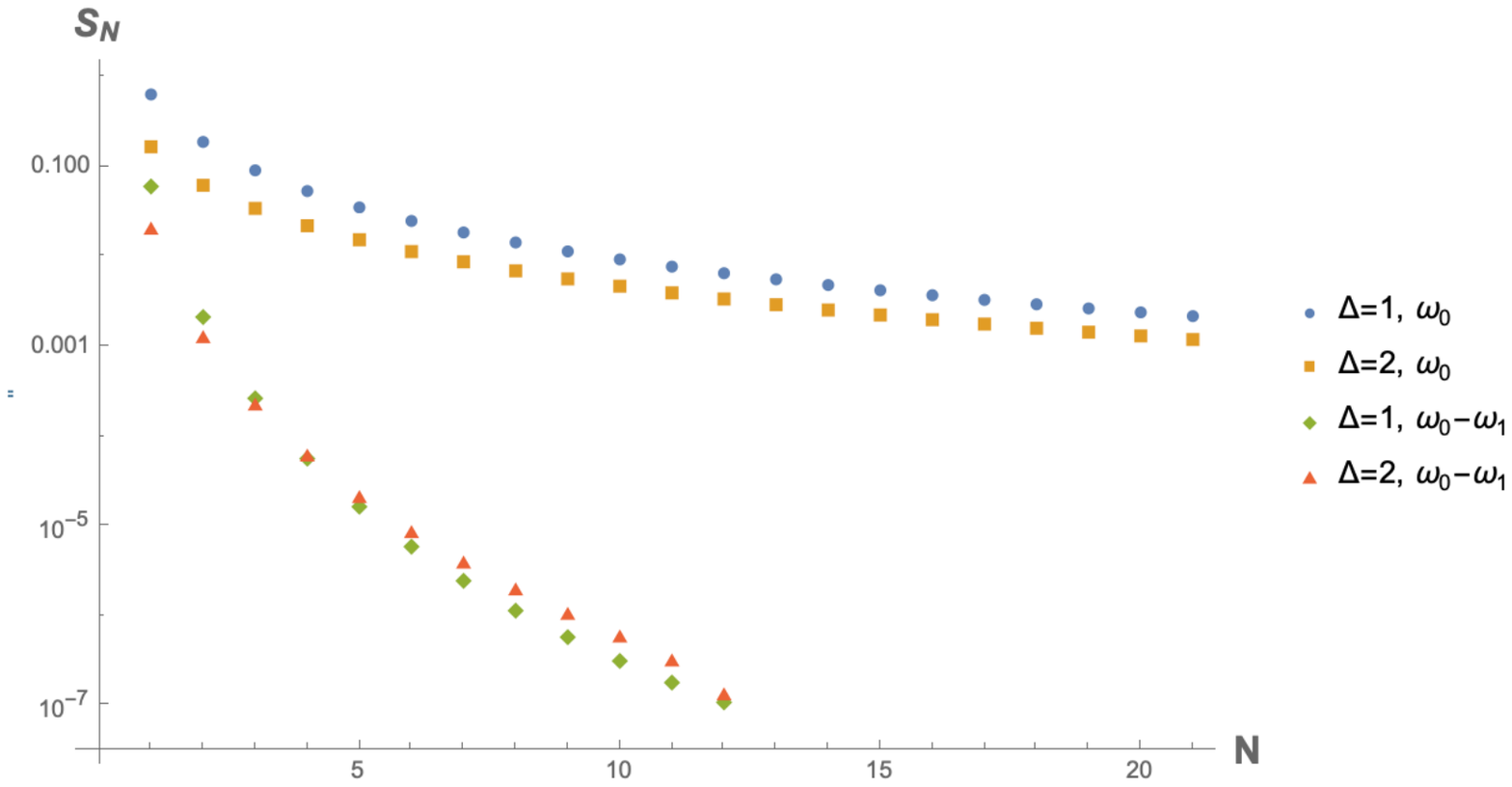}
	\caption{$L=0$ sum rules for external dimensions $\Delta_\phi=1,\Delta_\phi=2$ and different functionals $\omega_0$ and $\omega_0-\omega_1$. The plot shows the value of the finite sum truncated after $N$ terms in a logarithmic scale. The better convergence is seen for the $\omega_0-\omega_1$ functional, due to a better large $s$ behaviour.}\label{fig:sum-rule-phi4}
\end{figure}

\begin{figure}[htbp]
	\centering
	\includegraphics[width=0.9\linewidth]{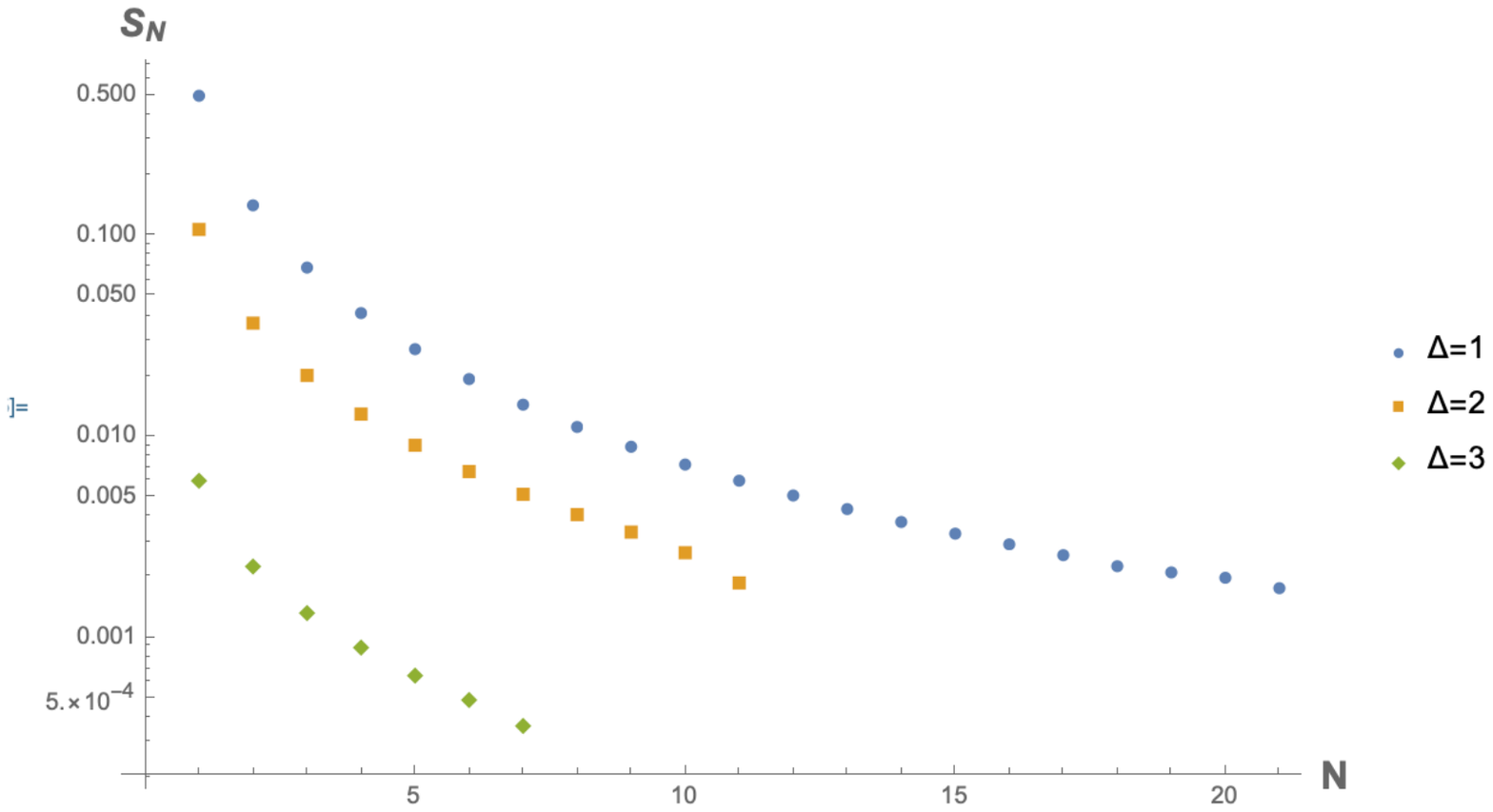}
	\caption{Sum rule applied for the $L=1$ case for $\Delta_\phi=1,2,3$, truncated after summing over N terms seen on a logarithmic scale. The slower convergence compared to $L=0$ is to be expected from the worse Regge behaviour.}
	\label{fig:sum-rule-dphi4}
\end{figure}

Let us start with the case $L=0$. In that case, the Regge behaviour is better than the GFF case, so we could even choose the function
\begin{equation}\label{functionforL0}
	F_p(s)=\frac{1}{s+p} \, ,
\end{equation}
which gives the sum rules
\begin{align}\label{sumruleL0}
	\sum_{\D,k} c_\D \frac{(-1)^{k+1}\Gamma(2\D)\Gamma(\D+k)}{\Gamma(\D)^2\Gamma(2\D+k)\Gamma(2\Delta_\phi-\D-k)\Gamma(k+1)(\D+k+p) (-2 \Delta +\D+k-p)}=0 \, .
\end{align}
Notice that we can also use the function \eqref{functionforL0} as a building block from which we can construct more suppressed functions of the class \eqref{Fp1p2}, for example
\begin{align}\label{combinationfunctions}
	F_{p,p+1}(s)=F_p(s)-F_{p+1}(s)=\frac{1}{(s+p)(s+p+1)} \, .
\end{align}
As we discussed below \eqref{Fp1p2}, the crossing-antisymmetric part of this function decays as $\frac{1}{s^3}$ at large $s$ and for this reason, we will also use it for the case $L=1$.
Inserting the expansions \eqref{expansionDelta} and \eqref{expansionc} into the sum rule \eqref{sumruleL0}, we get two contributions: a finite sum from the terms where $k=p-2n$
\begin{align}\label{finitesum}
	\tilde{\omega}_p&= \sum_{n=0}^{[\frac{p}{2}]}c_n^{(1)}\frac{\Gamma (p+1) \Gamma (4 (n+\Delta_\phi )) \Gamma (p+2 \Delta_\phi)}{(p-2 n)!\, \Gamma (2 (n+\Delta_\phi))^2 \Gamma (2 n+p+4 \Delta_\phi)}\\
	&+c_n^{(0)}\gamma_n^{(1)} \frac{ \Gamma (p+1) \Gamma (4 (n+\Delta_\phi)) \Gamma (p+2 \Delta_\phi)}{2 (\Delta_\phi+p) (p-2 n)! \,\Gamma (2 (n+\Delta_\phi))^2 \Gamma (2 n+p+4 \Delta_\phi)} \eta(\Delta_\phi,n,p)  \, ,
\end{align}
where 
\begin{align}
	\eta(\Delta_\phi,n,p) =1+2 (\Delta_\phi +p) (-2 \psi_{2 n+2\Delta_\phi }+2 \psi_{4 n+4\Delta_\phi}-2 \psi_{ 2 n+p+4 \Delta_\phi }+\psi _{p+2 \Delta_\phi}+\psi_{p+1}) 
\end{align}
with $\psi_n=\frac{\Gamma'(n)}{\Gamma(n)}$, and an infinite sum 
\begin{equation}
	\tilde{\omega}'_p=\sum_{n,k\neq p-2n}c_n^{(0)} \gamma_n^{(1)} \frac{2 (k+2 n)! (\Delta_\phi +k+2 n) \Gamma (4 (n+\Delta_\phi )) \Gamma (k+2 n+2 \Delta_\phi )}{k! (k+2 n-p) \Gamma (2 (n+\Delta_\phi ))^2 \Gamma (k+4 (n+\Delta_\phi )) (2 \Delta_\phi +k+2 n+p)} \, .
\end{equation}
This last term does not allow us to follow the same strategy we used for GFF to extract the value of $\gamma^{(1)}_{L,n}$. This is consistent with the fact that these sum rules are not as constraining as in the GFF case, and we may expect more than one solution. Therefore, instead of using the sum rules to derive the CFT data, we limit ourselves to numerically checking the validity of the equation
\begin{equation}
	\omega_p=\tilde{\omega}_p+\tilde{\omega}'_p=0
\end{equation}
by inserting the data \eqref{andimL0}, \eqref{andimL1} and \eqref{OPEcoeff1}. In Figure \ref{fig:sum-rule-phi4}, we show our results for the case $p=0$ both for the functional $\omega_0$ and for the combination \eqref{combinationfunctions}, i.e. the functional $\omega_0-\omega_1$. It is very clear from the plot that the latter shows a faster convergence reflecting the better asymptotic behaviour at $|s|\to \infty$.  In Figure \ref{fig:sum-rule-dphi4}, we show the analogous plot for the $L=1$ case, where we used the function \eqref{combinationfunctions} since we needed a large $s$ behaviour at least $\frac{1}{s^3}$.

We conclude with a couple of remarks. Since the finite sum in \eqref{finitesum} always ranges up to $[\frac{p}{2}]$, we can always find a combination of functionals $\omega_k$ such that the contributions proportional to $c^{(1)}_{L,n}$ are all cancelled, and we can write down a sum rule which only involves the anomalous dimensions $\gamma^{(1)}_{L,n}$
\begin{equation}\label{New sum rule}
	\sum_{k=0}^{2p+1}\frac{(-1)^{1+k}\,\Gamma (2 p+2)}{\Gamma (k+1) \,\Gamma (2 p+3-k)} \,\omega_k=0 \, .
\end{equation}
The simplest case is $p=0$ where \eqref{New sum rule} gives $\omega_0-2\omega_1=0$. Even after doing that, however, we are left with an infinite sum with alternating signs, which we did not manage to use constructively to extract a solution. Nevertheless, notice that once the anomalous dimensions are known, we can use our sum rules at fixed $p$ to determine recursively the value of the OPE coefficients $c^{(1)}_{L,n}$ following the same strategy as in the GFF case. This provides a very convoluted way to rederive the relation \eqref{OPEcoeff1}. 

\subsection*{Poles and Series}
\label{app:polesandseries}
In this Appendix, we address some subtleties in the perturbative expansion of the Mellin amplitudes in the context of the analytic sum rules. In all the following, we will consider a function $\hat{F}(s)$ defined nonperturbatively and its expansion in a small parameter $\lambda$, as well as the functional 
\begin{align}
	\omega_K[\hat{F}] = \oint_{\mathbb{C}|_{\infty}} \frac{ds}{2 \pi i} \hat{F}(s) K(s),
\end{align}
defined as the contour integral of $\hat{F}(s)$ over the circle at $\infty$ in the complex plane with an integration kernel $K(s)$. This functional is well defined and vanishing for meromorphic functions $\hat{F}(s)$ and $K(s)$ such that
\begin{align}\label{convergencecondition}
	\hat{F}(s)K(s)\underset{|s|\rightarrow\infty}{\sim} s^{-1-\alpha} \qquad \alpha>0.
\end{align}
\subsection*{Nonperturbative Zeros and Perturbative Poles}
The nonperturbative Mellin function $\hat{M}(s)$ has zeros at positions $s=2\Delta+n$, while the perturbative expansion has poles in those positions. The reason for this and the subtleties of evaluating the sum rule will be illustrated with an example.\par Let $f(s)$ be a well-behaved function\footnote{By this, we mean that $f$ is meromorphic, has no essential singularities and behaves as $s^{-\alpha}$ at large s, where $\alpha$ is a positive parameter. For simplicity, we take $f(s)$ to be regular without poles. However, including these poles (different to those of $\hat{F}$) is  straightforward and does not change the conclusions.}. We define the $\lambda$-dependent function
\begin{align}
	\hat{F}(s) = \frac{(s-1)f(s)}{s-1-\lambda}.
\end{align}
Upon expanding $F(s)$ in the parameter $\lambda$, we obtain the geometric series
\begin{align}
	\hat{F}(s) =\sum_{k=0}^\infty \left(\frac{\lambda}{s-1}\right)^k f(s),
\end{align}
whose radius of convergence is $\lambda<|s-1|$.
This perturbative series cannot be evaluated at finite $\lambda$ in $s=1$. However, several features are noteworthy. At $s=1$, we have poles of increasing degree at each order in the $\lambda$ expansion, while the nonperturbative expression has a simple zero at that point. Let us evaluate the contour integral
\begin{align}\label{Eq: Functional 1}
	\omega_1[\hat{F}] = \oint \frac{ds}{2\pi i} \frac{\hat{F}(s)}{s-1}
\end{align}
in three different ways. Firstly, by inserting the nonperturbative $\hat{F}$
\be
\oint \frac{ds}{2\pi i} \frac{\hat{F}(s)}{s-1} = 	\oint \frac{ds}{2\pi i} \frac{f(s)}{s-1-\lambda}
=f(1+\lambda).
\ee
This method is the most intuitive way of evaluating \eqref{Eq: Functional 1} and is perfectly well-defined. However, it requires the knowledge of the full nonperturbative function, which is generally unknown. \par Secondly, we can separately consider the poles of the integration kernel $\frac{1}{s-1}$ from those of the function $\hat{F}$:
\begin{align}
	\oint \frac{ds}{2\pi i} \frac{\hat{F}(s)}{s-1} &=\textrm{Res}(\frac{1}{s-1}) \hat{F}(1)+\sum_{s^*}\frac{\textrm{Res}(\hat{F})(s^*)}{s^*-1}=f(1+\lambda).
\end{align}
This is not identical to the previous method since it assumes more (notably that the poles of $K(s)$ and those of $\hat{F}$, $s^*$ above, are distinct) and requires less information from the nonperturbative function; only the residues of $\hat{F}(s)$ and the value of $\hat{F}(s_i)$ at a finite number of points are required to evaluate \eqref{Eq: Functional 1}.\par  Finally, one can use the series expansion of $\hat{F}(s)$
\begin{align}
	\omega_1[\hat{F}]&=\oint \frac{ds}{2\pi i}\left(\sum_{k=0}^\infty \frac{\lambda^k}{(s-1)^{k+1}} f(s)\right) =\sum_{k=0}^{\infty}\frac{\lambda^k f^{(k)}(1)}{k!},
\end{align}
which gives the $\lambda$-expansion of $f(1+\lambda)$. Under the assumption that $f$ is analytic near~1, we can then resum the Taylor series to obtain $f(1+\lambda)$ and obtain the same result as the previous two methods. In this specific example, the only operation which is not allowed is to evaluate $\hat{F}(1)$ using the nonperturbative series since the latter has a vanishing radius of convergence for $s=1$. In the next subsection, we will explore a setting where the perturbative evaluation of the functional is problematic and requires the truncation of the series at a fixed order.\par 
\subsection*{Bad Regge Behaviour}
We now consider another pathological case where the perturbative expansion of the correlator makes the large $s$  behaviour worse than that of the nonperturbative expression. To illustrate this, we take the example
\begin{align}\label{Fbad}
	\hat{F}(s) &= \frac{f(s)}{1-\lambda s^2},
\end{align}
and evaluate the functional
\begin{align}\label{Eq: Functional 2}
	\omega_3[\hat{F}] &= \oint_{\mathbb{C}|_{\infty}}\frac{ds}{2\pi i}\frac{\hat{F}(s)}{s(s^2-1)}.
\end{align}
The convergence of the integral~\eqref{Eq: Functional 2} imposes a bound on the large $s$ behaviour of $f(s)$ in~\eqref{Fbad}\footnote{As in the previous case, we consider that $f(s)$ is meromorphic in $s$ and ignore its poles since they act as spectators in the comparison between the perturbative and nonperturbative case. }
\begin{align}
	f(s) &\underset{s\rightarrow\infty}{\sim}s^{4-\alpha}& \alpha &>0.
\end{align}
The functional \eqref{Eq: Functional 2} can then be evaluated explicitly as
\begin{align}\label{Eq: functional non pert}
	\omega_3[\hat{F}]  =-f(0)+  \frac{f(1)+f(-1)}{2(1-\lambda) }-\frac{\lambda }{2(1-\lambda) } \left(f\left(\frac{1}{\sqrt{\lambda }}\right)+f\left(-\frac{1}{\sqrt{\lambda }}\right)\right).
\end{align}
The perturbative expansion of $\hat{F}(s)$ is 
\begin{align}
	\hat{F}(s)=f(s)+\lambda s^2 f(s)+\lambda^2 s^4 f(s)+O(\lambda^3).
\end{align}
In this perturbative setting, even with a stricter condition on the large $s$ behaviour of $f(s)$
\begin{align}\label{Eq: convergence Int}
	f(s)&\underset{s\rightarrow\infty}{\sim} s^{-\alpha}&\alpha&>0,
\end{align} 
we must truncate the series at order $\lambda$ in order to evaluate the functional 
\begin{align}\label{Eq functional Order lambda}
	\omega_3[f(s)+\lambda s^2 f(s)] &= -f(0)+\frac{f(1) +f(-1)}{2} + \lambda \frac{f(1)+f(-1)}{2}.
\end{align}
Comparing \eqref{Eq functional Order lambda} to the small-$\lambda$ expansion of \eqref{Eq: functional non pert}
\begin{align}
	\omega_3[\hat{F}]&=-f(0)+\frac{f(1) +f(-1)}{2} + \lambda \frac{f(1)+f(-1)}{2}-\frac{\lambda }{2}(f(\frac{1}{\sqrt{\lambda }})+f(-\frac{1}{\sqrt{\lambda }})) +o(\lambda),
\end{align}
we see that the condition of convergence \eqref{Eq: convergence Int} is exactly what is needed to get rid of the final terms and find agreement between the results
\begin{align}
	-\frac{\lambda }{2}(f(\frac{1}{\sqrt{\lambda }})+f(-\frac{1}{\sqrt{\lambda }}))& \underset{\lambda \rightarrow0}{\sim}\lambda^{1+\frac{\alpha}{2}}\underset{\lambda \rightarrow0}{=}  o(\lambda).
\end{align}
We therefore have agreement between the functional of the truncated $\lambda$-expansion of $\hat{F}(s)$ and the truncated $\lambda$-expansion of the functional of $\hat{F}(s)$
\begin{align}
	\omega_3[\hat{F}|_{\lambda}]  = \left(	\omega_3 [\hat{F}]\right)|_{\lambda},
\end{align}
where the order at which we are required to truncate is controlled by the large $s$ behaviour of $K$ (which can always be chosen to satisfy the convergence condition~\eqref{convergencecondition} at a given order of expansion).

\section{Anomalous Dimensions for Higher Derivative Interactions}\label{Ap: anomalous dimension}
This section lists the various results for the polynomial part of the anomalous dimension in equation \eqref{gammaLB2}. The \texttt{Mathematica} notebook attached to the arXiv submission of \cite{Bianchi:2021piu}  has values of $L$ ranging from $L=0$ to $L=8$ as well as a function~\texttt{FindBootstrapPolynomial[L,$\Delta$,n]} to compute $\mathcal{P}_{L,n}(\Delta_\phi)$ for arbitrary $L$ (the function gets slower and slower at higher $L$, but in principle, it works for any $L$).

\scriptsize
\begin{align}
	\mathcal{P}_{0,n}(\Delta)&=1\\
	\hspace{1.5cm}\nonumber \\
	\mathcal{P}_{1,n}(\Delta)&=8 (2 \Delta +1) n^4+8 \left(8 \Delta ^2+2 \Delta -1\right) n^3+2 (2 \Delta -1) (2 \Delta +1) (12 \Delta +1) n^2\nonumber \\
	&+\left(64 \Delta ^4-28 \Delta ^2-2 \Delta +2\right) n+\Delta ^2 \left(16 \Delta ^3-13 \Delta -3\right)\\
	\hspace{1.5cm}\nonumber \\
	\mathcal{P}_{2,n}(\Delta)&= 64 (2 \Delta +3) (2 \Delta +5) n^8+128 (2 \Delta +3) (2 \Delta +5) (4 \Delta -1) n^7\nonumber \\
	&+32 (2 \Delta +3) (2 \Delta +5) \left(56 \Delta ^2-22 \Delta -1\right) n^6\nonumber \\
	&+32 (2 \Delta +3) (2 \Delta +5) (4 \Delta -1) (28 \Delta ^2-5 \Delta -5) n^5\nonumber \\
	&+4 (2 \Delta +3) \left(2240 \Delta ^5+4800 \Delta ^4-2924 \Delta ^3-2156
	\Delta ^2+246 \Delta -415\right) n^4\nonumber \\
	&+8 (2 \Delta +3) (4 \Delta -1) (224 \Delta ^5+576 \Delta ^4-158 \Delta ^3-572 \Delta
	^2-243 \Delta -160) n^3\nonumber \\
	&+4 (2 \Delta -1) (2 \Delta +3) (448 \Delta ^6+1392 \Delta ^5+84 \Delta ^4-2183
	\Delta ^3-2091 \Delta ^2-1134 \Delta -105) n^2\nonumber \\
	&+4 (2 \Delta +3) (4 \Delta -1) (64 \Delta ^7+208 \Delta ^6-36 \Delta ^5-605 \Delta
	^4-554 \Delta ^3-30 \Delta ^2+243 \Delta +90) n\nonumber \\
	&+(\Delta -2) (\Delta -1) \Delta ^2 (\Delta +1)^2 (4 \Delta +3) (4 \Delta +5) (4 \Delta +7) (4 \Delta +9)\\
	\hspace{1.5cm}\nonumber \\
	\mathcal{P}_{3,n}(\Delta)&= 512 (2 \Delta +5) (2 \Delta +7) (2 \Delta +9) n^{12}\nonumber \\
	&+1536 (2 \Delta +5) (2 \Delta +7) (2 \Delta +9) (4 \Delta -1) n^{11}\nonumber \\
	&+128 (2 \Delta +5) (2 \Delta +7) (2 \Delta +9) (264 \Delta ^2-102 \Delta +5) n^{10}\nonumber \\
	&+640 (2 \Delta +5) (2 \Delta +7) (2 \Delta +9) (4 \Delta -1) (44 \Delta ^2-7 \Delta -3) n^9\nonumber \\
	&+96 (2 \Delta +5) (2 \Delta +7) (5280 \Delta ^5+22080 \Delta ^4-8610 \Delta ^3-4790 \Delta ^2-798 \Delta -2931) n^8\nonumber \\
	&+96 (2 \Delta +5) (2 \Delta +7) (4 \Delta -1) (2112 \Delta ^5+9792 \Delta ^4+268 \Delta ^3-5448 \Delta ^2-4628 \Delta -5493) n^7\nonumber \\
	&+8 (2 \Delta +5) (2 \Delta +7) \left(118272 \Delta ^7+556416 \Delta ^6-8736 \Delta ^5-656280 \Delta ^4-661308 \Delta ^3-560400 \Delta ^2 \right. \nonumber \\
	&\quad \left.+371392 \Delta +17415 \right) n^6\nonumber \\
	&+24 (2 \Delta +5) (2 \Delta +7) (4 \Delta -1) \left(8448 \Delta ^7+45504 \Delta ^6+22128 \Delta ^5-79708 \Delta ^4-143680 \Delta ^3-114082 \Delta ^2\right. \nonumber \\
	&\quad \left.+52985 \Delta +27645\right) n^5\nonumber \\
	&+4 (2 \Delta +5) \left( 253440 \Delta ^{10}+2327040 \Delta ^9+5816640 \Delta ^8-1506240 \Delta ^7-22985970 \Delta ^6-33151830 \Delta ^5 \right. \nonumber \\
	&\quad \left.-9079800 \Delta ^4+25792815 \Delta ^3+10370477 \Delta
	^2-446534 \Delta +2131794 \right) n^4\nonumber \\
	&+8 (2 \Delta +5) (4 \Delta -1)\left(14080 \Delta ^{10}+142080 \Delta ^9+423840 \Delta ^8+8160 \Delta ^7-2172753 \Delta ^6-4187481 \Delta ^5 \right. \nonumber \\
	&\quad \left.-1812050 \Delta ^4+3606930 \Delta ^3+3965596 \Delta ^2+1661325
	\Delta +791091 \right) n^3 \nonumber \\
	&+6 (2 \Delta -1) (2 \Delta +5) \left(11264 \Delta ^{11}+125184 \Delta ^{10}+437120 \Delta ^9+118880 \Delta ^8-2771604 \Delta ^7-6808095 \Delta ^6 \right. \nonumber \\
	&\quad \left.-4248981 \Delta ^5+6860955 \Delta ^4+13140919 \Delta
	^3+9496058 \Delta ^2+4002384 \Delta +360360\right) n^2\nonumber \\
	&+2 (2 \Delta +5) (4 \Delta -1) \left(3072 \Delta ^{12}+36096 \Delta ^{11}+132224 \Delta ^{10}-3360 \Delta ^9-1214676 \Delta ^8-2926395 \Delta ^7 \right. \nonumber \\
	&\quad \left.-970776 \Delta ^6+6196080 \Delta ^5+10143424 \Delta
	^4+5128059 \Delta ^3-1542528 \Delta ^2-3028860 \Delta -907200\right) n\nonumber \\
	&+(\Delta -3) (\Delta -2) (\Delta -1) \Delta ^2 (\Delta +1)^2 (\Delta +2)^2 (4 \Delta +5) (4 \Delta +7) (4 \Delta +9) (4 \Delta +11) (4 \Delta +13) (4 \Delta +15)
\end{align}
\normalsize

\chapter*{Translations}
\setcounter{chapter}{20}
\setcounter{equation}{0}
To pay hommage to the four languages I have used during this PhD, there is some text in English, French, Italian and German. The English translations are as follows:
\begin{align}\label{Heliane}
	&\text{`Will there be sky above me,//and will the stars} \non \\
&\text{// enfold my eyes//with their maternal light?'}\\ 
	&\non \\
		&\text{'By other routes, by other ports // you will reach the shore, not here for crossing}\non\\
	&\text{lighter ship must carry you'}\label{Dante 1}\\
	&\non \\
		&\text{`A thousand-beat waltz//alone gives lovers//}\non \\
	&\text{ three hundred and thirty-three times// the time to write a story.'}\label{Brel}
\end{align}

\bibliographystyle{nb}
\bibliography{RefPhD.bib,Ref2.bib}

\end{document}